\documentclass[11pt]{amsart} 
\usepackage{amsmath}
\usepackage{amssymb}
\usepackage{verbatim}
\usepackage{amsthm}
\usepackage{mathrsfs}
\usepackage[top=3cm, bottom=3cm, right=3cm, left=3cm]{geometry}
\usepackage{graphicx}
\usepackage[utf8]{inputenc}
\usepackage{mathtools}
\usepackage[colorlinks,pdfdisplaydoctitle,linkcolor=blue,citecolor=red]{hyperref}
\usepackage{caption}
\usepackage{subcaption}
\usepackage{url}
\usepackage{epstopdf}
\usepackage{pgf,tikz}
\usepackage{todonotes}
\usepackage{bbm}
\usepackage{cleveref}
\usepackage{extarrows}
\usetikzlibrary{arrows}
\usepackage{algorithm}
\usepackage{algpseudocode}
\usepackage{placeins}
\usepackage[style=authoryear,natbib=true,bibstyle=numeric]{biblatex}
\usepackage{amssymb,amsthm}    
\usepackage{graphicx}          
\usepackage{amsmath}         
\usepackage{caption}
\usepackage{subcaption}
\usepackage{dsfont}
\usepackage{listings}
\usepackage{bbm}
\usepackage{color}
\usepackage{ dsfont }
\usepackage{enumerate}
\usepackage{relsize}
\usepackage{xfrac}
\usepackage{bbm}
\usepackage{tabularx}

\definecolor{darkblue}{rgb}{0,0,0.7}

\definecolor{darkgreen}{rgb}{0.01,0.75,0.24}

\newcommand{\be}{\begin{equation}}
\newcommand{\en}{\end{equation}}
\newcommand{\ben}{\begin{equation*}}
\newcommand{\enn}{\end{equation*}}
\newcommand{\bea}{\begin{aligned}}
    \newcommand{\ena}{\end{aligned}}

\def\ba#1\ena{\begin{align}#1\end{align}}
\def\ban#1\enan{\begin{align*}#1\end{align*}}

\usepackage{mathtools}
\mathtoolsset{centercolon}  

\renewcommand{\phi}{\varphi}

\definecolor{mygreen}{rgb}{0.1,0.75,0.2}

\theoremstyle{plain}
\newtheorem{thm}{Theorem}[section]
\newtheorem{defn}[thm]{Definition}

\newtheorem{remark}[thm]{Remark}

\numberwithin{equation}{section}
\renewcommand{\theequation}{\thesection .\arabic{equation}}

\begin{document}
    
    \title[Cauchy Markov random field priors for Bayesian inversion]{Cauchy Markov random field priors for Bayesian inversion}

    \author[J. Suuronen] {Jarkko Suuronen}
    \address{School of Engineering Science, Lappeenranta-Lahti University of Technology, PO Box 20, FI-53851 Lappeenranta, Finland.}
    \email{Jarkko.Suuronen@lut.fi}

    \author[N. K. Chada] {Neil K. Chada}
    \address{Applied Mathematics and Computational Science Program, King Abdullah University of Science and Technology, Thuwal, 23955, KSA}
    \email{neilchada123@gmail.com}

    \author[L. Roininen] {Lassi Roininen}
    \address{School of Engineering Science, Lappeenranta-Lahti University of Technology, PO Box 20, FI-53851 Lappeenranta, Finland.}
    \email{lassi.roininen@lut.fi}

    \begin{abstract}
        The use of Cauchy Markov random field priors in statistical inverse problems can potentially lead to posterior distributions which are non-Gaussian, high-dimensional, multimodal and heavy-tailed. 
        In order to use such priors successfully, sophisticated optimization and Markov chain Monte Carlo (MCMC) methods are usually required.
        In this paper, our focus is largely on reviewing recently developed Cauchy difference priors, while introducing interesting new variants, whilst providing a comparison.
        We firstly propose a one-dimensional second order Cauchy difference prior, and construct new first and second order two-dimensional isotropic Cauchy difference priors. Another new Cauchy prior is based on the stochastic partial differential equation approach, derived from Mat\'ern type Gaussian presentation.
        The comparison also includes Cauchy sheets.
        Our numerical computations are based on both maximum a posteriori and conditional mean estimation.
        We exploit state-of-the-art MCMC methodologies such as Metropolis-within-Gibbs, Repelling-Attracting Metropolis, and No-U-Turn sampler variant of Hamiltonian Monte Carlo.
        We demonstrate the models and methods constructed for one-dimensional and two-dimensional deconvolution problems.
        Thorough MCMC statistics are provided for all test cases, including potential scale reduction factors.
        %
        %
        %
        %
    \end{abstract}
    
    \maketitle


    \section{Introduction}
    \label{sec:intro}
    
    The focus of this paper is in quantifying and learning continuous-parameter models in Bayesian settings, which can be considered as Bayesian non-parametric or continuous-parameter Bayesian inversion  \citep{KS04,AMS10,AT87}.
    Specifically, we are interested in quantifying an unknown, in the form of an input to a model, from unobservable data of the model solution. 
    Mathematically, these are referred  as inverse problems, where an unknown, or quantity of interest $\textcolor{black}{\mathbf u \in \mathbb{R}^n}$ is aimed to recover from noisy measurement $\textcolor{black}{\mathbf y \in \mathbb{R}^m}$, such that their relationship is defined as
    \begin{equation}
    \label{eq:ip}
    \mathbf y = \mathbf{G}(\mathbf u) + \eta, \quad \eta \sim \mathcal{N}(\mathbf 0,\Gamma),
    \end{equation}
    where $\mathbf G$ is a potentially non-linear mapping, and $\eta$ is Gaussian zero-mean noise with covariance matrix $\Gamma$.
    
    The solution of a \textcolor{black}{finite-dimensional} inverse problem in a Bayesian approach is a probability distribution, given as, by Bayes' Theorem
    \begin{equation*}
    \pi(\mathbf u|\mathbf y) = \frac{\pi(\mathbf y|\mathbf u) \pi(\mathbf u)}{\pi(\mathbf y)} \propto \pi(\mathbf y|\mathbf u) \pi(\mathbf u),
    \end{equation*}
    which is known as the posterior distribution \citep{KS04,AMS10}. 
    %
    The term $\pi(\mathbf y|\mathbf u)$ refers to data-likelihood, in which where the data comes  via Equation \eqref{eq:ip}, and $\pi(\mathbf u)$ denotes the prior distribution, which acts as an initial belief of what the posterior may be. 
    
    \textcolor{black}{Often, in Bayesian statistical inverse problems, one starts by constructing infinite-dimensional prior, and then finite-dimensional approximations for computational purposes, and finally studying the posterior consistency, that is, the convergence of the finite-dimensional approximations to infinite-dimensional models. }
    \textcolor{black}{
    One way is to give the posterior of the unknown $u$ (defined in Banach spaces) as the Radon--Nikodym derivative
    $$
    \frac{d \mu^y}{d \mu_0} = \frac{\exp(-\Phi(u;y))}{Z} \propto \exp(-\Phi(u;y)),
    $$
    where $\mu^y$ denotes the posterior measure, $\mu_0$ the prior measure, and $\Phi(u;y) = \frac{1}{2}|y - \mathcal{G}(u)|^2_{\Gamma}$ is the misfit functional with $\mathcal G$ being continuous-parameter to observation mapping. }
    \textcolor{black}{While this analysis is mathematically sound, it typically requires well-defined continuous-parameter priors, and here, as our specific interest lies in constructing non-Gaussian Cauchy priors by starting from discrete presentations, the continuous prior models do not exist per se. 
    Thus, from now on, we take a computational point of view for studying the priors in finite-dimensional setting, and leave the rigorous consistency studies to later papers.
    }
    . 
    
\begin{figure}[ht!]      
        \centering      
        \begin{subfigure}[b]{4.5cm}
            \includegraphics[width=\linewidth]{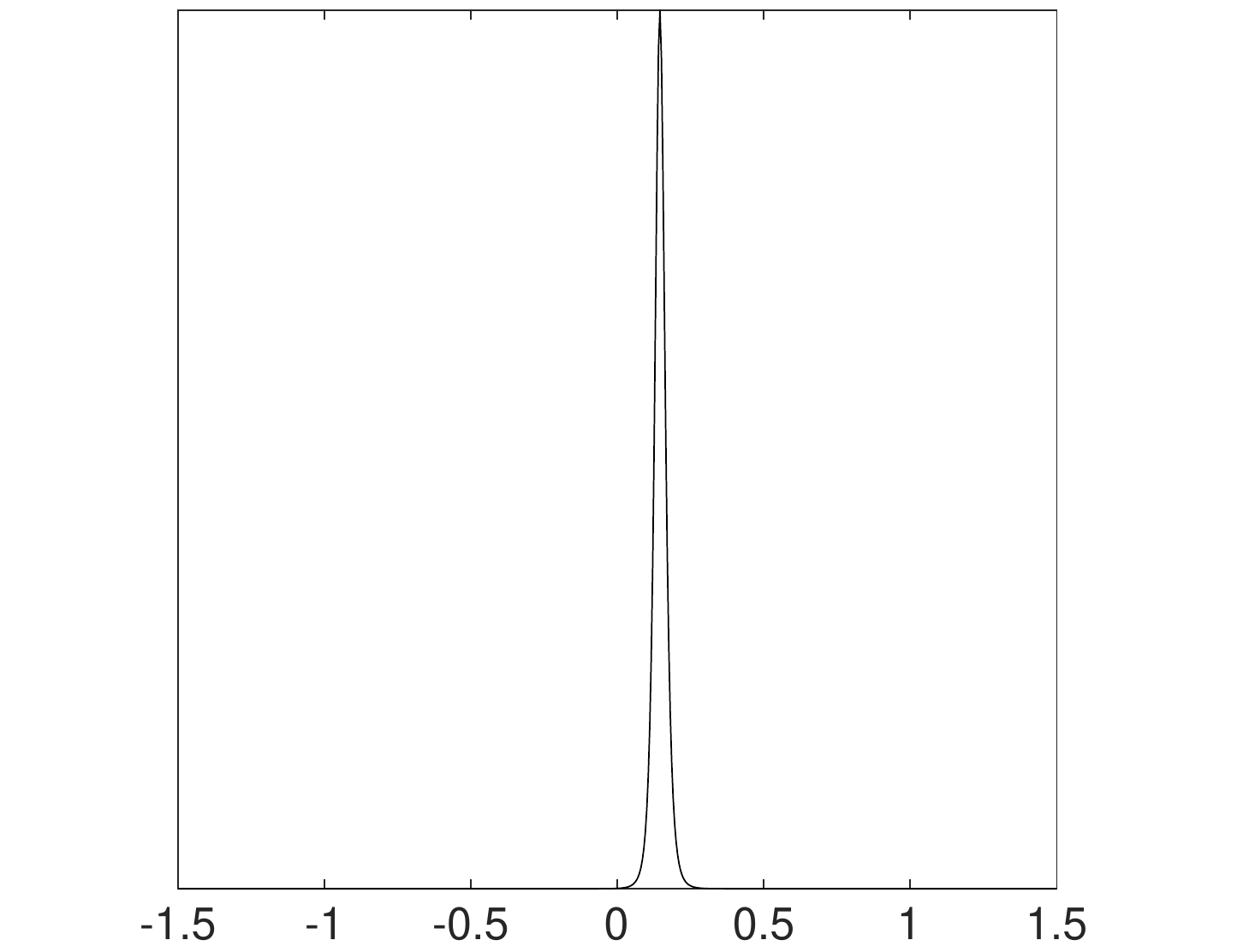}
            \caption{ }\label{condispde}\end{subfigure}
        \begin{subfigure}[b]{4.5cm}
            \includegraphics[width=\linewidth]{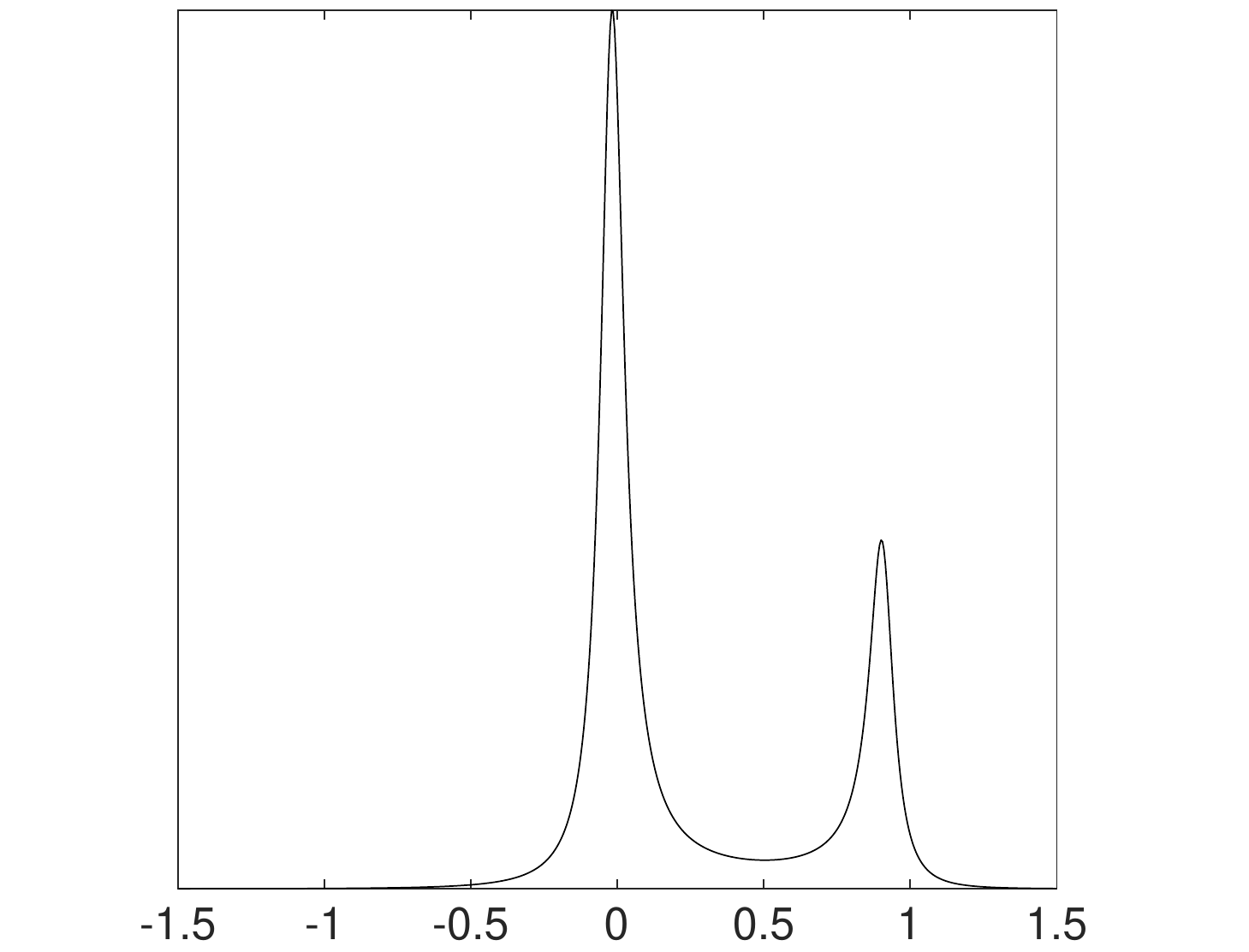}
            \caption{ }\label{condidiff}\end{subfigure}
        \begin{subfigure}[b]{4.5cm}
            \includegraphics[width=\linewidth]{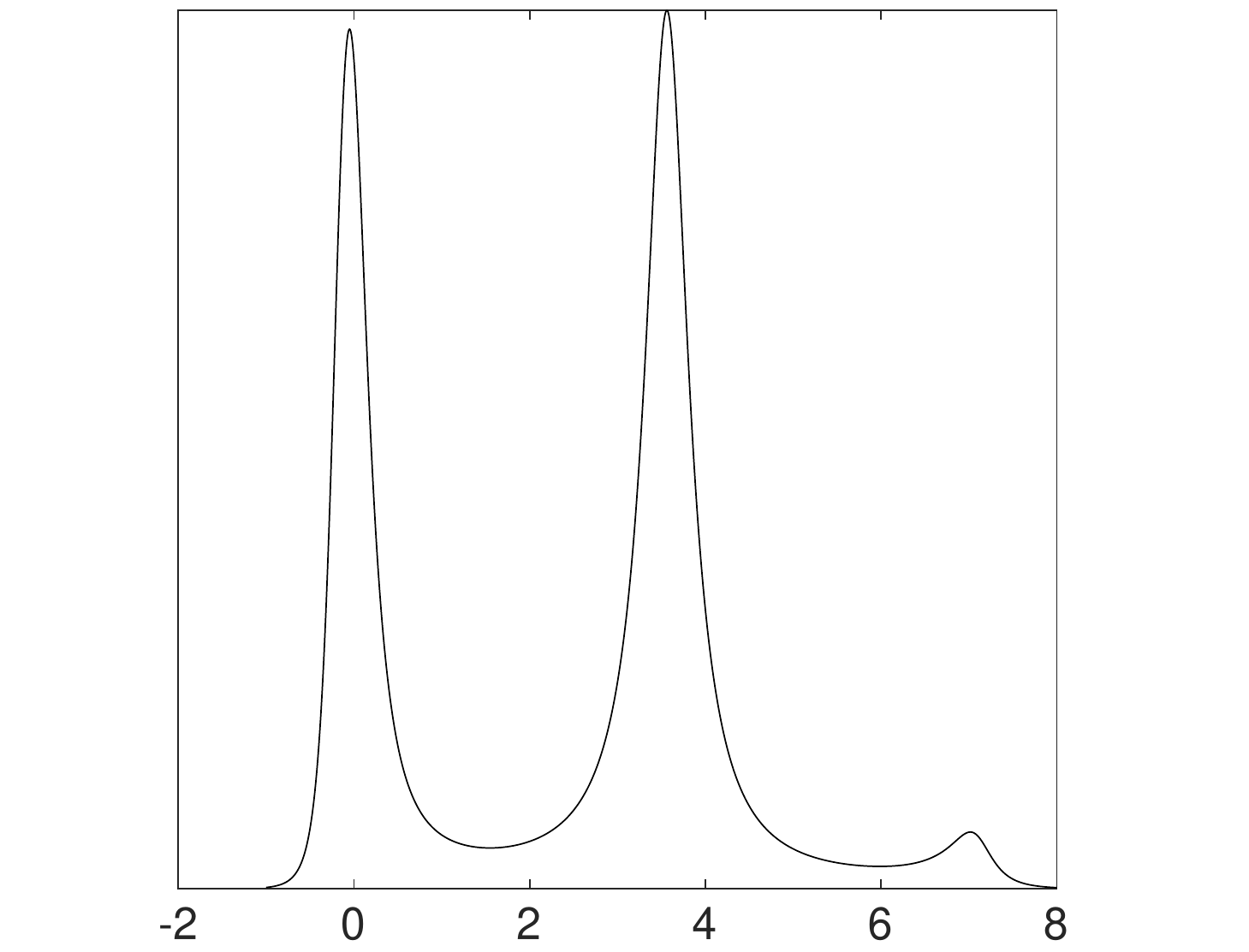}
            \caption{}\label{condi2d}\end{subfigure}
        \caption{Three examples of posterior distributions with  Cauchy priors. Left: uni-modal, middle: bi-modal, right: tri-modal.}     
        \label{fig:cond} 
    \end{figure}

    \Cref{fig:cond} shows three examples of conditional posterior distributions,  based on Cauchy priors. 
    These probability density functions are obtained when  all, expect one, of the  components of the posterior  have been fixed to their maximum a posteriori estimates.  
    As  can be observed, using such priors can result in exotic conditional and marginal distributions, which are spiky, or highly multimodal. The examples in \Cref{fig:cond} demonstrate that doing inference with such distributions is far from straightforward.
    
    %
    An important question is how to develop priors used for particular applications and models. 
    A common prior choice, due to theoretical properties and computational benefits, is the Gaussian prior \citep{VIB98,DGST19, KVZ11,RW06,RHL14}. 
    They have been widely used for inverse problems, and other applications as well, since they are easy to implement, have often desirable properties and are suitable at modelling various phenomenon of real-world models. 
    However, for certain problems, the Gaussian priors have been shown not to be the optimal choice. 
    A common issue with the use of Gaussian priors is their inability to recover rough features or edges.
    Although Gaussian priors are widely used, they are, however, deemed a poor choice in situations where the point of  interest is non-smooth reconstruction such as reconstruction of edges. 
    If the posterior favors such an edge-preserving form, sophisticated priors which can exhibit such phenomenon, can aid the problem significantly. 

    Numerous priors have been promoted and suggested to alleviate the issues that arise with Gaussian priors. 
    In the context of Bayesian inversion, this was firstly proposed by  \citet{LS04}, who conjectured whether the total variation prior might be a good choice. 
    However, the results of their work suggested the prior degenerated as the mesh was refined. 
    Other priors based on similar ideas include, but are not limited to, the Besov prior \citep{DHS12,LSS09}, geometric priors \citep{CIRL17,ILS14} and Laplace priors \citep{BH17,HN17}. 
    Besov priors are based on wavelet expansions and require a dyadic construction, which suggests they can be difficult to use in practice. 
    Geometric priors are designed based on random geometries of known-knowledge of the underlying unknown. As a result, these priors can be model-specific, dependent on the application.
    The Laplace priors  are also based on a similar representation of wavelet expansion, that is, a modified Karhunen-Lo\'{e}ve expansion.

    An  interesting direction has  been the exploitation of neural network and deep learning machinery \citep{AMOS19,RMN12}. This approach has included data-driven priors based on neural network architecture \citep{HPZ21,SS21}, as well as the promotion of deep learning frameworks for Gaussian processes, known as deep Gaussian processes \citep{DL13,DGST19}. These priors have been proven to be successful and capable of quantifying deeper structures of images, but it is important to note that much of the work presented has not been tailored for non-Gaussian priors.

    In this context, it is worth mentioning  the recent progress in neural network -based learned priors. 
    The rationale of underlying development of (Generative Adversarial Network) (GAN) priors \citep{PO19} like the conditioned Wasserstein-GAN \citep{AO18} is the desire to attain the best possible posterior inference performance by ensuring  flexiblity and the ability to capture complex features of the training data, for which  a prior  cannot be  manually crafted. 
    However, neural network -based approaches rely on learning a latent vector representation of the training data distribution, and hence, they often exhibit inadequate  robustness against training data inconsistencies and have unsatisfactory generalization capability.  These deficiencies provide further motivation for non-Gaussian random field priors, which can be constructed for invariant to translation, and preferably rotation, and which generate predictable qualitative features, including the edge-preserving property, of the resulting posterior distributions.

    A more recent class of priors which have been developed are Cauchy and $\alpha$-stable priors \citep{CLR19,RGLM16,TJS16}.
    These priors are known to encapsulate more edge-like features, where their advantage is that they have been analyzed both theoretically and numerically. 
    Such priors include Cauchy-difference and $\alpha$-stable process priors.  Cauchy difference priors, which are based on  coupling of Cauchy random walks, were first proposed by \citet{MRHL19}. These priors were constructed for multi-dimensional problems. Recent work has extended this approach for both hyper-prior estimation of hyperparameters, and demonstrating benefits over other non-Gaussian priors, such as Besov priors. Posterior convergence analysis of probability has been analyzed for $\alpha$-stable priors \citep{CLR19}. The discretization of such priors was constructed using hypercubes. As well known, $\alpha$-stable priors encapsulate numerous processes, such as Cauchy processes, which can be viewed as a special case.

    Bayesian inference with posteriors with random field priors is challenging due to their high dimensionality. 
    While  point estimates like maximum a posteriori (MAP) estimate can be derived analytically for posteriors with Gaussian random field prior and Gaussian likelihood with a linear direct mapping $\eqref{eq:ip}$, this is not the case with Cauchy priors. MAP estimates require the use of a numerical optimization tool to find  the estimate. 
    Other estimators, like conditional mean and marginal variance estimators require evaluation of intractable integrals. Due to the dimensionality, the estimators cannot be evaluated analytically, but have to be calculated using Markov Chain Monte Carlo (MCMC) methods by drawing samples from the posteriors.

    In order to accommodate the use of such non-Gaussian priors, we use a variety of Markov chain Monte Carlo methods. However, efficient sampling is computationally  difficult  with the Cauchy random field priors due to possible heavy-tailedness and multimodality  of the posteriors. Consequently, traditional MCMC methods such as Metropolis-Hastings may not perform well. We discuss and test a number of MCMC algorithms, which may prove to be useful in various scenarios. These algorithms include the NUTS algorithm, proposed by \citet{HG14,RMN03}, and the Repelling-Attracting Metropolis method, which are considered in addition to the blockwise Metropolis-Hastings adaptive algorithm \citep{TMD18}.

    The motivation for comparing the three aforementioned MCMC algorithms comes  from the high-dimensional posteriors with Cauchy random-field priors, that can have multimodal marginal or conditional distributions. On the other hand, multimodality of the conditional distributions of the posteriors are ingredient of the edge-preserving property of the priors, and usage of biased inference methods, such as the Laplace approximation, for the inference would ruin the desired properties of the priors.  Importance sampling and other particle-based Monte Carlo algorithms deal well with multimodality, but the Curse of Dimensionality renders them impractical for approximating high-dimensional distributions \citep{QML10}. Certain adaptive rejection Metropolis samplers are viable choices for sampling multimodal and even heavy-tailed distributions \citep{MRL15}, but they are computationally infeasible for utilization within-Gibbs sampling to sample distributions with a few hundred dimensions or more. Modern MCMCs like the Hamiltonian Monte Carlo \citep{N12} deal well with high-dimensionality, but not so well with multimodality and heavy-tailedness. Finally, most MCMC algorithms that are targeted for multimodal distributions, such as various multi-chain MCMC that are run in different virtual temperatures \citep{GT95} and JAMS \citep{PHL20}, are not designed for use within-Gibbs sampling  and cannot be applied for high-dimensional sampling.

    \subsection{Our contributions}
    
    The promotion of non-Gaussian priors, \textcolor{black}{raises} the question how these Cauchy priors compare with each another. The formulation of such priors is inherently different and yet it is of interest to see, which variant is most suitable for particular setups. Thus, the aim of this work is to systematically understand these varying Cauchy random field priors for Bayesian inverse problems. To aid development of this understanding, we are also motivated to formulate two new prior forms,  based on a modification of the Cauchy difference prior and a reformulation of the stochastic partial differential equation (SPDE) approach \citep{LRL11,YAR77} for non-Gaussian priors. We, however, emphasize with this work that we are not concerned with comparing our Cauchy priors with other non-Gaussian priors. This study has been done extensively in the work of  \citet{SELS20}.

    The contributions of this work are summarized below.
    \begin{itemize}
        \item We provide a \textit{cookbook} overview of the various Cauchy priors that have been introduced in the literature, and  introduce new variants. Existing priors include one-dimensional Cauchy difference prior and two-dimensional anisotropic Cauchy difference priors.
        We generalise the models with second order differences providing continuous, but non-smooth reconstructions suitable e.g.\ for detecting non-differentiable functions,  that is,  promoting piecewise linear reconstructions.
        We provide new two-dimensional first and second order isotropic difference priors.
        For advocating detection of spikes and similar features, we propose to use SPDE based Cauchy priors, which have been previously studied in Bayesian statistics \citep{WB15}.
        We also study Cauchy sheet priors and their validity in inversion.
        \item We propose to use Metropolis-within-Gibbs (MwG),  the Repelling-Attracting Metropolis (RAM) and  the No-U-turn sampler (NUTS), a variant of Hamiltonian Monte Carlo, and show their validity on sampling resulting high-dimensional, multimodal and heavy-tailed posteriors.
        \item We contrast and compare these newly developed and existing priors with the three sampling methods for one-dimensional and two-dimensional deconvolution problems, as we are interested in reconstructing an image containing both rough edges and non-Gaussian spikes. \textcolor{black}{We also compare our Cauchy priors to both total variation and Gaussian priors.}
    \end{itemize}
    
    \subsection{Outline of the paper}
    The outline of this work is as follows. 
    In Section \ref{sec:priors}, we provide an overview of existing Cauchy priors used for inverse problems, and construct the new variants. 
    New prior forms require also a discussion on the SPDE approach for random fields and general alpha-stable random fields. 
    In Section \ref{section:Methods}, we discuss existing optimization and MCMC methods, and make notes on how we apply them to the problem at hand.
    This discussion leads to Section \ref{sec:num}, where we provide  numerical examples comparing all priors on different sets of experiments.
    Finally, we summarize our work, and suggest further directions of research regarding the Cauchy priors in Section \ref{sec:conc}.

    \section{Cauchy Markov Random Field Priors}
    \label{sec:priors}
    
    First-order one-dimensional difference priors and two-dimensional anisotropic difference priors were originally studied by \citet{MRHL19}, with applications to subsurface imaging by \citet{Muhumuza_2020} and to X-ray tomography by \citet{mendoza:2019}. 
    An extension to Cauchy sheets, and especially to posterior consistency results are discussed by \citet{CLR19}.
    In this section, we give all the aforementioned priors in a systematic manner in conjunction with new second-order difference priors, two-dimensional isotropic priors, and SPDE priors.
    All the constructions are based on finite difference methods, since the discussion on finite element methods and series expansions require different kind of techniques, they are left for future studies.
    
    \subsection{One-dimensional Cauchy  difference priors}
    
    Cauchy random walks are well-known objects and they have been extensively studied \citep[see e.g.][]{TS94}.
    Cauchy random walks are special cases of L\'{e}vy $\alpha$-stable random walks $u_i$, $i \in \mathbb{Z}^+$, and the corresponding continuous-time stochastic processes $u(t)$, $t\in\mathbb{R}^+$.
    Let us denote by $S_\alpha(\lambda^2,\beta,\mu)$ a stable distribution with  stability parameter $0<\alpha \leq 2$, scale parameter $\lambda^2\in \mathbb{R}^+$, skewness parameter $\beta\in[-1,1]$, and location parameter $\mu \in \mathbb{R}$.
    For simplicity, we choose $\beta = 0$ and $\mu = 0$.
    We note that Cauchy and Gaussian distributions are both part of the family of $\alpha$-stable distributions, with stability parameters $\alpha=1$ and $\alpha=2$, respectively.     
    
    Now, we can define L\'{e}vy $\alpha$-stable stochastic process and random walk  via increments
    \begin{equation*}
    \begin{split}
    u(t)-u(s) &\sim S_{\alpha}\left((t-s)^{1/\alpha}, \beta,0\right), \\
    u_{i+1} - u_{i} &\sim S_{\alpha}\left(h^{1/\alpha}, \beta,0\right),
    \end{split}
    \end{equation*}
    for any $0 \leq s <t<\infty$, and discretization $t=jh$ with $h>0$ is the discretization step. 
    Cauchy random walk $u_i$ has a well-defined limit  $h \rightarrow 0$, that is, $u_i$  converges to $u(t)$ \citep[for details, see][]{TS94}. 
    Cauchy random walks, similarly to Gaussian random walks, are Markovian by construction, that is $u_{i+1}$ depends on only $u_i$.  
    
    It is worth mentioning that the probability density functions of a $\alpha$-stable distribution with arbitrary stability parameter $\alpha$, do not have an analytical formula, with a few exceptions, including Cauchy and Gaussian distributions. 
    However, integral expressions  for the univariate and elliptically contoured multivariate $\alpha$-stable  distributions do exist \citep{VMZ81, JN13}, which would possibly enable numerical applications of such priors using  computationally feasible approximations.

    For Bayesian inversion, \citet{MRHL19} used Cauchy random walks as difference priors.
    %


    \begin{defn}[First order Cauchy difference prior]
        A stochastic process $\mathbf{u} := (u_1,\dots, u_N)^T$ is called first order Cauchy difference prior if it can be presented as an unnormalized probability distribution
        \begin{equation}
        \label{eq:cd1}
        \pi(\mathbf{u}) \propto \frac{1}{ \gamma^2 + u_{1} ^2} \prod_{\textcolor{black}{i}=1}^{N-1} \frac{ 1}{ \lambda^2 + (u_{i+1} - u_{i})^2},
        \end{equation}
        where $\gamma,\lambda>0$ are strictly positive  parameters, acting as an initial condition and the scale parameter, respectively. 
    \end{defn}

    We note that by choosing $\lambda=h$, then the scale parameter is chosen in such a way the random walk converges to its continuous counterpart.
    \citet{CLR19} showed that when using this kind of a prior in a Bayesian inverse problem, the posterior behaves well in the discretization limit. That is, the posterior is consistent with respect to refining the computational grid.
    In Bayesian inversion, this property is called discretization invariance, which guarantees all the estimators are essentially the same on dense enough meshes.
    
    In the definition of the first order Cauchy difference prior, we use an initial condition that guarantees the prior is proper. 
    We could also utilize Cauchy bridges, in the same sense as Brownian bridges, as we could impose boundary conditions for $u_1$ and $u_N$.

    Similarly, we can use second order differences, in Bayesian inversion, that is, to use integrated  L\'{e}vy flights.
    \begin{defn}[Second order Cauchy difference prior]
        A stochastic process $\mathbf{u}$ is  called second order Cauchy difference prior if it can be presented as a probability distribution
        \begin{equation*}
        \label{eq:cd2}
        \pi(\mathbf{u}) \propto  \frac{ 1}{\gamma^2 + u_{1} ^2} \frac{1}{ \gamma'^2 + (u_{2} - u_{1})^2}  \prod_{i=2}^{N-1} \frac{ 1}{ \lambda^2 + (u_{i+1} - 2u_i + u_{i-1})^2} ,
        \end{equation*}
        with $\gamma,\gamma',\lambda>0$.
    \end{defn}
    
    Again, the left boundary conditions are given in order to guarantee properness of the prior probability distribution.
    To the best of our knowledge, this kind of prior has not been reported in inverse problems literature earlier.
    

    \subsection{Two-dimensional Cauchy difference priors}
    
    
    %
    

    Let $\mathbf u$ be a two-dimensional discrete random field with elements $u_{i,j}$.
    We define  horizontal differences $d_h := u_{i+1,j}-u_{i,j}$ and vertical differences $d_v := u_{i,j+1}-u_{i,j}$.
    Then, related to each pixel $(i,j)$, we model increments with a bivariate Cauchy distribution
    \begin{equation}\label{formula2i}
    \pi(d_h,d_v) \propto  \frac{1}{\left(c^2 + d_h^2+d_v^2\right)^{3/2}}.
    \end{equation}   
    That is, the increments to the vertical and horizontal direction at each grid node are set to follow the introduced multivariate radially symmetric Cauchy distribution.
    We note that all univariate, bivariate and multivariate Cauchy distributions can be obtained from univariate, bivariate and multivariate Student distribution with one degree of freedom.
    The construction of the isotropic Cauchy difference prior is  analogous to the construction of isotropic total variation prior \citep{GKS17}. 
    
    \begin{defn}[Isotropic first order Cauchy difference prior]
        A random field $\mathbf{u}$ on domain $\Omega\in \mathbb{R}^2$ with approximation on lattice $(i,j)$ is called isotropic first order Cauchy difference prior, if it can be written as a probability distribution
        \begin{equation}
        \label{eq:ci1}
        \pi(\mathbf{u}) \propto 
        \pi_{\partial \Omega}(\mathbf{u})
        \prod_{i=1}^{N-1} \prod_{j=1}^{N-1} \frac{ 1}{ \left( \lambda^2 + (u_{i+1,j} - u_{i,j})^2 + (u_{i,j+1} - u_{i,j})^2 \right)^{3/2}} 
        \end{equation} 
        with 
        \begin{equation*}
        \pi_{\partial \Omega}(\mathbf{u}) = \prod_{(i,j) \in \partial \Omega} \ \frac{ 1}{ \gamma^2 + u_{i,j} ^2},
        \end{equation*}
        where $\partial \Omega$ denotes the grid nodes on the boundary of the lattice $(i,j)$, and $\lambda,\gamma>0$. 
    \end{defn}
    
    This construction differs from the previously studied difference priors by \citep{MRHL19} in the sense that the earlier construction used products of independent increments of \Cref{eq:cd1} in horizontal and vertical directions   
    \begin{equation*}
    \pi(\mathbf{u}) \propto 
    \pi_{\partial \Omega}(\mathbf{u})
    \prod_{i=1}^{N-1} \prod_{j=1}^{N-1} \frac{ 1}{  \lambda^2 + (u_{i+1,j} - u_{i,j})^2  }  \frac{ 1}{  \lambda^2 + (u_{i,j+1} - u_{i,j})^2 }      .   
    \end{equation*}
    We call this existing construction of the prior as an anisotropic first-order Cauchy difference prior.
    We note that the prior parameter $\lambda$ is not necessarily consistent between the isotropic and anisotropic formulations. For example, even if the unknown object consists of features that are perfectly aligned with the coordinate axis, the ratio of the priors might not be constant when changing the parameter.
    
    Similarly to the one-dimensional difference priors, we aim to obtain a second-order difference prior in spatial dimensions of two.
    There are two alternative paths, the first alternative is to consider a stochastic partial differential equation $\Delta u = \mathcal{M}$, where $\mathcal{M}$ is Cauchy noise field. 
    \begin{equation*}
    \pi(\mathbf{u}) \propto 
    \pi_{\partial \Omega}(\mathbf{u})
    \prod_{i=2}^{N-1} \prod_{j=2}^{N-1} \frac{ 1}{  \lambda^2 + ( - 4u_{i,j} + u_{i+1,j} + u_{i-1,j} + u_{i,j+1} + u_{i,j-1})^2 } ,
    \end{equation*}
    We come back to this construction later in this section. However, we want to include another construction of the prior which mimics more the properties of the one-dimensional second order prior, as we wish to have properties which promote piecewise linear type reconstructions. 
    Based on Equation \eqref{eq:ci1}, we construct a higher order version by using bivariate Cauchy distribution.
    \begin{defn}[Isotropic second order Cauchy difference prior]
        A random field $\mathbf{u}$ on domain $\Omega\in \mathbb{R}^2$ with approximation on lattice $(i,j)$ is called isotropic second order Cauchy difference prior, if it can be written as a probability distribution        
        \begin{equation*}
        \pi(\mathbf{u}) \propto  
        \pi_{\partial \Omega}(\mathbf{u})
        \prod_{i=2}^{N-1} \prod_{j=2}^{N-1} \frac{ 1}{ \left( \lambda^2 + (u_{i+1,j} - 2u_{i,j} + u_{i-1,j})^2 + (u_{i,j+1} - 2u_{i,j} + u_{i,j-1})^2 \right)^{3/2}}  ,
        \end{equation*}
        with 
        \begin{equation*}
        \pi_{\partial \Omega}(\mathbf{u}) = \prod_{(i,j) \in \partial \Omega}   \frac{1}{ \gamma^2 + (u_{i,j} - u_{i-\cdot,j-\cdot})^2}  \frac{ 1}{ \gamma'^2 + u_{i,j} ^2},
        \end{equation*}
        \textcolor{black}{where by $u_{i-\cdot,j-\cdot}$ we denote the closest lattice point inside the domain $\Omega$ of the pixel $(i,j)$, and $\lambda, \gamma,\gamma' >0$}.
    \end{defn}
    
    \subsection{Two-dimensional Cauchy sheets}   
    
    Brownian sheets can be generalized to $d$-dimensional $\alpha$-stable sheets.
    We are interested on the Cauchy sheets, and they can be given similar Markovian structure as all the presentations above:
    \begin{equation*}
    \pi_\textrm{sh}(\mathbf{u}) \propto \pi_{\partial \Omega}(\mathbf{u})  \prod_{i=1}^{N-1} \prod_{j=1}^{N-1} \frac{ 1}{ \lambda^2 + (u_{i+1,j+1} - u_{i+1,j} - u_{i,j+1} + u_{i,j})^2}.
    \end{equation*}
    A discussion on  $\alpha$-stable sheets in Bayesian inversion with posterior consistency can be found from \citet{CLR19}.
    
    However, a disadvantage of the sheet prior compared to the proposed isotropic priors \eqref{eq:ci1}, is that the Cauchy sheet prior is radially asymmetric, this is similar to the anisotropic difference priors. This means the prior favors features that are aligned with the coordinate axis, therefore  the prior is affected by the underlying geometry and chosen discretization lattice. 
    
    \subsection{Cauchy priors via SPDEs}
    
    The stochastic partial differential equation (SPDE) approach is a  powerful methodology for simulating Gaussian random fields \citep{LRL11}. 
    In the context of inverse problems, the SPDE approach for Gaussian priors has been used, for example, in applications such as imaging and geophysical sciences \citep{RHL14,CIRL17,RGLM16}.
    The SPDE approach for Gaussian random fields deals with solving the following linear elliptic SPDE
    \begin{equation}
    \label{eq:matern}
    (I - \ell^2 \Delta)^{(\nu+d/2)/2}u = \phi \mathcal{W}, \quad \phi  =  \sigma^2 \frac{2^d\pi^{d/2}\Gamma(\nu + d/2)}{\Gamma(\textcolor{black}{\nu})},
    \end{equation}
    where $\Delta$ is the Laplace operator, $\ell \in \mathbb{R}^+$ is the length-scale, $\nu \in \mathbb{R}^+$ denotes the regularity or smoothness, $\sigma \in \mathbb{R}$ denotes the amplitude. \textcolor{black}{Furthermore, $\mathcal{W}$ is Gaussian white noise, and $\Gamma(\cdot)$ is a gamma function}.
    The solution of \eqref{eq:matern} is $d$-dimensional stationary Gaussian random field $u$ with Mat\'{e}rn covariance function, of the form
    \begin{equation*}
    C(x,x') = \sigma^2\frac{2^{1-\nu}}{\Gamma(\nu)}\left(\frac{|x-x'|}{\ell}\right)^{\nu}K_{\nu}\left(\frac{|x-x'|}{\ell}\right),
    \end{equation*}
    where $K_{\nu}(\cdot)$ is a modified Bessel function of the second kind of order $\nu>\alpha+d/2$. Motivated by the Gaussian case, our next new Cauchy prior we introduce, effectively replaces the Gaussian white noise $\mathcal W$ with Cauchy noise $\mathcal{M}$, resulting in a Cauchy SPDE prior.
    As an example, let us consider a two-dimensional random field.  At first, we  drop the constant $\phi$ and assume that $(\nu+d/2)/2 = 1$, and hence obtain
    \begin{equation}
    \label{eq:spde}
    (I - \ell \Delta)u =  \mathcal{M}.
    \end{equation}
    \textcolor{black}{To numerically solve \eqref{eq:spde}}, we can approximate the operator part, the right hand of the equation, using a canonical finite-difference approximation of the Laplace operator as follows:
     \begin{equation*}
     \Delta)u  \approx  \frac{4 u_{i,j}-u_{i+1,j}-u_{i,j+1}-u_{i-1,j}-u_{i,j-1}}{ h^2},
    \end{equation*} where $h$ is the node interval within the discretized domain. We obtain an approximation
    \begin{equation*}
    (I - \ell \Delta)u  \approx   u_{i,j} - \ell \frac{4 u_{i,j}-u_{i+1,j}-u_{i,j+1}-u_{i-1,j}-u_{i,j-1}}{ h^2} =: p_{i,j} = m_{i,j},
    \end{equation*}
    where $m_{i,j}$ is independent and identically distributed discrete Cauchy noise $\mathcal{M}$.
    Finally, the resulting prior is of the form
    \begin{equation*}
    \pi_\textrm{p}(\mathbf{u}) \propto \prod_{(i,j)} \frac{ 1}{ \xi^2 + p_{i,j}^2},
    \end{equation*}
    where $\xi^2$ is the scale parameter of the Cauchy noise. In other words, we employ the finite-difference stiffness matrix of the SPDE by setting the entries of the  product of the stiffness matrix and the discretized random field  vector to follow Cauchy distribution. 
    
    \textcolor{black}{As we have introduced various Cauchy random field priors priors in this section, we provide a comparison related to their advantages of disadvantages. This can be found in Table \ref{table:priors}, which highlights key differences between priors, and what we can expect in the numerical experiments section.
    Based on our understanding, we initially would expect that the best performing priors would be the isotropic Cauchy priors.}
    
    \begin{remark}
        \textcolor{black}{It is important to note that for the numerical computation of $u$ in \eqref{eq:spde}}, we could also resort to other known numerical methods such as spectral Galerkin method, or a finite element method. However, for simplicity and computational purposes, we stick to finite-differences.
    \end{remark}
    
     \begin{remark}
        \textcolor{black}{Discretization invariant scaling of  $\xi^2$ as $h\rightarrow \infty$ in \eqref{eq:spde} is yet to be verified for Cauchy and other non-Gaussian $\alpha$-stable noise. Mat\'ern SPDE with Gaussian noise converges to the continuous limit when a nominal variance $\sigma_0^2$ is scaled as $\sigma^2 = \frac{\sigma_0^2}{h}$ in one-dimensional domain \citep{RHL14}. 
        The discretization of the SPDE via finite elements would offer a straightforward way to obtain the proper scaling of the discretized Cauchy or $\alpha$-stable noise in the SPDE prior through stochastic integrals. If the domain $\Omega$ of the SPDE is divided in disjoint elements $E_i$, and the discretized noise $\mathcal{M}$ is independently scattered in the elements, then the proper scaling of the discretized element-wise noise is to set}
        $$\int_{E_i} h(x) \mathcal{M}(dx),$$ 
   to follow $\alpha$-stable distribution with scale $ \left( \int_{E_i} |h(x)|^{\alpha} dx\right)^{1/\alpha}$ \citep{CLR19}. However, a solid theoretical discussion regarding the  infinite-dimensional SPDE random fields with $\alpha$-stable noise is missing for us to use and exploit.
    \end{remark}
    
    \subsection{One-dimensional realizations}
    
\begin{figure}
        \centering
        \includegraphics[width=0.6\linewidth]{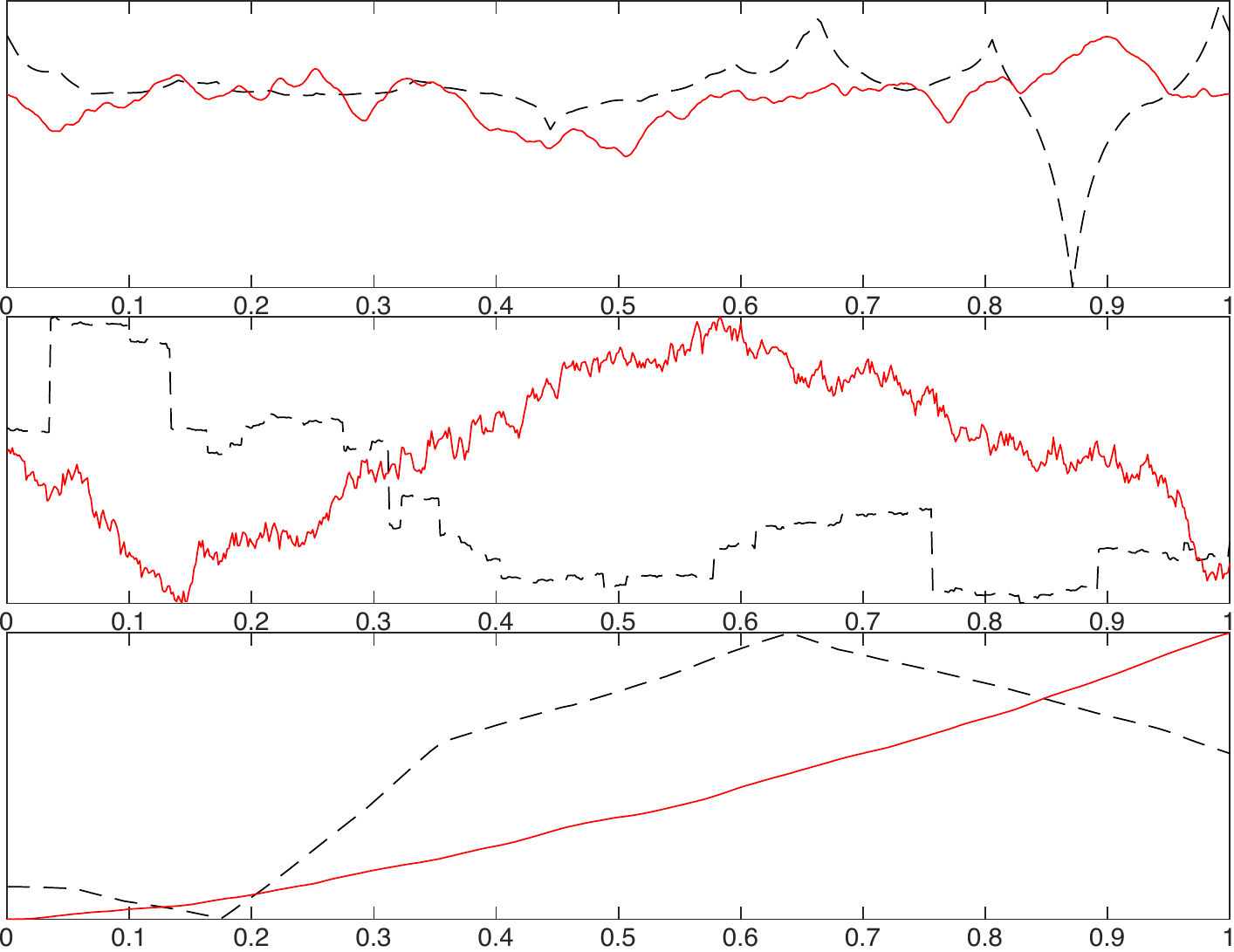}
        \caption{   \textcolor{black}{Realizations of 1st and 2nd order $\alpha$-stable random walks and SPDEs. Top image: A solution for the Mat\'{e}rn SPDE with Cauchy noise (black) and Gaussian noise (red). Middle image: First order Cauchy random walk (black) and Gaussian random walk (red). Bottom image: Second order  Cauchy random walk (black) and Gaussian random walk (red). Scales of the solutions  have been normalized on purpose for comparison. }}
        \label{fig:real}
    \end{figure}

    Realizations of both Cauchy and Gaussian random walks and Matérn SPDEs are illustrated in Figure \ref{fig:real}. \textcolor{black}{The samples were obtained by interpreting the priors as stochastic differential equations (SDE), which noise process is either independent and identical Gaussian or Cauchy noise. Doing so is possible  in the one-dimensional case, since there is a linear one-to-one correspondence between the discretized random walk processes and their increments, so realizations of the processes can be obtained by solving the corresponding systems of SDEs.  However, in dimensions more than one, only the SPDE priors can be sampled efficiently, since the difference priors cannot be interpreted as SPDEs.} 
    A characteristic feature of solutions  of Matérn SPDEs with Cauchy noise are exponential spikes. However, they seek to center back to zero outside the peaks. Solutions of the Matérn SPDE with Gaussian noise do not have such strong peaky features. Naturally, the mean of the SPDE with Gaussian noise is zero, but its realizations do not tend to revert close to zero as strongly as the solutions of Cauchy noisy SPDE do.
 
    \begin{table}
        \textcolor{black}{
    \centering
    \begin{tabularx}{1.0\textwidth} { 
  | >{\centering\arraybackslash}X 
  | >{\centering\arraybackslash}X 
  | >{\centering\arraybackslash}X | }
 \hline
 \textbf{Prior} & \textbf{Advantages} & \textbf{Disadvantages} \\
 \hline
Isotropic first order difference prior  & Piecewise-constant coordinate-axis independent reconstruction  &  Lack of convergence proofs, difficult to sample from \\
\hline
Anisotropic first order difference prior  &  Piecewise-constant reconstruction, tentative results for convergence in the discretization limit exist  & Favors features aligned with coordinate-axis, difficult to sample from (in two and higher dimensions)  \\
\hline
Isotropic second order difference prior  & Piecewise-constant derivative coordinate-axis independent reconstruction  &  Not based on a SPDE so lacks convergence proofs, difficult to sample from \\
\hline
Anisotropic second order difference prior  &  Piecewise-constant derivative reconstruction & Not based on a SPDE so lacks convergence proofs, favors features aligned with coordinate-axis, difficult to sample from\\
\hline
Cauchy sheet prior  & Easily generalized and sampled, convergence proofs exist  & Radially asymmetric, i.e. affected by local geometry  \\
\hline
Cauchy SPDE prior  & Reconstructing exponential and non-Gaussian peaks, convergence results likely available thanks to the SPDE construction   & Does not favor piecewise-constant features, very sensitive to the average level (median) of the fields  \\
\hline 
\end{tabularx} \\ \bigskip
\caption{\textcolor{black}{Comparison of the different Cauchy priors, in terms of both advantages and disadvantages of what they can help reconstruct.}} 
\label{table:priors}}
\end{table}

\subsection{Analytical properties of Cauchy priors}
\label{ssec:analysis}

\textcolor{black}{
%
Out
of the priors to be numerically tested, only two of them have been introduced prior to this paper; (i) the 
anisotropic Cauchy-difference priors \citet{MRHL19} and (ii) the Cauchy sheet prior \citet{CLR19}. For the former mainly numerical experiments have been provided related to x-ray tomography. However \citet{CLR19} provide a first result related to difference priors which showcase an $L^p$-convergence result  (Theorem 4.1.),for $1 \leq p< \infty$. For the later the same work of \citet{CLR19} showed that $\alpha$-stable Cauchy sheet priors, one could
prove both well-posedness and well-defindness defined as
\begin{itemize}
    \item \textit{Well-posedness 1}: There exists a set $D \subset \mathcal{Y}$ such that $y \mapsto \int \exp(-\Phi(u,y)\mu(du)$ is continuous.
    \item \textit{Well-posedness 2}: The function $y\mapsto \Phi(u{,}y)$ is continuous on $\mathcal{Y}$ for every 
$u\in \mathcal{X}$.
\item \textit{Well-definedness}: The function $\Phi$ is bounded on bounded subsets of $\mathcal{X} \times \mathcal{Y}$.
    \end{itemize}
    In particular the former has been proven in the total-variation metric, which is different to the usual
    Hellinger metric. This is due to the complexity induced by the Cauchy sheet prior. Other such results, in function-space settings can be found related to the convergence for $\alpha$-stable Cauchy sheets, for $0 < \alpha \leq 2$.\\
Given the current analysis discussed above, the derivation of new results related to well-posedness and
well-definiteness are beyond the scope of such a paper, as our intention is primarily numerical based. The
difficulty with analzying such priors is related to the non-Gaussian phenomenon and thus we leave this
direction for future work.}
    
    \section{Posterior estimates and uncertainty quantification}
    \label{section:Methods}
    
    Sampling from high-dimensional, multimodal and heavy-tailed posteriors is far from trivial. 
    Hence, in this section we discuss in detail the benefits of different methods, and specific adaptations to the algorithms deployed.
    For conditional mean estimation we use three Markov Chain Monte Carlo methods (MCMC): the adaptive Metropolis-within-Gibbs (MwG) algorithm  which is also called the Single-Component Adaptive Metropolis-Hastings (SCAM) \citep{HST05}, the Repelling-Attracting Metropolis (RAM) algorithm \citep{TMD18}, and the No-U-Turn Sampler variant of the Hamiltonian Monte Carlo (NUTS). The respective algorithmic forms can be found in the Appendix. 
    Additionally, we briefly discuss the role of MAP estimation. 
    
    \subsection{MAP-estimation}
    The maximum a posteriori estimator (MAP) is one of the most widely used estimators in Bayesian inference, and it is given as an optimization problem of the posterior 
    \begin{equation*}
    \mathbf{u}_\text{MAP} = \arg \max_\mathbf{u} \pi (\mathbf{u}|\mathbf{y}). 
    \end{equation*}
    MAP estimator a point-estimate and  therefore provides no information regarding the spread of the posterior, however, it is often sufficient for many applications. 
    
    MAP estimator can be obtained by minimizing the negative posterior probability density function or its logarithm.
    In our cases it cannot be analytically evaluated, but we can deploy existing numerical optimization algorithms.
    Here, we choose the Limited-memory Broyden–Fletcher–Goldfarb–Shannon (L-BFGS) algorithm implemented by \citet{MR18}. While  L-BFGS is not a global optimization algorithm, it is a  quasi-Newton method and therefore  applicable to  minimization of high-dimensional differentiable functions.

    \subsection{{CM-estimation}}
    The conditional mean (CM) is another common estimator for posteriors. 
    It is  the expectation of the posterior
    \begin{equation*}
    \mathbf{u}_\text{CM} = \int_{\mathbb{R}^d} \pi(\mathbf{u}|\mathbf{y})\mathbf{u} \text{d}\mathbf{u}. 
    \end{equation*}
    The advantage of the CM estimator -- when compared to MAP -- is that the CM estimator takes the whole distribution into account. 
    \textcolor{black}{If the marginal distributions of the posterior are multimodal like in \Cref{fig:cond}, the MAP and CM estimators can be significantly different.}
    \textcolor{black}{In general, the CM estimator is typically preferred over the MAP estimator in parameter and state estimation of various hidden Markov models, especially if the transition density of the model is intractable \citep{L15}. The MAP estimator is more common in Bayesian continuous-space parameter estimation, such as estimating the electrical conductivity in electrical impedance tomography \citep{GKS17}. That is because the CM estimator requires high-dimensional integration, and is thus computationally more demanding in models where the dimensionality of the posterior is proportional to the discretized dimensionality of the continuous space. If the computational issues are not taken into account, the question of whether the CM or the MAP estimator should be preferred cannot be answered universally.  In our Cauchy random field models, the MAP estimator can be considered somewhat more natural than the CM, since the MAP estimator promotes the existence of discontinuous features. For instance, the CM estimators of the multimodal conditional distributions in \Cref{fig:cond} would not be close to either of the modes, and the CM estimator of the node would be a compromise of the values of the adjacent nodes, what is sub-optimal for discontinuity.}
    Hence, in CM estimation we need to resort to MCMC \citep{KS04, LMBES20}. In brief, the common principle of different MCMC methods is  to implement a  transition kernel $\text{k}(\cdot)$ that is used to propose new samples from the target distribution. 
    The transition kernel has to be able to generate an ergodic chain, that is, the distribution of the generated chain converges to the target distribution in total variation distance.

    For the above reasons, we select three MCMC algorithms  each having their own advantages and disadvantages. In the sections below, we  briefly explain the algorithms and their core principles.  
    
    \subsubsection{Adaptive Metropolis-within-Gibbs}
    
    Metropolis-Hastings is one of the oldest and most common MCMC methods \citep{M53}, and it is the vanilla MCMC for many target distributions with typically up to a few tens of parameters.  
    We use the Metropolis-within-Gibbs (MwG) scheme instead of the basic Metropolis-Hastings with a transition kernel
    \begin{equation*}
    \text{k}_\text{\textbf{j}}\left(\mathbf{u}^*_\mathbf{j}|\mathbf{u}_\mathbf{j}^\text{p}\right) = \text{h}\left(\mathbf{u}_\mathbf{j}^*|\mathbf{u}_\mathbf{j}^\text{p}\right)\alpha_\mathbf{j}\left(\mathbf{u}^*_\mathbf{j}|\mathbf{u}_\mathbf{j}^\text{p}\right)\text{d}\mathbf{u}_\mathbf{j}^* + \delta_{\mathbf{u}^{\text{p}}_\text{\textbf{j}}}\left(\mathbf{u}^*_\mathbf{j}\right)\left(1-Z_{\mathbf{j}}\left(\mathbf{u}_\mathbf{j}^\text{p} \right)\right),
    \end{equation*}
    and acceptance probability
    \begin{equation}\label{m01}
    \alpha_\mathbf{j}\left(\mathbf{u}^*_\mathbf{j}|\mathbf{u}_\mathbf{j}^\text{p}\right) = \min \left (1, \frac{\pi(\mathbf{u}_\mathbf{j}^*|\mathbf{u}^\text{p}_{-\textbf{j}})}{\pi(\mathbf{u}_\mathbf{j}^\text{p}|\mathbf{u}^\text{p}_{-\textbf{j}})} \right),
    \end{equation}
      \textcolor{black}{where $\mathbf{u}^*_{\mathbf{j}}$ denotes the proposed and $\mathbf{u}^p_{\mathbf{j}}$ the previous values for the indices $\mathbf{j}$ of $\pi$. } 
    We propose new values for index set $\mathbf{j}$ using the proposal distribution $\text{h}_\mathbf{j}$.
    For high-dimensional distributions, drawing samples from the subsets of the target distribution components  improves the mixing of the marginal chains. 
    
    In the adaptive random walk MwG \citep{HST01}, a Gaussian distribution centered at the previous sample is used as the proposal kernel $\text{h}(\cdot)$, and  the proposal covariances $\mathbf{C}_\mathbf{j}$ are tuned automatically during the burn-up period \citep{HST05}. The adaptation for the proposal covariances is calculated from the empirical covariance estimates  using the chain generated so far. The adaptation is stopped after a specified number of  iterations has been reached, so that we guarantee the ergodicity of the algorithm during the actual sampling. A pseudocode for the random walk adaptive MwG using a lexicographic scan of the target distribution component subsets is given in \Cref{scam}. \textcolor{black}{We intentionally select adaptive MwG instead of an adaptive Metropolis-Hastings algorithm that uses global proposals. That is because the chains an algorithm like adaptive MwG generates have been shown to have better autocorrelation, as they are able to propose samples that are further from the previous sample, at least component-wise \Cref{scam}. In MwG, we evaluate only the components of the matrix-vector products of the likelihood function and the prior function that refer to the current index $\mathbf{j}$ in the Metropolis-within-Gibbs algorithm. Hence, by leveraging the linearity of the forward model and the likelihood function, and by comparing the difference of the partially evaluated log-posterior densities, we avoid the increased total cost of generating a new sample to be added in the chain.}
    
    For heavy-tailed distributions, the random walk MwG and its adaptive variants do not  generate geometrically ergodic chains without special treatment of the target distribution, like transformations \citep{JT03}. 
    For sampling multimodal distributions, the only way  to improve MCMC capability is to tune the proposal covariance.  
    \textcolor{black}{In general, chain convergence issues in of sampling non-Gaussian posteriors are mostly caused by heavy-tailedness of the conditional distributions. That is because  make most of the MCMC algorithms  generate only ergodic but not geometrically ergodic chain for  heavy-tailed posterior distributions \citep{JT03, RR04}. The issue shows up as slowly mixing chains which cannot be really improved by tuning the parameters of the MCMC algorithm altering its transition kernel. On the other hand, the multimodality or spikiness of the  target distributions that is not heavy-tailed may appear as chains that mix well only part of the time -- when the algorithm proposes and accepts a move that escapes an old mode or a peak, the chain can get stuck at the new mode although the chain seems to mix well around the new mode. Such a phenomenon can be seen in sampling mixtures of Gaussian distributions. }
    We note that, in adaptive Metropolis algorithms, the optimal scaling of the proposal covariance  is admissible for Gaussian-like target distributions only. 
    While non-Gaussian proposal distributions have been studied \citep{Y13}, at least the  applicability of heavy-tailed proposal distributions is limited, as no proper adaptation scheme exists for such distributions  \citep{SFR10}.
    In the following, we consider adaptive random walk MwG as our reference method when comparing the other two MCMC algorithms. 
    \begin{remark}\textcolor{black}{
    An important, and perhaps obvious question from the discussion above, is why not simply
    implement adaptive Metropolis-Hastings. We initially tested this out, for which the performance was not adequate, related to the mixing of chains, and the choice of step-size. Therefore, the implementation of the Gibbs sampling was required for us.}
    \end{remark}


    \subsubsection{{Repelling-Attracting Metropolis (RAM)}}
    The RAM algorithm \citep{TMD18} uses Gaussian random walk to generate new proposals, and hence it resembles the random walk Metropolis-Hastings algorithm. 
    The  advantage of RAM compared to the Metropolis-Hastings algorithm is the multimodality awareness of RAM, as RAM is specifically tailored for sampling multimodal distributions. 
    With more than a few dozen variables, we deploy the RAM property enabling using it the within-Gibbs sampling.
    %
    %
    The performance of RAM in sampling of heavy-tailed distributions has not been studied, but we make the assumption that it generates ergodic chains instead of geometrically ergodic ones  when using Gaussian proposal kernels, like the random walk Metropolis-Hastings does. 
     %
    %
    %
    %
    For generating RAM proposal moves, we first force a move away from the current position.
    This implies that the proposal prefers moves towards regions of lower probability densities. 
    Thus, the transition kernel to generate a mode-repelling move to $\mathbf{u}^*$ using the proposal kernel $\text{h}(\cdot)$ and the previous sample $\mathbf{u}^p$ is
    \begin{equation}
    \label{eq:s1}
    \text{k}_\text{R}(\mathbf{u}^*|\mathbf{u}^\text{p}) = \frac{\text{h}(\mathbf{u}^*|\mathbf{u}^\text{p}) \alpha_\text{R}(\mathbf{u}^*|\mathbf{u}^\text{p}) \text{d}\mathbf{u}^*}{Z_\text{R}(\mathbf{u}^\text{p})},
    \end{equation}
    with acceptance probability
    \begin{equation*}
    \alpha_\text{R}(\mathbf{u}^*|\mathbf{u}^\text{p}) = \min\left(1,\frac{\pi(\mathbf{u}^\text{p})}{\pi(\mathbf{u}^*)}\right),
    \end{equation*}
    and normalization constant
    \begin{equation*}
    Z_{\text{R}}(\mathbf{u}^\text{p}) = \int \text{h}(\mathbf{u}|\mathbf{u}^\text{p}) \alpha_\text{R}(\mathbf{u}|\mathbf{u}^\text{p}) \text{d}\mathbf{u}.
    \end{equation*} 
    Since this transition kernel does not include a Dirac measure term as an option to reject the proposed value, the proposals are generated repeatedly until a mode-repelling move is accepted \citep{TMD18}. 
    What follows is the mode-attracting move, which uses the intermediate sample from the previous repelling stage \eqref{eq:s1}, and  has the transition kernel 
    \begin{equation*}
    \text{k}_\text{A}(\mathbf{u}^{**}|\mathbf{u}^*) = \frac{\text{h}(\mathbf{u}^{**}|\mathbf{u}^*) \alpha_\text{R}(\mathbf{u}^{**}|\mathbf{u}^*)\text{d}\mathbf{u}^{**}}{Z_\text{A}(\mathbf{u}^*)},
    \end{equation*}
    where the acceptance probability has an inverted logic compared to that of the repelling stage
    \begin{equation*}
    \alpha_\text{A}(\mathbf{u}^{**}|\mathbf{u}^*) = \min\left(1,\frac{\pi(\mathbf{u}^{**})}{\pi(\mathbf{u}^*)}\right).
    \end{equation*}
    The overall transition kernel is of form
    \begin{equation*}
    \text{k}_\text{RA}(\mathbf{u}^{**}|\mathbf{u}^\text{p}) = \int  \text{k}_\text{R}(\text{d}\mathbf{u}^{**}|\mathbf{u}) \text{k}_\text{A}(\mathbf{u}|\mathbf{u}^\text{p}) \text{d}\mathbf{u},
    \end{equation*}
    The algorithm assumes that the proposal kernel $\text{h}(\cdot)$ is symmetric and used in both stages. It can be proven \citep{TMD18} that  after a proposal $\mathbf{u}^{**}$ is obtained, it is accepted for insertion in the chain with a probability 
    \begin{equation*}
    \alpha_\text{RA}(\mathbf{u}^{**}|\mathbf{u}^\text{p}) = \min\left(1,\frac{\pi(\mathbf{u}^{**}) Z_\text{R}(\mathbf{u}^\text{p})}{\pi(\mathbf{u}^\text{p})Z_\text{R}(\mathbf{u}^{**})}\right).
    \end{equation*}
    Because of the normalization constants, the proposal acceptance probability of the generated sample is intractable.
    However, we can use an auxiliary variable approach \citep{M06} to estimate the probability in an unbiased manner. 
    This is done by sampling in an augmented space for which, if $\mathbf{w}$ is marginalized out from the augmented joint distribution $\hat{\pi}(\mathbf{u},\mathbf{w})$, then the marginal distribution should equal to  $\pi(\mathbf{\mathbf{u}})$. 
    For simplicity, the joint distribution can be selected as  $\hat{\pi}(\mathbf{u},\mathbf{w}) = \text{h}(\mathbf{w}|\mathbf{u}) \pi(\mathbf{\mathbf{u}})$. 
    More specifically, $\mathbf{w}^*$ is generated using $\text{k}_\text{R}(\mathbf{w}^*|\mathbf{u}^{**})$.
    Then the overall sample $(\mathbf{u}^{**},\mathbf{w}^*)$, in the augmented space, is accepted with probability
    \begin{equation*}
    \begin{split}
    \alpha_\text{final}(\mathbf{u}^{**},\mathbf{w}^{*}|\mathbf{u}^\text{p},\mathbf{w}^\text{p}) &= \min\left(1,\frac{\text{h}(\mathbf{w}^*|\mathbf{u}^{**})\pi(\mathbf{u}^{**})Z_\text{R}(\mathbf{u}^\text{p}) \text{k}_\text{R}(\mathbf{u}^\text{p}|\mathbf{w}^\text{p})  }{\text{h}(\mathbf{w}^\text{p}|\mathbf{u}^\text{p})\pi(\mathbf{u}_\text{p}) Z_\text{R}(\mathbf{u}^{**}) \text{k}_\text{R}(\mathbf{u}^{**}|\mathbf{w}^{*}) }\right)\\
    &  = \min\left(1,\frac{\pi(\mathbf{u}^{**}) \min\left({1,\frac{\pi(\mathbf{u}^\text{p})}{\pi(\mathbf{w}^\text{p})}} \right)}{\pi(\mathbf{u}^\text{p})  \min\left({1,\frac{\pi(\mathbf{u}^{**})}{\pi(\mathbf{w}^{*})}} \right) }\right),
    \end{split}    
    \end{equation*}  
    after which either $\mathbf{u}^{**}$ or $\mathbf{u}^\text{p}$ is inserted in the chain.
    For technical details, see \Cref{ram}. 
    Similar to the adaptive MwG, we use a proposal covariance adaptation in RAM.

    \subsubsection{HMC-NUTS}
    
    As a third alternative, we use the implementation by  \citet{GXG18} of the No-U-Turn sampler (NUTS) \citep{HG14}. 
    NUTS is an MCMC that is based on the Hamiltonian Monte Carlo (HMC). 
    HMC adapts a mechanistic interpretation of a $d$-dimensional target distribution by construing its log-density  as a potential energy function. 
    HMC also requires introducing an additional momentum vector $\mathbf{p}\in \mathbf{R}^d$ and a kinetic energy function $K(\mathbf{u,p})$ for the system. The total energy function $H(\mathbf{u},\mathbf{p})$ of the virtual system is called as Hamiltonian, which is merely a sum of the potential and kinetic energies in the HMC. 
    
    The principle of the HMC is to integrate the trajectories defined by the Hamiltonian in virtual time, and proposing a new sample from the trajectory to be inserted in the chain.  For a detailed description of the HMC, and numerical integration of the trajectories of the system,  please see \citet{N12} and \citet{BC17}.  
    
    Although HMC generates an ergodic chain \citep{DMS20}, it does not generate a geometrically ergodic chain when the target distribution is heavy-tailed \citep{LV17}.   When dealing with heavy-tailed distributions, the chain mixing can be improved  with a careful selection of the kinetic energy, as discussed by \citet{LV17}. Still, employing other than Gaussian kinetic energies is rare due to the lack of adaptation schemes, and that is why Gaussian kinetic energy functions are nearly always used. \citet{LV17} also suggest that  utilizing NUTS instead of a basic HMC can be beneficial for sampling of heavy-tailed distributions. Dealing with multimodality is, however, where NUTS and HMC are likely to struggle more than random walk MCMC methods. This issue has been  discussed in great details by \citet{MPS18}. There is some existing evidence that NUTS may indeed underperform with   field posteriors having Cauchy random field priors  \citep{SELS20}.   Hence, we seek to compare the NUTS algorithm to the  two MCMC methods presented earlier.

    Like the basic HMC algorithm, the NUTS uses  gradient information of the target distribution induced by the Hamiltonian function, and  should hence be able to generate chains with adequate  mixing, even if the target distribution is nonlinearly correlated.  However unlike  HMC, where a fixed  trajectory length is used for integrating the Hamiltonian equations, and a proposal sample is selected  from that trajectory using a rule that keeps the chain ergodic \citep{BC17}, NUTS  does not use such static trajectory generation procedure.  Instead, NUTS keeps doubling the trajectory  randomly in the forwards (+) and backwards (-) directions in virtual time  until a stopping condition of a U-turn is fulfilled. The generalized stopping rule can be  expressed as \citet{BC17}
    \begin{equation}
    \label{eq:stop}
    \left( \mathbf{M}^{-1}\mathbf{p}^-\cdot \left(\sum_{\mathbf{p}_j \in \text{trajectory}} \mathbf{p}_j \right) <0 \right) \lor \left( \mathbf{M}^{-1}\mathbf{p}^+ \cdot \left(\sum_{\mathbf{p}_j \in \text{trajectory}} \mathbf{p}_j \right) <0 \right),
    \end{equation}
    where $\mathbf{M}$ is a constant metric tensor for the system, and $\mathbf{p}^-$ and $\mathbf{p}^+$ denote  the momentum vectors at the left and right end of the trajectory, respectively.
    The generalized stopping rule is slightly modified from the original stopping rule of \citet{HG14}, which uses a criterion of 
    \begin{equation}
    \label{eq:stop2}
    \left(\left( \mathbf{u}^+ -  \mathbf{u}^-\right) \cdot \mathbf{p}^-  <0 \right) \lor \left(\left( \mathbf{u}^+ -  \mathbf{u}^- \right) \cdot \mathbf{p}^+ <0 \right)
    \end{equation}
    as a stopping rule. Both forms check  whether a trajectory of the system has made a U-turn and is about to turn back to the initial point. The generalized rule is tailored for general kinetic distributions whose marginal distribution is not necessarily a non-correlated Gaussian distribution.  
    The doubling of the trajectory in both directions is needed to keep the algorithm reversible for ergodicity \citep{HG14}.   
    
    In addition to the generalized stopping rule, AdvancedHMC employs multinomial sampling to the generated trajectory for drawing samples from the generated Hamiltonian trajectories. That is, a sample with index $i$ is selected from a trajectory $[\mathbf{u}_{-m}, \mathbf{u}_{-m+1} \cdots \mathbf{u}_{n-1},\mathbf{u}_{n}]$ with convention $\mathbf{u}^- = \mathbf{u}_{-m} $ and $ \mathbf{u}^+ = \mathbf{u}_{n}$ with probability
    \begin{equation*}
    w_i = \frac{\exp\left( -H({\mathbf{u}_i},  {\mathbf{p}_i} )\right)}{\sum_i\exp\left( -H({\mathbf{u}_i},  {\mathbf{p}_i} )\right)}.
    \end{equation*}    
    The multinomial sampling strategy differs from the original NUTS \citep{HG14}, which uses  slice sampling to draw a  sample from the overall trajectory by setting the auxiliary slice variable to the probability density of the initial position $\mathbf{u}_0$ of the trajectory. The NUTS combined with the generalized stopping rule, multinomial sampling, and adaptation of the metric tensor and step size render the resulting NUTS efficient for sampling high-dimensional probability distributions \citep{BC17}. 
    
    A pseudocode for the multinomial sampling based NUTS with Gaussian kinetic energy function is given in \Cref{nuts}.  AdvancedHMC and Stan software employ a maximum tree doubling parameter in their NUTS implementations. That is, doubling of the trajectory is stopped and a sample is proposed from the trajectory if a specified number of doublings is exceeded, what practically prevents using too small step size for integration of the trajectories of the Hamiltonian function.  
    

    \section{Numerical experiments}
    For this section  we demonstrate the capabilities of the Cauchy Markov random field priors in one-dimensional and two-dimensional  deconvolution problems. 
    In the one-dimensional deconvolution, we calculate MAP estimators, and kernel density estimators of the marginals of the posteriors.  In the two-dimensional deconvolution, we calculate MAP, conditional mean  and marginal variance estimators.  \textcolor{black}{To aid our computations, we use present MCMC methods for the inversion in \Cref{ssec:1d} - \ref{ssec:2d}}.
    We use an implementation of \citep{GXG18} for NUTS, and we programmed the other two MCMC algorithms by ourselves. Sample codes for the experiments can be found from a \href{https://github.com/suurj/markovcauchy}{ GibHub repository}. 
    \label{sec:num}
    
    \subsection{Potential Scale Reduction Factor}
    
    We need metrics to assess the sampling efficiency and sample quality of  MCMC chains. 
    The so-called  potential scale reduction factor (PSRF)  \citep{BG96,GR92}  is  one of the most widely used. Unlike other metrics, PSRF requires the running of multiple independently seeded MCMC chains. After removing the burn-in period from the chains, the diagnostic is calculated from intra-chain and inter-chain variances. 
    
    Let us denote the number of samples in each chain  as $N_s$, the number of chains as $N_c$,  $u_i^j$ the $j^{th}$ sample of $i^{th}$ chain,  \textcolor{black}{$\bar{u}_j$ the mean of the $j^{th}$ chain, and $\bar{u}$ the mean of means of all the chains}. Then, if the intra-chain variance is denoted by
    \begin{equation*}
    V = \frac{1}{N_c} \sum_{i=1}^{N_c} \left(\frac{1}{N_s -1}  \sum_{j=1}^{N_s} ( u_i^j - \bar{u_i} )^2\right)^2, \end{equation*}
    and the inter-chain variance by
    \begin{equation*}
    K = \frac{N_s}{N_c-1} \sum_{i=1}^{N_c} \left( \bar{u_i} - \bar{u} \right)^2,
    \end{equation*}     
    and the PSRF is defined as
    \begin{equation}
    \label{eq:rhat}
    \hat{R} = \left(\frac{ \frac{N_s -1}{N_s} V + \frac{1}{N_s}K} {V}\right)^{1/2}.
    \end{equation}
    
    PSRF allows improper chain mixing and convergence to be detected in a more robust manner than the metrics that evaluate the chain quality using one chain only. For instance, the so-called Geweke's diagnostic compares the start and end parts of a chain to each other  \citep{GW92} and the autocorrelation function is used to calculate effective sample size  (ESS). While there is no perfect MCMC convergence metric,  diagnostics that try to estimate the degree convergence out of one chain can be  sensitive to burn-in period and hence adaptation of the MCMC.     As a rule of thumb, PSRF considers the  MCMC chain as having converged, if  the value of the diagnostics \eqref{eq:rhat} is  $\hat{R} < 1.2$ \citep{BG96}.  
    
    It is not, however, clear what is a good value for PSRF diagnostic if the target distribution is very high-dimensional and multimodal, like the posteriors we are considering. Furthermore, the diagnostic assumes that the target distribution is approximately Gaussian, which may be not the case.  \citet{BG96}  suggest similar diagnostics  based on comparing inter-chain and intra-chain interval length estimators instead of the variance estimators.  A common practice is also to initialize the chains from different points so that the inter-chain variance is not underestimated in the case of severe multimodality, for instance. 
    
    In our numerical experiments below, we start all the chains from the MAP estimate with different seeds and calculate the canonical PSRF diagnostics for convergence evaluation. This approach is used because we are dealing with high-dimensional distributions and want to minimize the burn-in period. \textcolor{black}{Moreover, our posteriors have a Gaussian likelihood function. Despite this the posteriors are still non-Gaussian, because we are using Cauchy priors, combined with Gaussian likelihoods}.  Finally, the objective for using PSRF is to detect sampling issues and assess their severity, rather than evaluate the exact diagnostic values and tell whether the estimators calculated from the chains are reliable.

    \subsection{One-dimensional deconvolution}
    \label{ssec:1d}
    
    As the first numerical experiment, we demonstrate properties of the priors in a one-dimensional deconvolution problem
    \begin{equation} \label{eqn:convolution}
    y = k(r) * u(r) + \eta, \quad \eta \sim \mathcal{N}(0,\sigma^2I ),
    \end{equation}
    where by $*$ we denote convolution operation with a convolution kernel being Gaussian-shaped
    \begin{equation}\label{kernel}
    k(r) = \frac{1}{\sqrt{\pi s}}\exp\left(-\frac{r^2}{s}\right),
    \end{equation} 
    and $r\in\mathbb{R}^d$ with $d=1,2$ in our cases. For simulations, we chose $s=1/500$,  $\sigma=0.01$.
    To avoid committing an inverse crime of reusing the computational grid used in the simulation in the inversion also, \textcolor{black}{ we simulated the convolutions of the test function 
    \begin{equation} \label{eqn:test}
    \begin{split}
    u(x) =& H(x-0.75)H(-x+0.9)+ \Lambda\big(10(x-0.15)\big) \\  &+ \Lambda\big(10(x-0.55)\big)H(x-0.55) + \exp\big(-70|x-0.4|\big),
    \end{split}
    \end{equation}
    using an adaptive Gauss-Kronrod quadrature integration. In \Cref{eqn:test},  $H(\cdot)$ denotes the Heaviside step function and $\Lambda(x) = \max(0, 1- |x|)$. We evaluated the convolutions at 67 equispaced points within the domain. }
    The reconstruction grid consists of $200$ equispaced nodes. Therefore the posterior then is
    \begin{equation*}
    \pi(\mathbf{u}|\mathbf{y}) \propto \exp \left( -\frac{1}{2\sigma^2} (\mathbf{y}-\mathbf{Fu})^T(\mathbf{y}-\mathbf{Fu}) \right)\pi(\mathbf{u}),
    \end{equation*}
    where $\mathbf{F}$ is a matrix that approximates the convolution in the reconstruction grid.
    
    For the first and second order Cauchy priors, we choose $\lambda =  \lambda' = 0.01$.
    The parameter $\ell$ of the SPDE matrix was set to $0.015^2$, and the scale of the Cauchy noise of the SPDE prior was set to $\lambda = 0.01$. The parameter $h$ of the SPDE prior discretization was set to ${1}/{199}$. 

    We ran adaptive MwG and RAM for 250\,000 adaptation iterations. Both of the algorithms generated proposals for one component of the posteriors at time, so the overall number of proposed samples was 50\,000\,000. After that, the proposal covariances were fixed and another 250\,000 total iterations of the algorithms were performed to obtain samples for the inference. Similarly, NUTS was let to generate 40\,000 samples, of which 20\,000 were used for only adaptation of the step size and the diagonal metric tensor  to achieve an average proposal acceptance rate of 0.8. The maximum tree doubling parameter was set to 12. All the adaptation and sampling procedures using the tree algorithms were repeated ten times for calculating the PSRF diagnostics. The chains of RAM and MwG were thinned by a factor of 10 in order to save memory. \textcolor{black}{Although the thinning improves the autocorrelation of the resulting chains, it is usually performed indeed as  measure to reduce the need for central memory and disk space, as information is lost in the process. However, thinning is still commonly employed, and the convergence diagnostics as PSRF can be applied upon the thinned chains \citep{YS19}. We compute and plot all the Monte Carlo estimators and chain convergence diagnostics, including the PSRF values, trace plots and autocorrelation estimates, on the basis of the thinned chains.}

    The test function, the noisy measurements and the MAP estimates are plotted in upper part of  \Cref{fig:meas}, where the characteristics of each prior can be seen. We observe that the first order Cauchy prior leads to MAP estimates that are  piecewise nearly constant, while the second order favors functions whose first derivative is piecewise constant. The second order prior is the best to reconstruct the triangle part of the test function. The SPDE prior  seems to reconstruct the exponential peak well around, but other parts of the test function are more difficult for the prior. The reconstructions amplify our hypothesis that the SPDE prior is suitable for exponential peaks, with the expense of reconstruction ability  of other kinds of  features. 
    
    One of the generated chains of all the MCMC method and prior combinations was used for kernel density estimation of the marginal distributions, while the other chains of were employed to assess the chain convergence. Using only one chain for inference was a conscious selection, since sampling  high-dimensional distributions is numerically very burdensome, and  in  many applications, it is not realistic to run several chains for  inference due to time or memory constraints.
    The kernel estimators of the marginal distributions are plotted in \Cref{oned} with different priors and methods. While the posteriors with the first and the second order Cauchy priors  have reasonable PSRF values over all the domain, the posterior with the SPDE prior has serious convergence issues at some of the marginal chains. Trace plots for two different nodes in \Cref{t1d} reveal the SPDE prior equipped posteriors are very difficult for all the MCMC methods when the expectation value of the node differs noticeably from zero. The MCMC chains seem to slowly drift around the mean like the chain generated by MwG, or to switch between the modes of the marginal with inadequate mixing like the NUTS does.  The kernel density estimators of the SPDE posteriors have merely visible width at all. NUTS seems to perform better than RAM and especially MwG, which struggle in the sampling since they sample the conditional distributions sequentially. If  a conservative metric like PSRF is used to assess the convergence, the estimators calculated from MCMC chains should be considered suspicious when treating Cauchy noisy SPDE priors. We claim that unless the uncertainty quantification of posteriors with the SPDE priors is not necessary, sticking with the MAP estimators should be preferred instead.

    \newcommand{\fhei}{5.9cm}
    \newcommand{\fspa}{0.3cm}

    \newcommand{\fheis}{4.4cm}
    \newcommand{\fspas}{0.4cm}
    
    \begin{figure}[ht!]
        \centering
        \begin{subfigure}[ht!]{\fheis}
            \includegraphics[height=\linewidth]{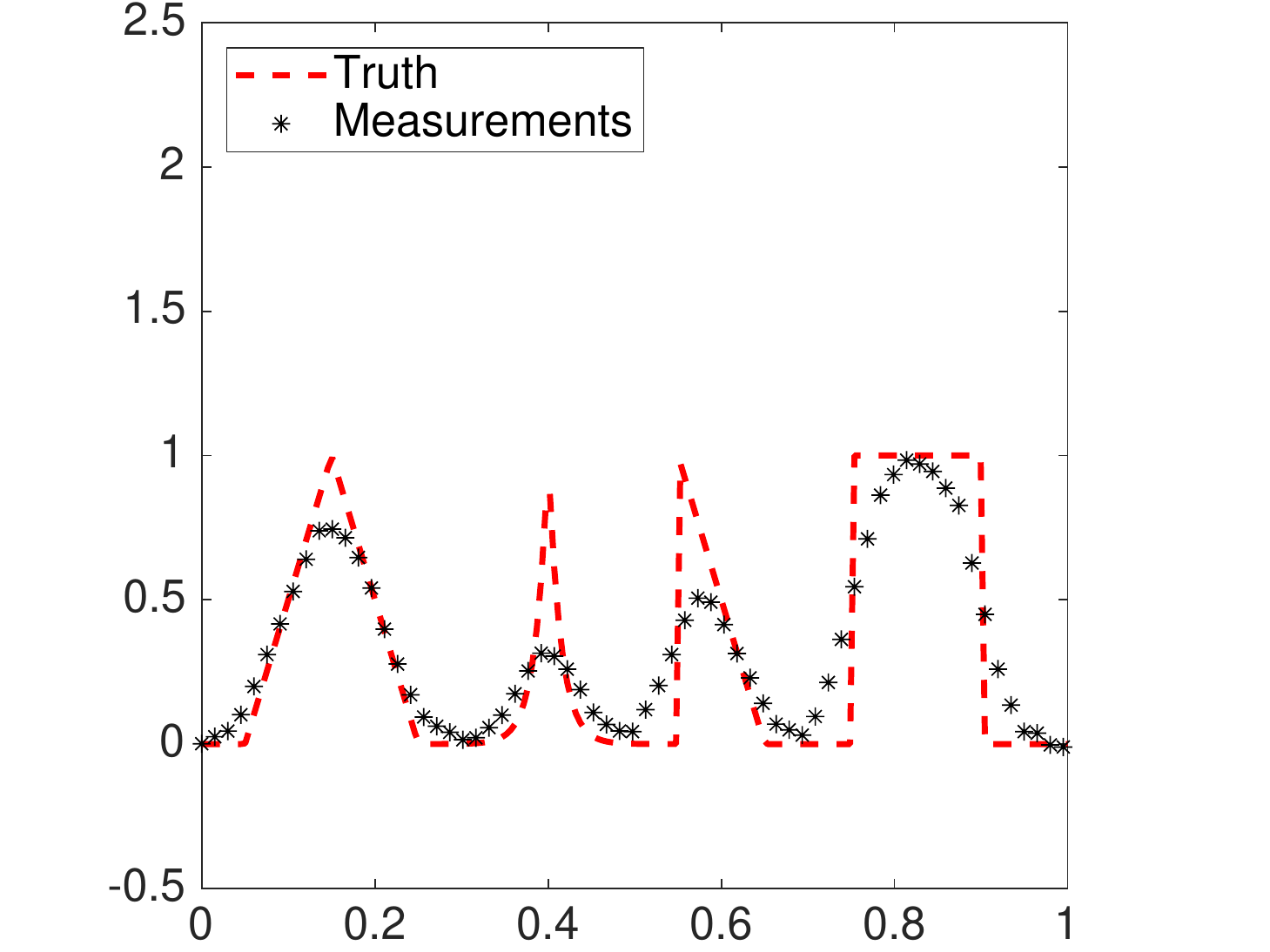}
            \caption{Measurements. }
            \label{fig:meas}\end{subfigure} 
        \begin{subfigure}[ht!]{\fheis}
            \includegraphics[height=\linewidth]{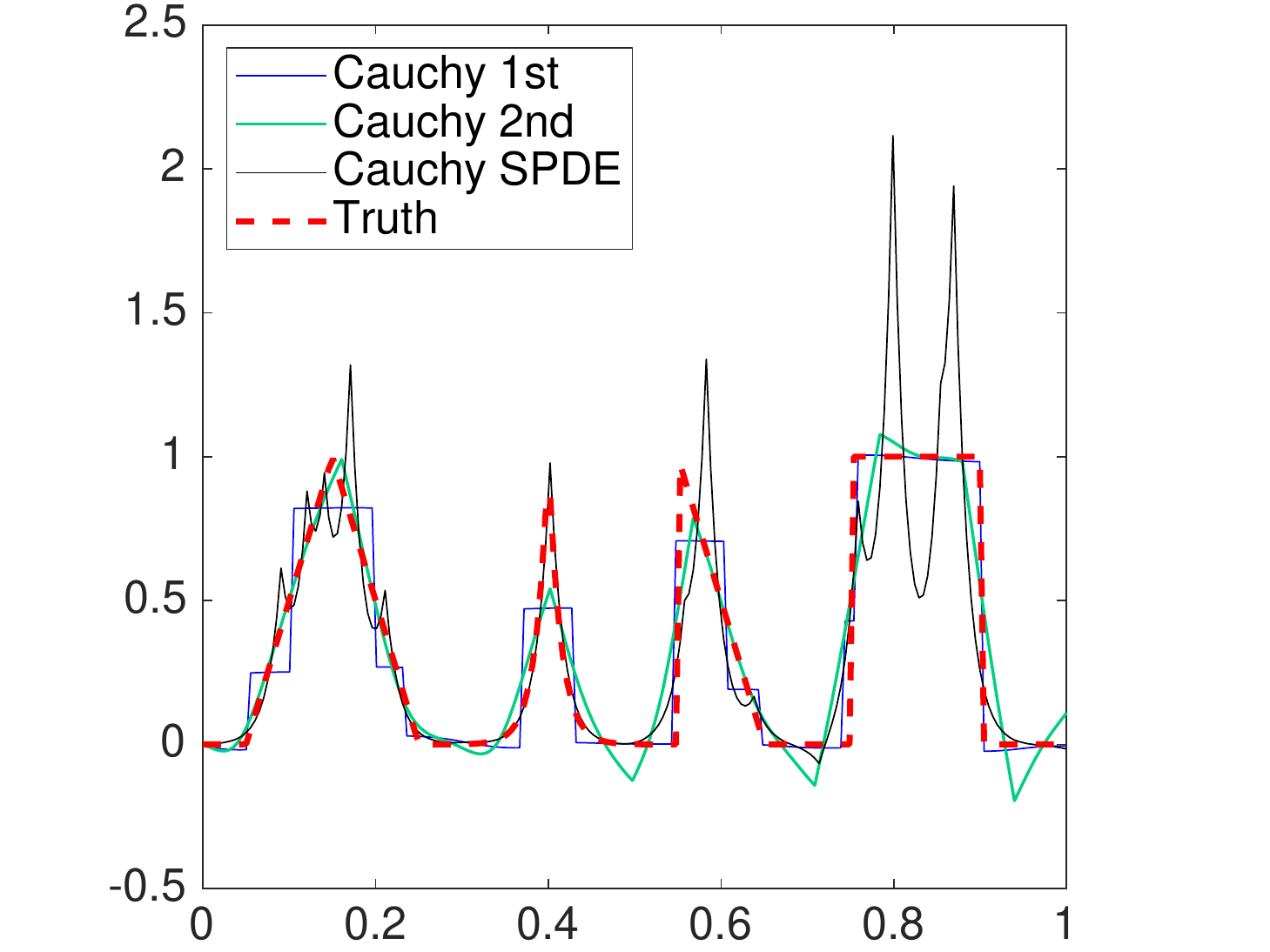}
            \caption{MAP estimates.}
            \label{fig:maps}\end{subfigure}\hspace{\fspas}

        \begin{subfigure}[ht!]{\fheis}
            \includegraphics[height=\linewidth]{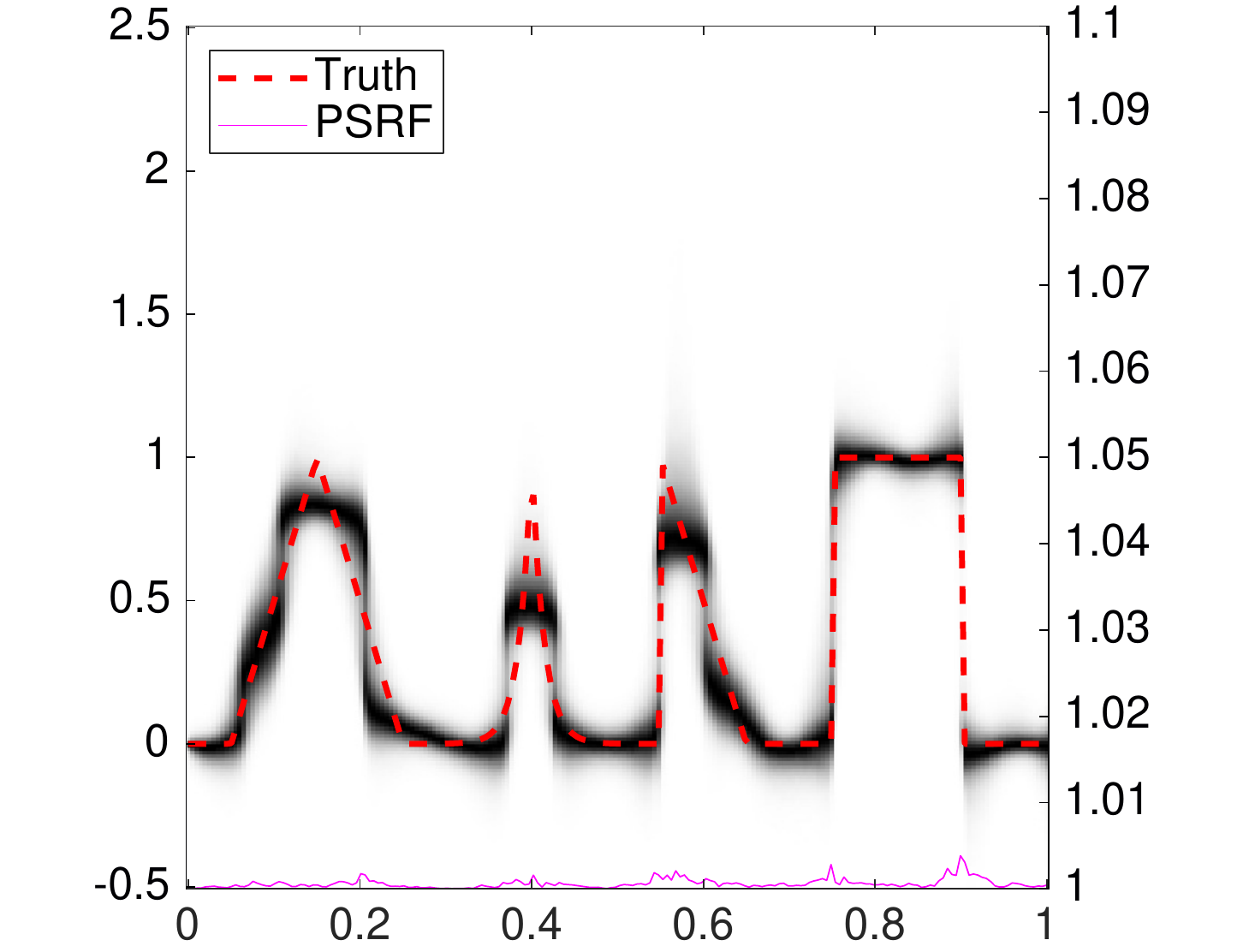}
            \caption{ NUTS }\label{nutsd}\end{subfigure}\hspace{\fspas}
        \begin{subfigure}[ht!]{\fheis}
            \includegraphics[height=\linewidth]{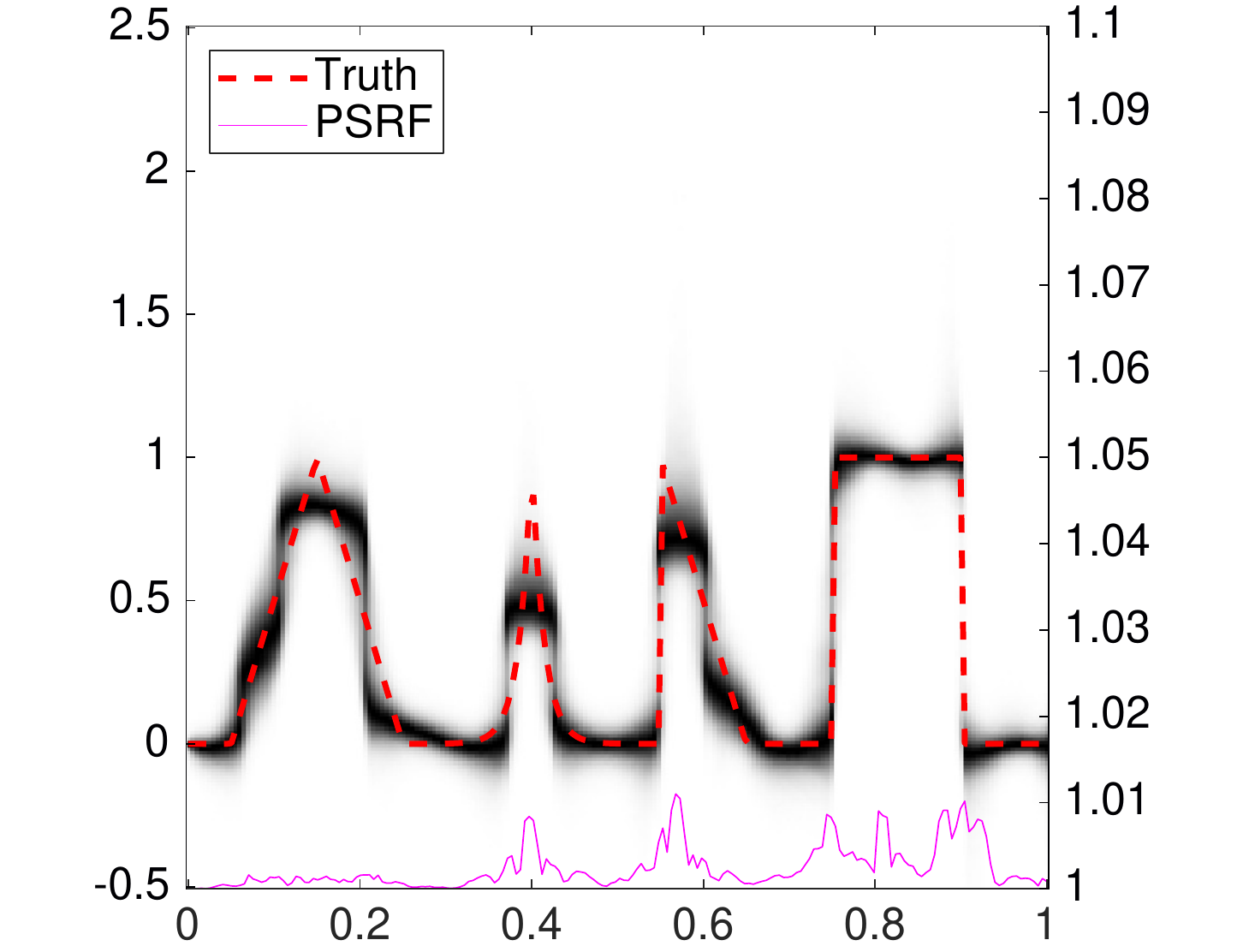}
            \caption{MwG  }\label{mwgd}\end{subfigure}\hspace{\fspas}
        \begin{subfigure}[ht!]{\fheis}
            \includegraphics[height=\linewidth]{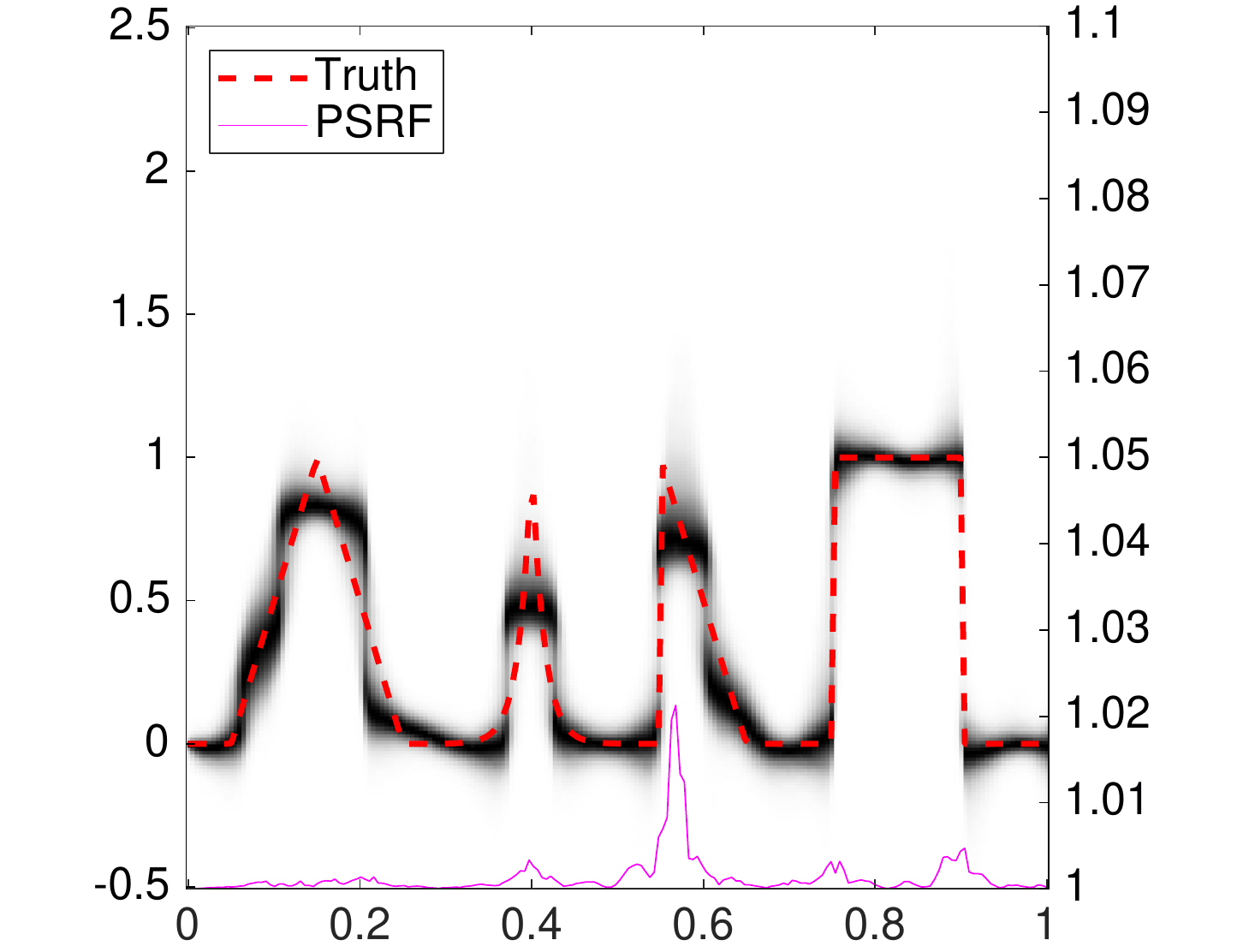}
            \caption{RAM  }\label{ramd}\end{subfigure}
        
        \begin{subfigure}[ht!]{\fheis}
            \includegraphics[height=\linewidth]{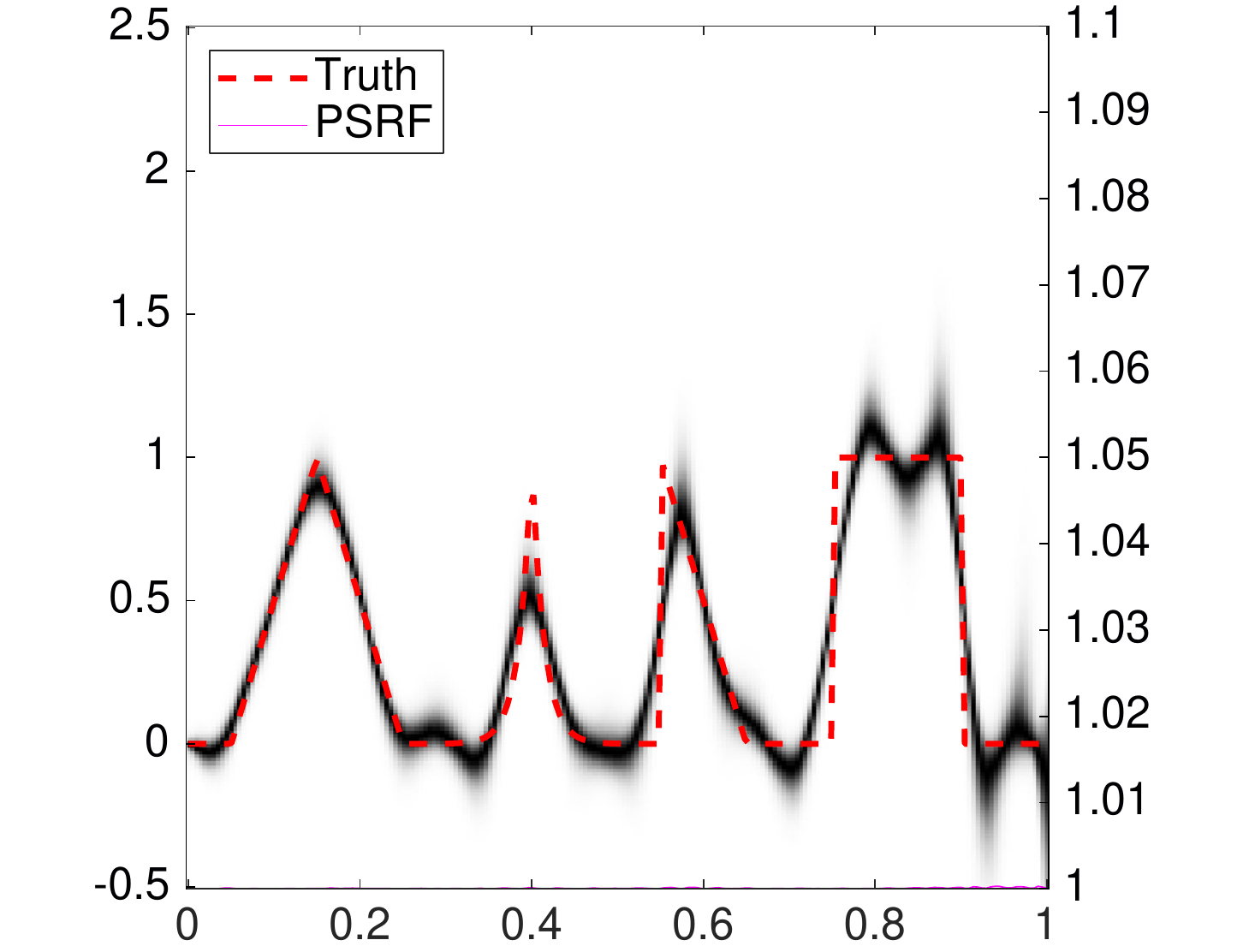}
            \caption{ NUTS }\label{nusd2}\end{subfigure}\hspace{\fspas}
        \begin{subfigure}[ht!]{\fheis}
            \includegraphics[height=\linewidth]{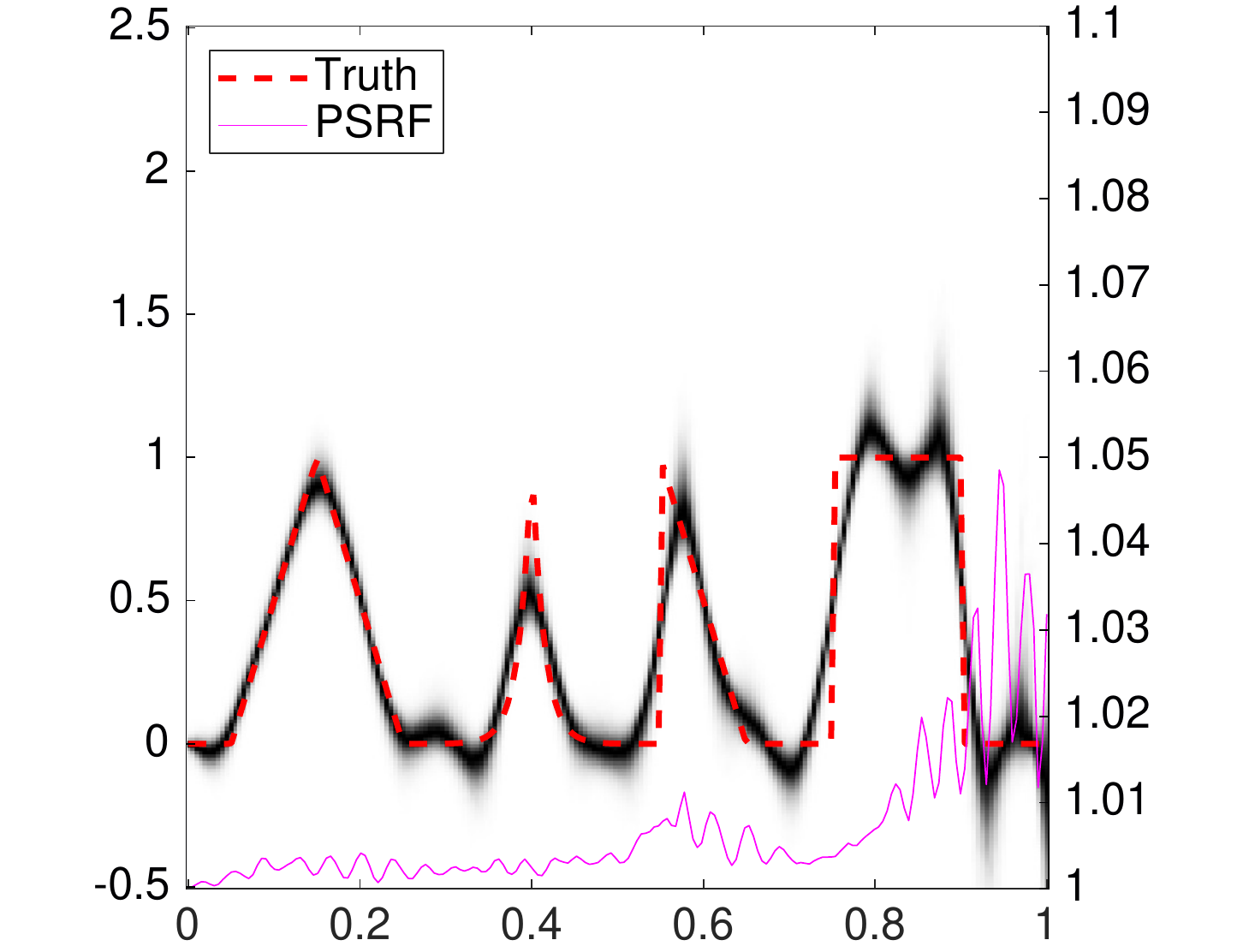}
            \caption{MwG  }\label{mwgd2}\end{subfigure}\hspace{\fspas}
        \begin{subfigure}[ht!]{\fheis}
            \includegraphics[height=\linewidth]{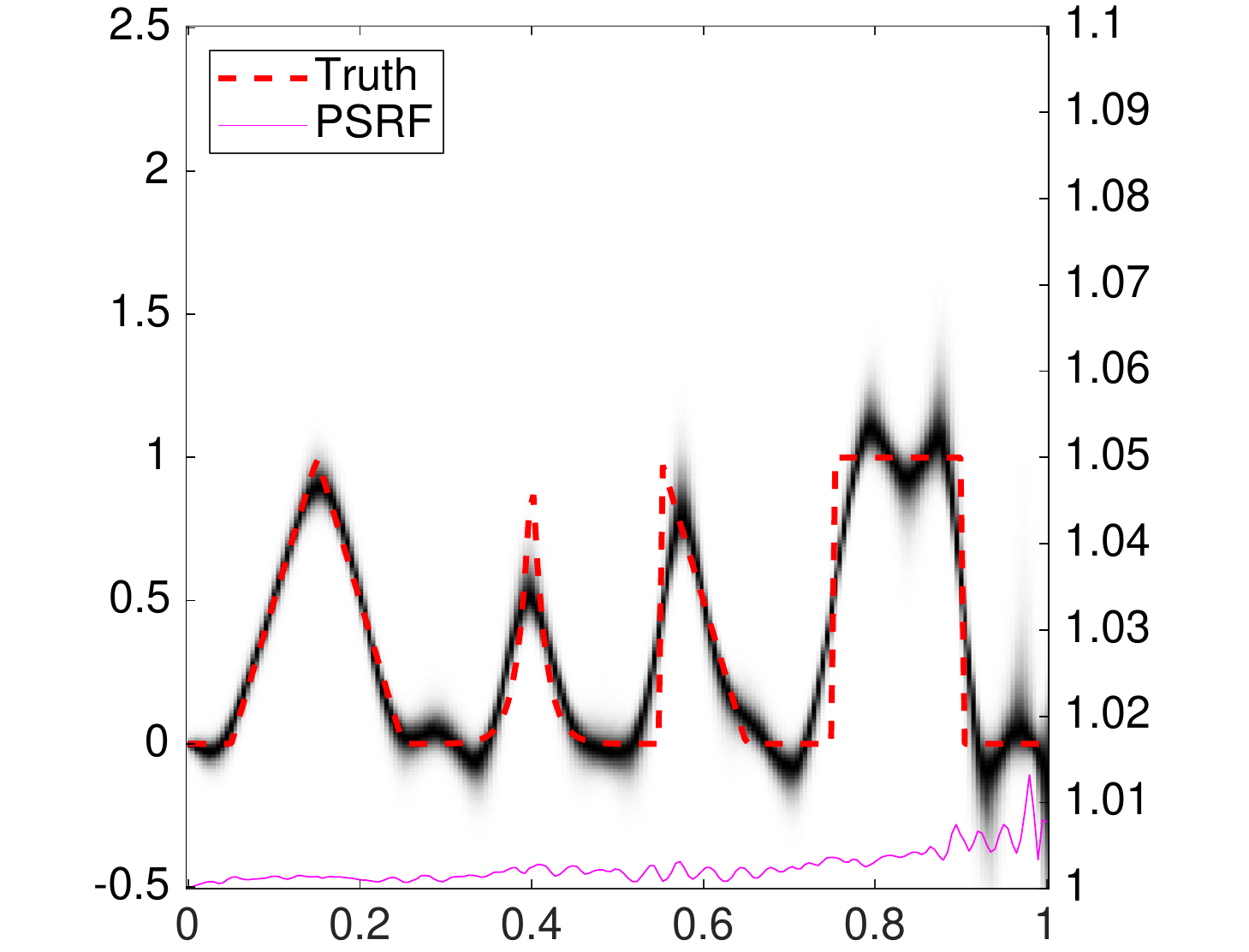}
            \caption{RAM  }\label{ramd2}\end{subfigure} 
        
        \begin{subfigure}[ht!]{\fheis}
            \includegraphics[height=\linewidth]{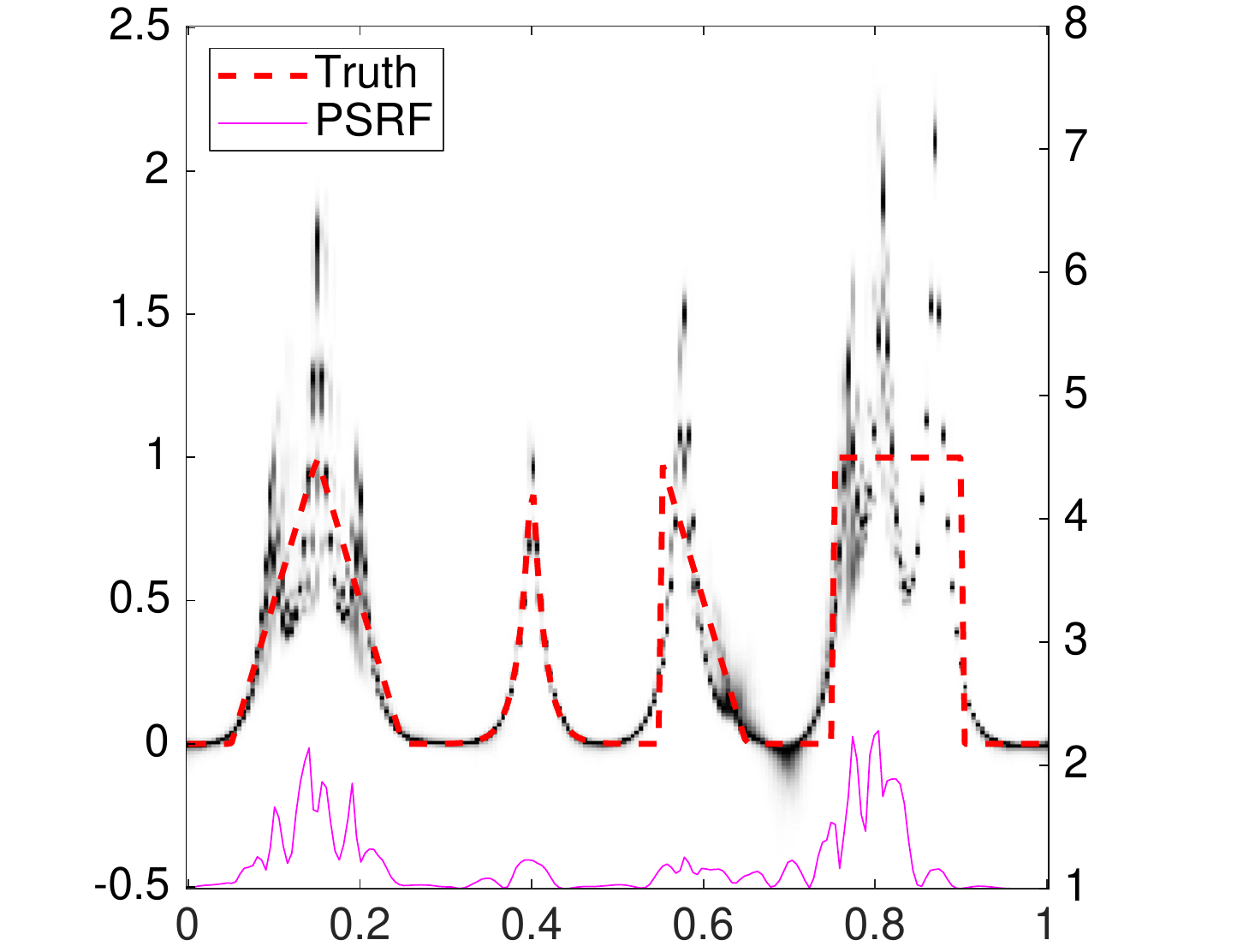}
            \caption{ NUTS }\label{nutsp}\end{subfigure}\hspace{\fspas}
        \begin{subfigure}[ht!]{\fheis}
            \includegraphics[height=\linewidth]{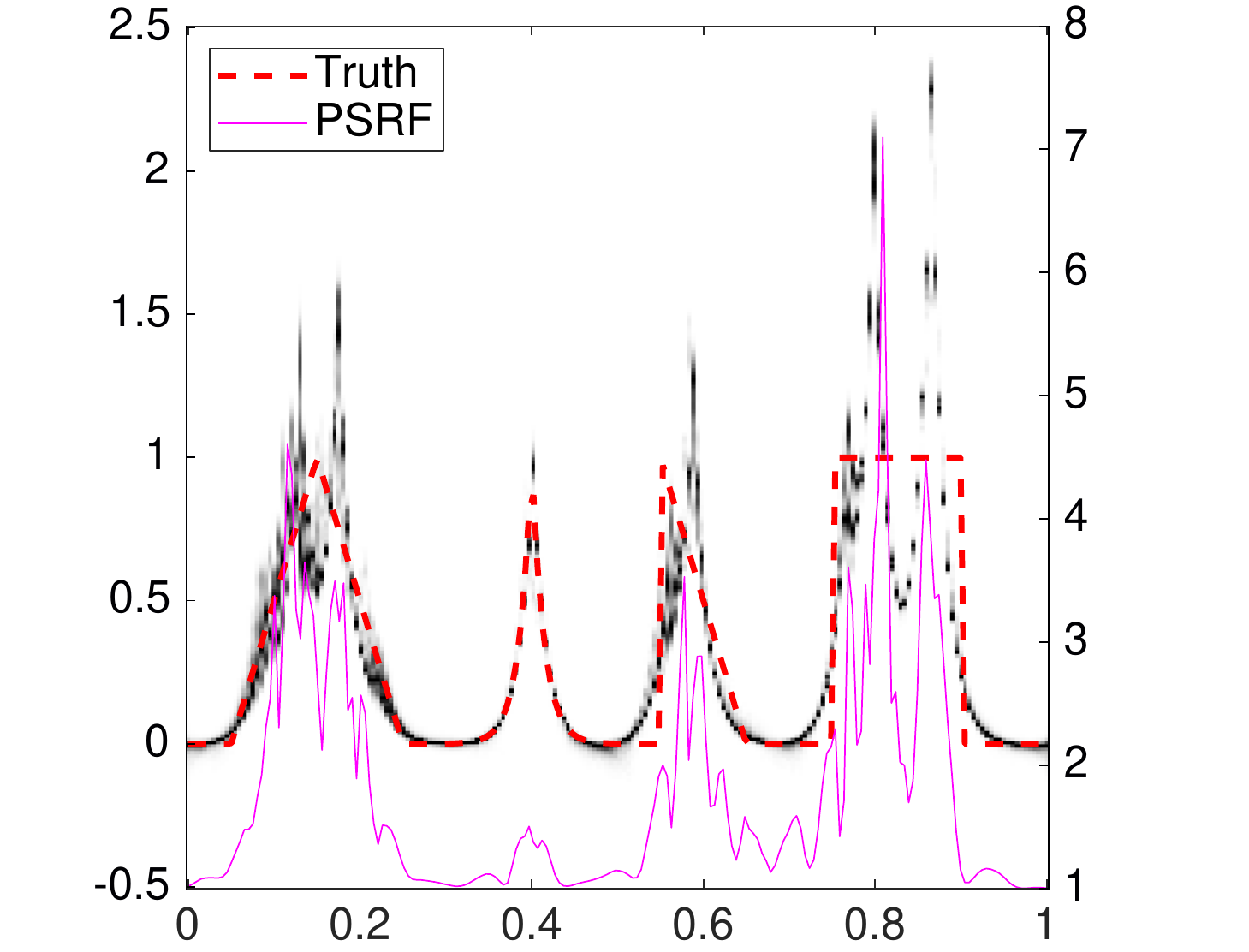}
            \caption{MwG  }\label{mwgp}\end{subfigure}\hspace{\fspas}
        \begin{subfigure}[ht!]{\fheis}
            \includegraphics[height=\linewidth]{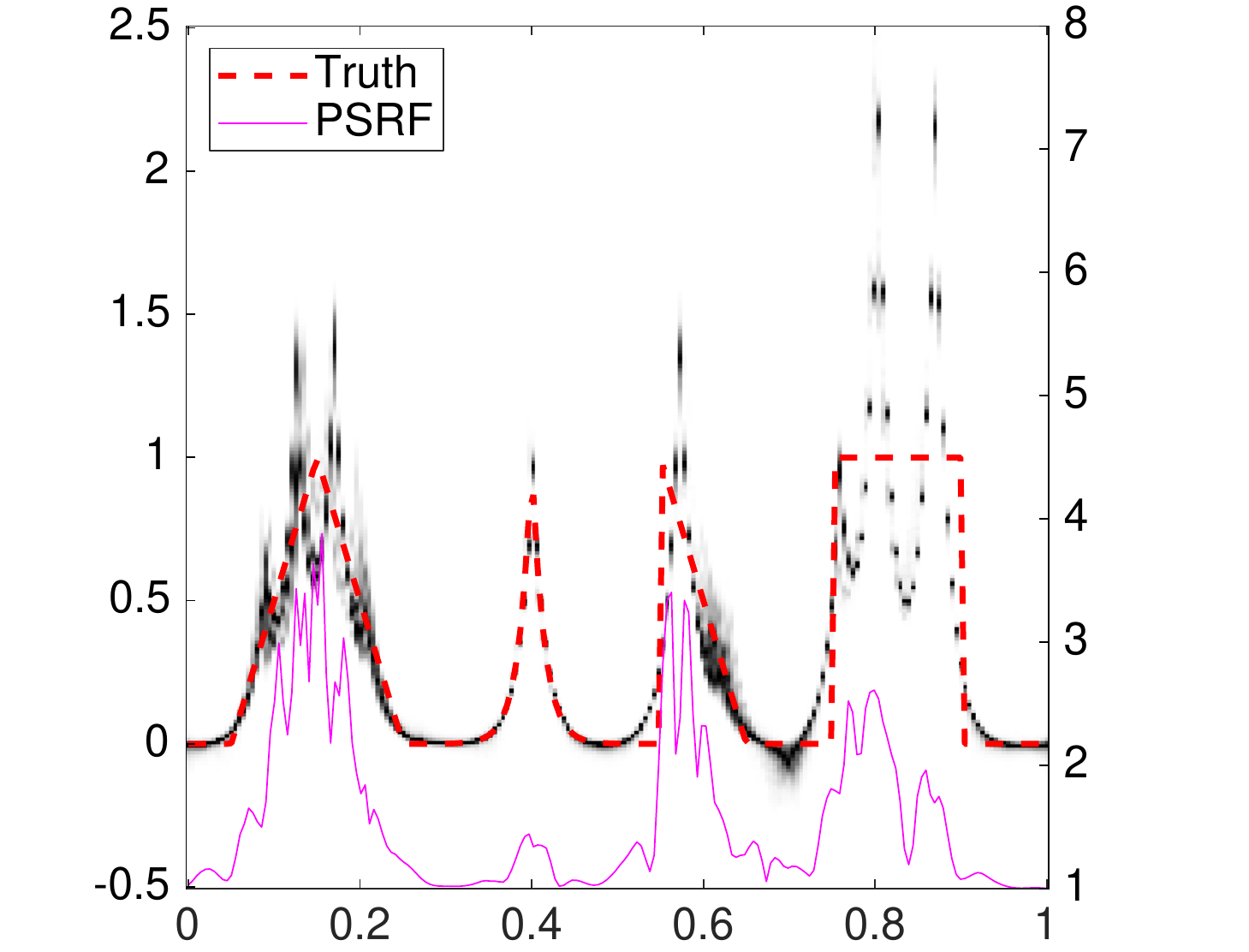}
            \caption{RAM  }\label{ramp}\end{subfigure}
        \caption{ Kernel density estimates and PSRF convergence diagnostics (right Y-axis) for the posteriors. \textcolor{black}{The 2nd row:  the first order Cauchy difference prior. The  3rd row:  the second order Cauchy difference prior. The 4th row: SPDE prior with Cauchy noise.} }
        \label{oned}
        
    \end{figure}

    \newcommand{\qhei}{2.8cm}
    \newcommand{\phei}{2.8cm}
    \newcommand{\qspa}{-0.3cm}
    \newcommand{\pspa}{-0.4cm}
    
    \newcommand{\qheis}{4cm}
    \newcommand{\qspas}{0.0cm}
    
    \begin{figure}
        \centering
        \begin{subfigure}[b]{\phei}
            \includegraphics[width=\linewidth]{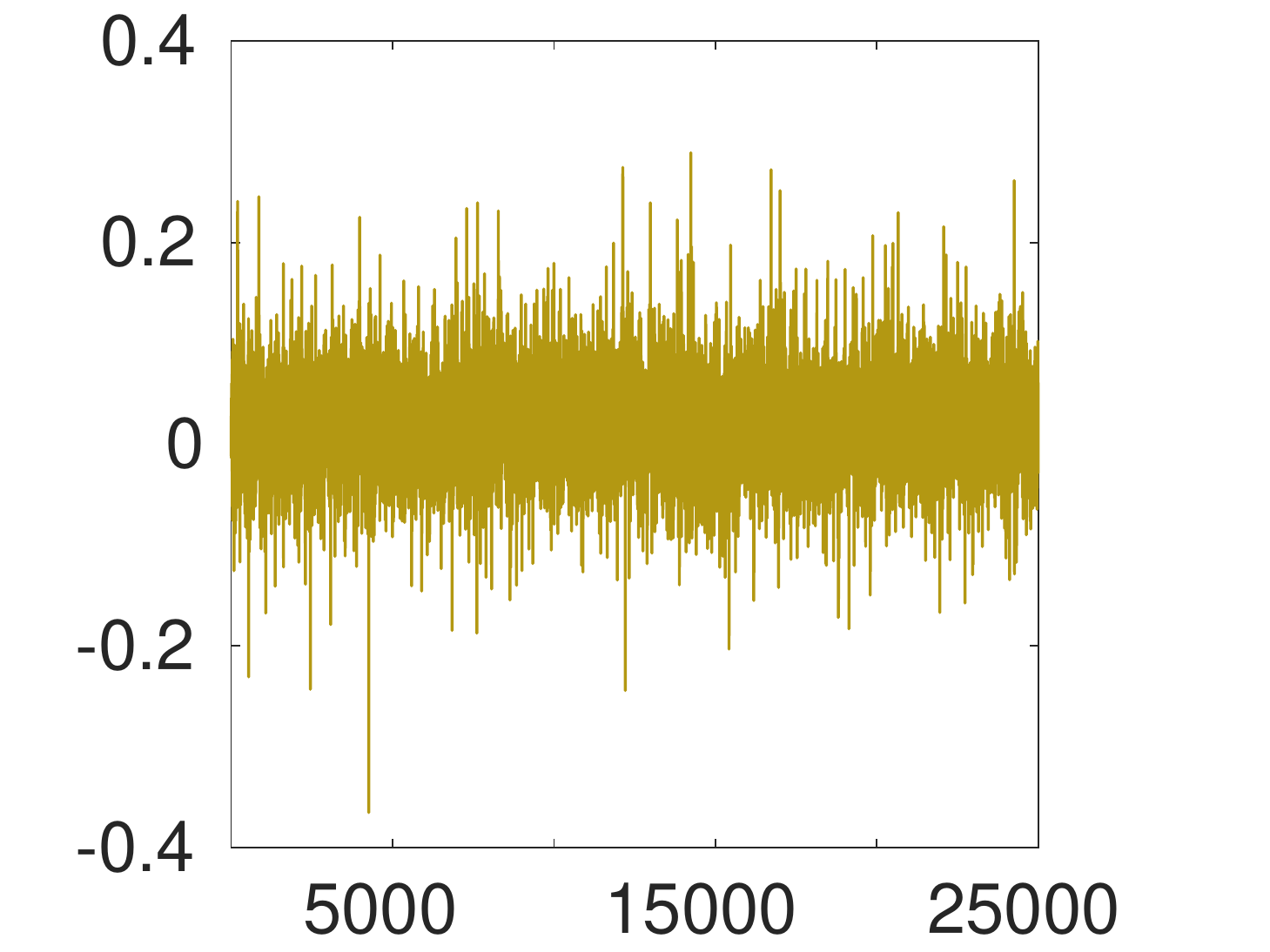}
            \includegraphics[width=\linewidth]{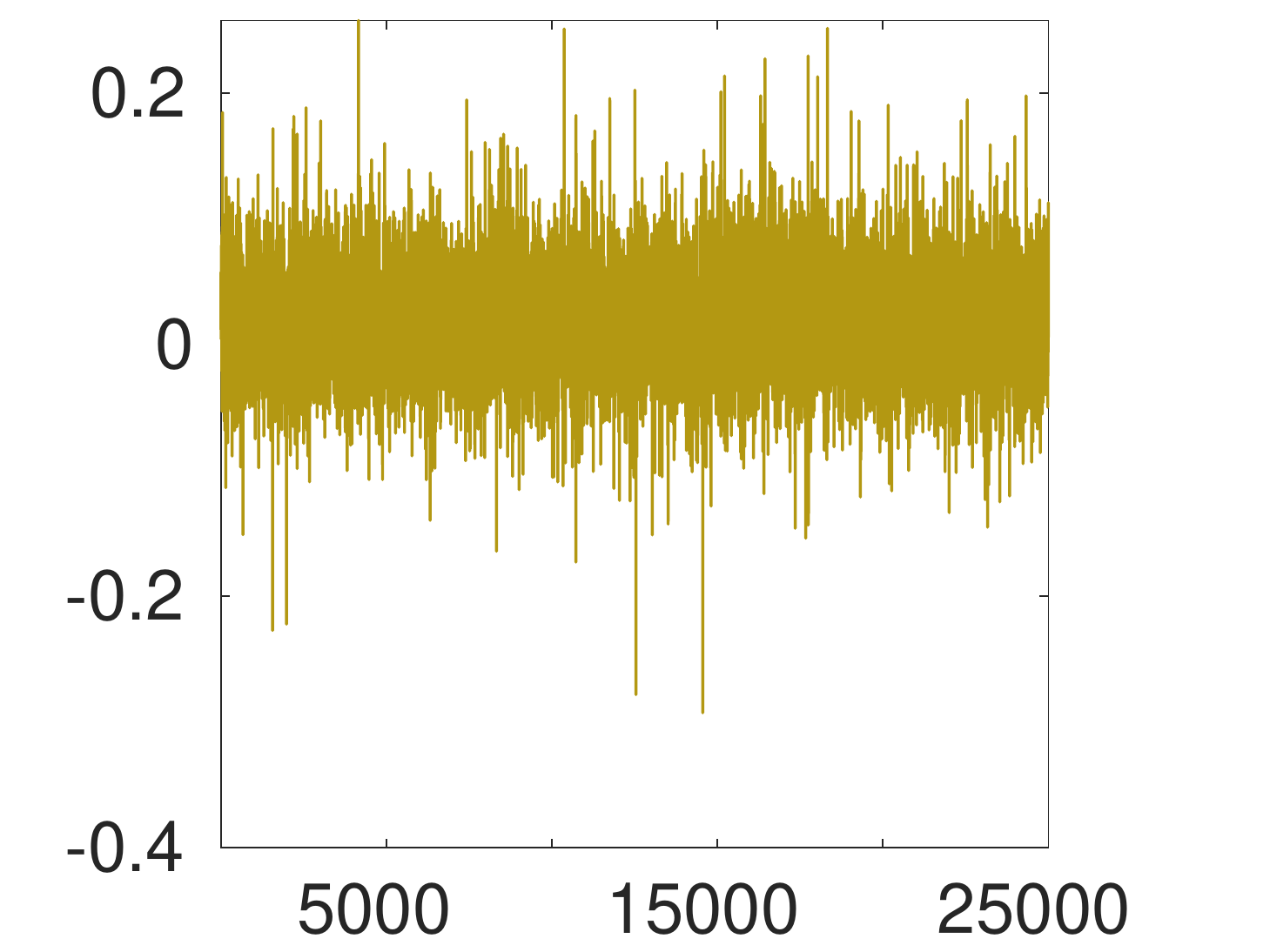}
            \includegraphics[width=\linewidth]{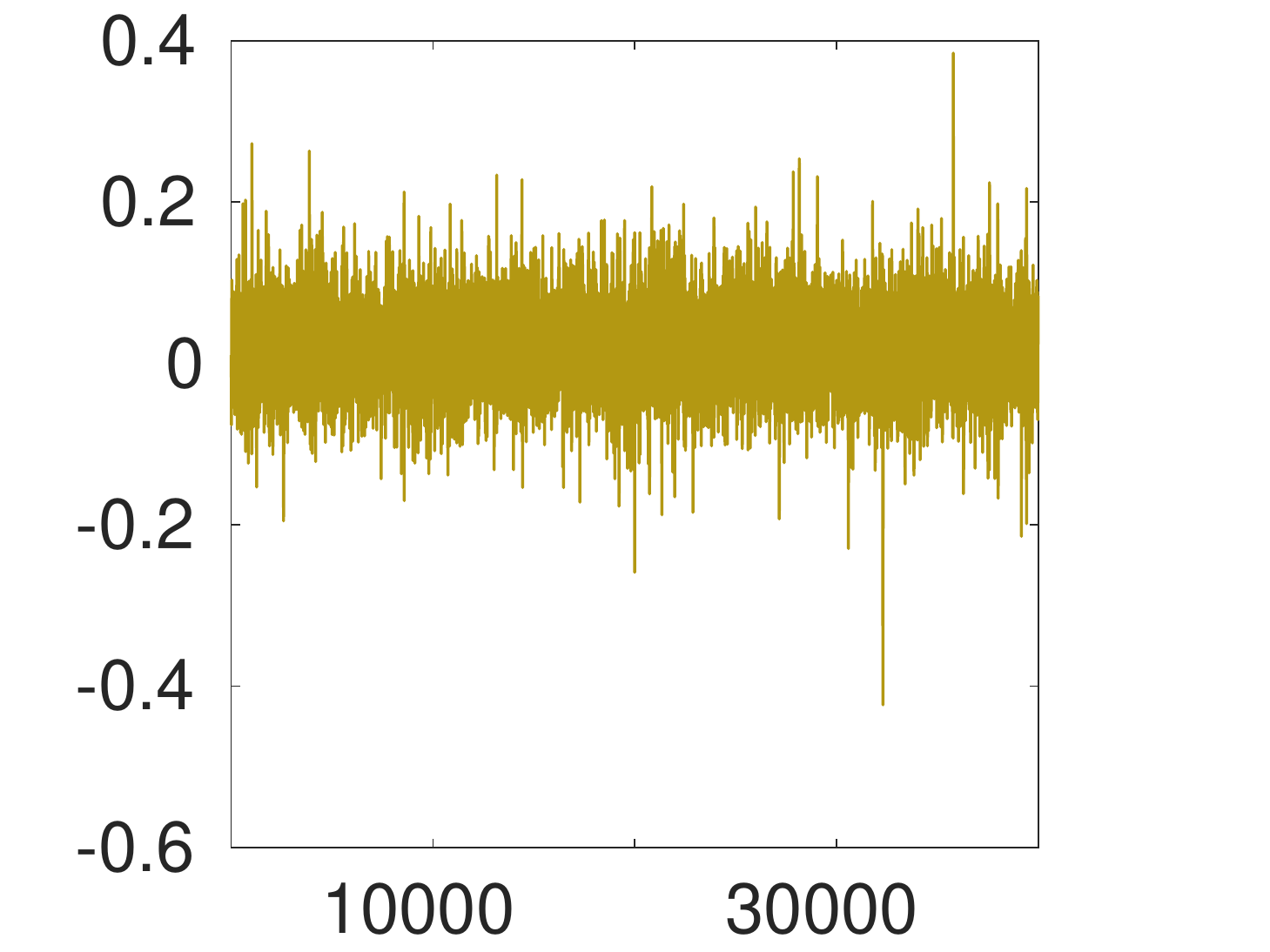}
            \caption{1st at 0.3 }\end{subfigure}\hspace{\pspa}
        \begin{subfigure}[b]{\phei}
            \includegraphics[width=\linewidth]{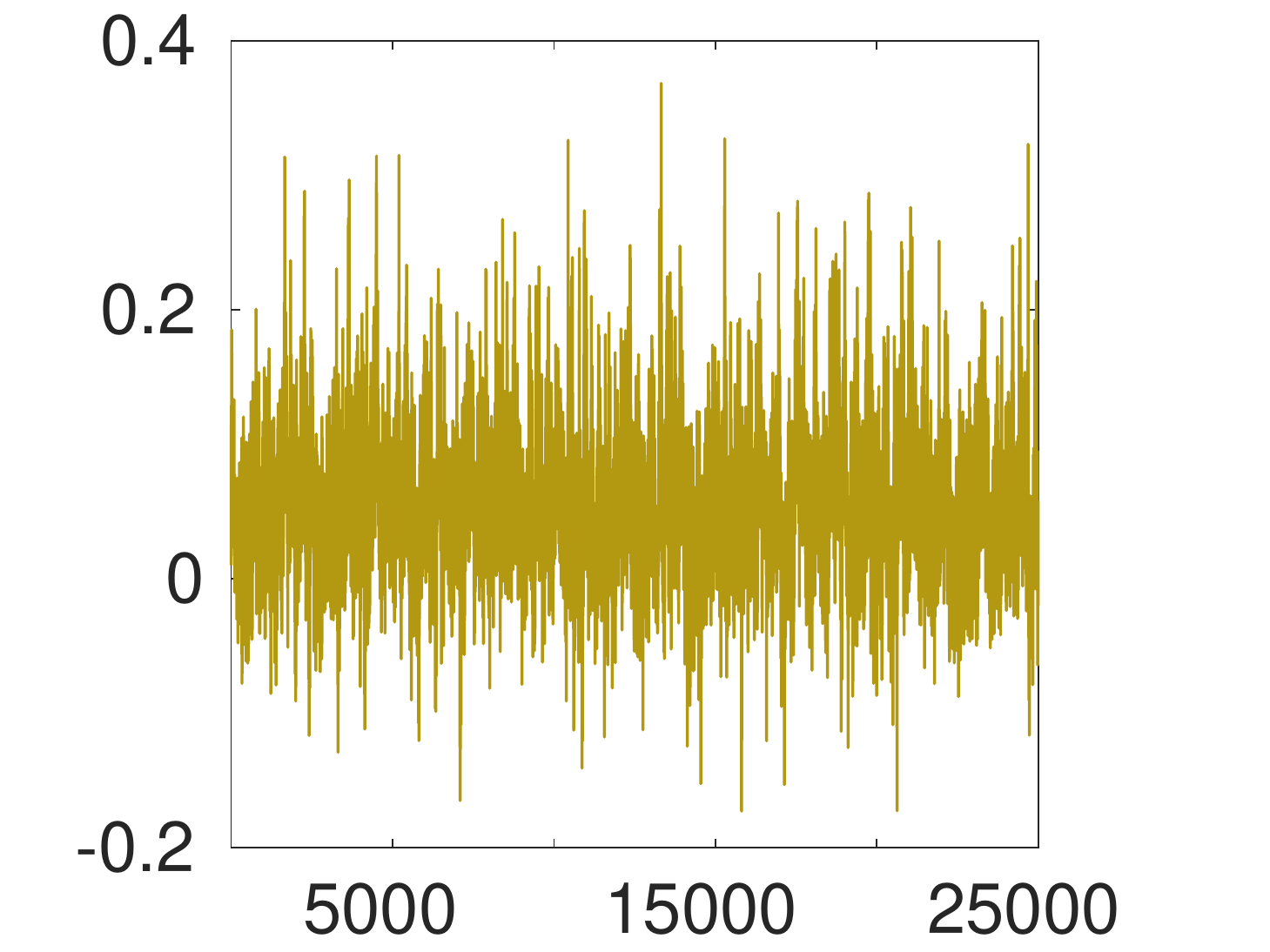}
            \includegraphics[width=\linewidth]{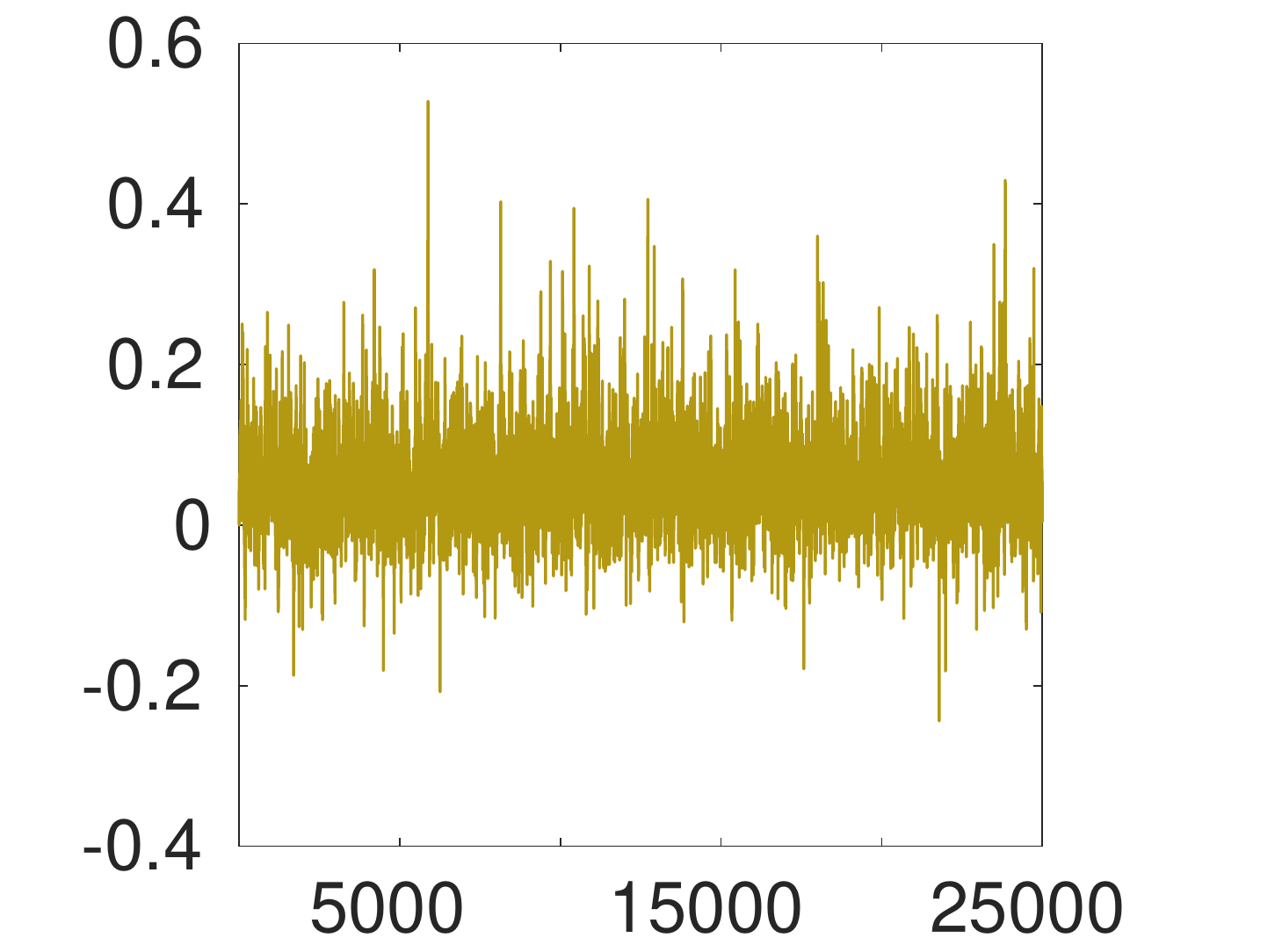}
            \includegraphics[width=\linewidth]{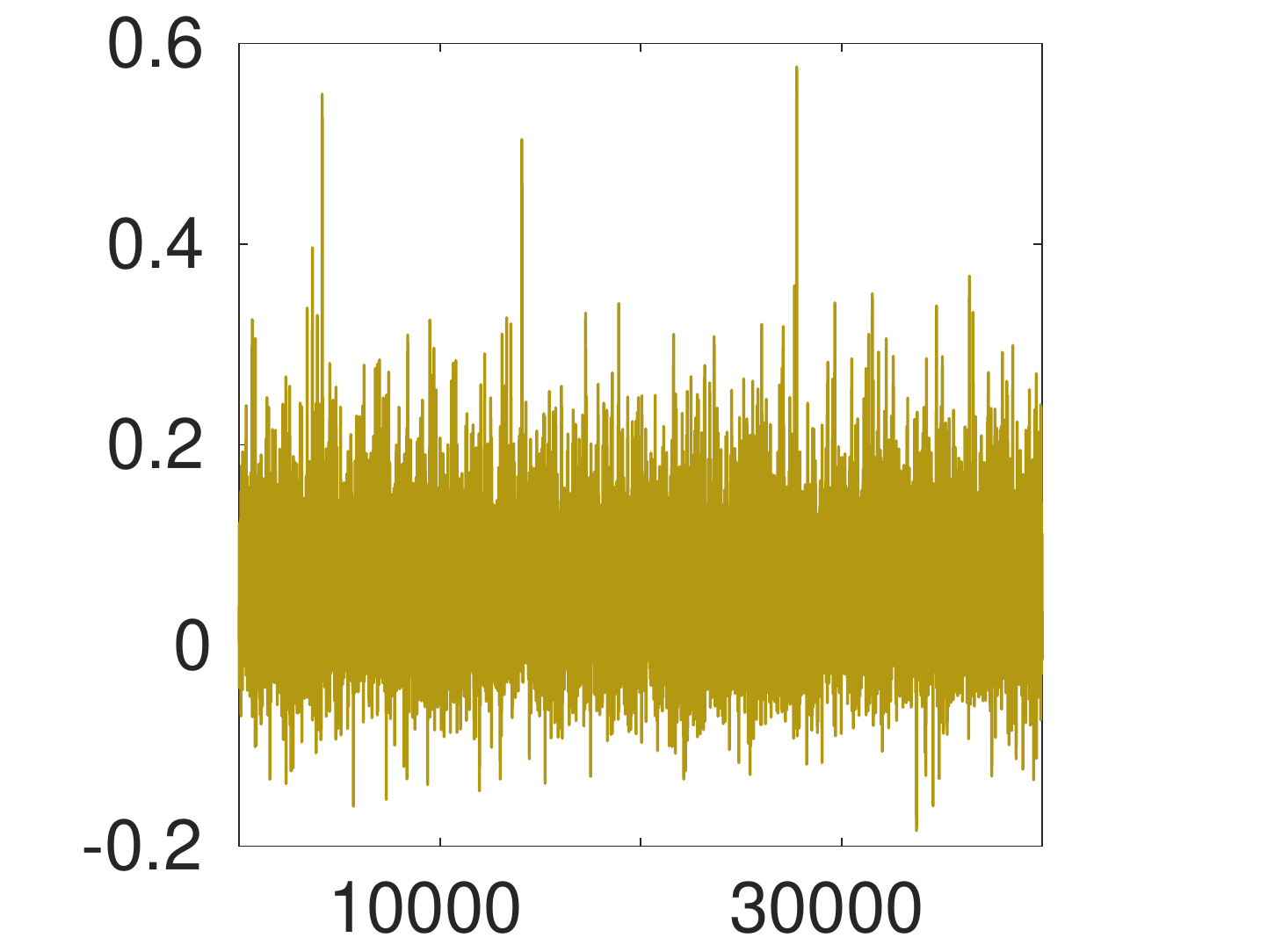}
            \caption{ 2nd at 0.3 }\end{subfigure}\hspace{\pspa}
        \begin{subfigure}[b]{\phei}
            \includegraphics[width=\linewidth]{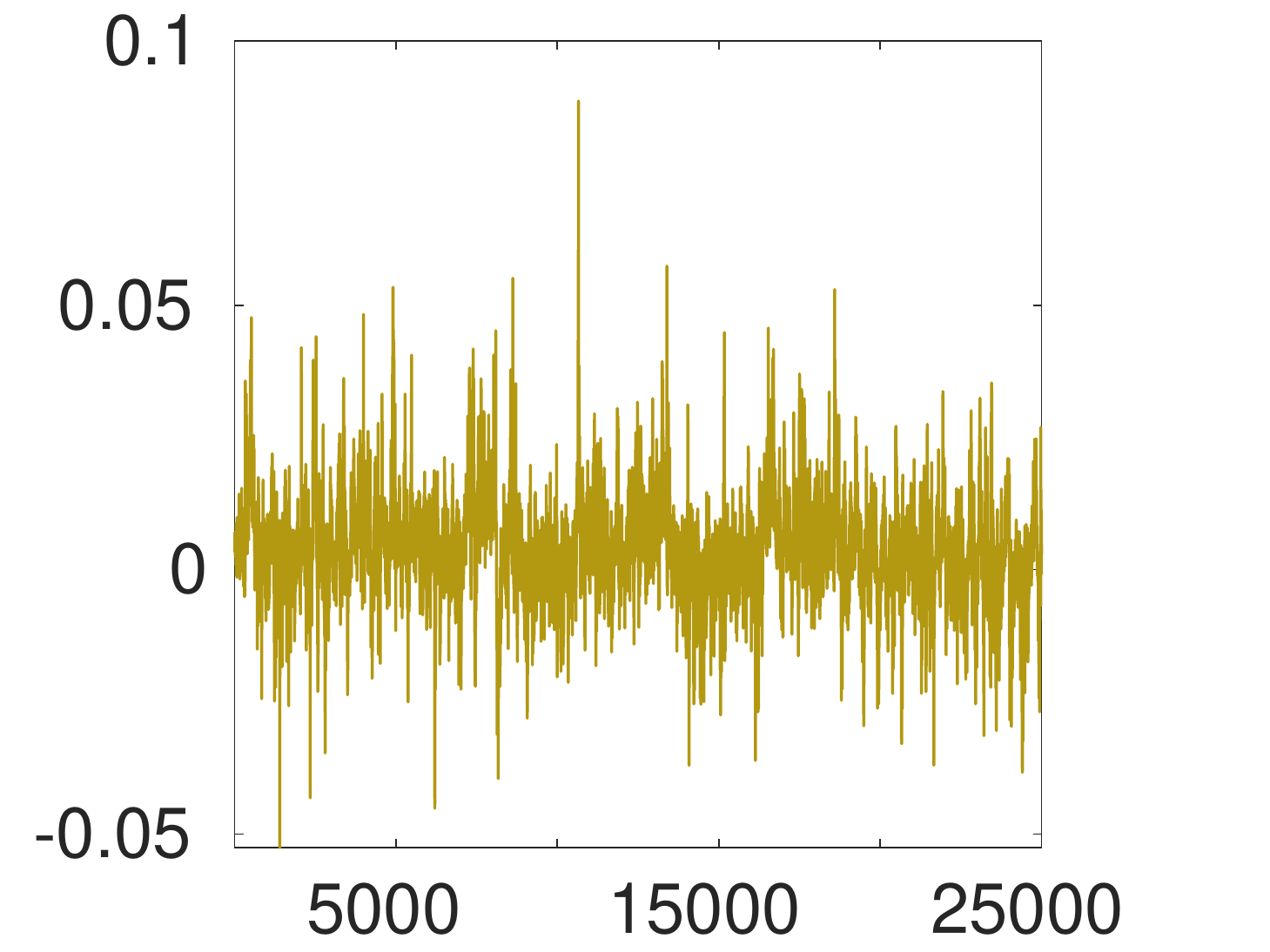}
            \includegraphics[width=\linewidth]{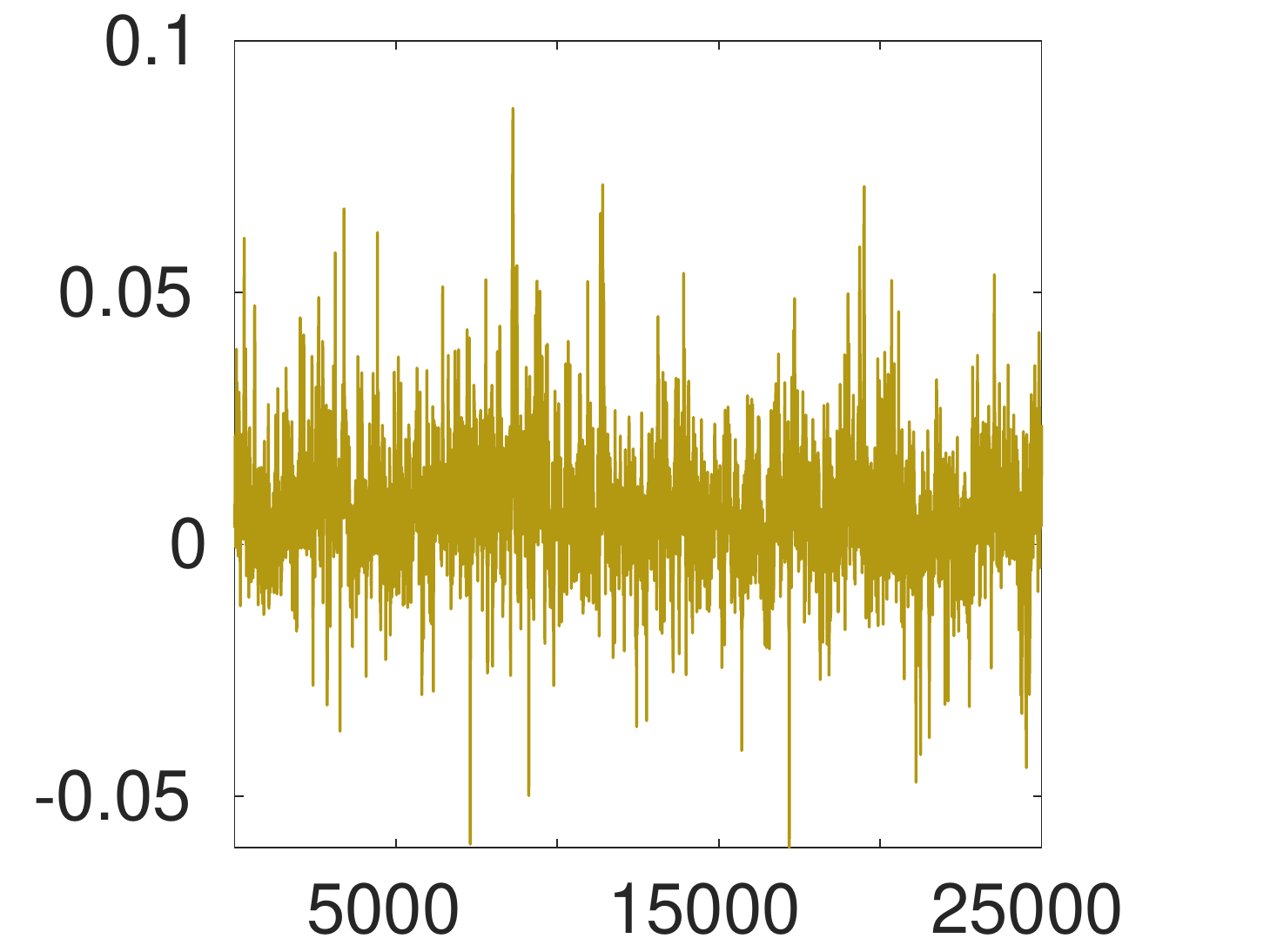}
            \includegraphics[width=\linewidth]{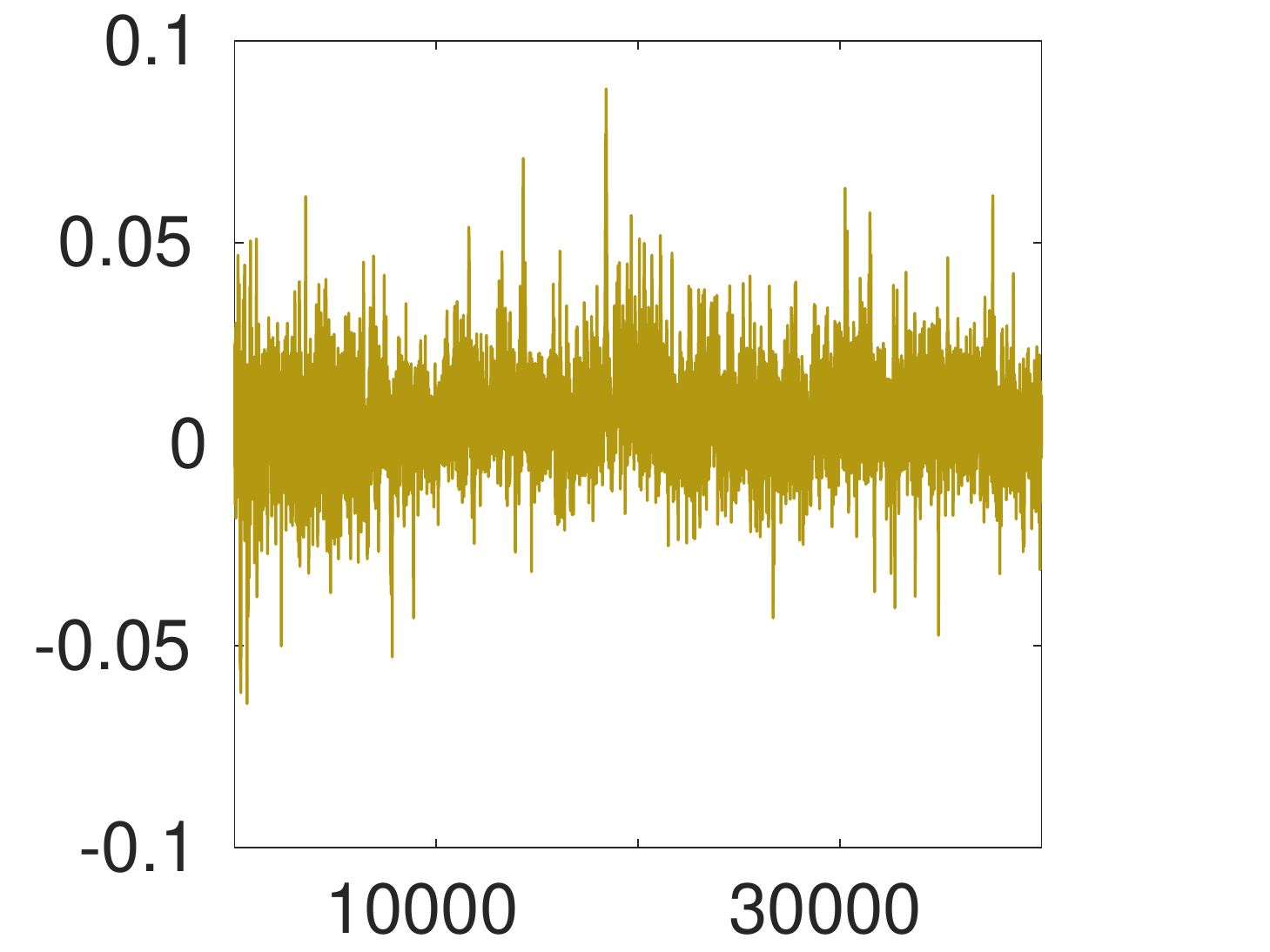}
            \caption{SPDE at 0.3 }\end{subfigure}\hspace{\pspa}
        \begin{subfigure}[b]{\phei}
            \includegraphics[width=\linewidth]{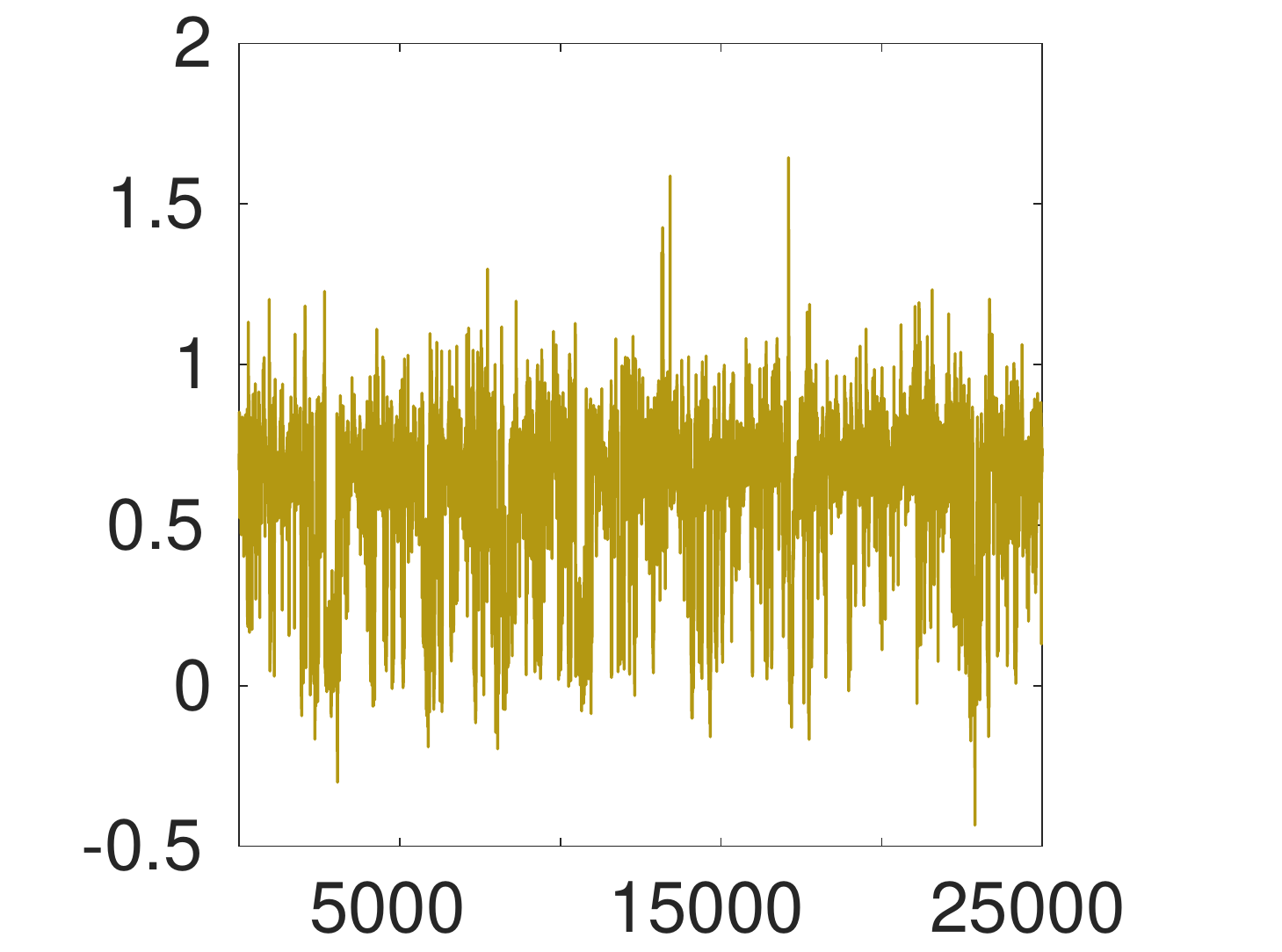}
            \includegraphics[width=\linewidth]{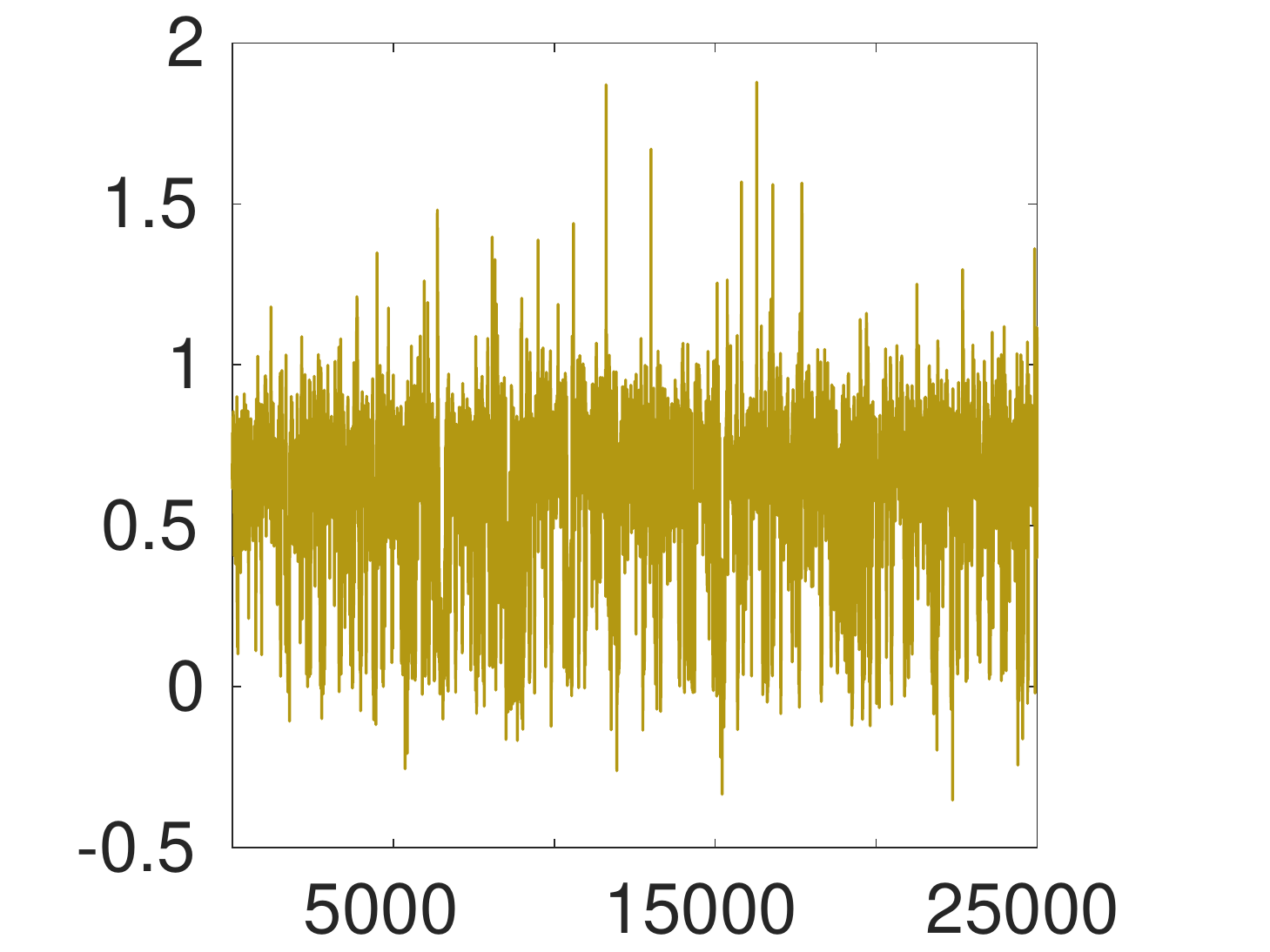}
            \includegraphics[width=\linewidth]{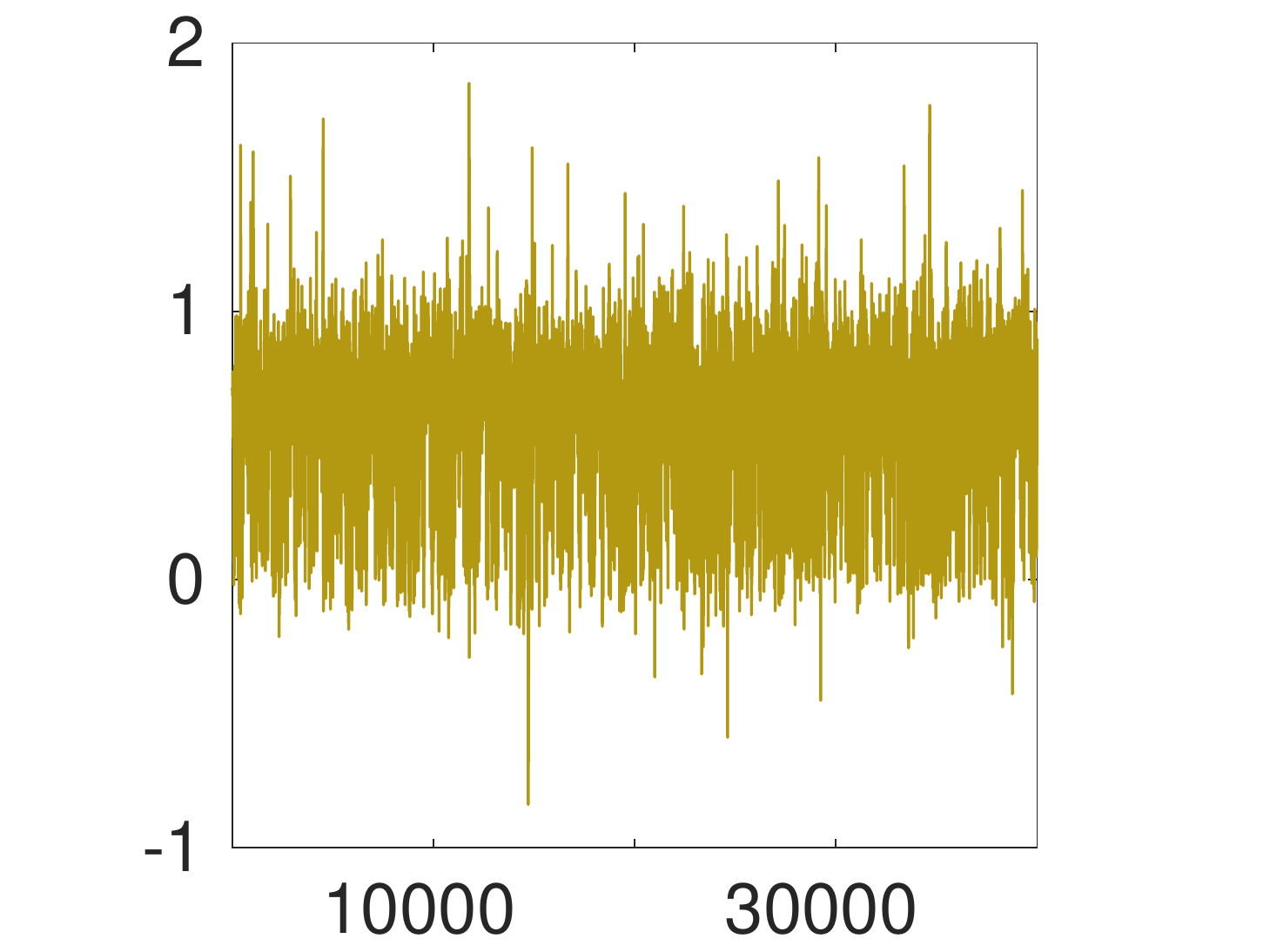}
            \caption{ 1s, at 0.56}\end{subfigure}\hspace{\pspa}
        \begin{subfigure}[b]{\phei}
            \includegraphics[width=\linewidth]{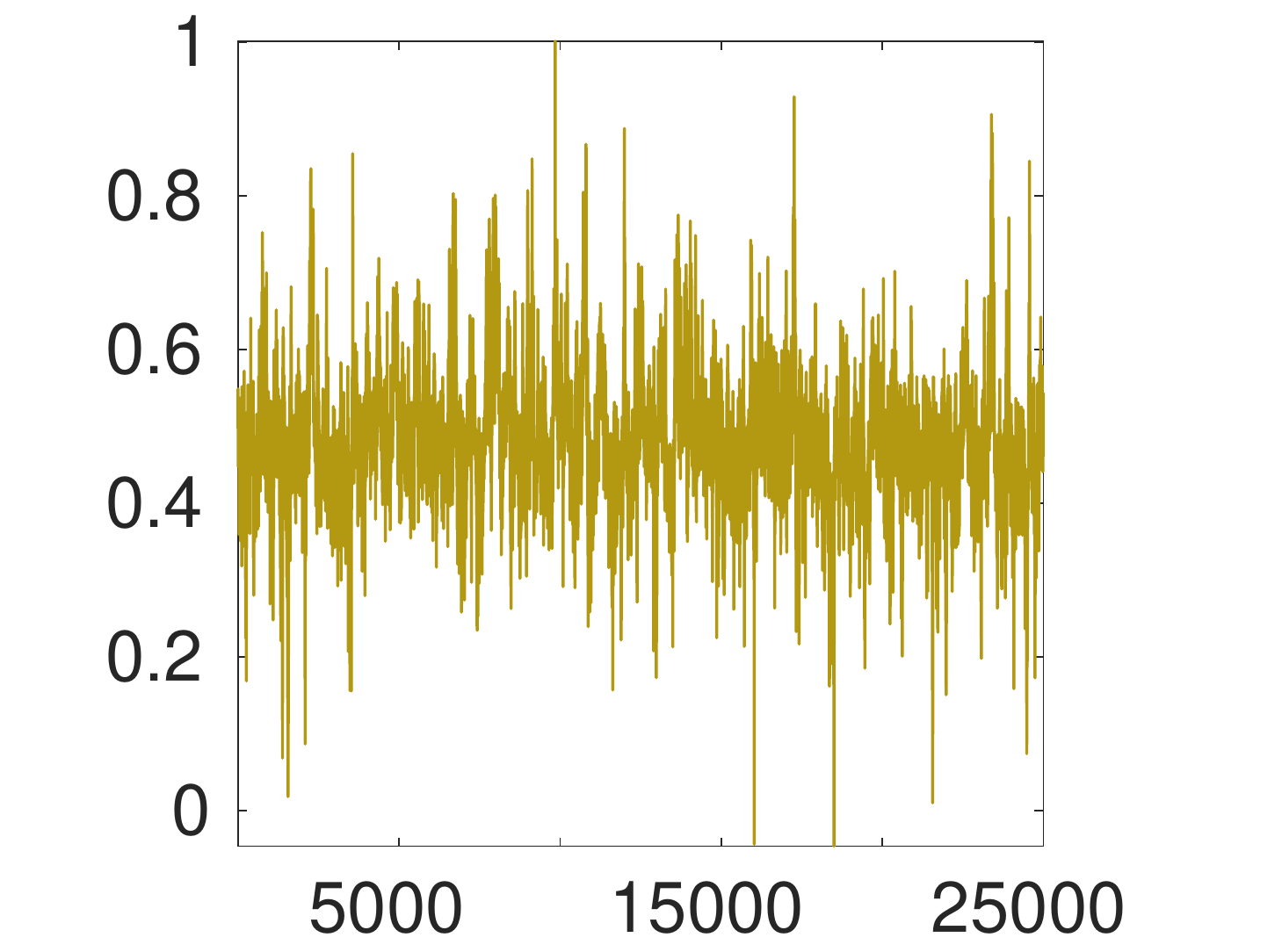}
            \includegraphics[width=\linewidth]{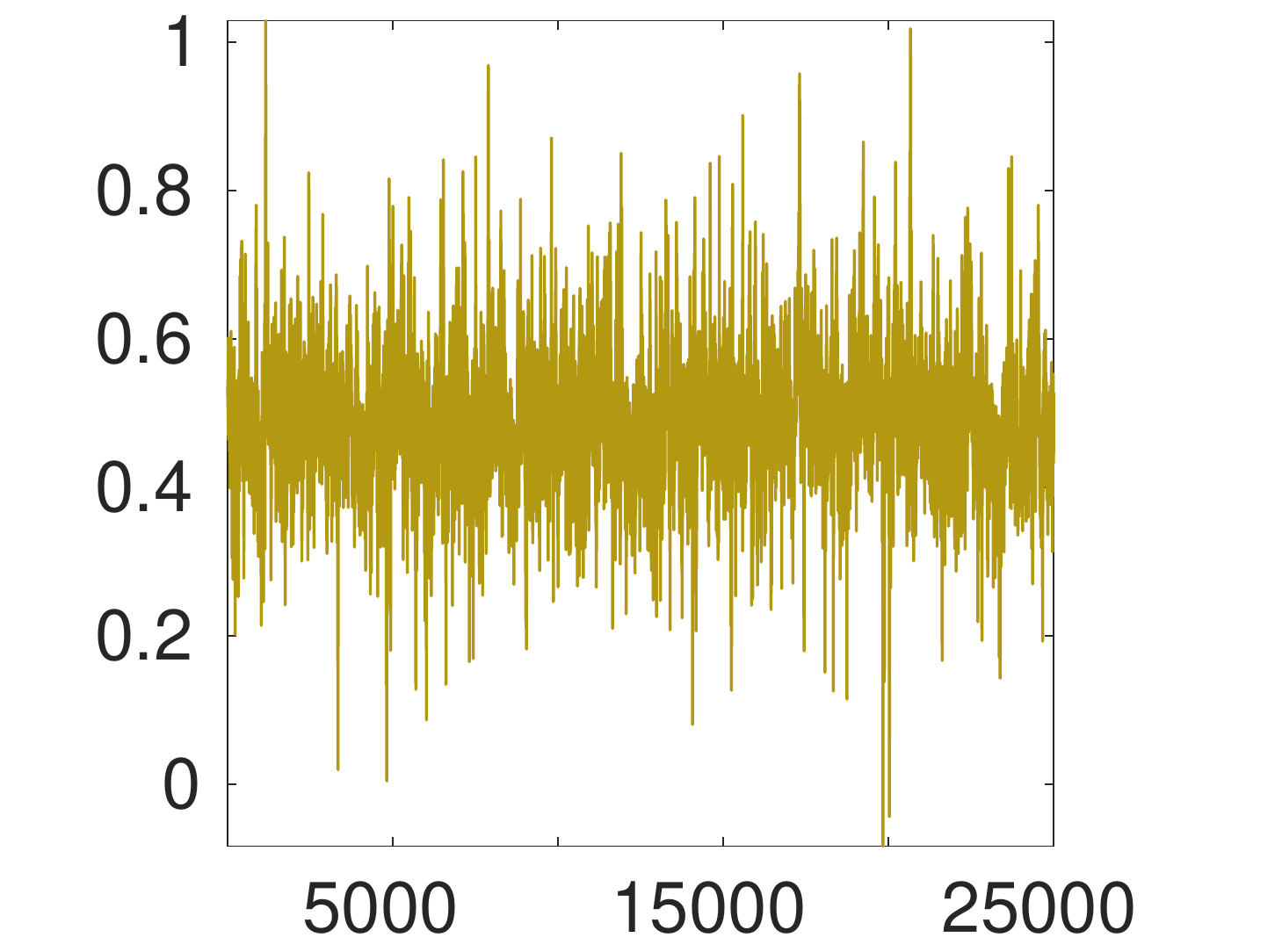}
            \includegraphics[width=\linewidth]{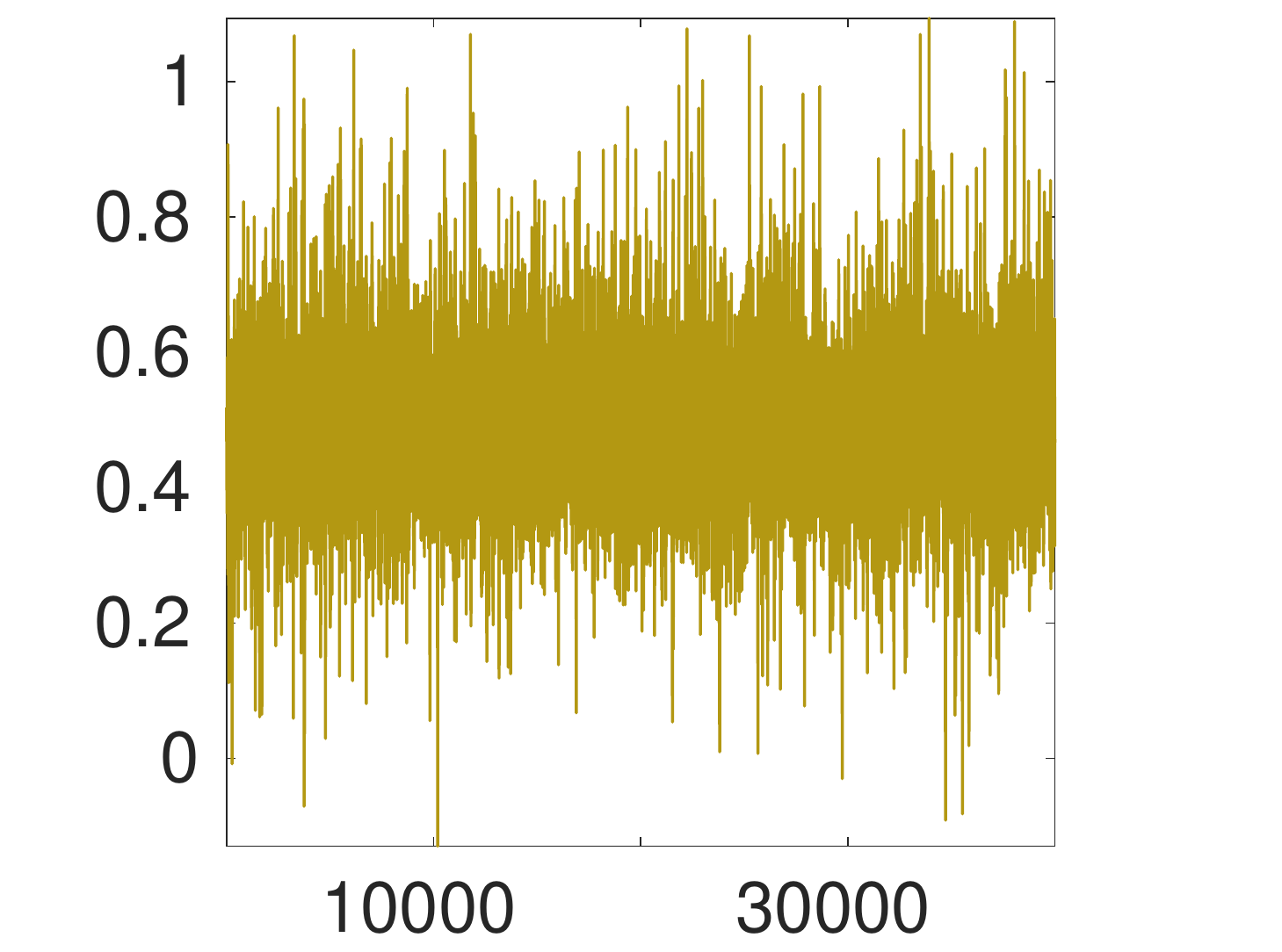}
            \caption{2nd at 0.56}\end{subfigure}\hspace{\pspa}
        \begin{subfigure}[b]{\phei}
            \includegraphics[width=\linewidth]{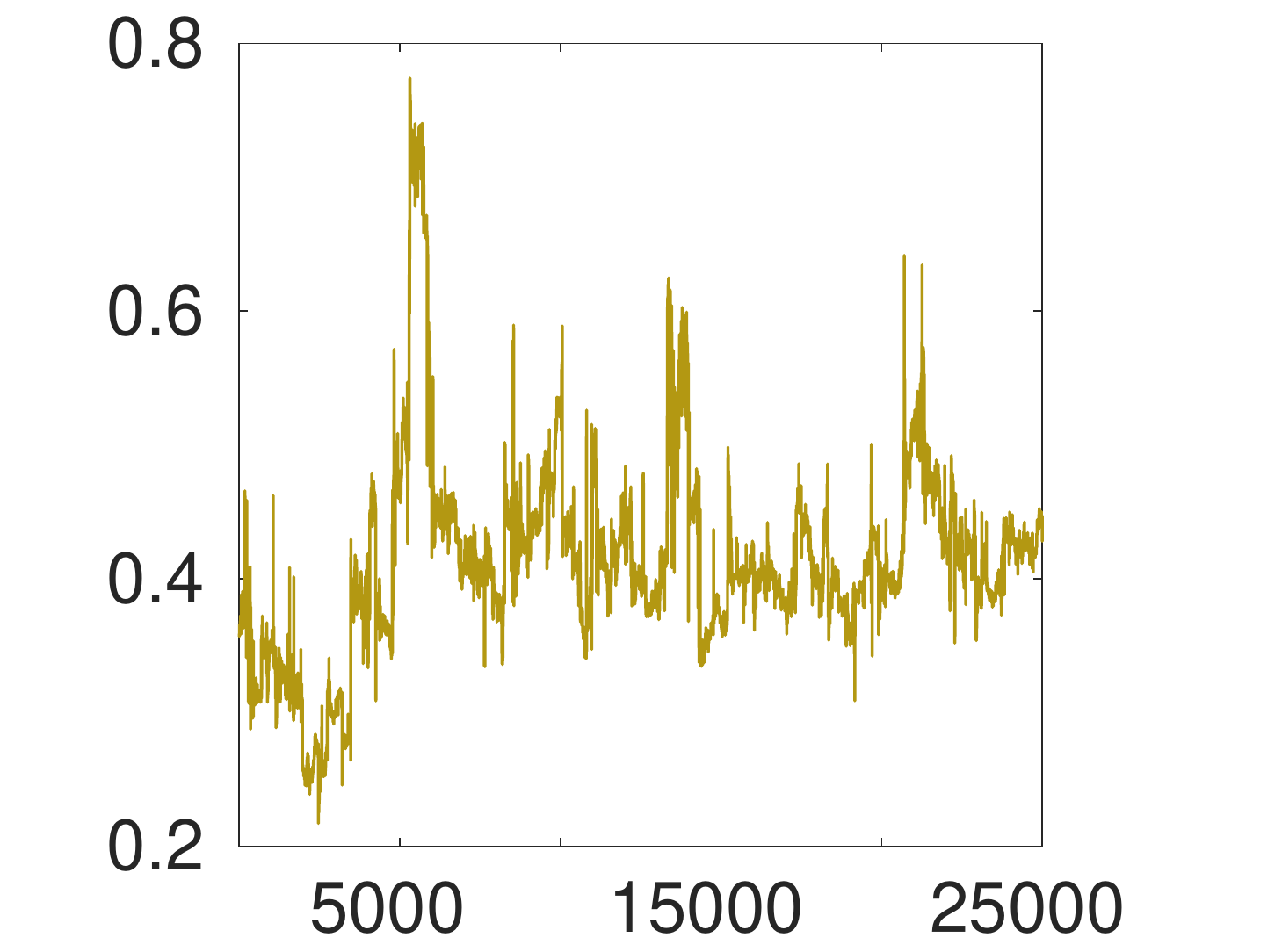}
            \includegraphics[width=\linewidth]{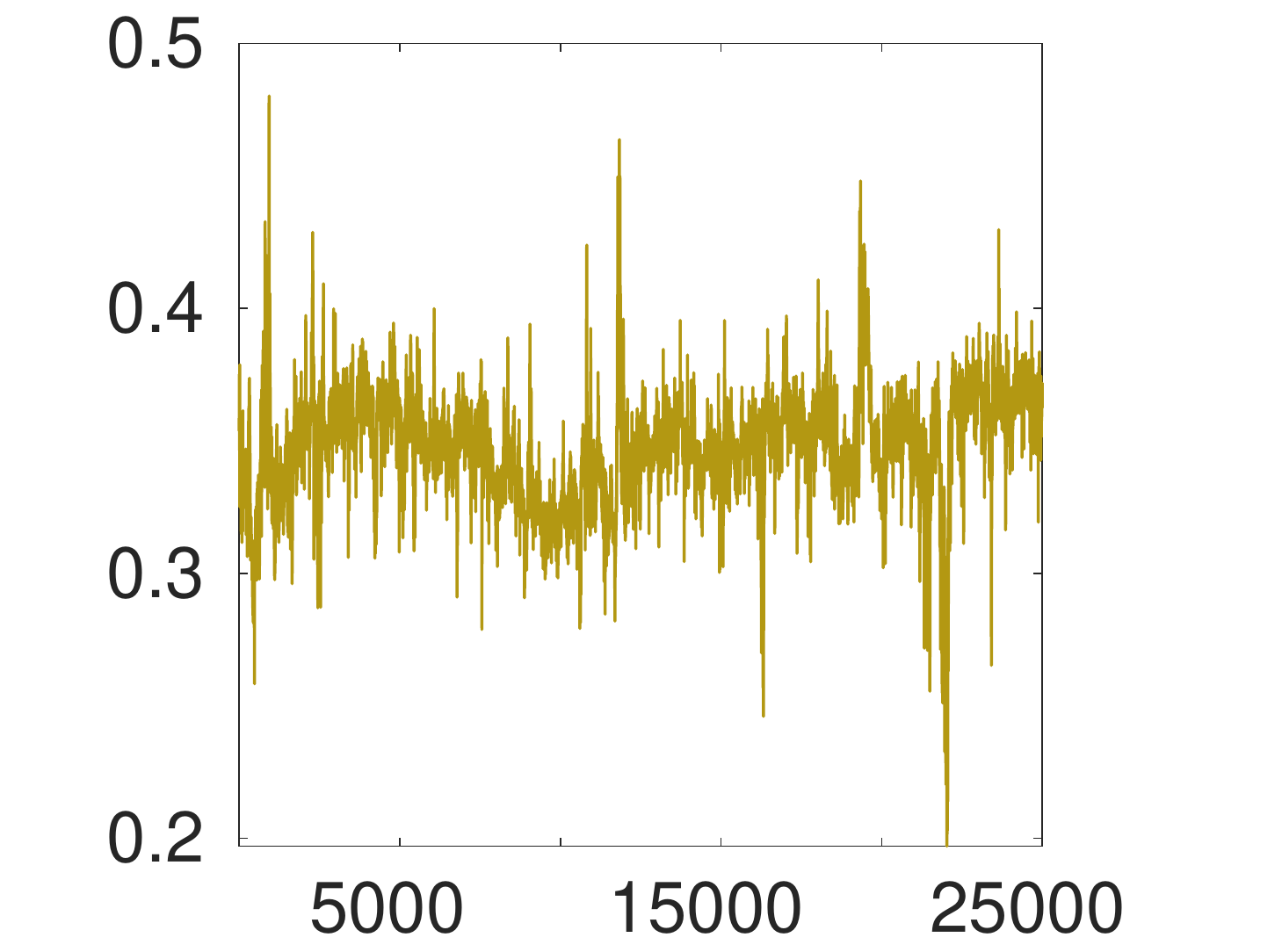}
            \includegraphics[width=\linewidth]{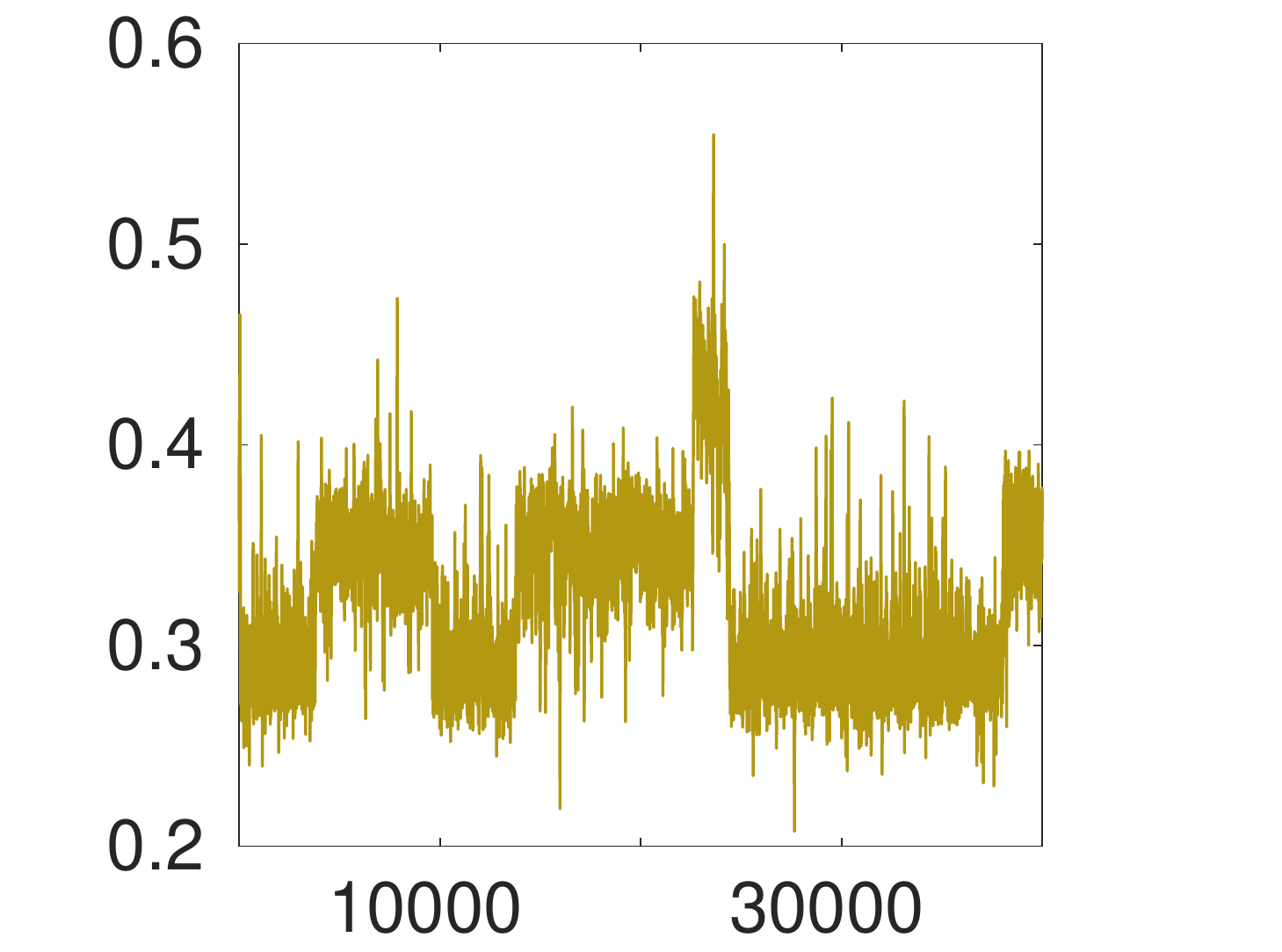}
            \caption{ SPDE at 0.56}\end{subfigure}
        
        \begin{subfigure}[b]{\phei}
            \includegraphics[width=\linewidth]{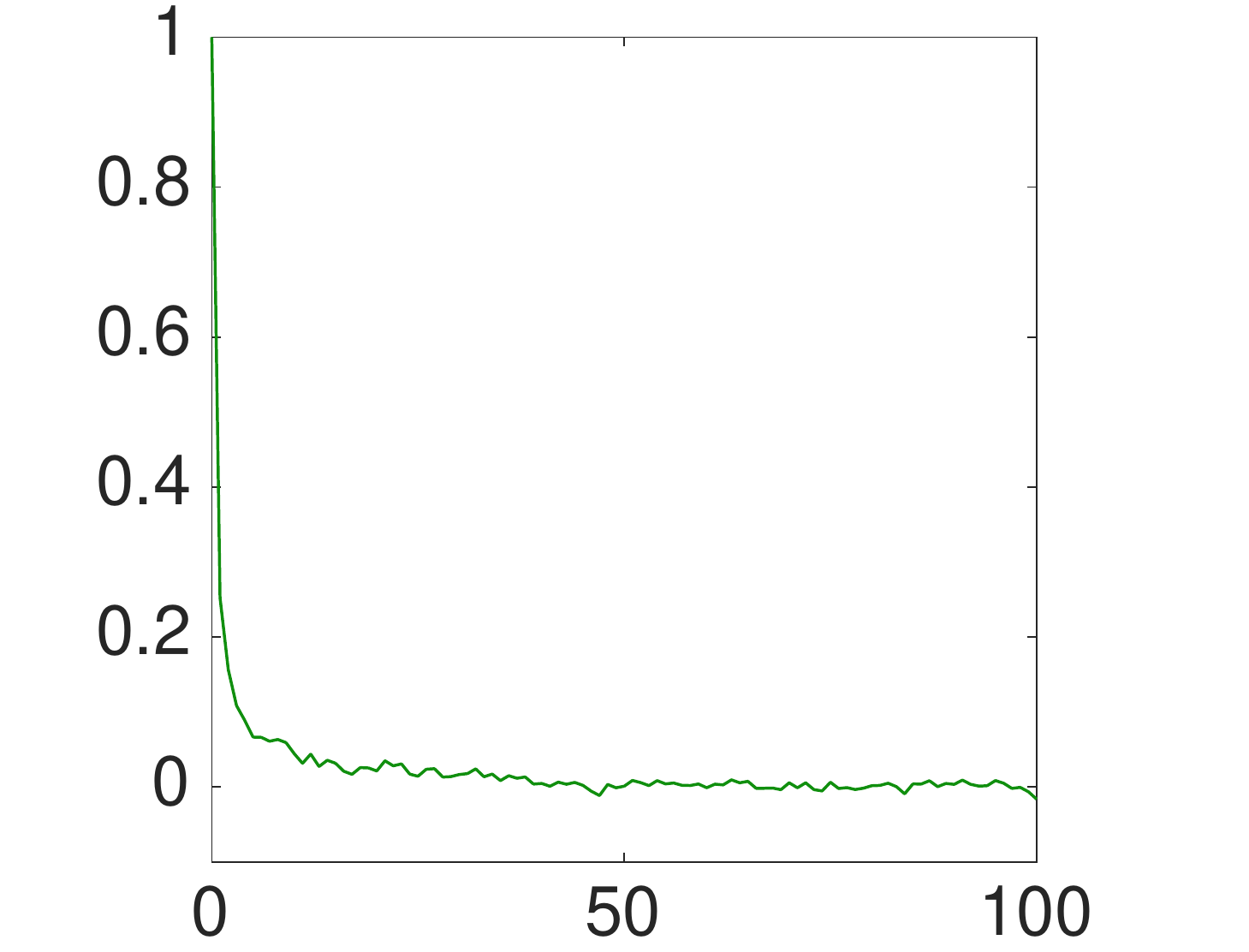}
            \includegraphics[width=\linewidth]{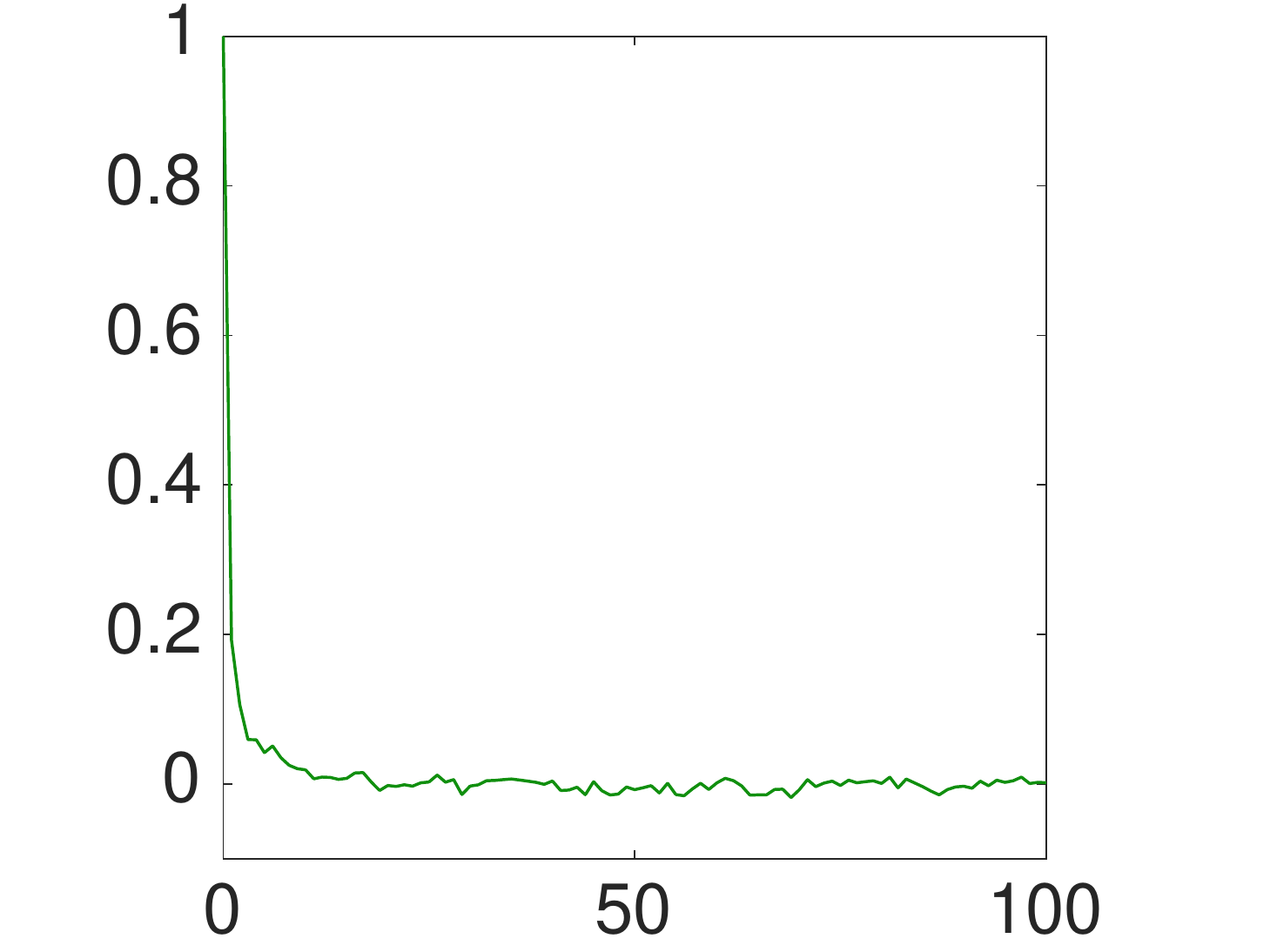}
            \includegraphics[width=\linewidth]{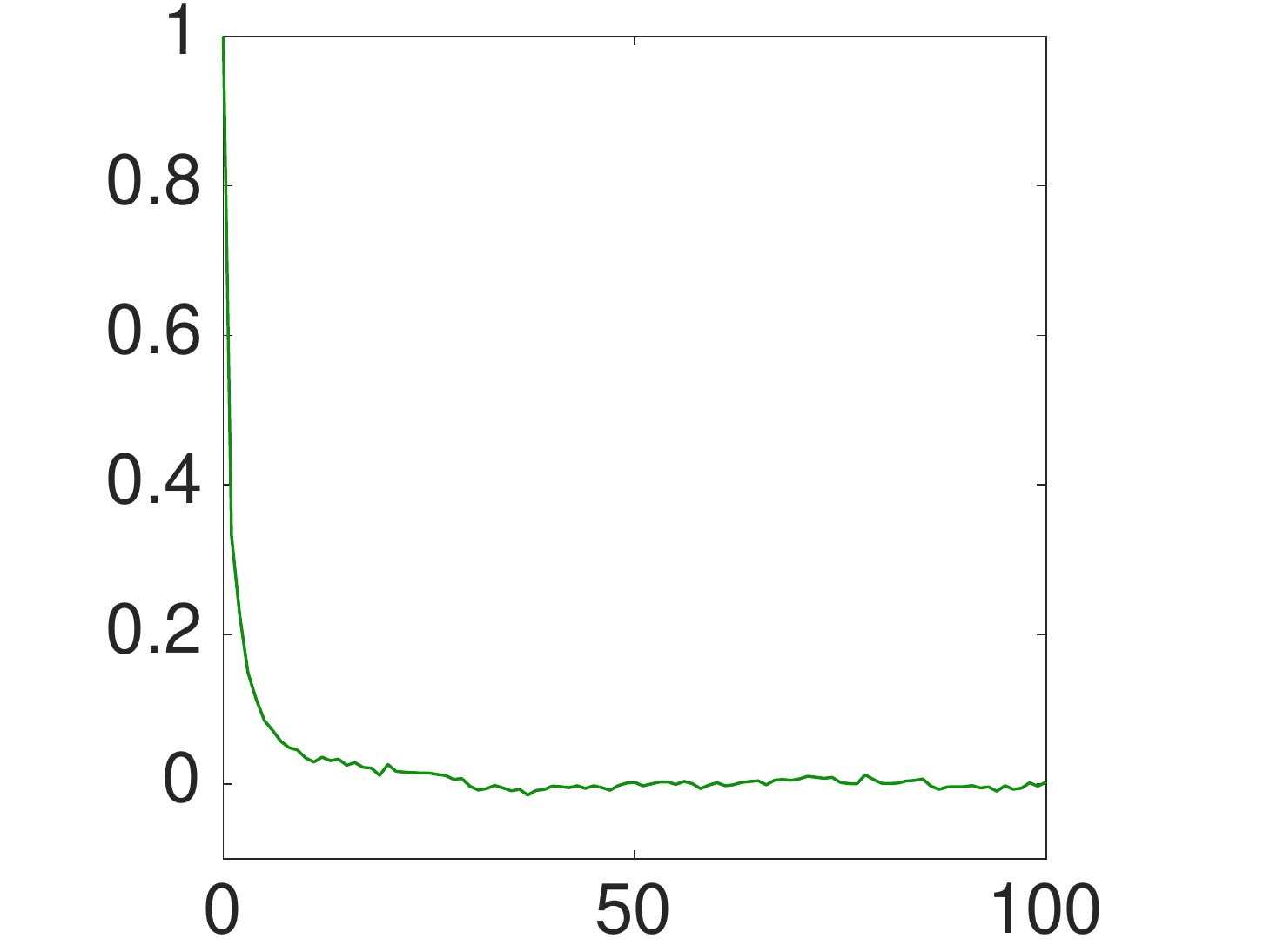}
            \caption{1st at 0.3 }\end{subfigure}\hspace{\pspa}
        \begin{subfigure}[b]{\phei}
            \includegraphics[width=\linewidth]{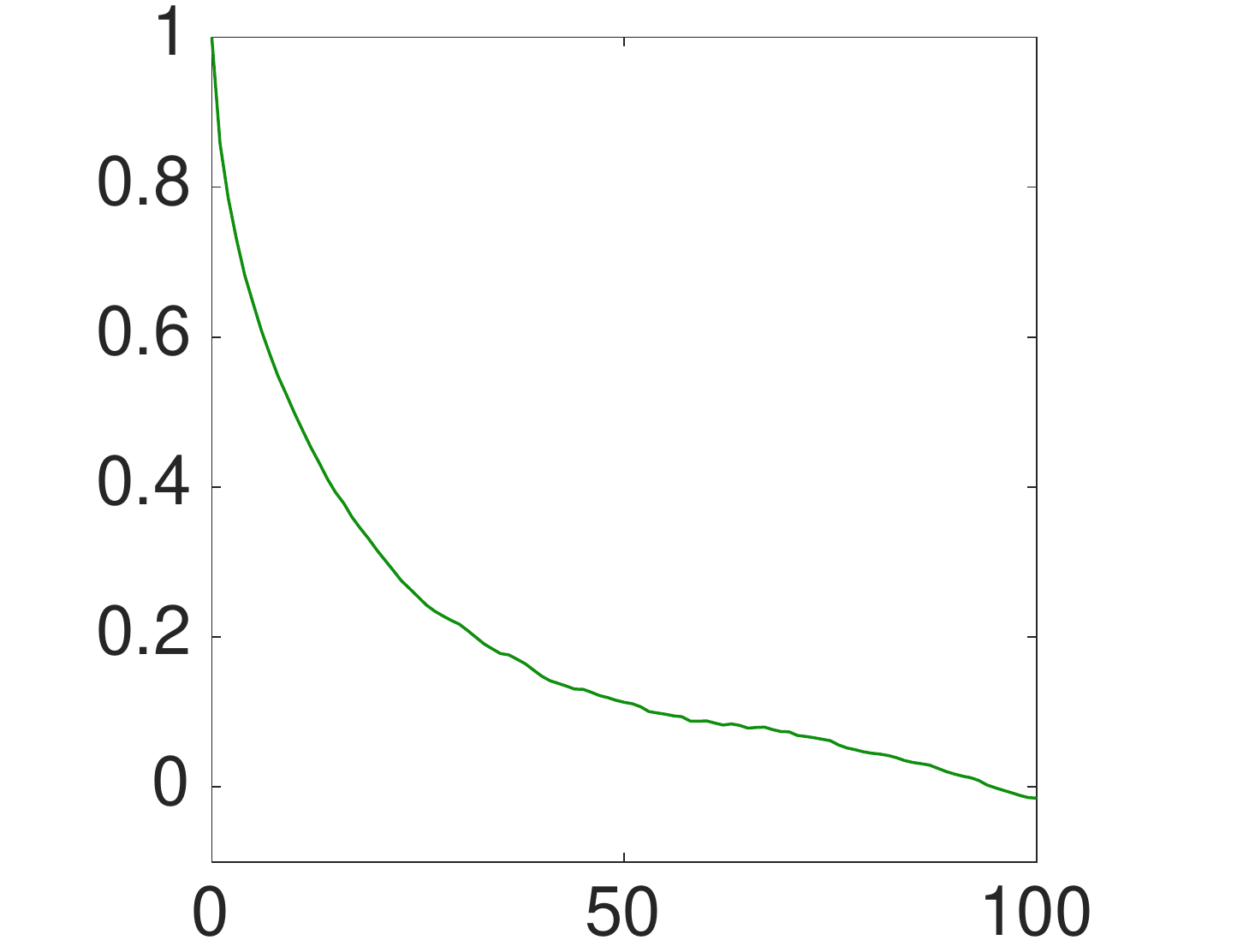}
            \includegraphics[width=\linewidth]{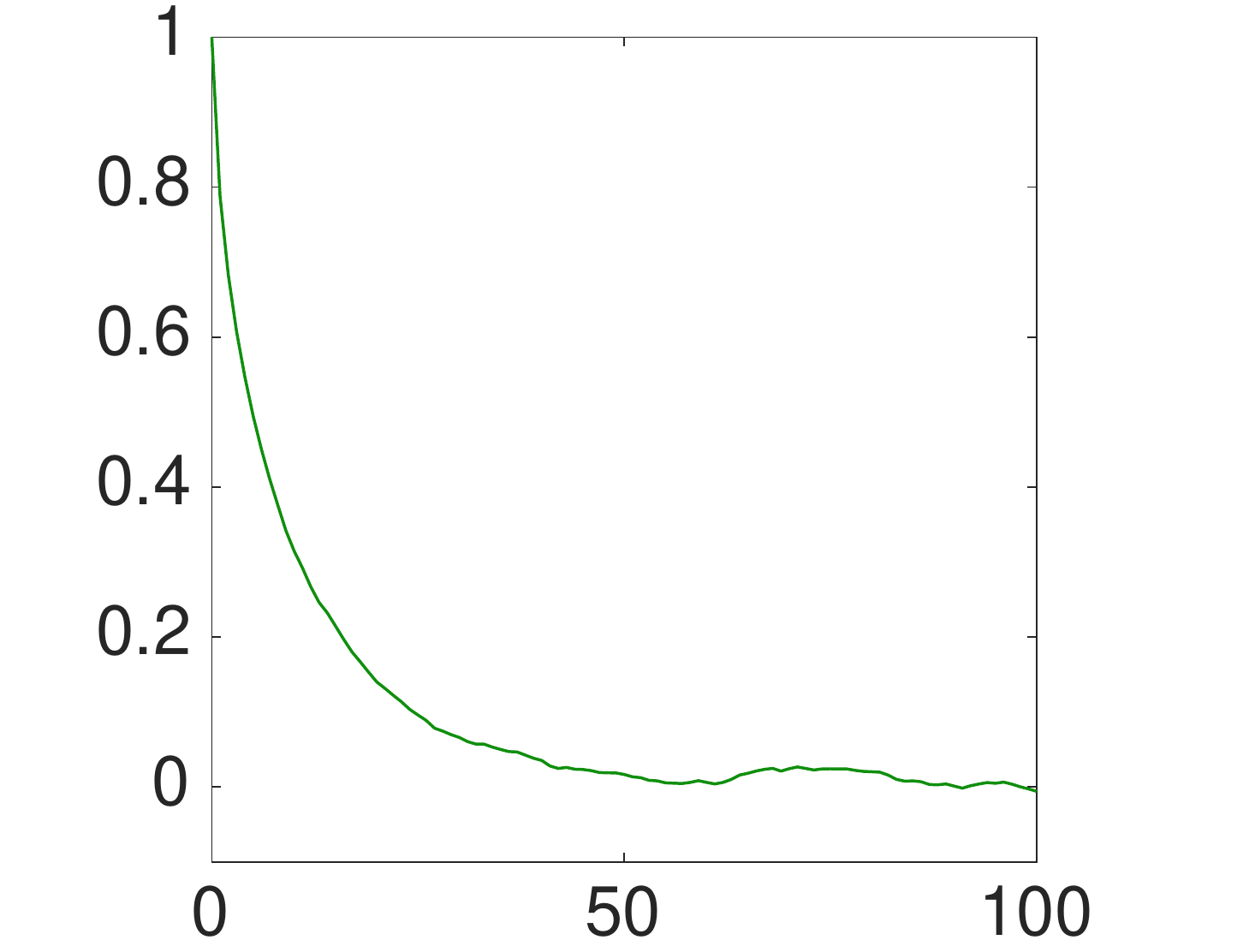}
            \includegraphics[width=\linewidth]{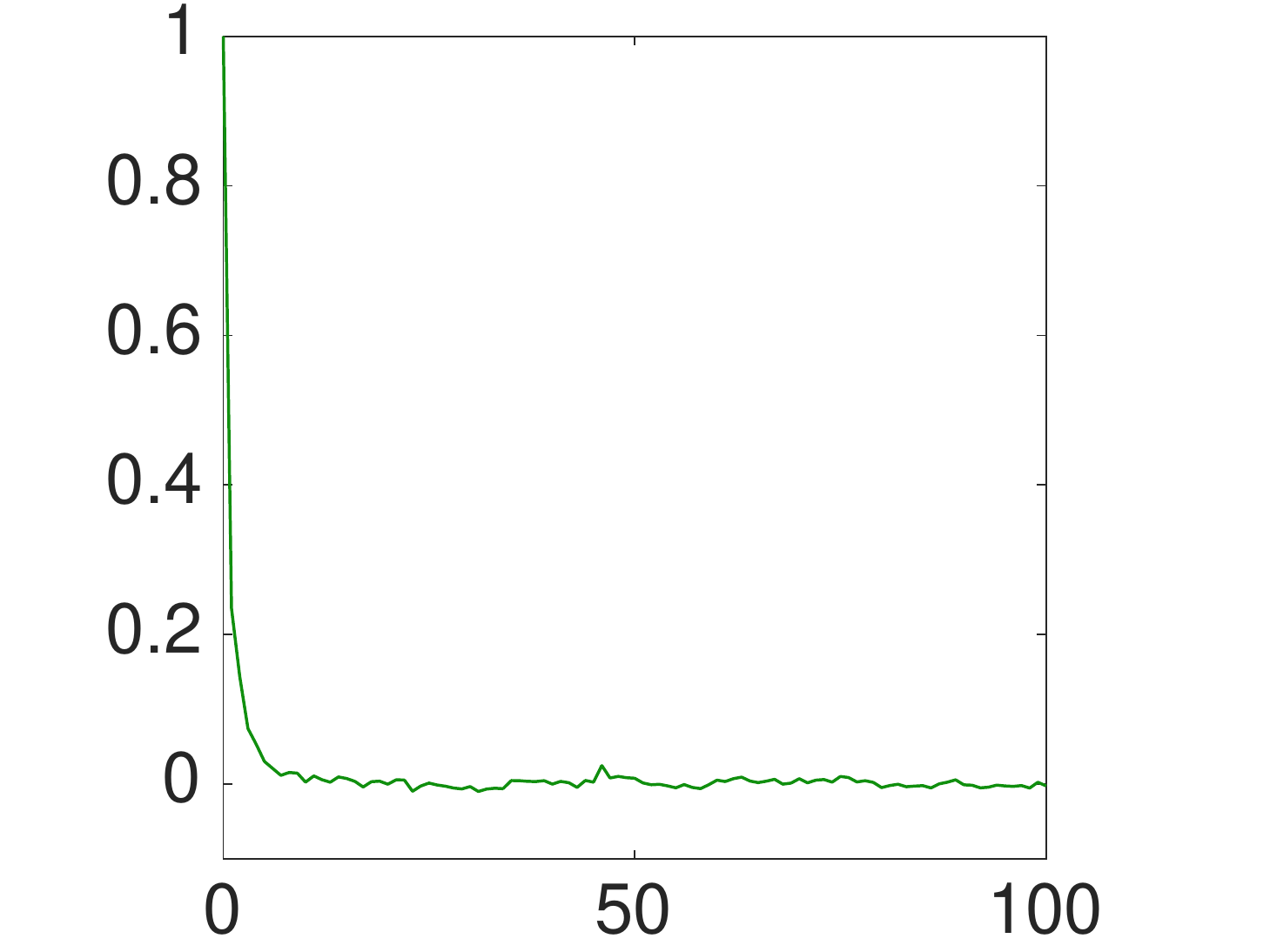}
            \caption{ 2nd at 0.3 }\end{subfigure}\hspace{\pspa}
        \begin{subfigure}[b]{\phei}
            \includegraphics[width=\linewidth]{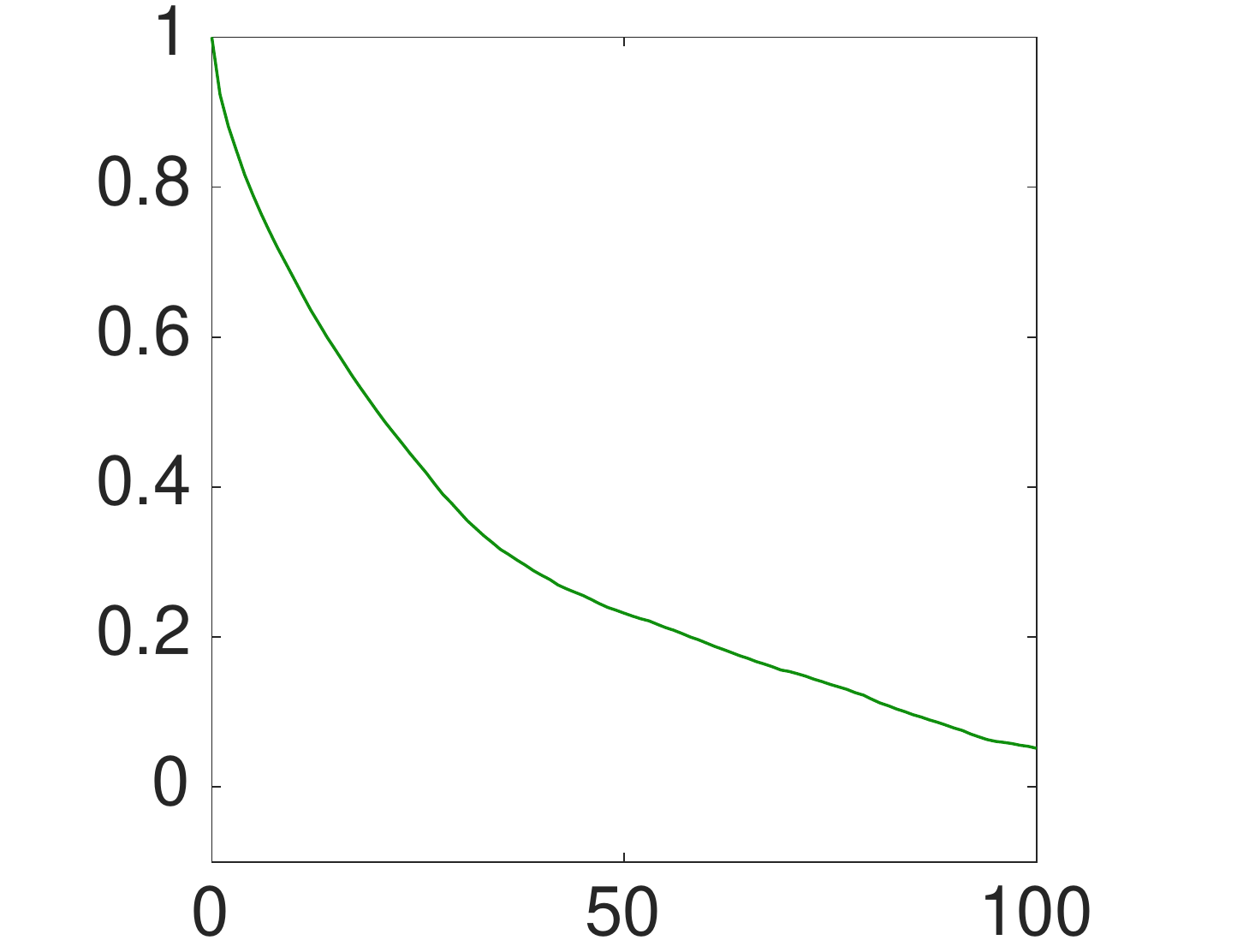}
            \includegraphics[width=\linewidth]{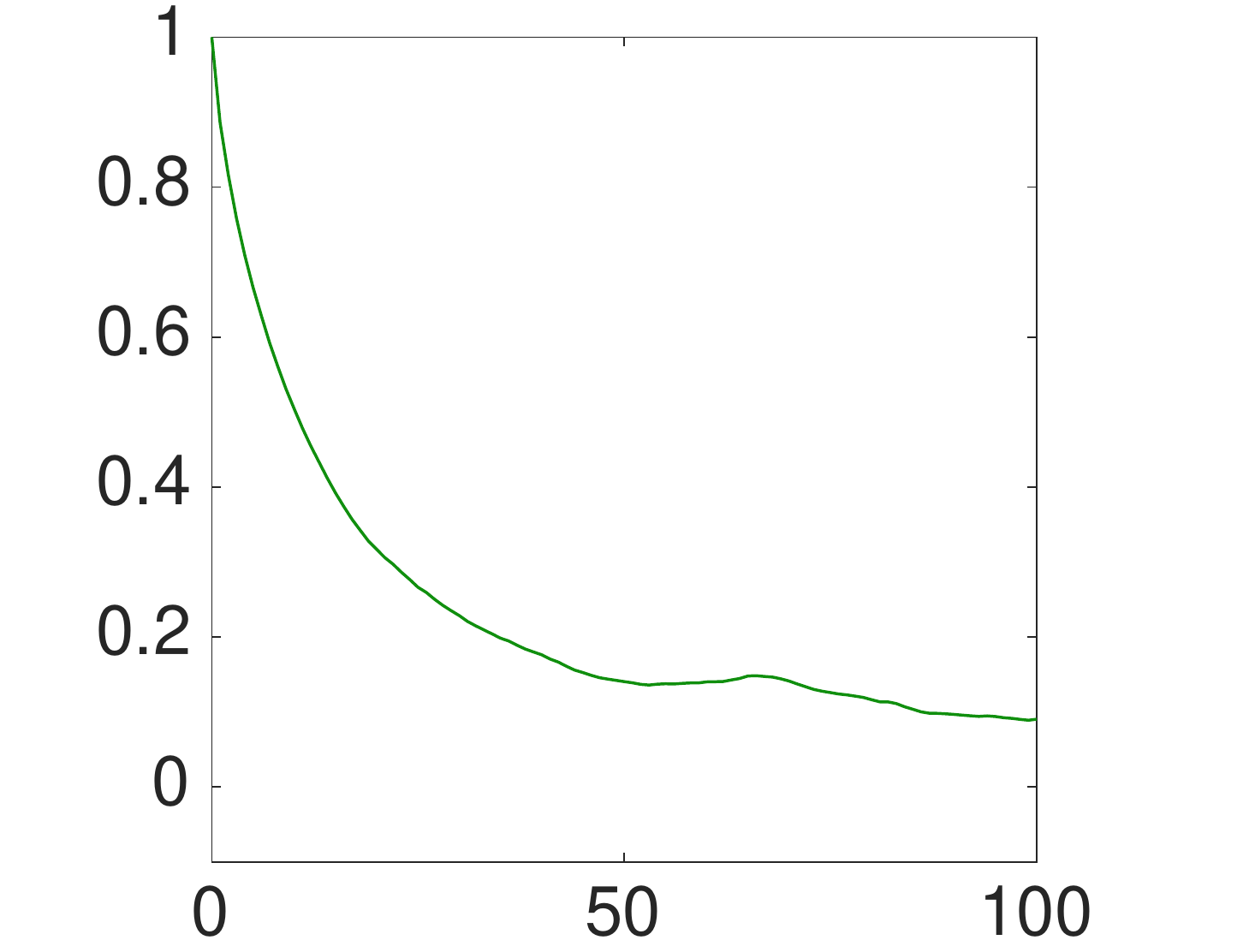}
            \includegraphics[width=\linewidth]{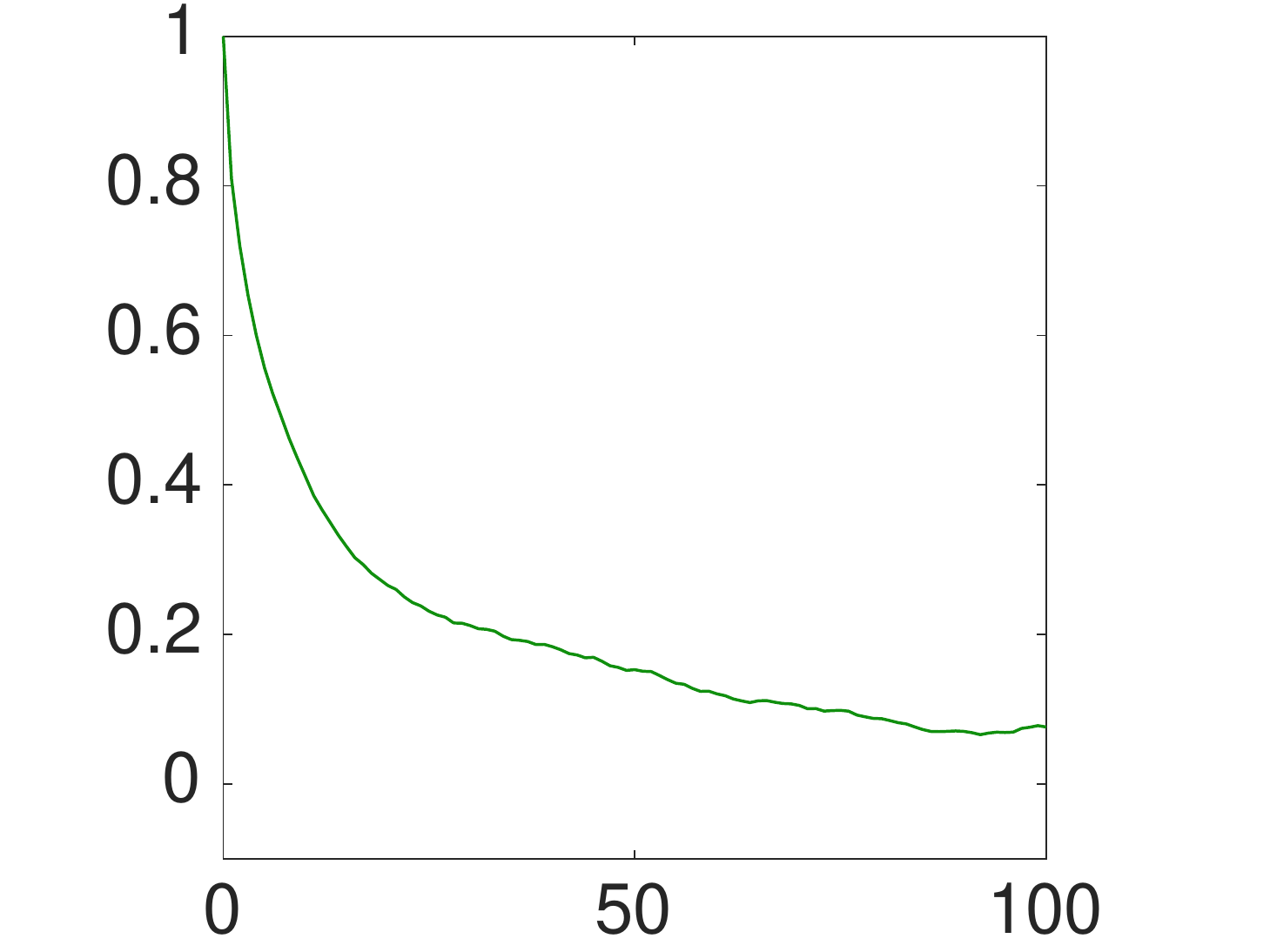}
            \caption{SPDE at 0.3 }\end{subfigure}\hspace{\pspa}
        \begin{subfigure}[b]{\phei}
            \includegraphics[width=\linewidth]{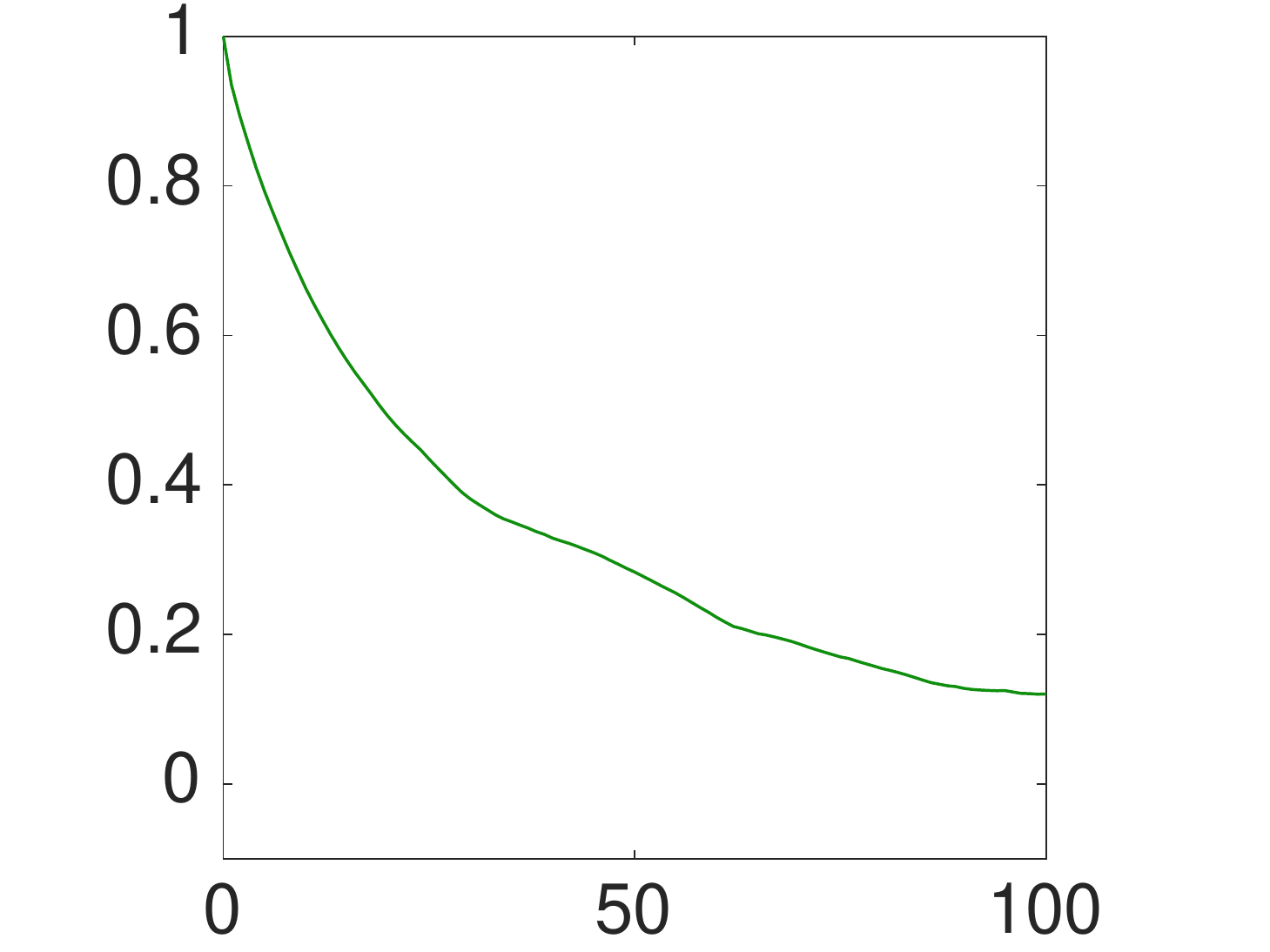}
            \includegraphics[width=\linewidth]{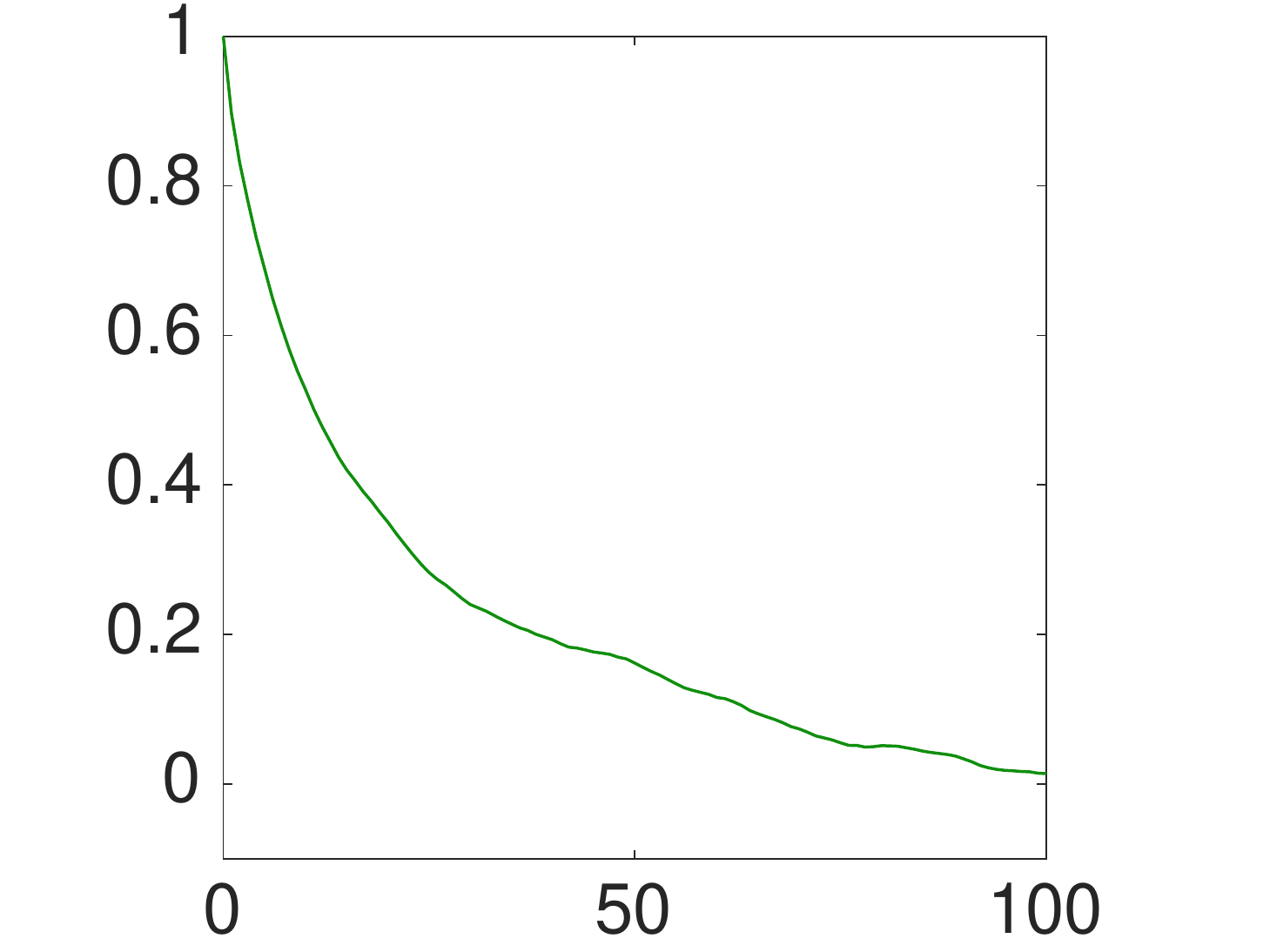}
            \includegraphics[width=\linewidth]{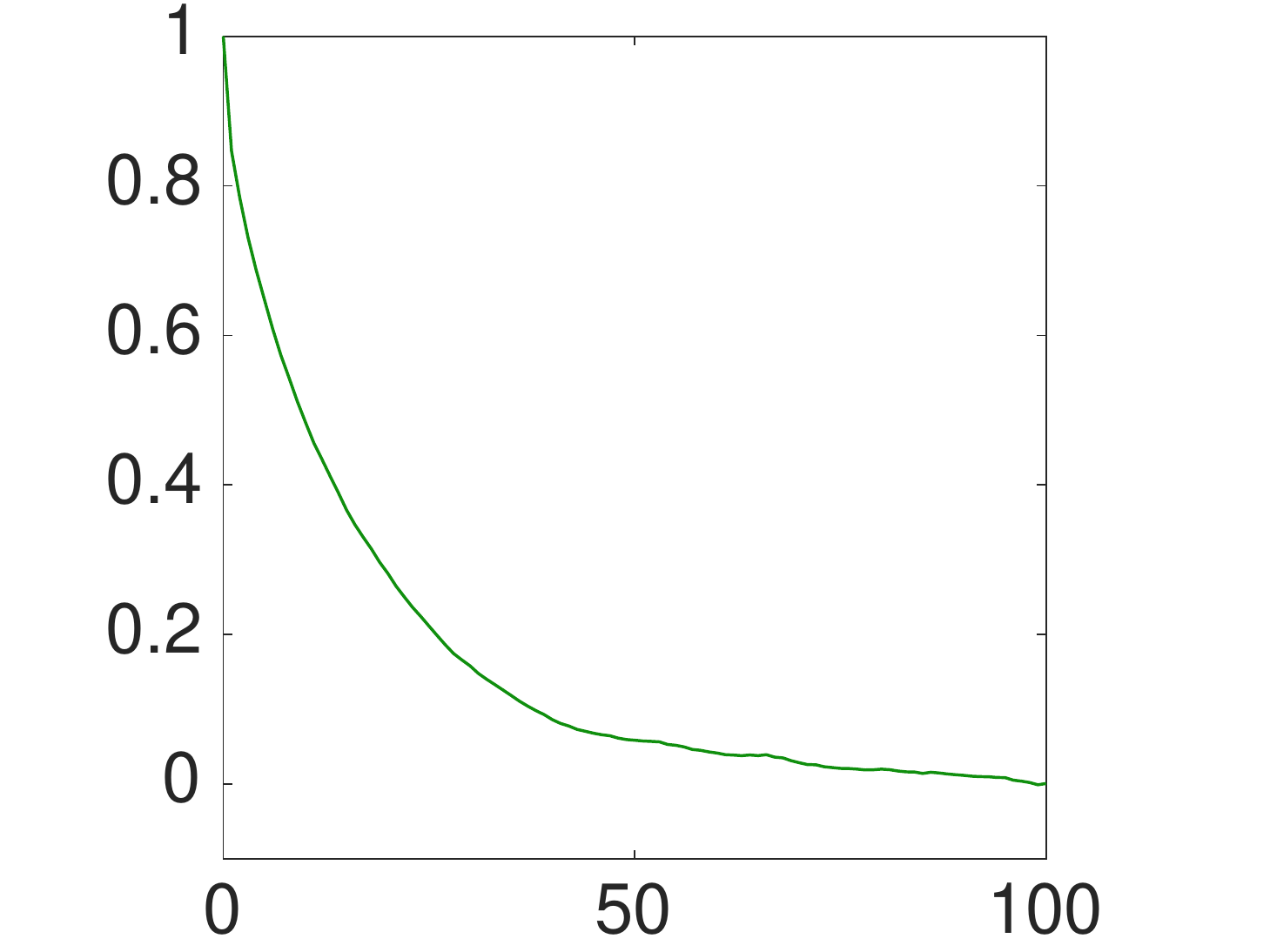}
            \caption{ 1st, at 0.56}\end{subfigure}\hspace{\pspa}
        \begin{subfigure}[b]{\phei}
            \includegraphics[width=\linewidth]{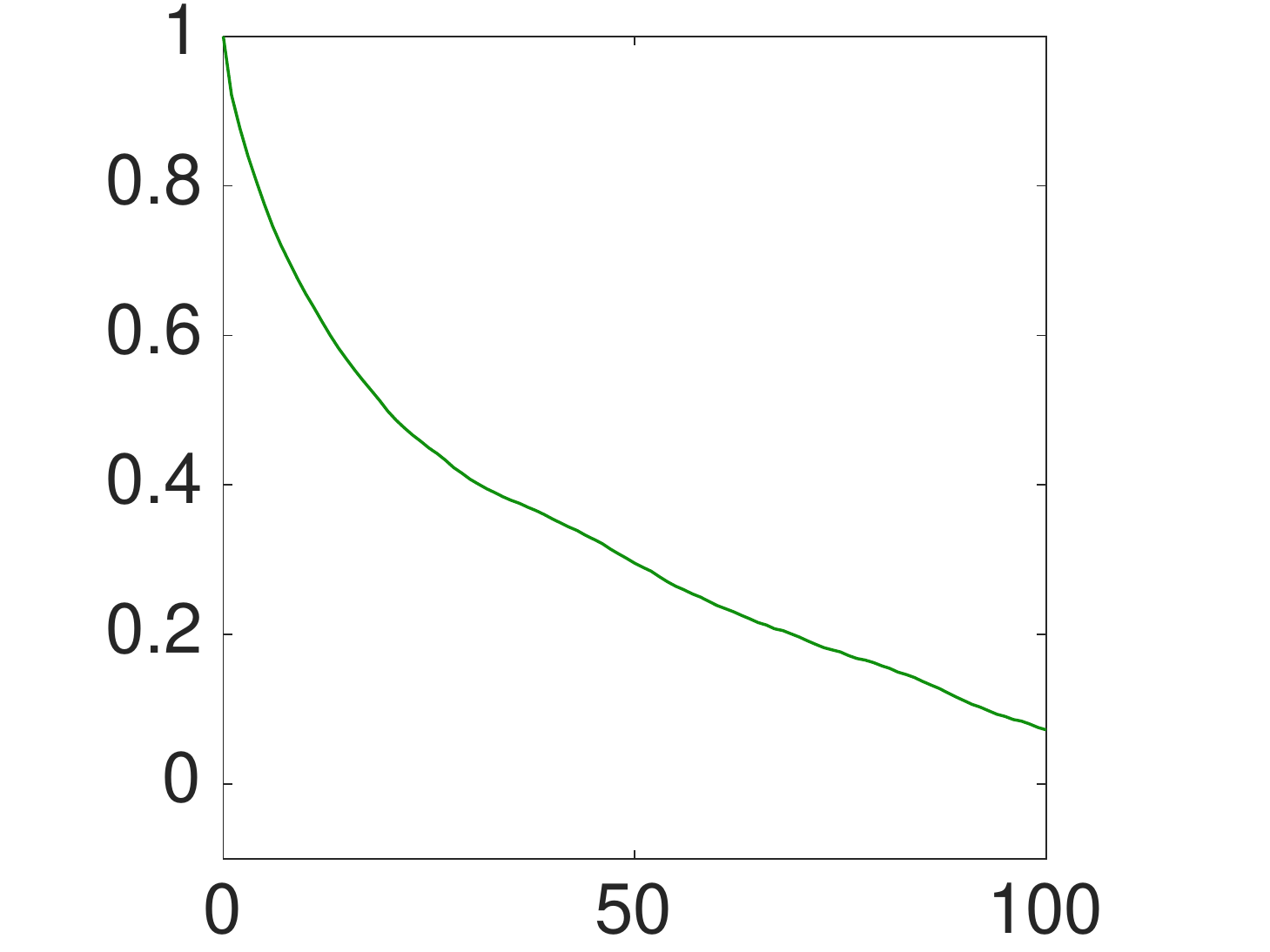}
            \includegraphics[width=\linewidth]{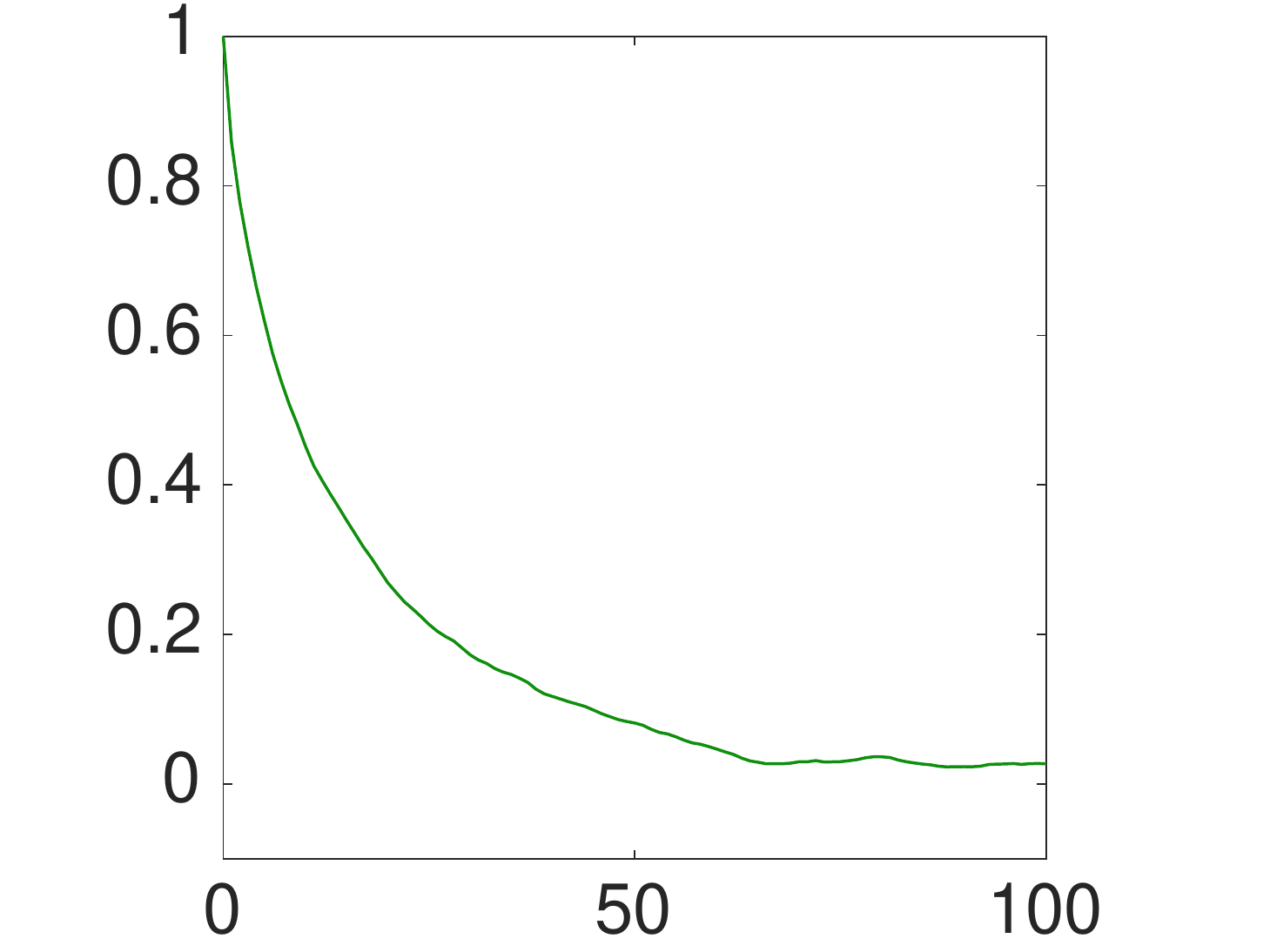}
            \includegraphics[width=\linewidth]{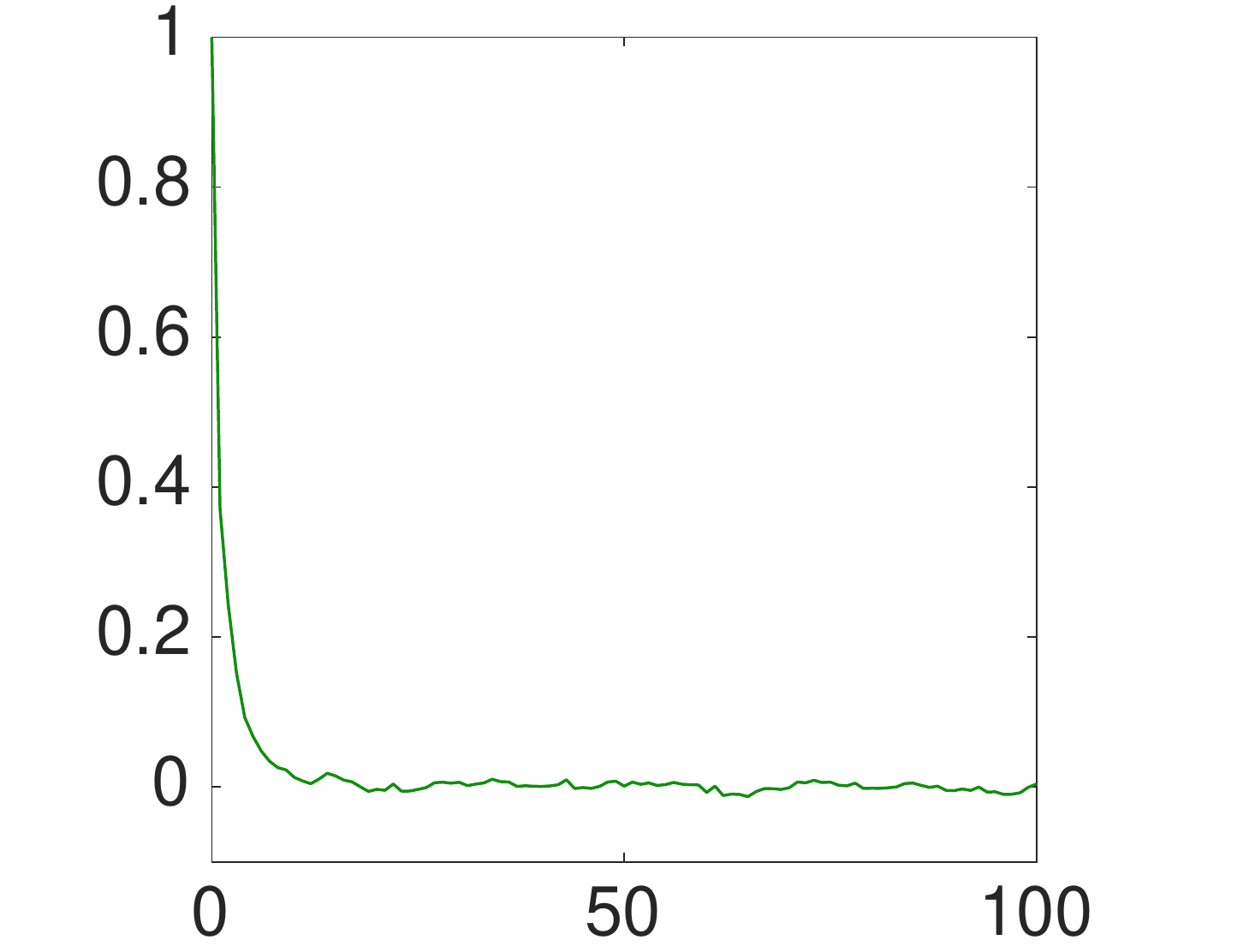}
            \caption{2nd at 0.56}\end{subfigure}\hspace{\pspa}
        \begin{subfigure}[b]{\phei}
            \includegraphics[width=\linewidth]{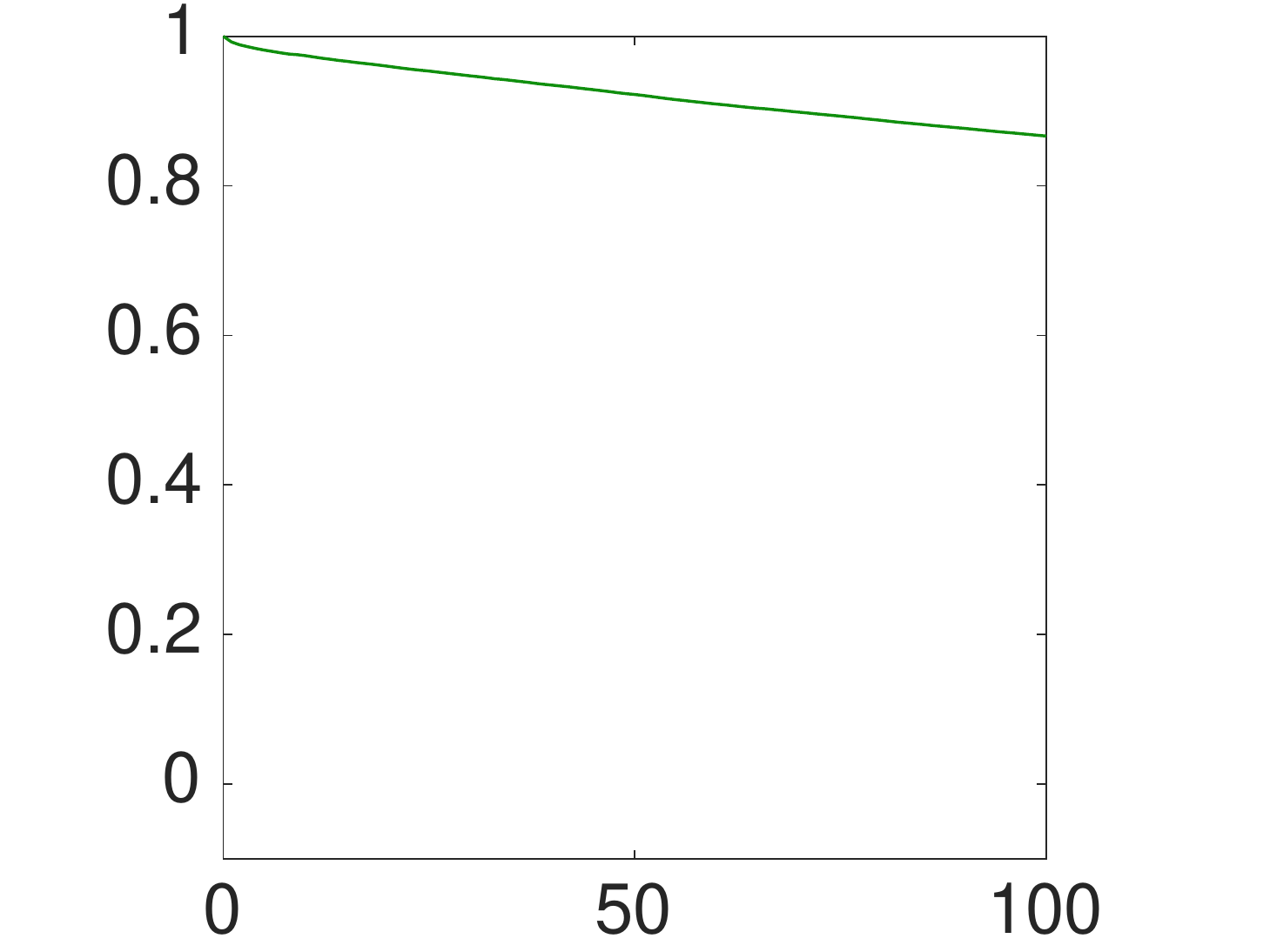}
            \includegraphics[width=\linewidth]{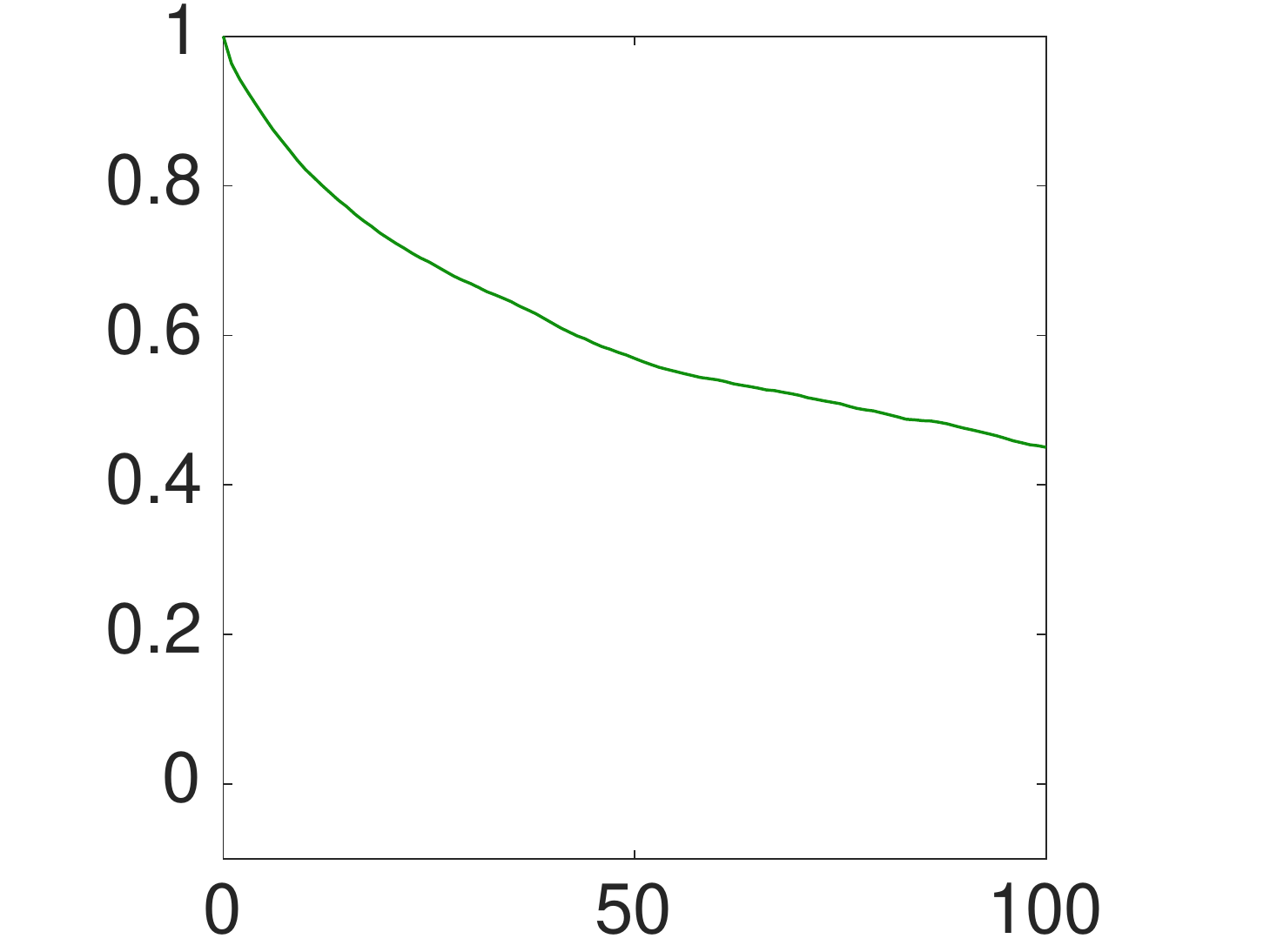}
            \includegraphics[width=\linewidth]{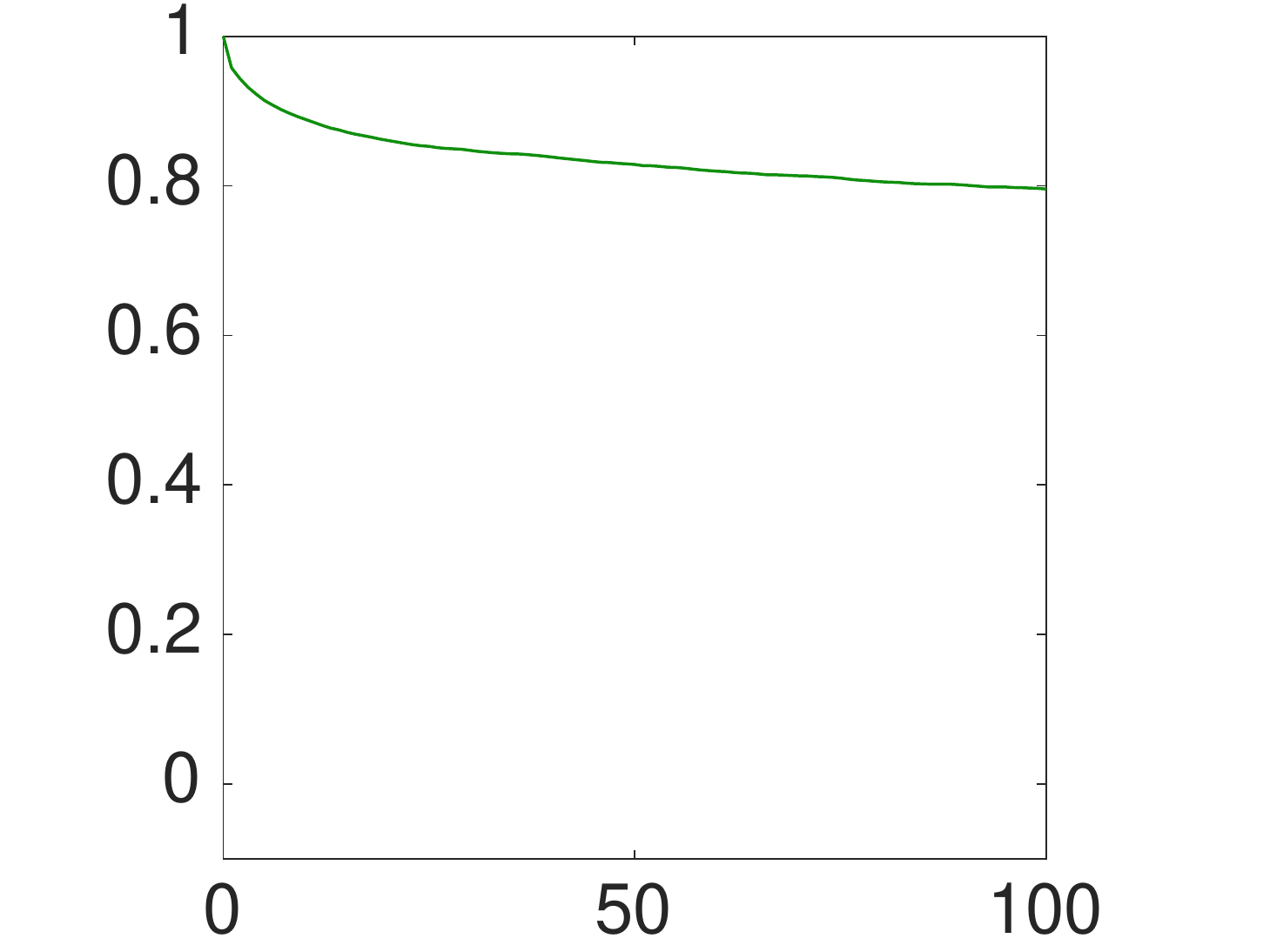}
            \caption{ SPDE at 0.56}\end{subfigure}
        \caption{  Trace plots and autocorrelation functions at two different locations. Top rows: MwG. Middle rows: RAM. Bottom rows: HMC. }
        \label{t1d}
    \end{figure}

    \subsection{Two-dimensional deconvolution}
        \label{ssec:2d}
    
\begin{figure}[!ht]
        \centering
        \begin{subfigure}[b]{\qheis}
            \includegraphics[height=\qheis]{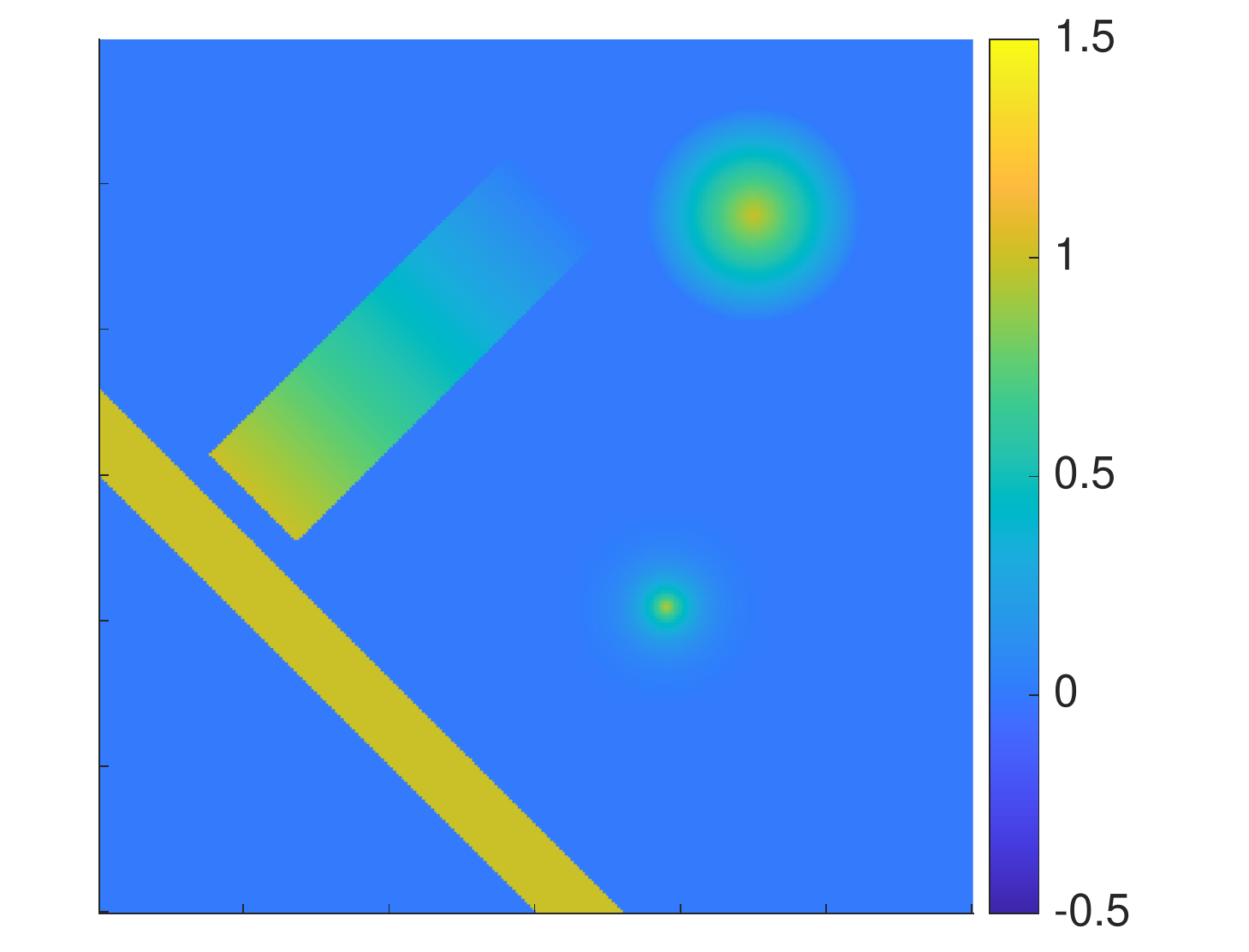}
            \caption{ Ground truth. }\end{subfigure}\hspace{0.5cm}
        \begin{subfigure}[b]{\qheis}
            \includegraphics[height=\qheis]{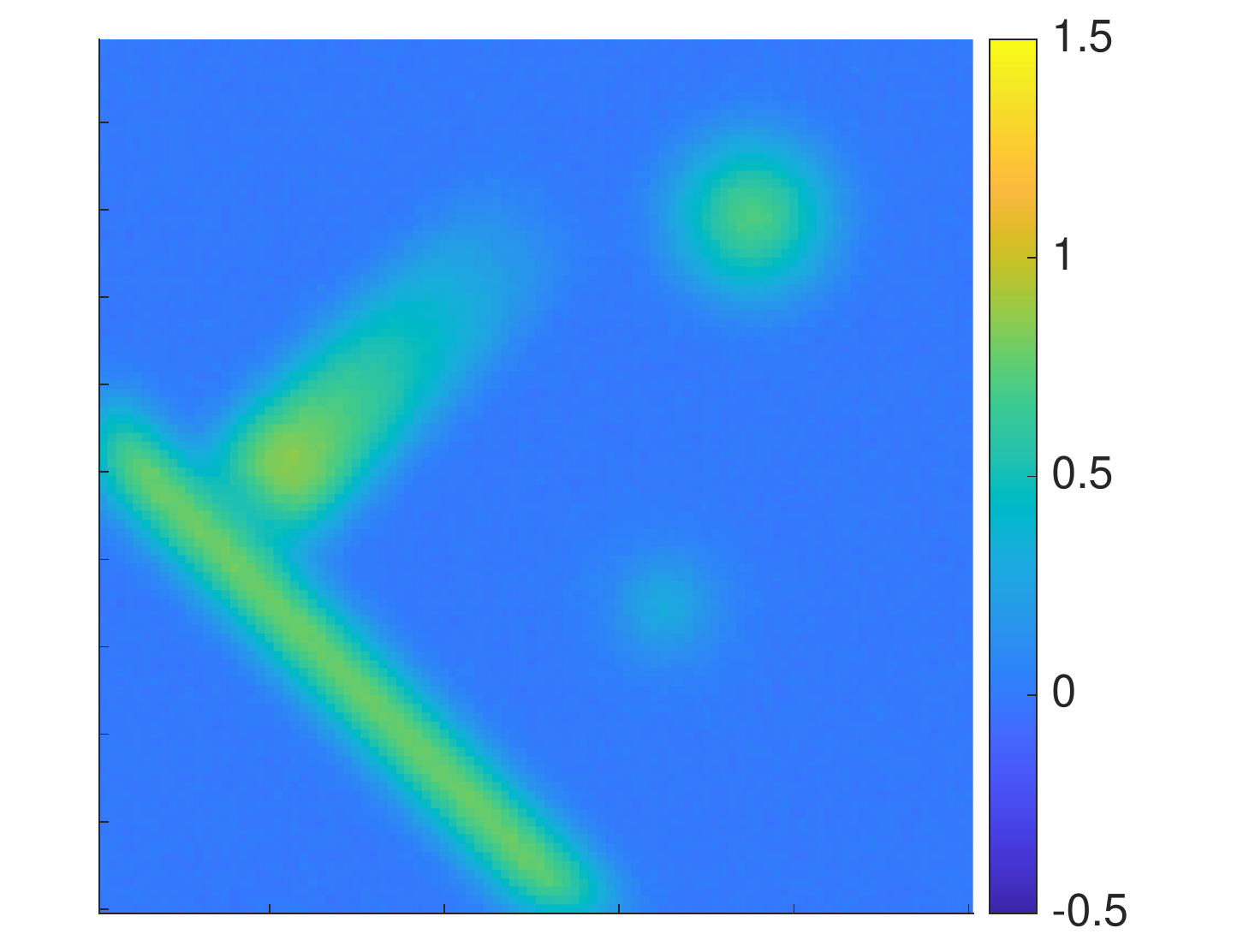}
            \caption{ Measurement data. }\end{subfigure}  
        
        \begin{subfigure}[b]{\qheis}
            \includegraphics[height=\linewidth]{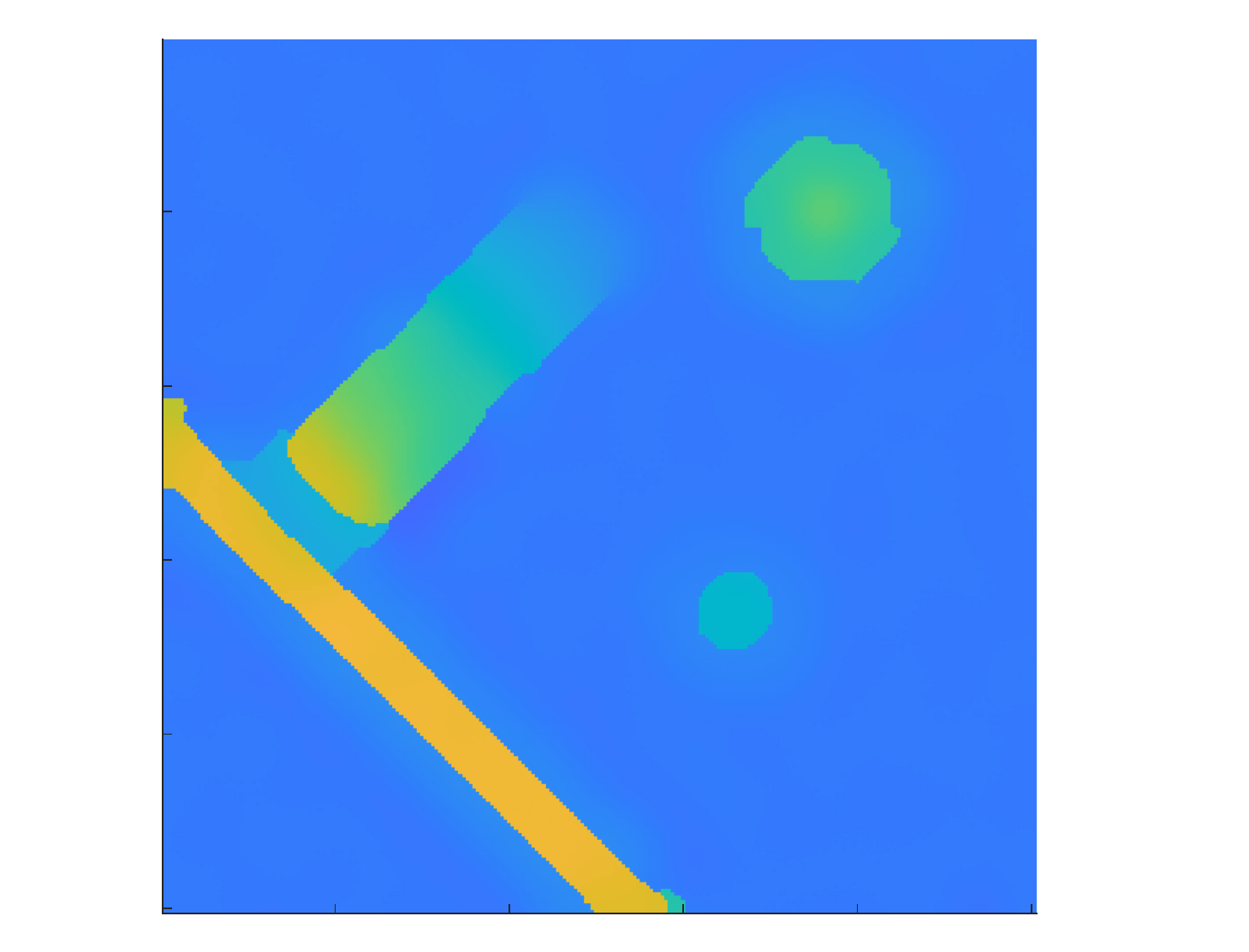}
            \caption{ Isot. 1st order diff. }\end{subfigure}
        \begin{subfigure}[b]{\qheis}
            \includegraphics[height=\linewidth]{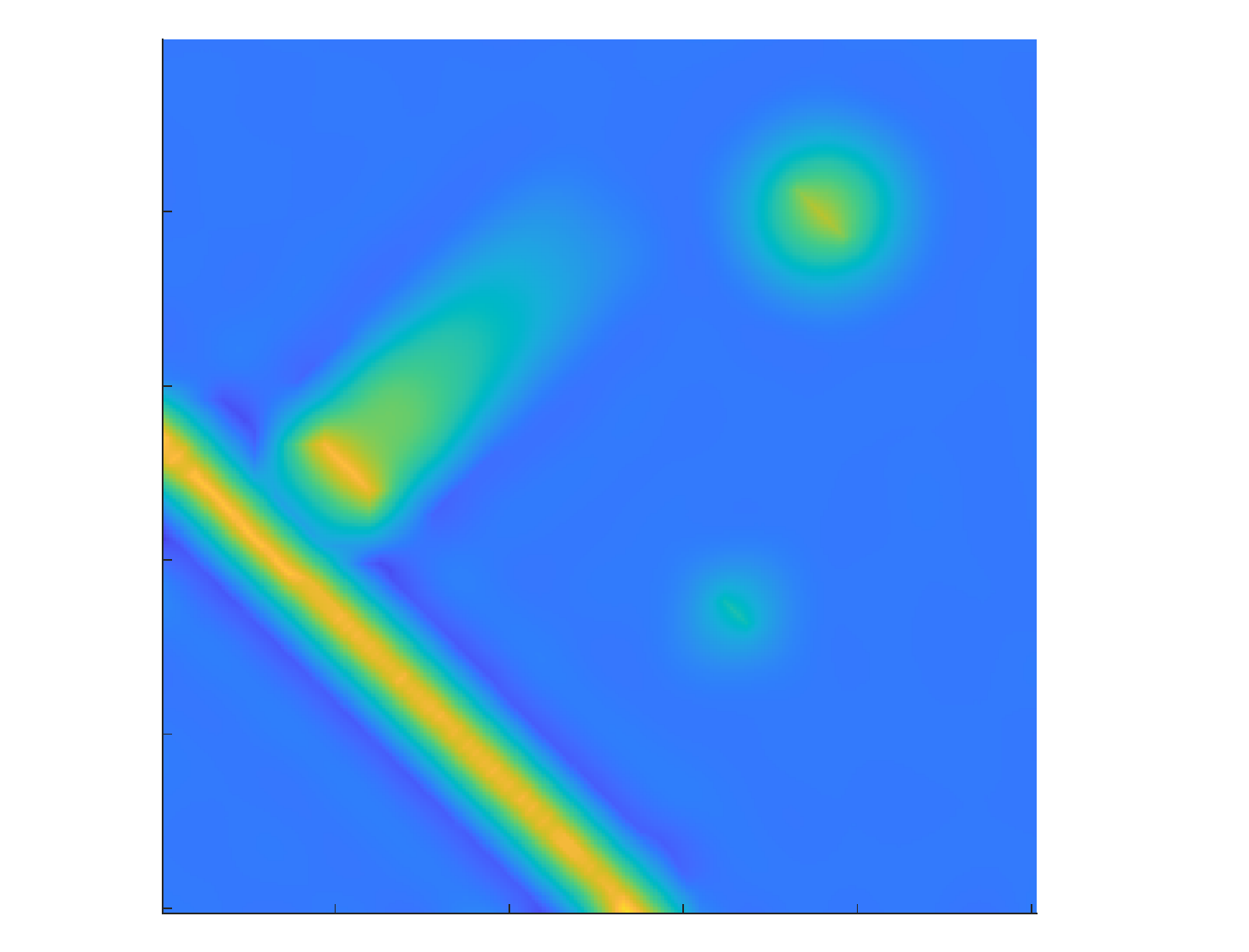}
            \caption{ Isot. 2nd order diff. }\end{subfigure}
        \begin{subfigure}[b]{\qheis}
            \includegraphics[height=\linewidth]{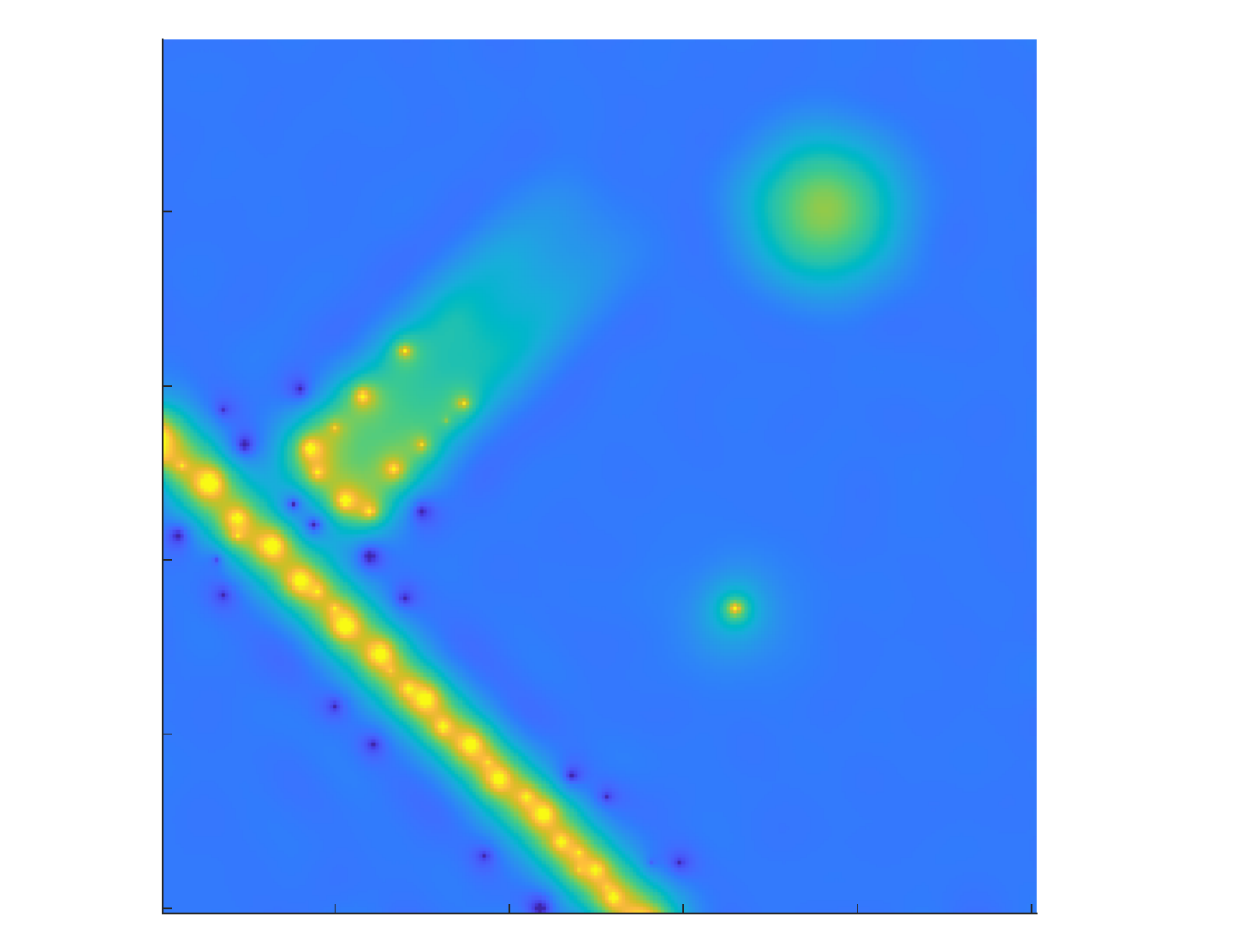}
            \caption{ SPDE. }\end{subfigure}
        
        \begin{subfigure}[b]{\qheis}
            \includegraphics[height=\linewidth]{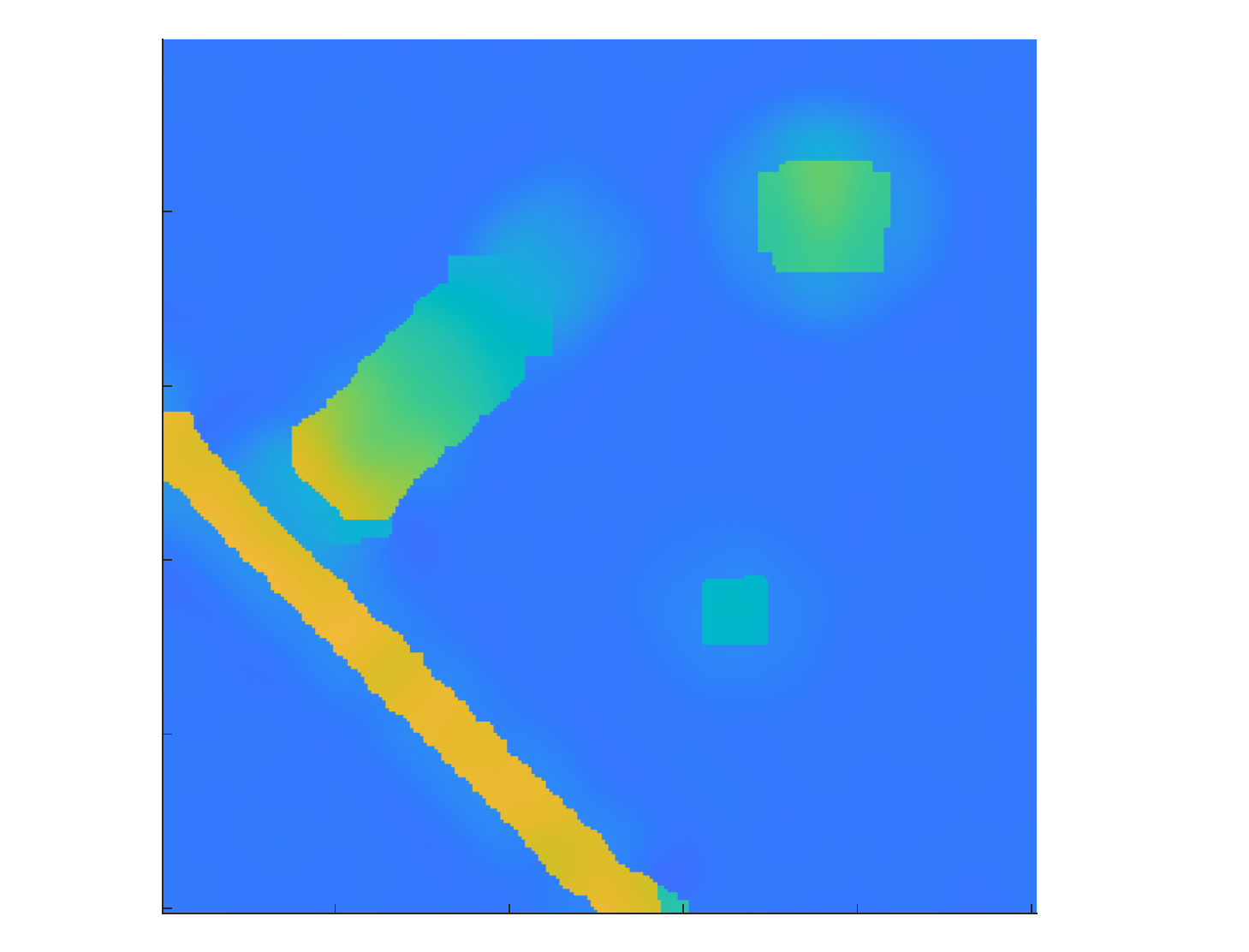}
            \caption{ Anisot. 1st order diff. }\end{subfigure}
        \begin{subfigure}[b]{\qheis}
            \includegraphics[height=\linewidth]{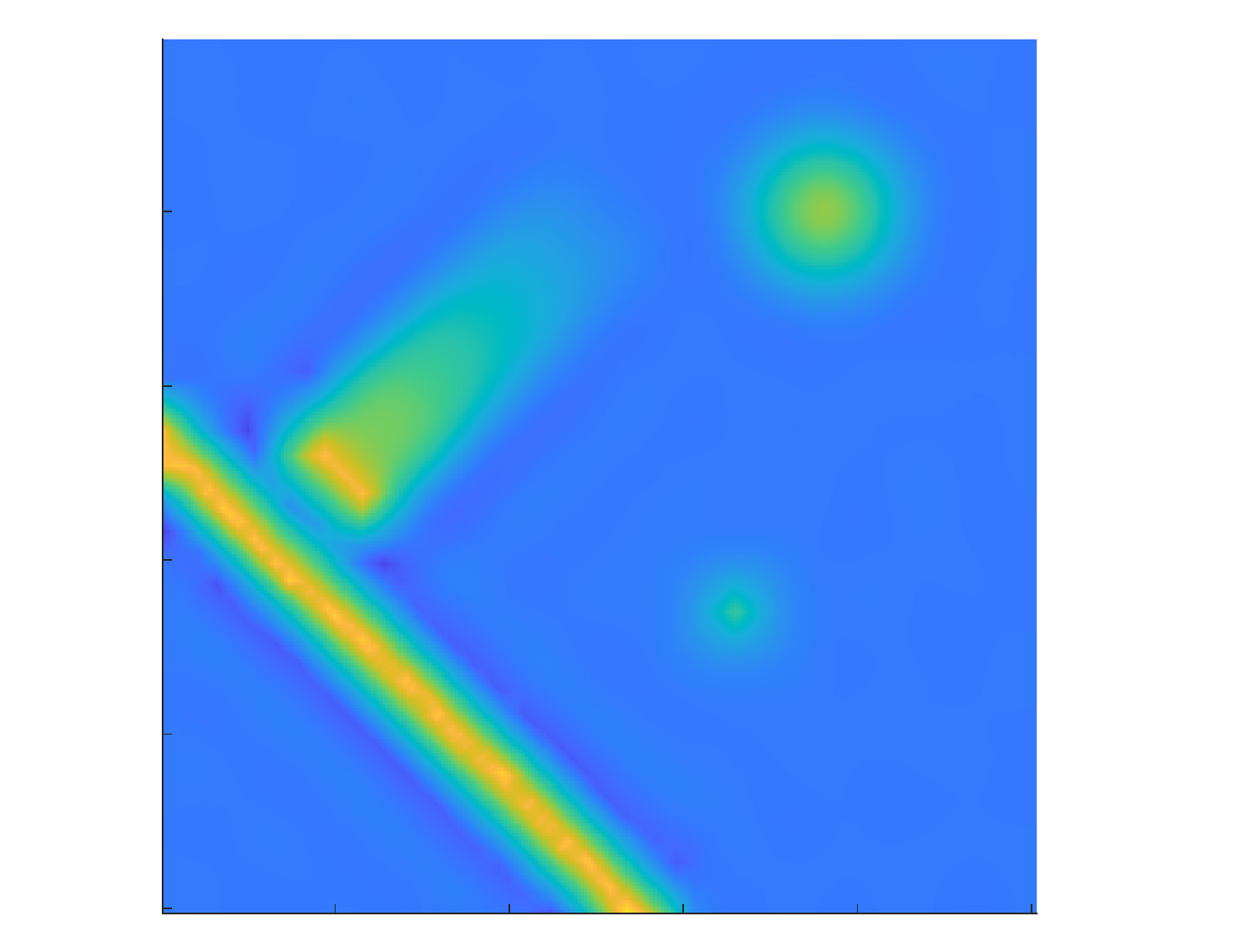}
            \caption{ Anisot. 2nd order diff. }\end{subfigure}
        \begin{subfigure}[b]{\qheis}
            \includegraphics[height=\linewidth]{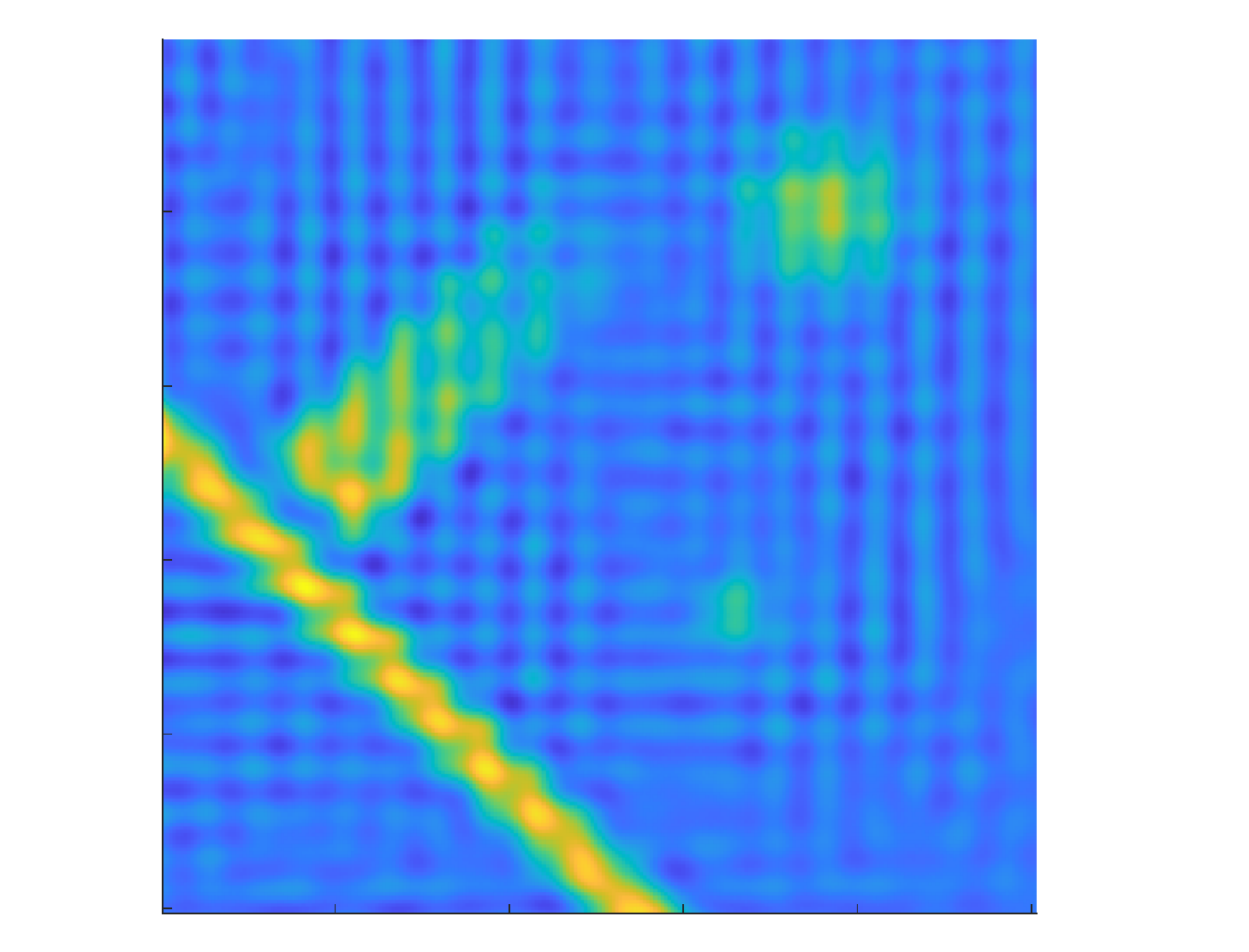}
            \caption{ Sheet. }\end{subfigure}
        
        \begin{subfigure}[b]{\qheis}
            \includegraphics[height=\linewidth]{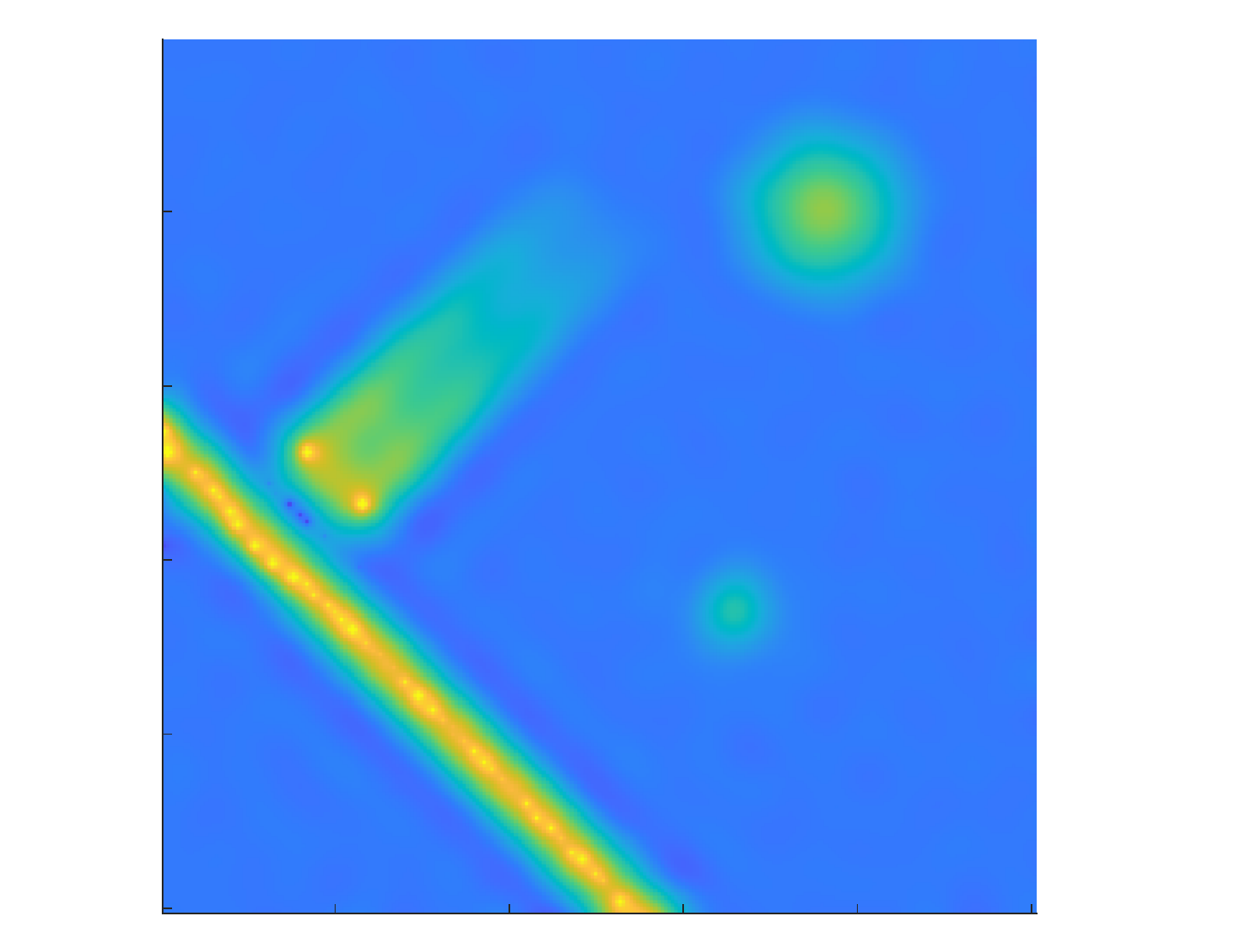}
            \caption{SPDE, Laplace only.}\end{subfigure}
        \caption{ MAP estimates with different priors. }
        \label{twomap}
    \end{figure}

    The second numerical example deals with a deconvolution problem in spatial dimensions of two, that is, it is deconvolution example with $r\in\mathbb{R}^2$.
    We compare six different Cauchy priors: anisotropic and isotropic first and second order difference priors, the SPDE prior with Cauchy noise and the Cauchy sheet prior. 
    The test function of the experiment consists of a diagonal strip with constant value, a diagonally decaying rectangle, an exponential peak and a cone. The test function, measurement data and the MAP estimators are plotted in \Cref{twomap}.
    
    The measurement data was generated using a grid of equispaced $300\times300$ pixels,  and employing  a  convolution matrix corresponding to the grid. As the convolution kernel, a Gaussian kernel in \eqref{kernel} with $s=1/150$. The convoluted  data was extracted in an equispaced grid of $100\times100$  pixels, and the data was added Gaussian white noise with variance of $\sigma^2=0.01^2$. Finally, the reconstruction grid size was set to $256\times256$.
    
    We set the parameter $\lambda =0.03$ for both the anisotropic and the isotropic first order Cauchy difference priors. Likewise, the prior parameter was given value of $\lambda=0.005$ for both of the second order difference priors. The parameter $\lambda$ of the Cauchy sheet prior was set to $0.07$.  The SPDE prior parameters were specified as  $\ell=0.01$, and  the Cauchy noise was set to have scale of $3.0$. The boundary term parameter of all the difference  priors was set to  $\gamma'=1.0$, as well as the parameter $\gamma$ of the first order differences in the boundary terms of the second order priors. The discretization parameter $h$ of the SPDE prior matrix was set to $1/128$.  We also setup a Laplace-only variant of the SPDE prior in the MAP estimation for illustration purposes by removing the identity operator from the SPDE, and setting the  parameter  $\ell=-6.1\cdot10^{-3}$, and the Cauchy noise scale to $3.0$. 
    
    RAM and adaptive MwG were ran to generate a chain of 200\,000 samples, of which half were used for adaptation and discarded from inference. In other words, 13,107,200,000 overall proposals were generated in a full run of both algorithms thanks to the Gibbs sampling. NUTS was ran for 2000 iterations, of which 1000 were discarded as adaptation and burn-in. The maximum tree doubling parameter of NUTS was set to 8. We tested using the parameter value of 10 in sampling of the first order difference priors, but mixing of the chains was almost identical between the chains generated with different maximum tree doublings, and hence we stick with the lower value for the final experiments. Chains generated by RAM and MwG  were thinned by a factor of 200.  
    We ran 5 chains with each method for PSRF diagnostics, and used the first chain of  each of the MCMC algorithm and posterior combinations for the CM and marginal variance estimation.

    The first and second order difference priors produce qualitatively similar estimates as they do in the one-dimensional case. The first order difference priors excel at reconstructing the constant valued stripe at the bottom left corner It it not a surprise that the isotropic variant of the prior is better to the anisotropic one. The anisotropic prior suffers from clearly coordinate-aligned features, what can be seen well from the reconstruction of the exponential peak.
    
    On the other hand, there is not so great distinction between the second order difference priors, although the MAP estimator of the anisotropic prior seems to have more cross-shaped details near the stripe. The MAP estimator of the SPDE prior behaves qualitatively the same as it does in the one-dimensional case, and it recovers the exponential peak well, again. The Laplace-only variant of the SPDE prior is isotropic, but does not favor features having constant first derivatives, like the second order difference priors do. Instead, the MAP estimator of the Laplace-only variant  of the SPDE prior looks a bit same as the posterior of the SPDE prior with both identity and Laplace operators, but the lack of the identity part in the variant makes the spikes in the estimator less prevalent.
    The MAP estimator of the Cauchy sheet prior is corrupted with wave artifacts that are aligned to the coordinate axis. It is known that the sheet priors are not isotropic, but this major artifacts are excessive. 
    
\begin{figure}
        \centering
        \begin{subfigure}[b]{\qhei}
            \includegraphics[width=\linewidth]{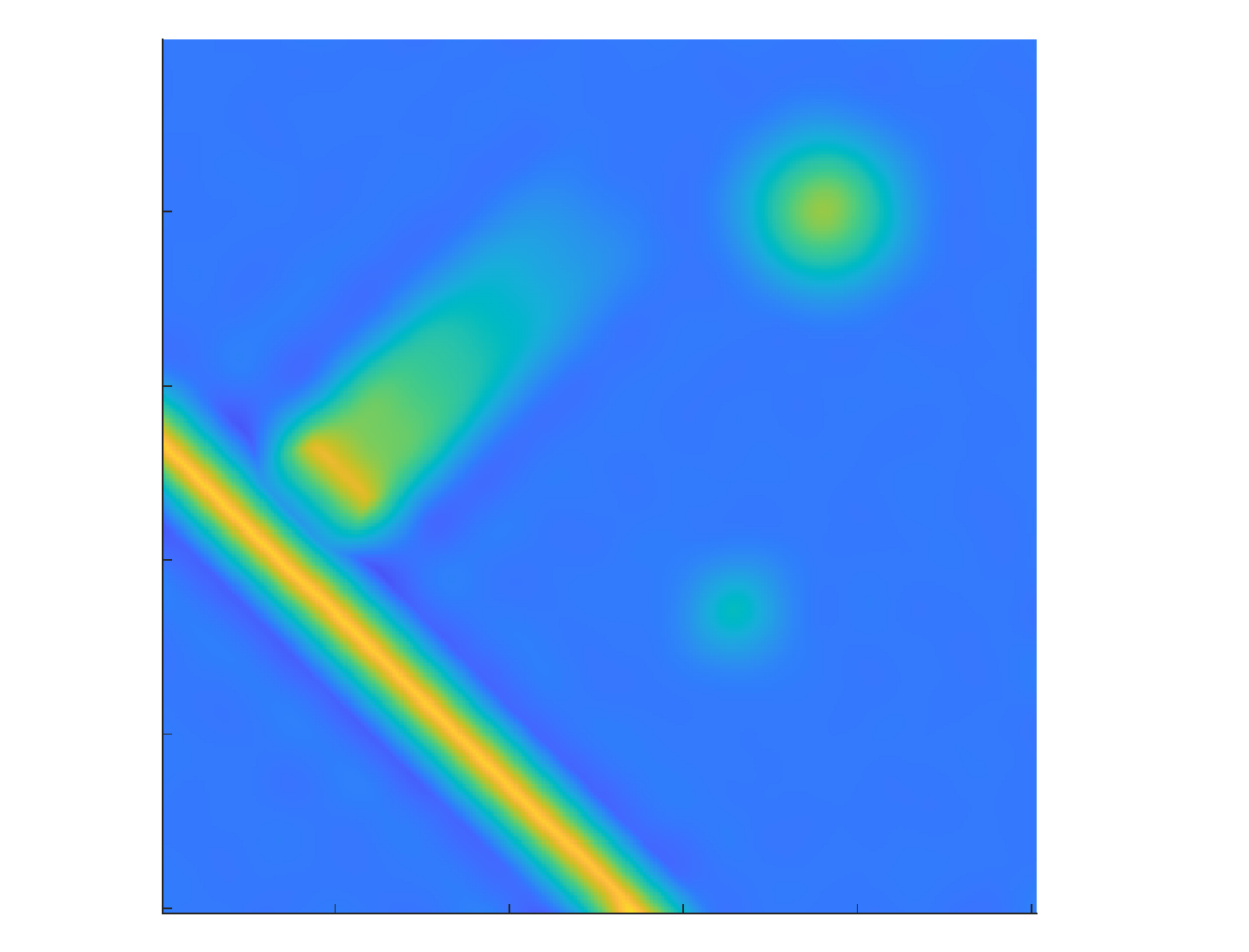}
            \includegraphics[width=\linewidth]{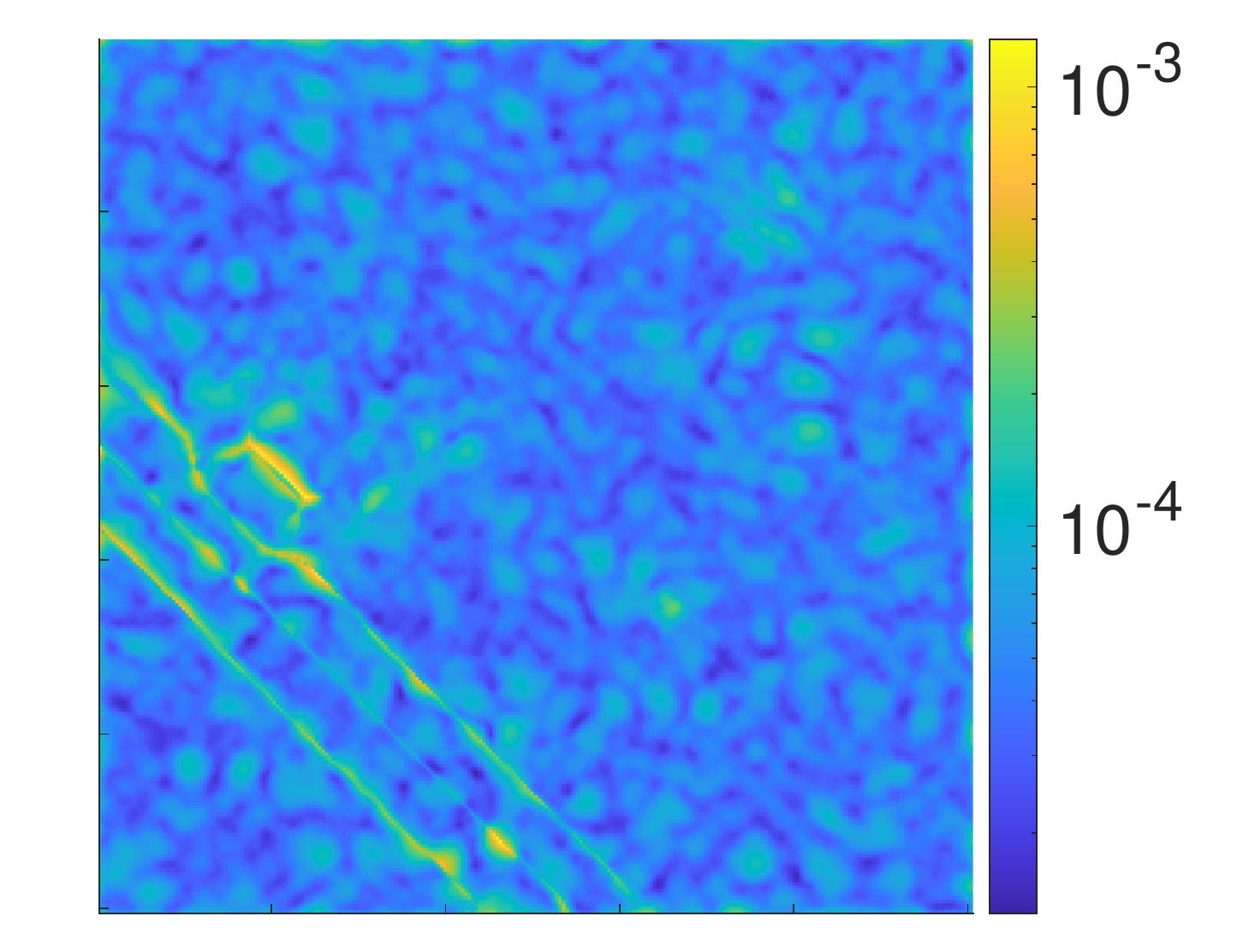}
            \includegraphics[width=\linewidth]{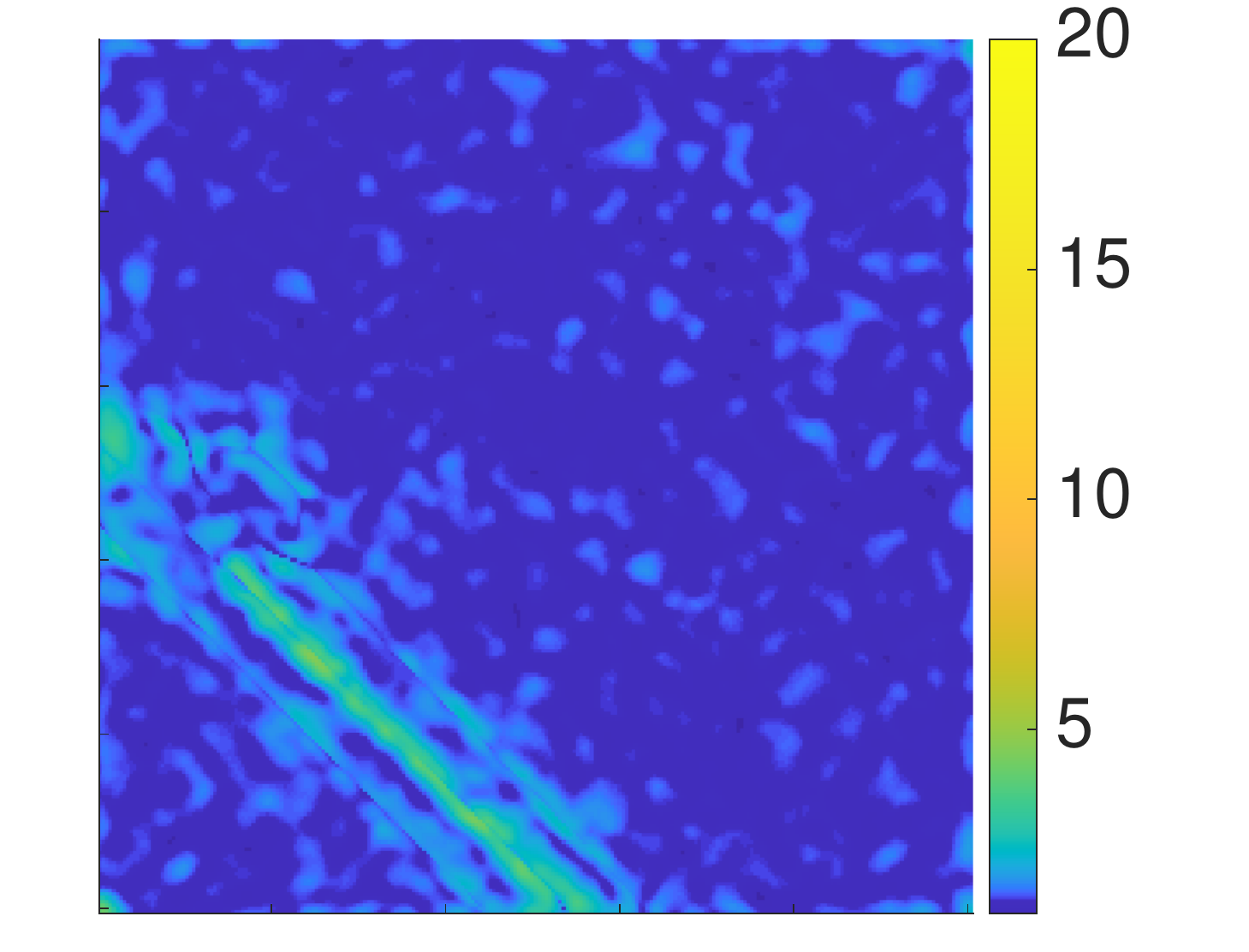}
            \caption{Isotropic,  MwG}\label{isod2_mwg_mean}\end{subfigure}
        \begin{subfigure}[b]{\qhei}
            \includegraphics[width=\linewidth]{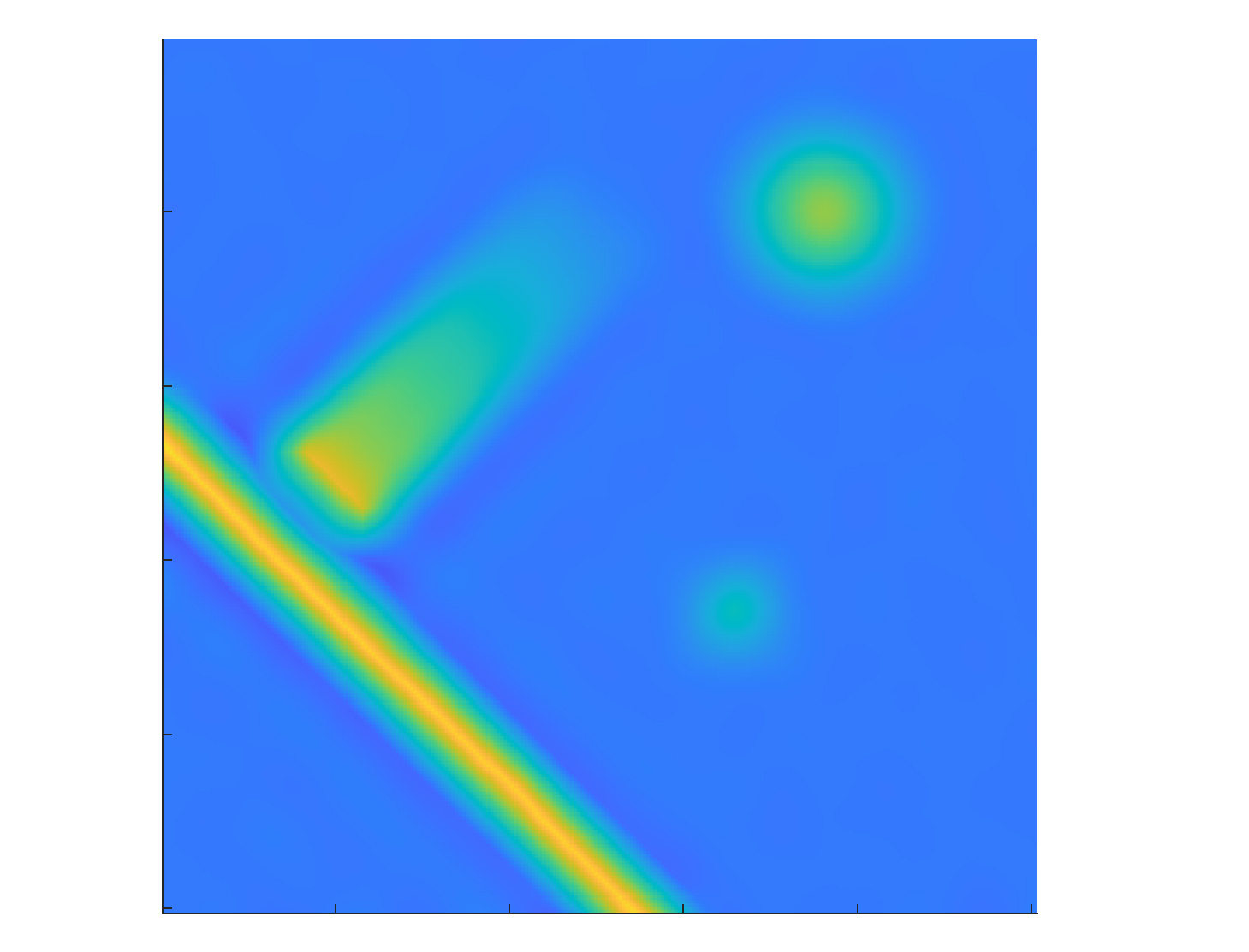}
            \includegraphics[width=\linewidth]{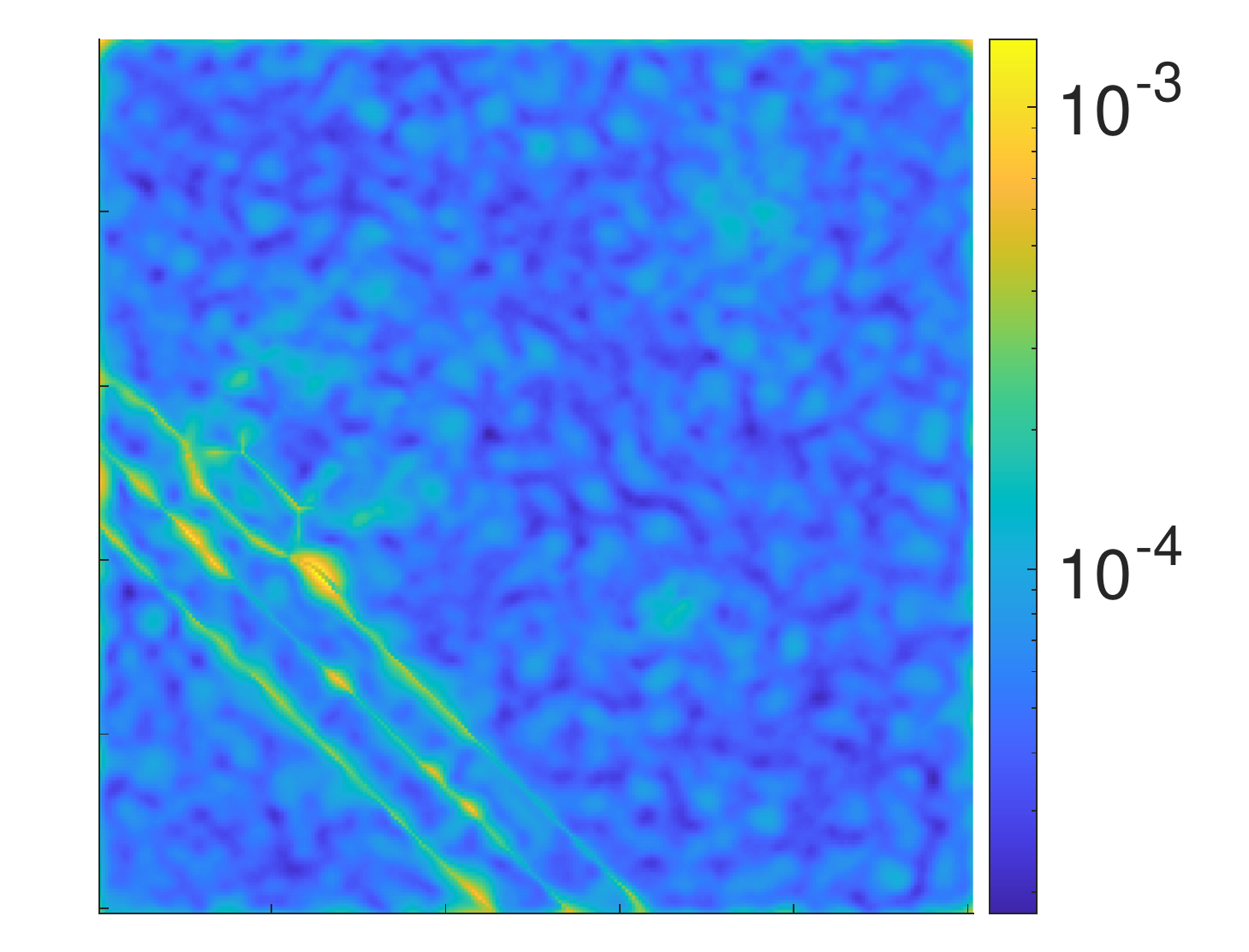}
            \includegraphics[width=\linewidth]{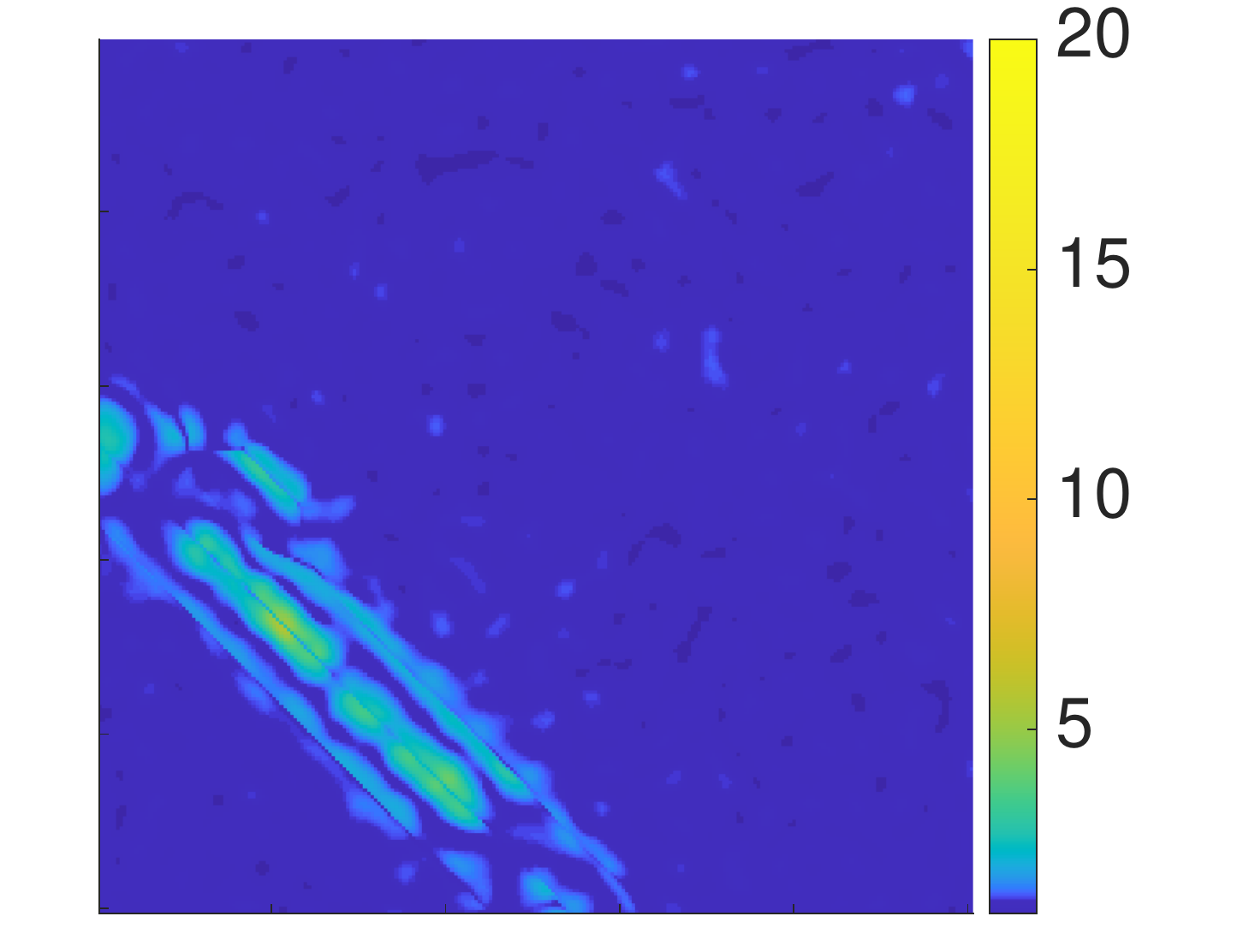}
            \caption{Isotropic, NUTS }\label{isod2_nuts_mean}\end{subfigure}
        \begin{subfigure}[b]{\qhei}
            \includegraphics[width=\linewidth]{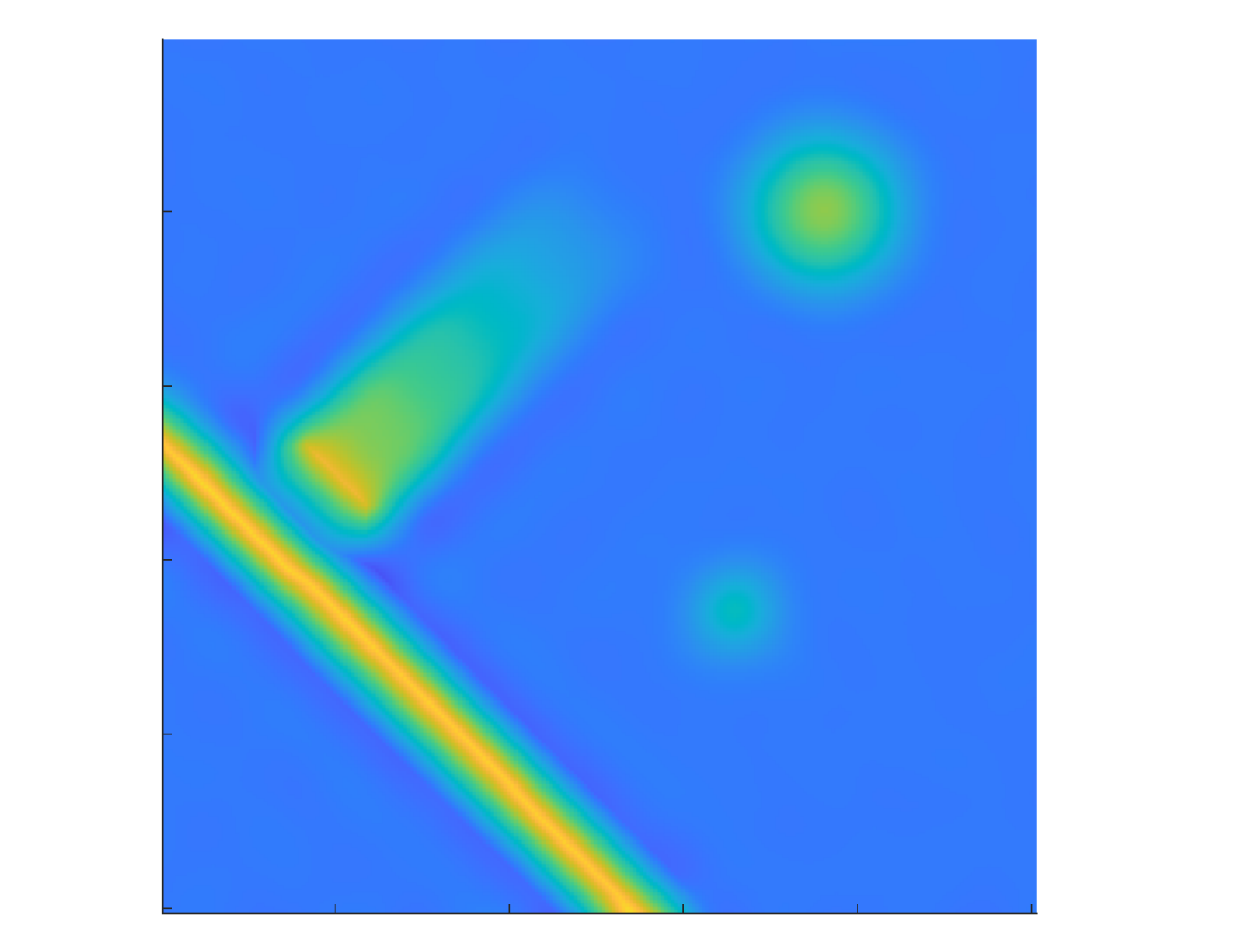}
            \includegraphics[width=\linewidth]{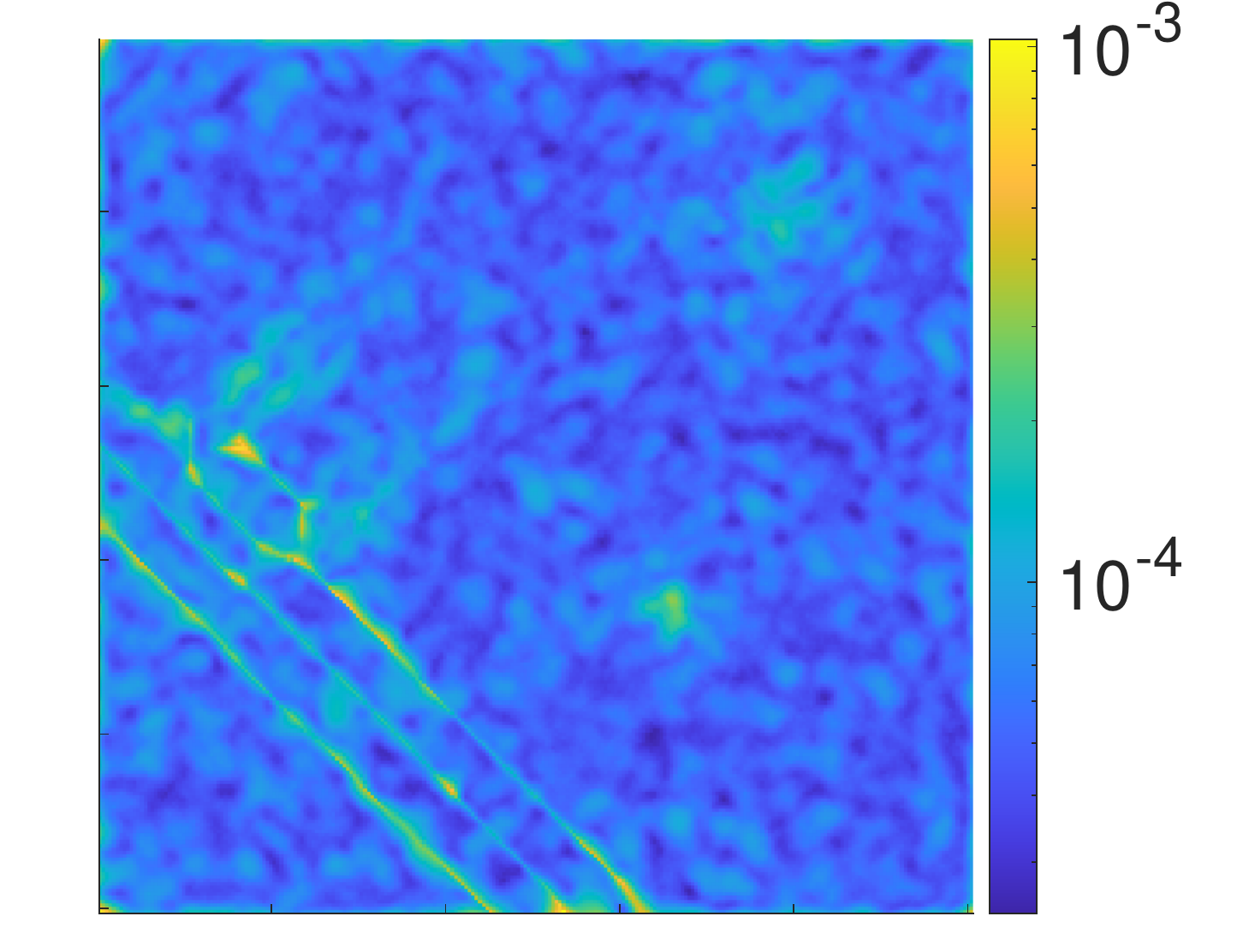}
            \includegraphics[width=\linewidth]{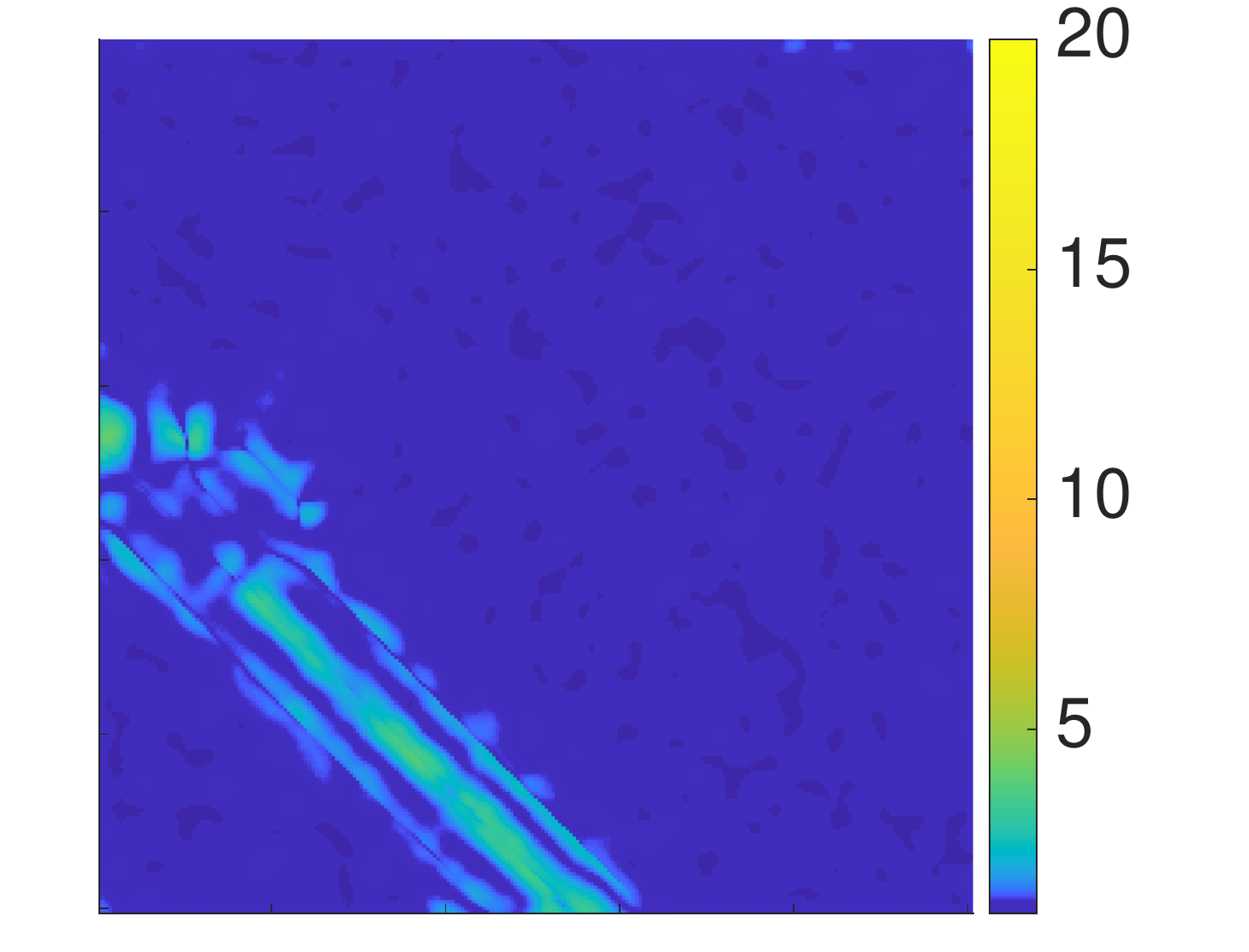}
            \caption{Isotropic, RAM}\label{isod2_ram_mean}\end{subfigure}
        \begin{subfigure}[b]{\qhei}
            \includegraphics[width=\linewidth]{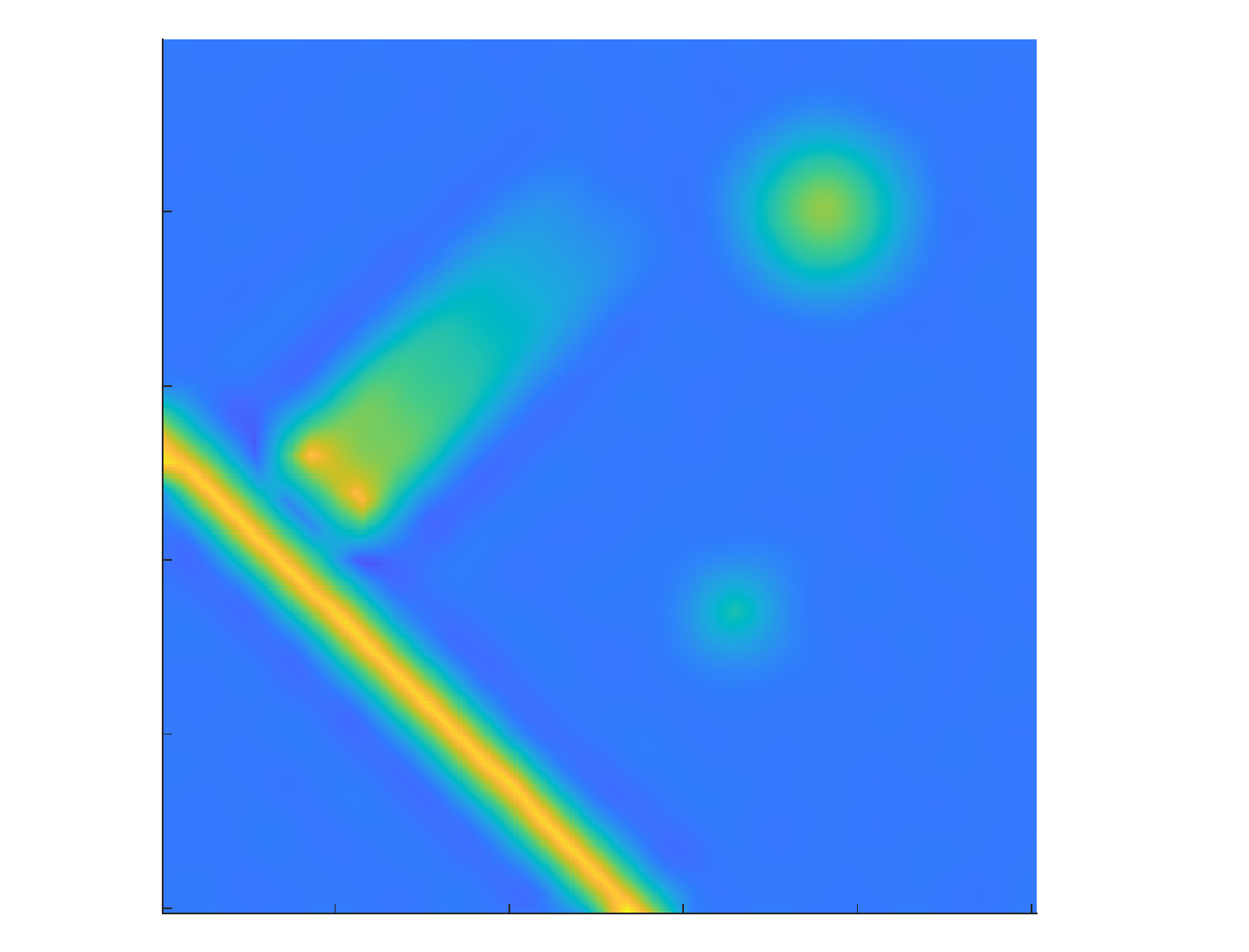}
            \includegraphics[width=\linewidth]{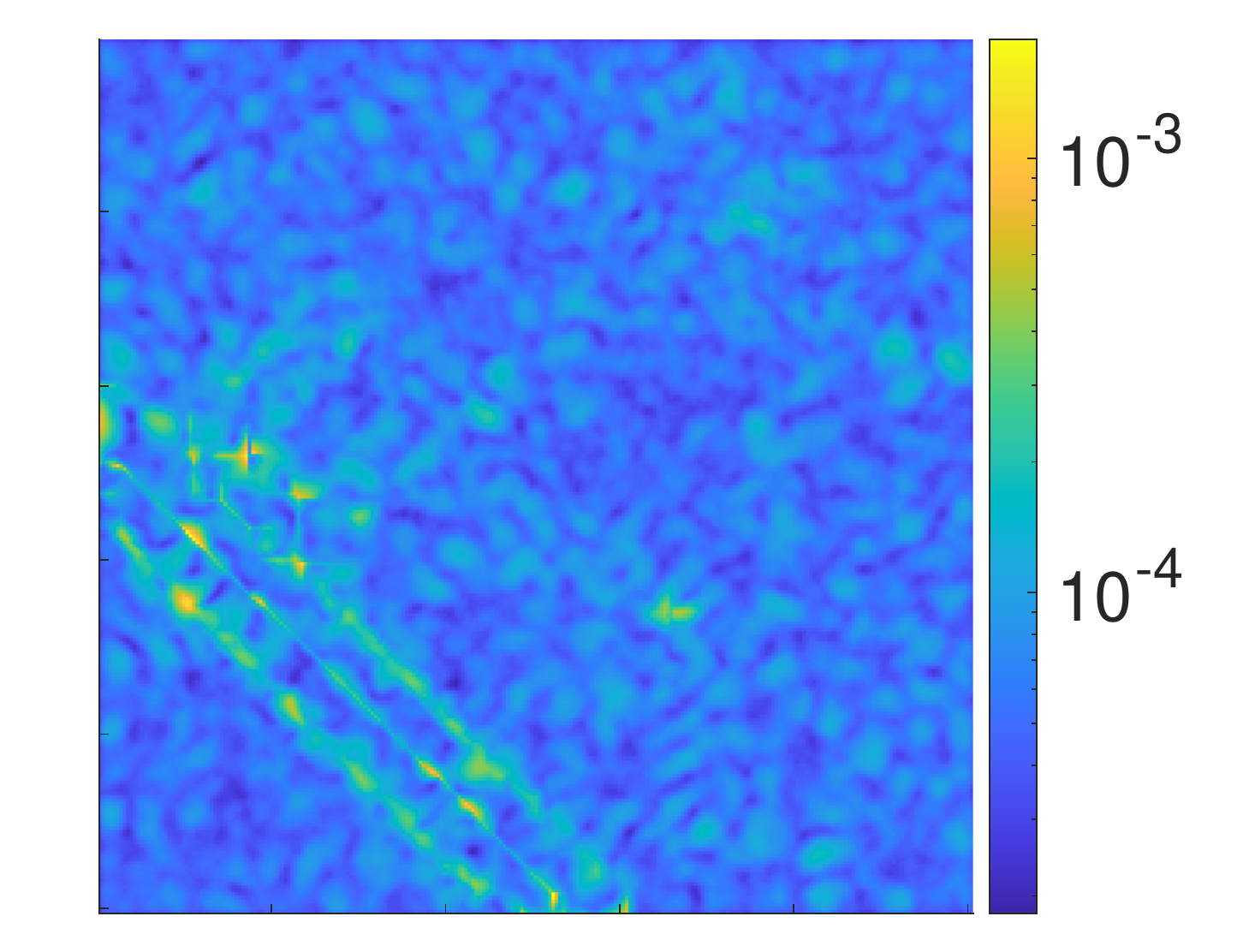}
            \includegraphics[width=\linewidth]{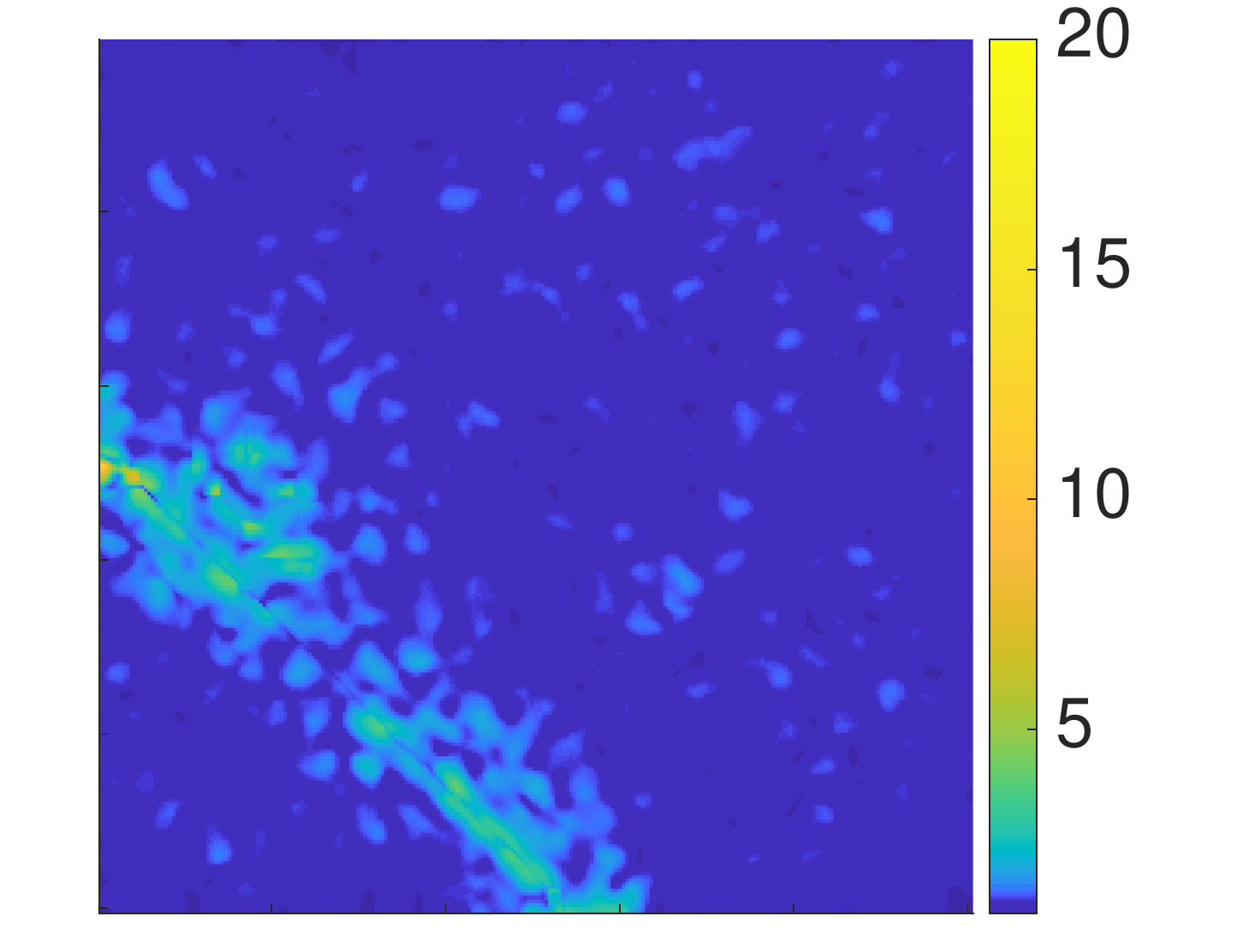}
            \caption{Anisotropic, MwG }\label{anisod2_mwg_mean}\end{subfigure}
        \begin{subfigure}[b]{\qhei}
            \includegraphics[width=\linewidth]{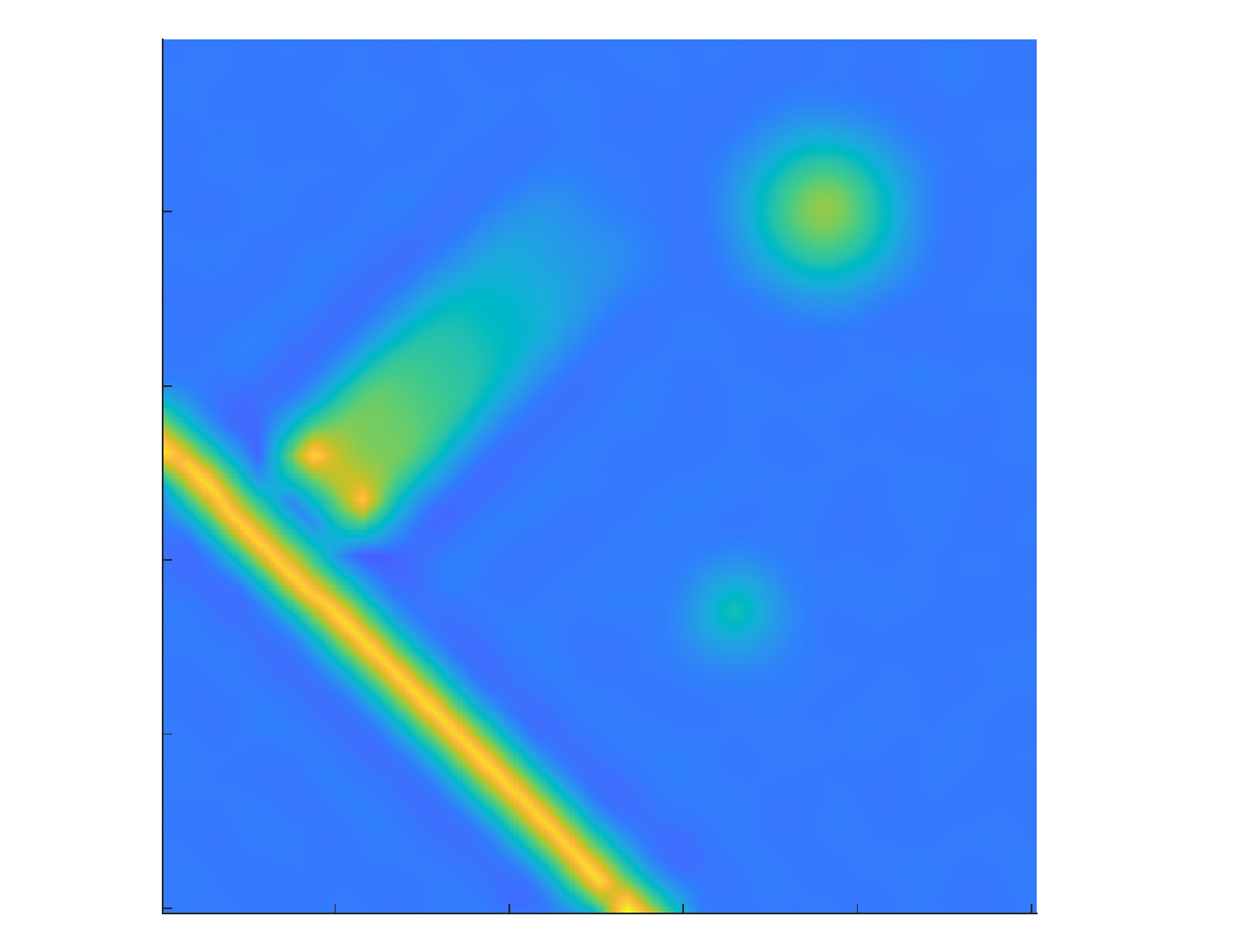}
            \includegraphics[width=\linewidth]{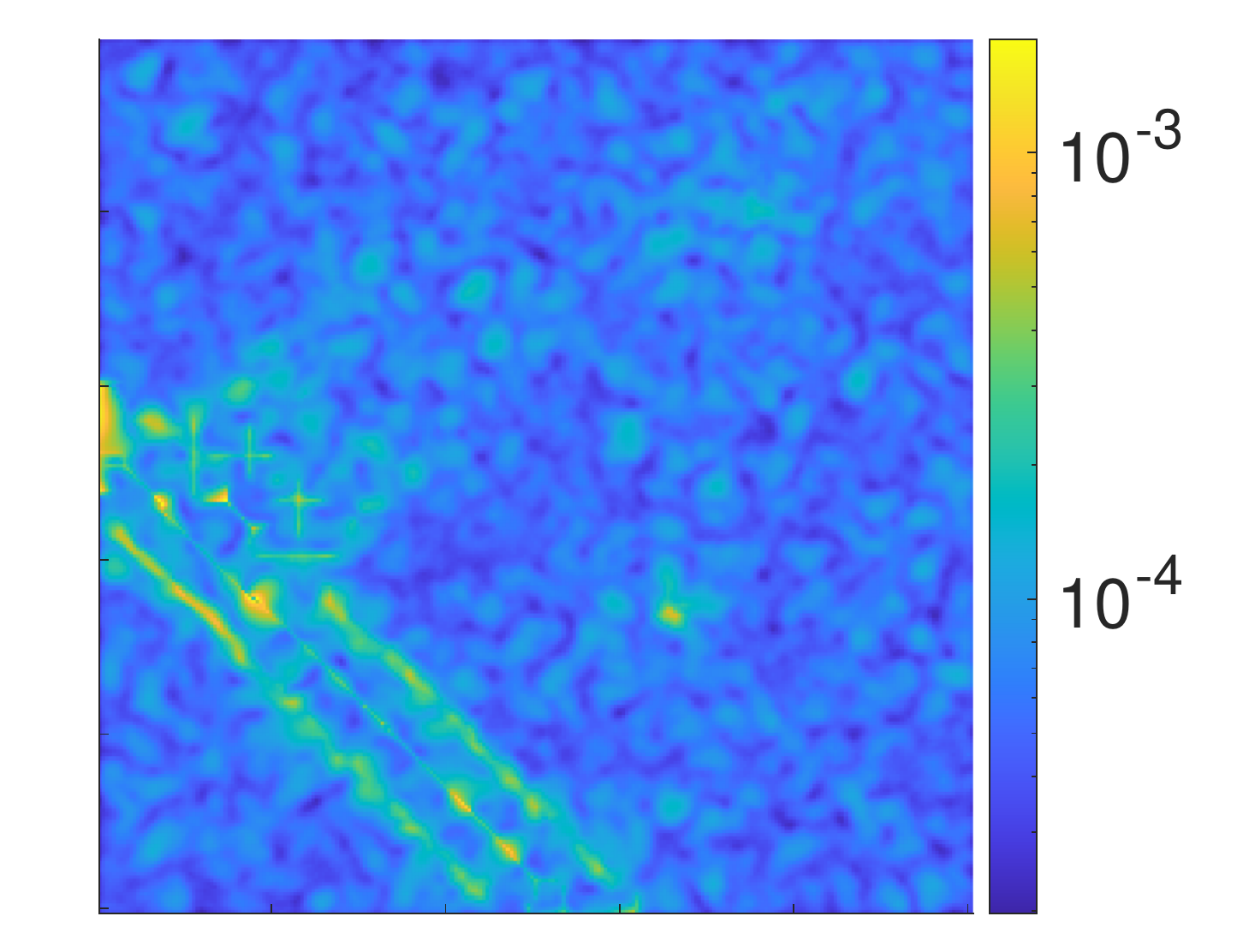}
            \includegraphics[width=\linewidth]{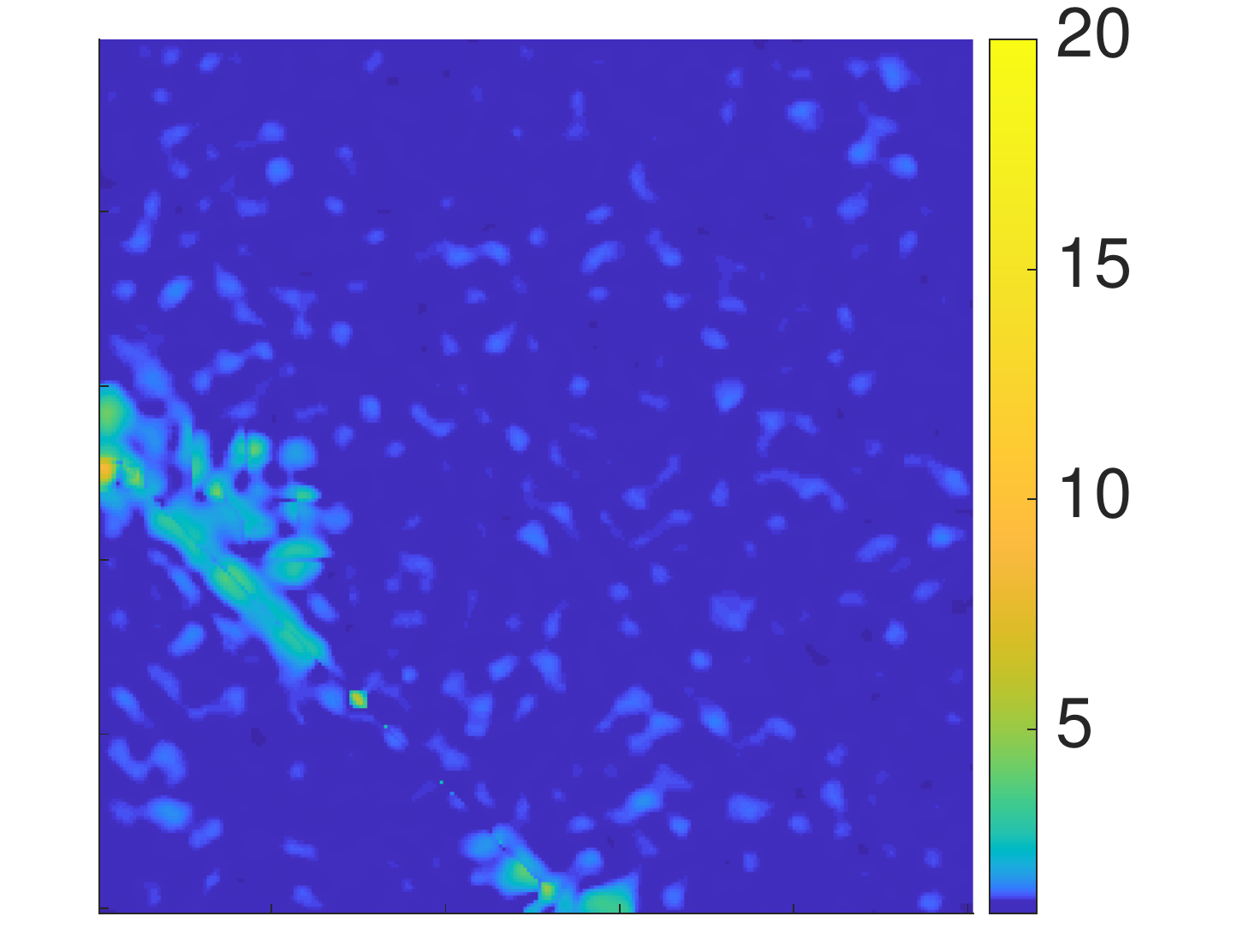}
            \caption{Anisotropic, NUTS }\label{anisod2_nuts_mean}\end{subfigure}
        \begin{subfigure}[b]{\qhei}
            \includegraphics[width=\linewidth]{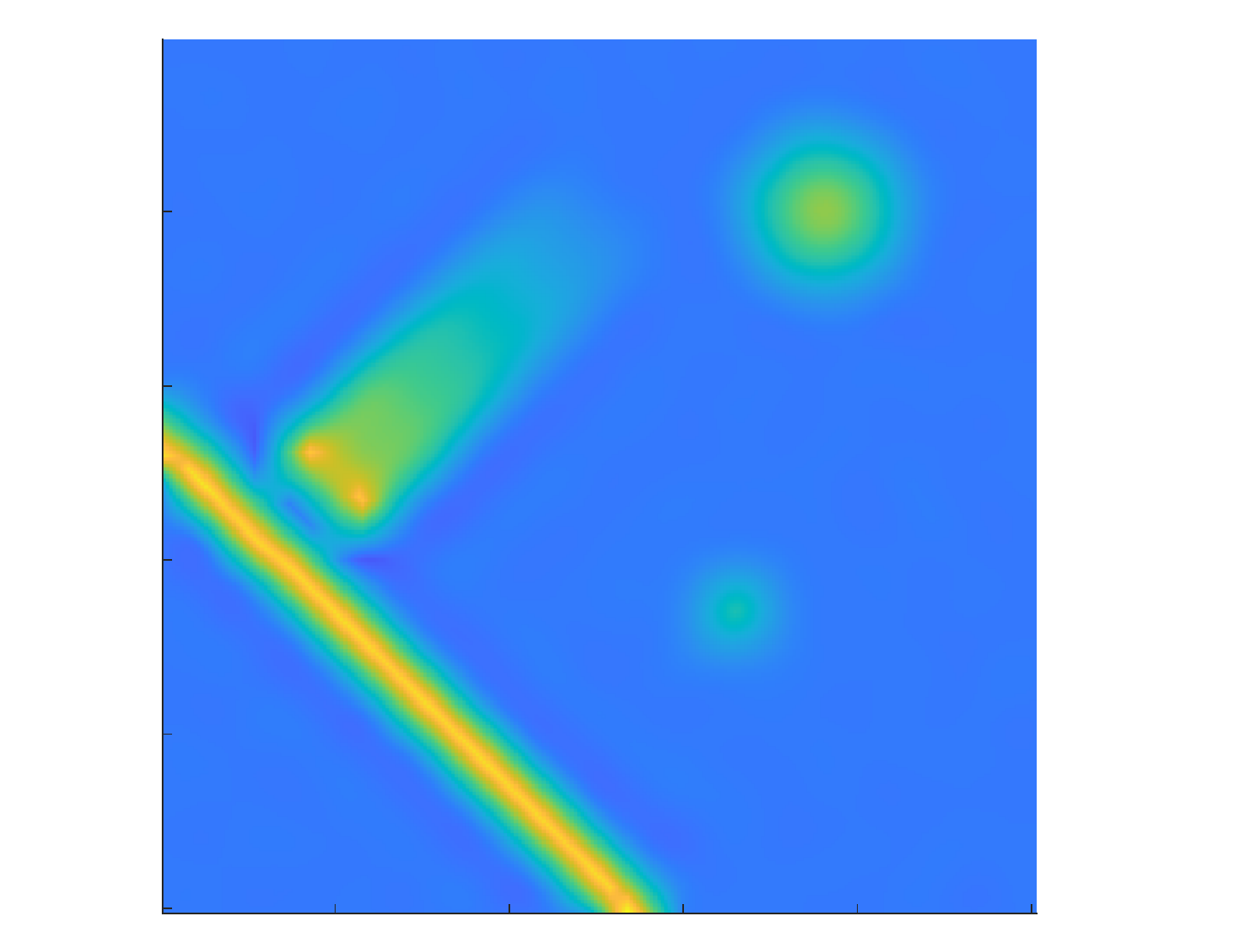}
            \includegraphics[width=\linewidth]{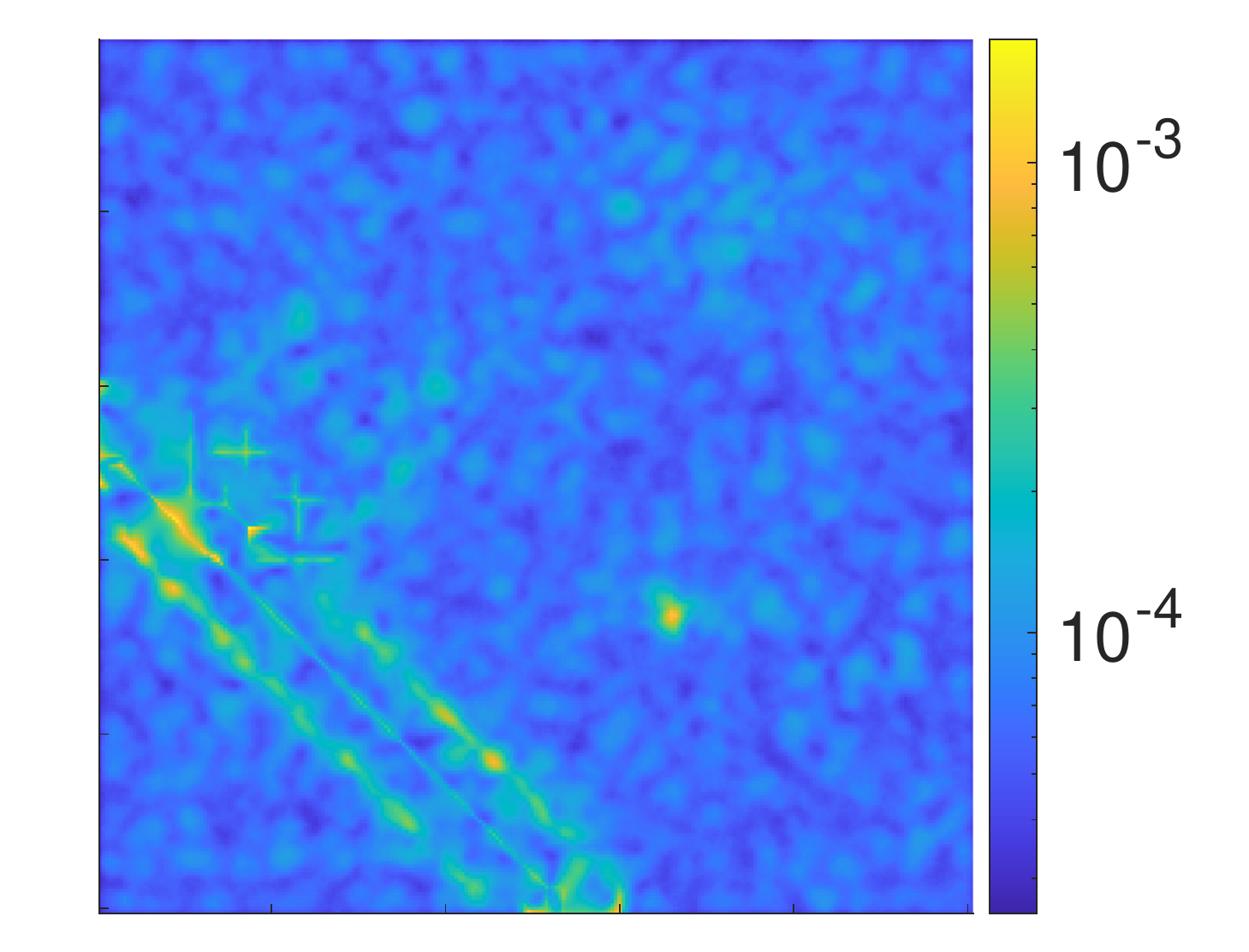}
            \includegraphics[width=\linewidth]{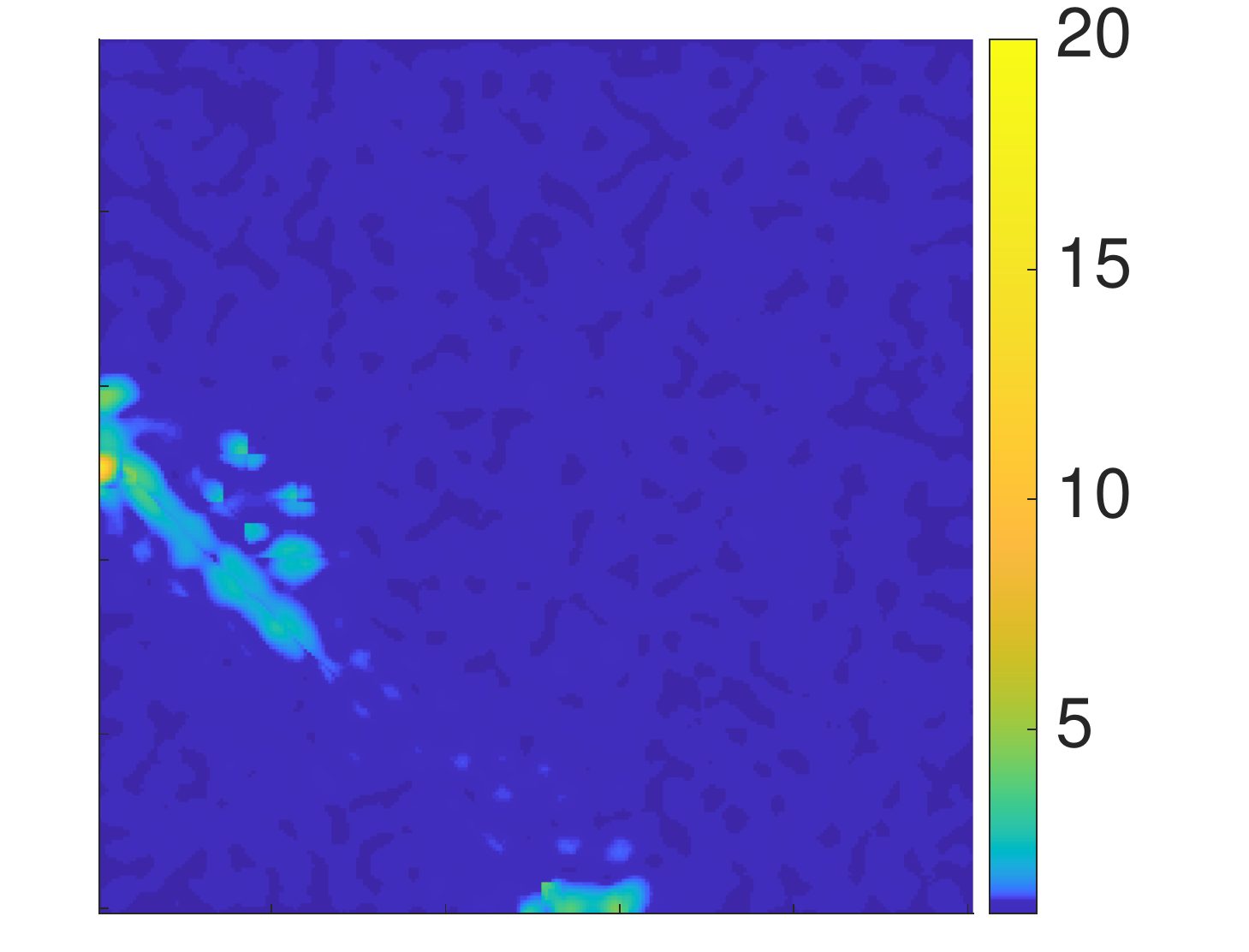}
            \caption{Anisotropic, RAM }\label{anisod2_ram_mean}\end{subfigure}

        \caption{ Second order Cauchy priors -- Top row:  mean estimates. Middle row: Variance estimates. Bottom row: PSRF convergence  diagnostics. }
        \label{andiff2d}
    \end{figure}

\begin{figure}
        \centering
        \begin{subfigure}[b]{\qhei}
            \includegraphics[width=\linewidth]{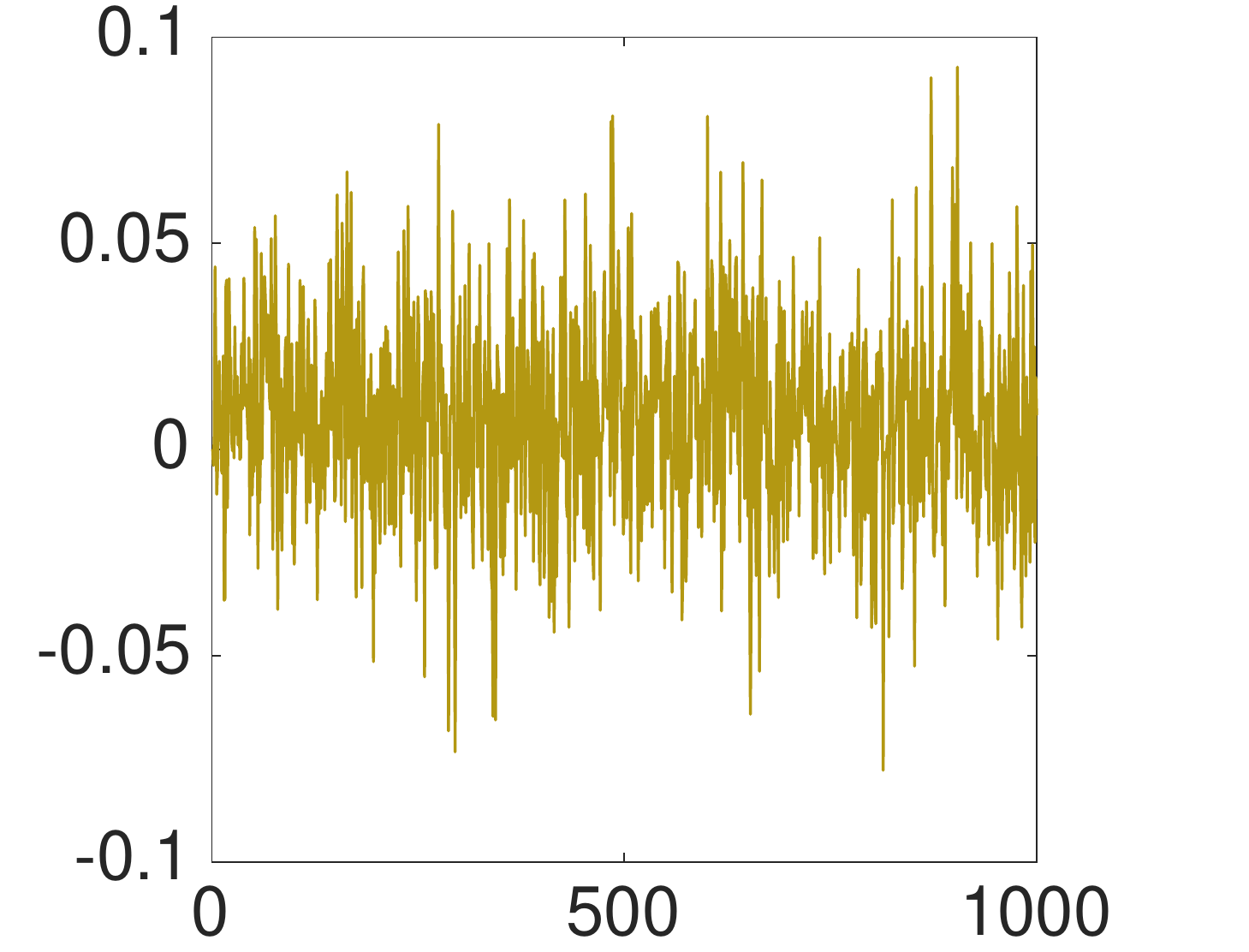}
            \includegraphics[width=\linewidth]{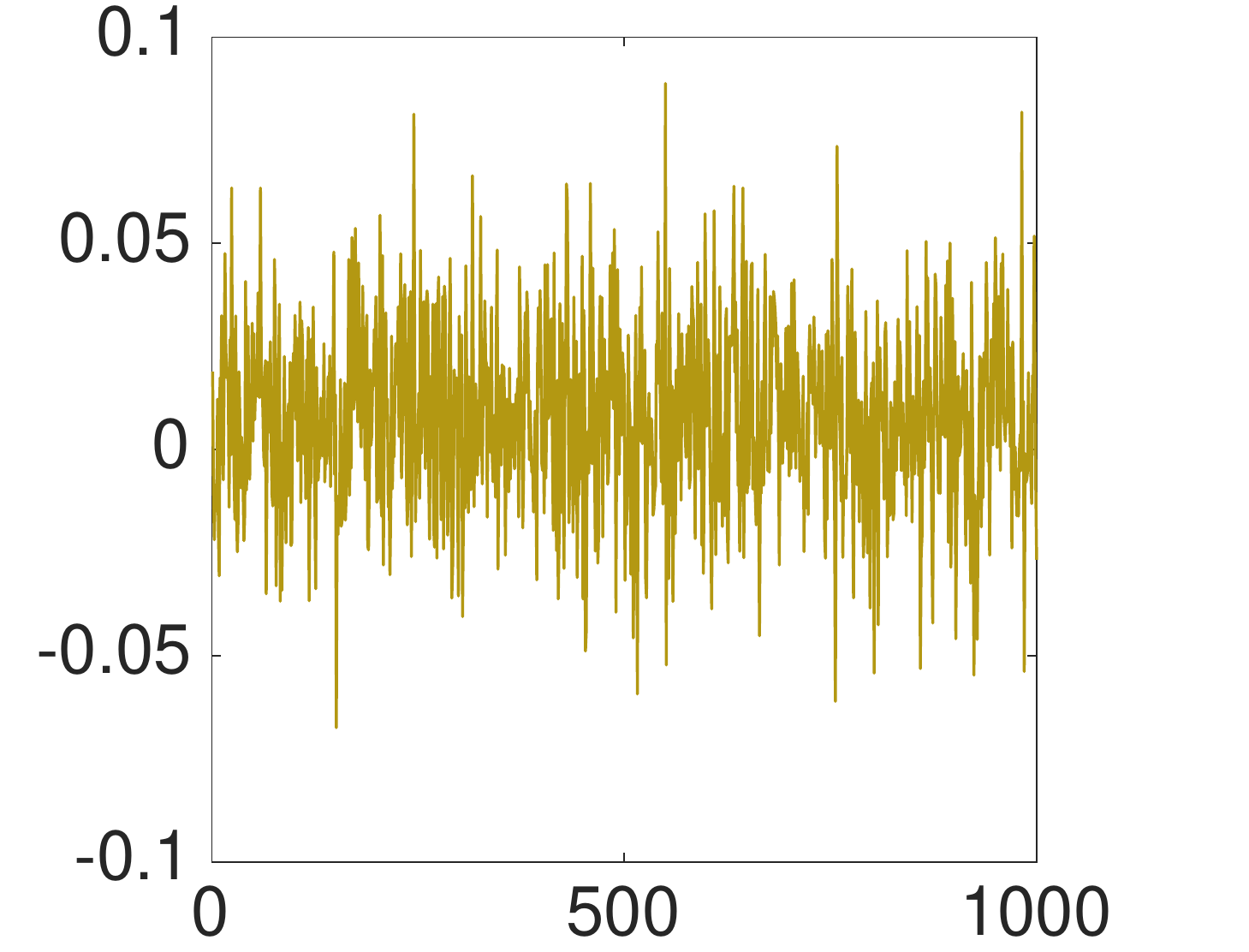}
            \includegraphics[width=\linewidth]{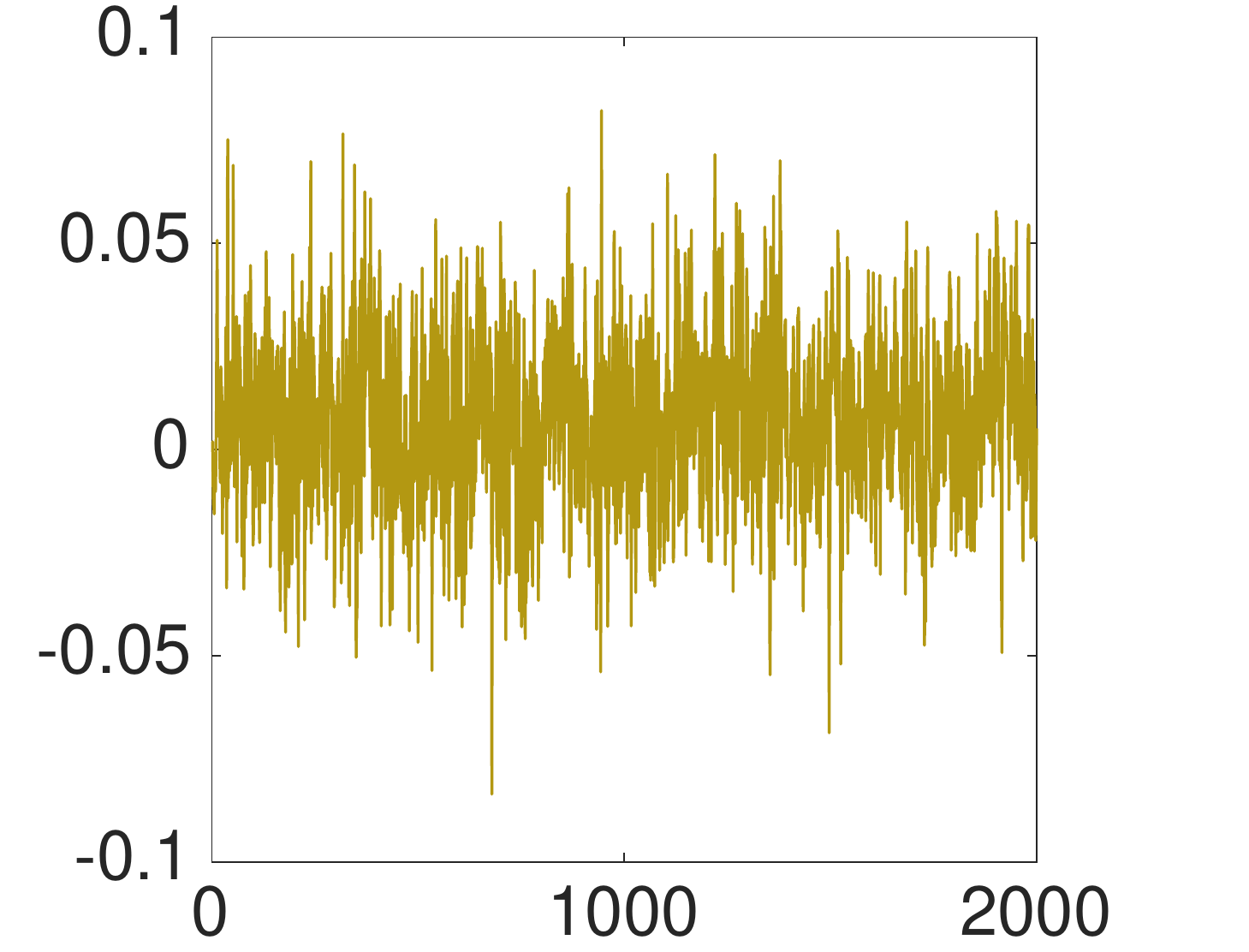}
            \caption{Aniso 1st }\end{subfigure}
        \begin{subfigure}[b]{\qhei}
            \includegraphics[width=\linewidth]{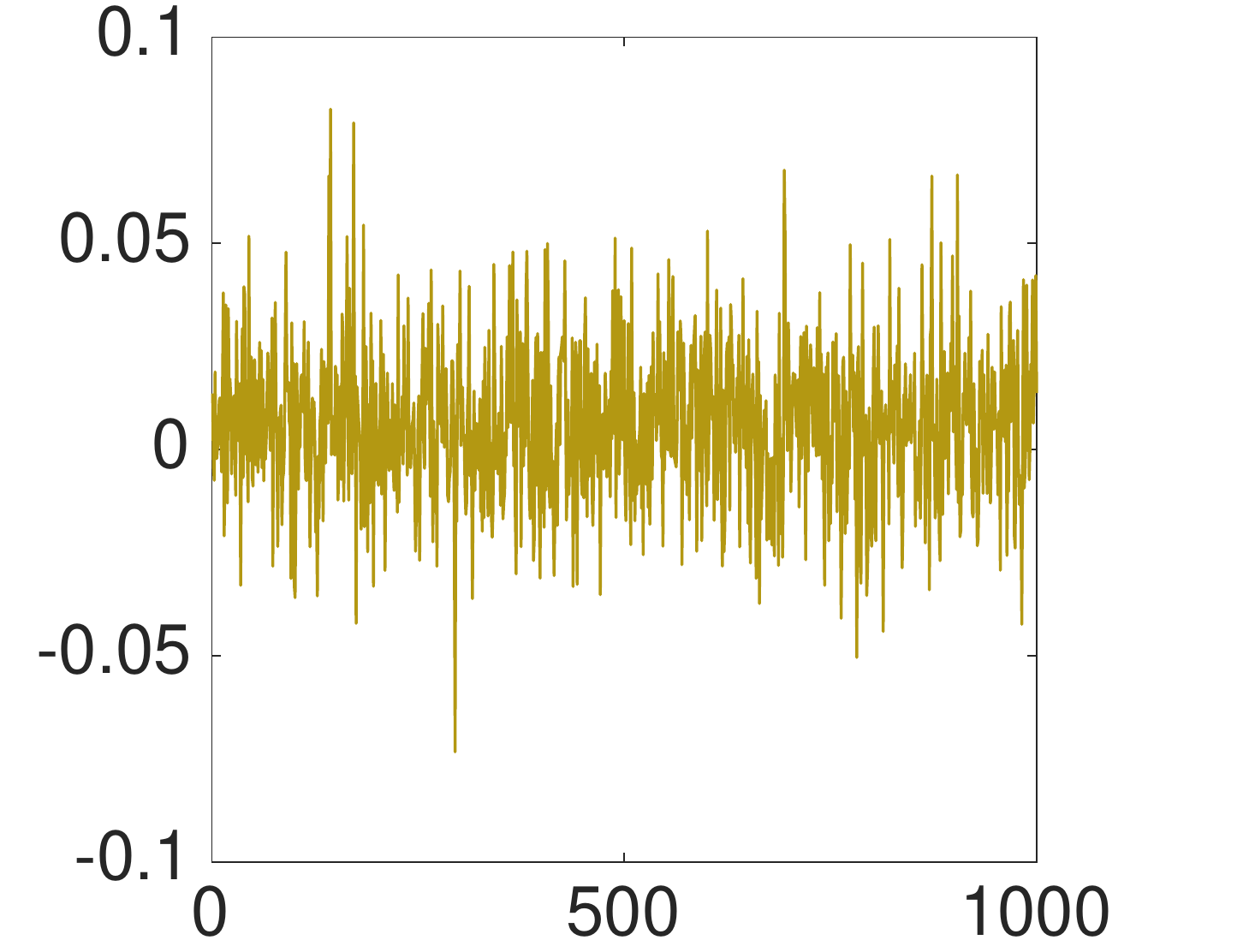}
            \includegraphics[width=\linewidth]{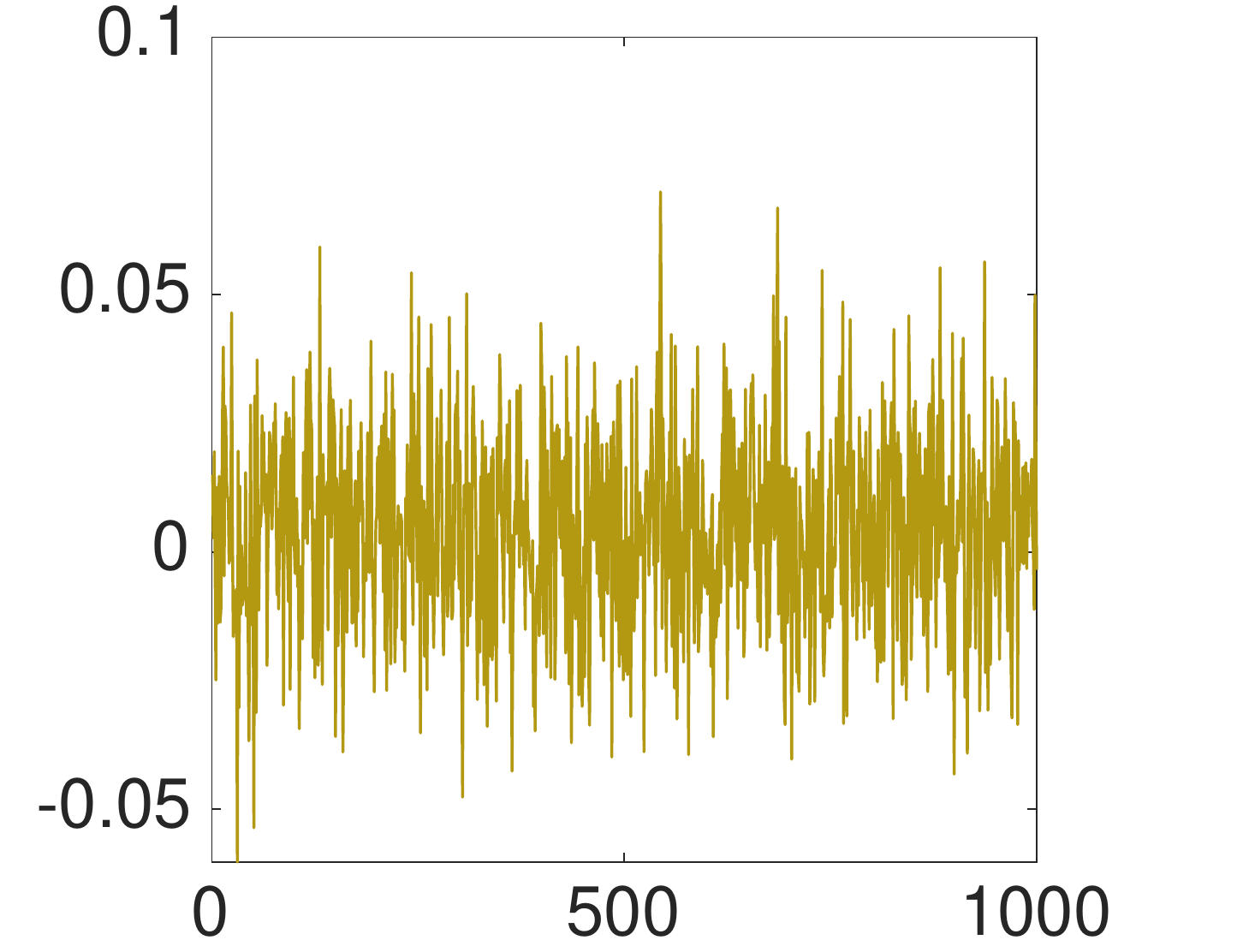}
            \includegraphics[width=\linewidth]{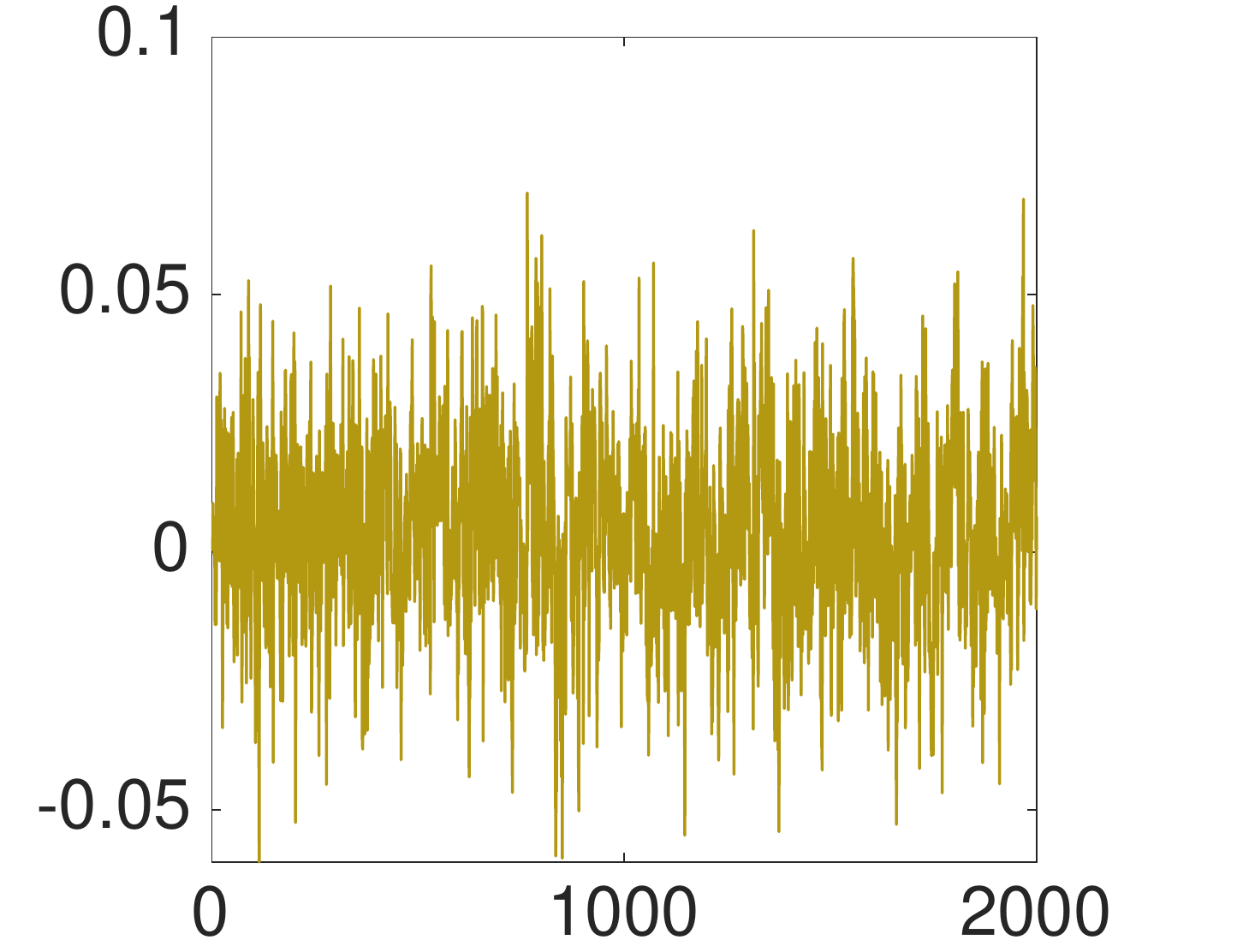}
            \caption{Iso 1st }\end{subfigure}
        \begin{subfigure}[b]{\qhei}
            \includegraphics[width=\linewidth]{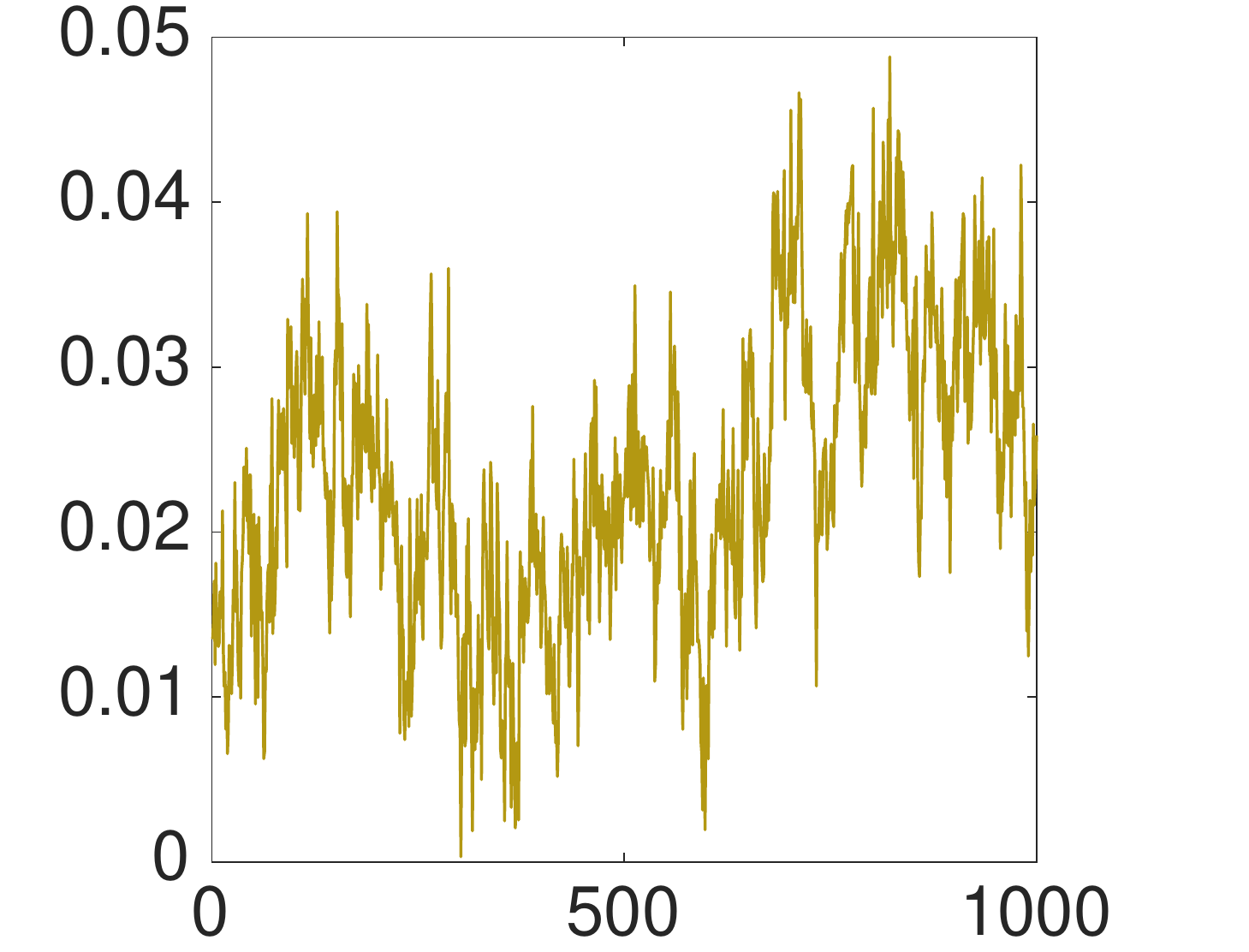}
            \includegraphics[width=\linewidth]{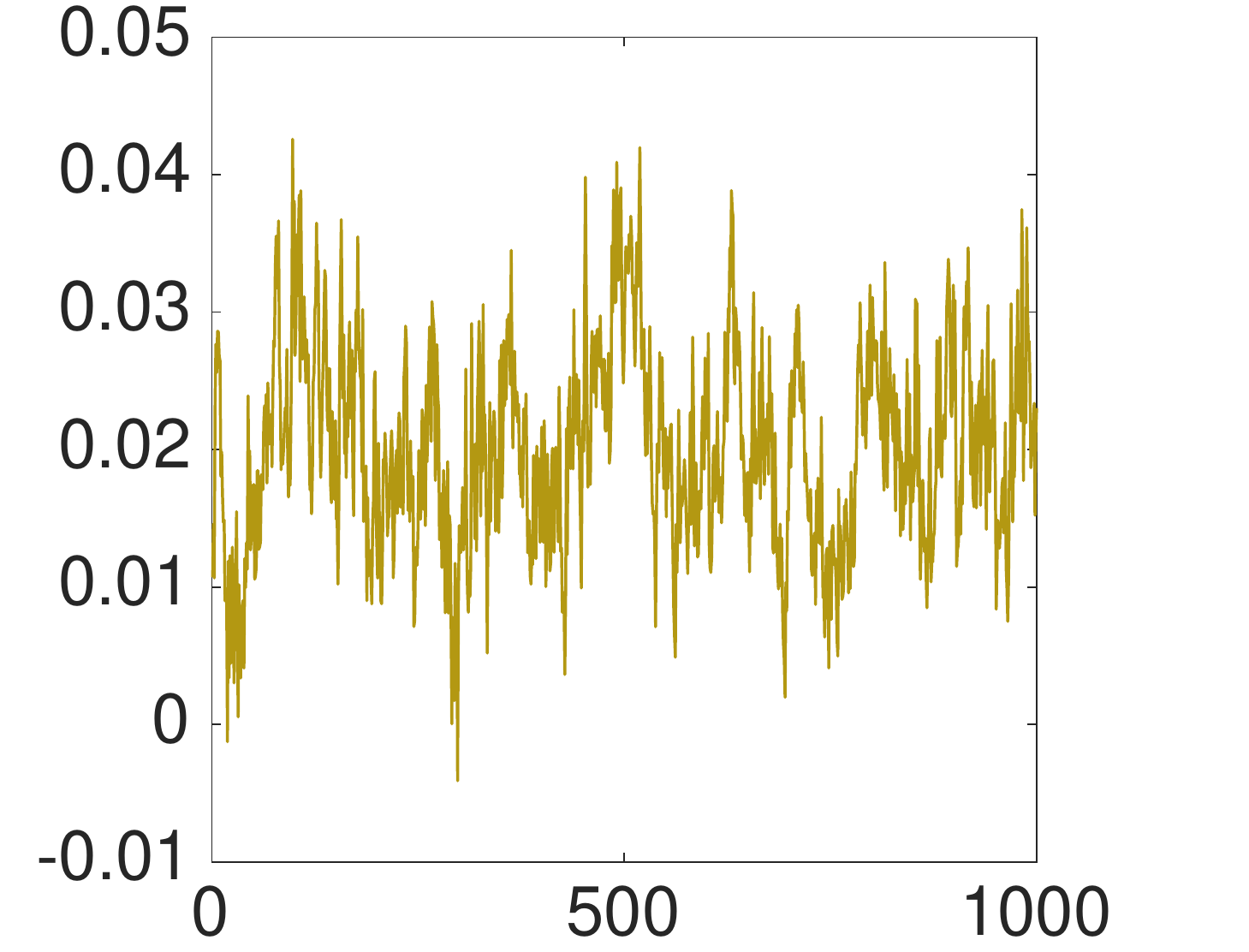}
            \includegraphics[width=\linewidth]{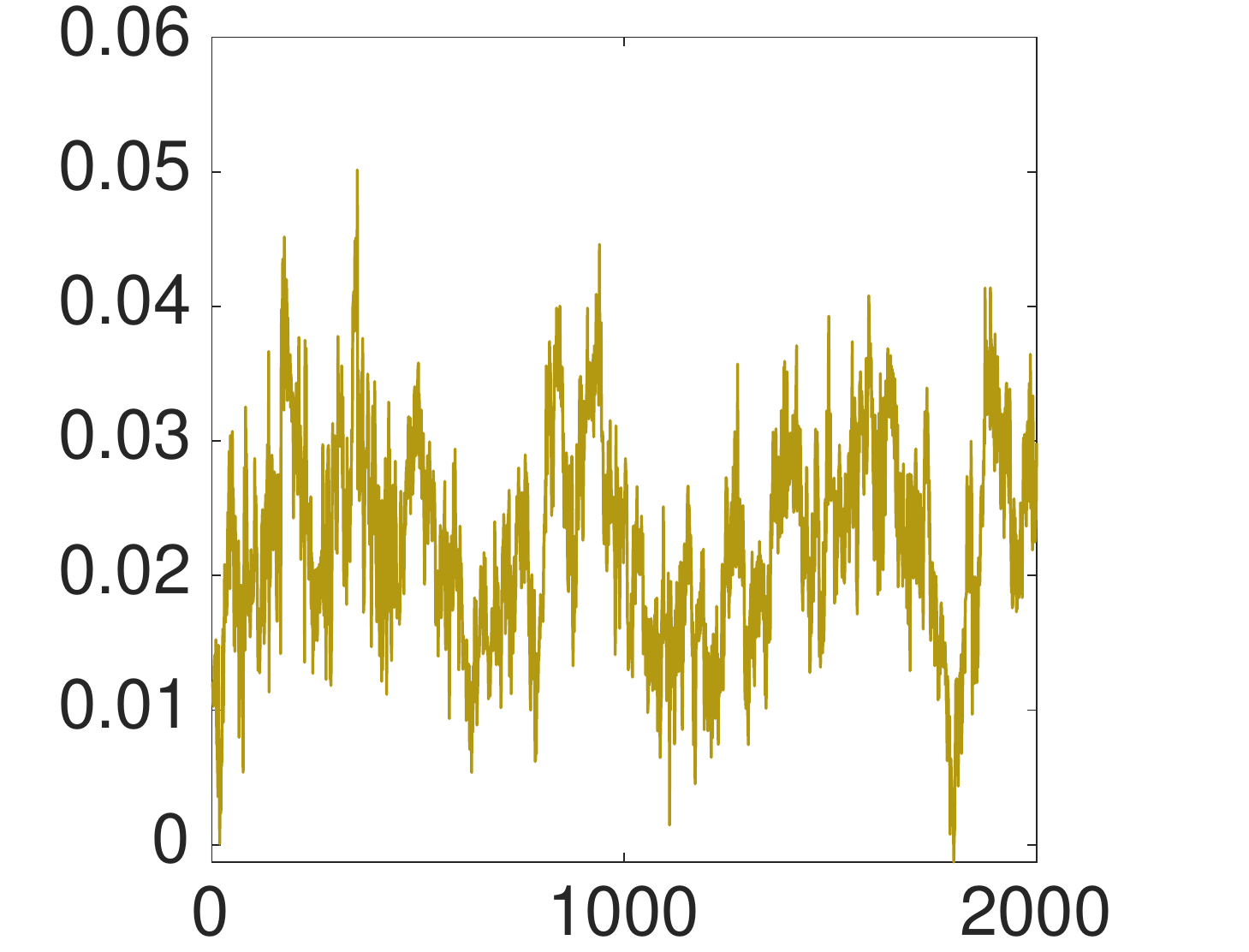}
            \caption{Aniso 2nd }\end{subfigure}
        \begin{subfigure}[b]{\qhei}
            \includegraphics[width=\linewidth]{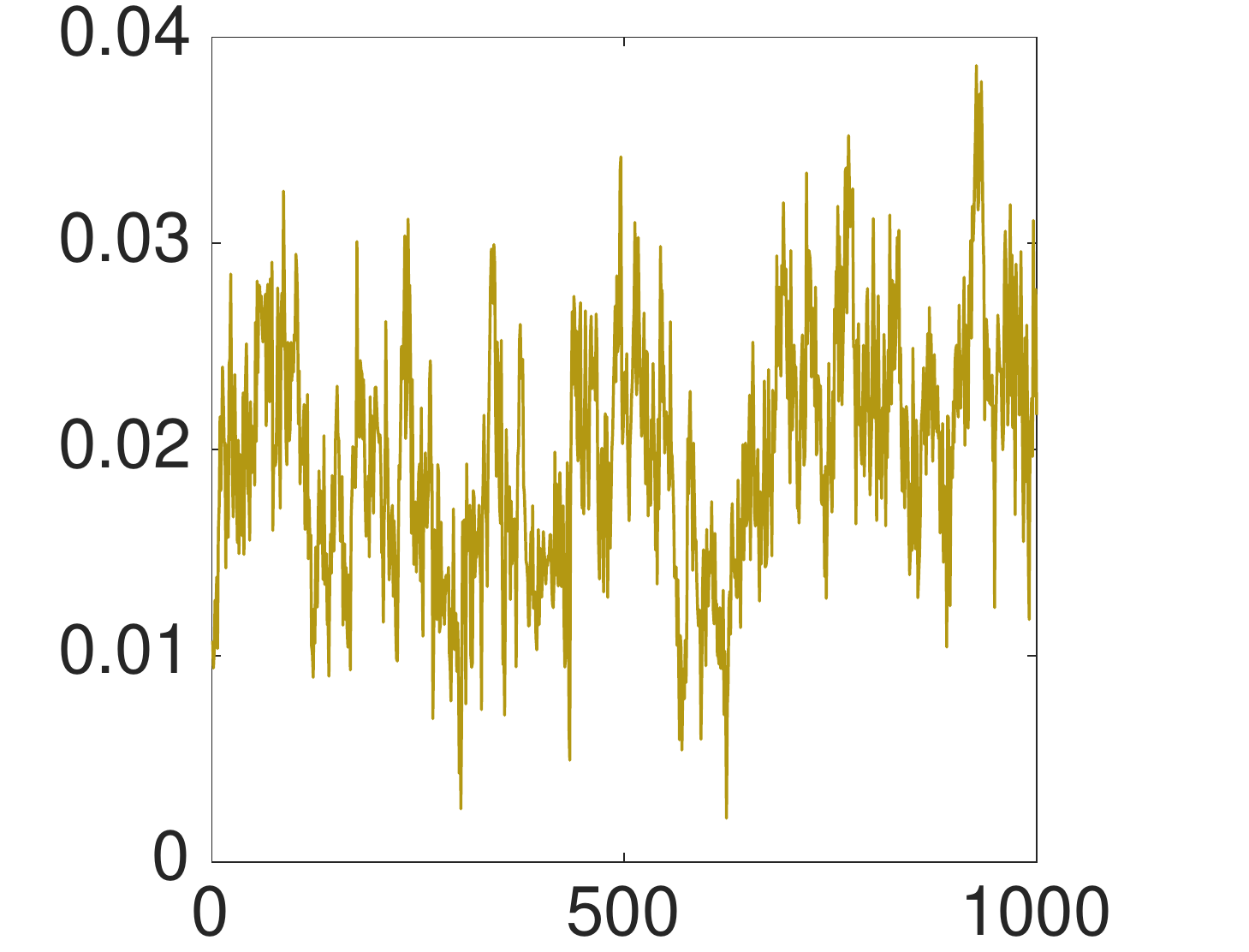}
            \includegraphics[width=\linewidth]{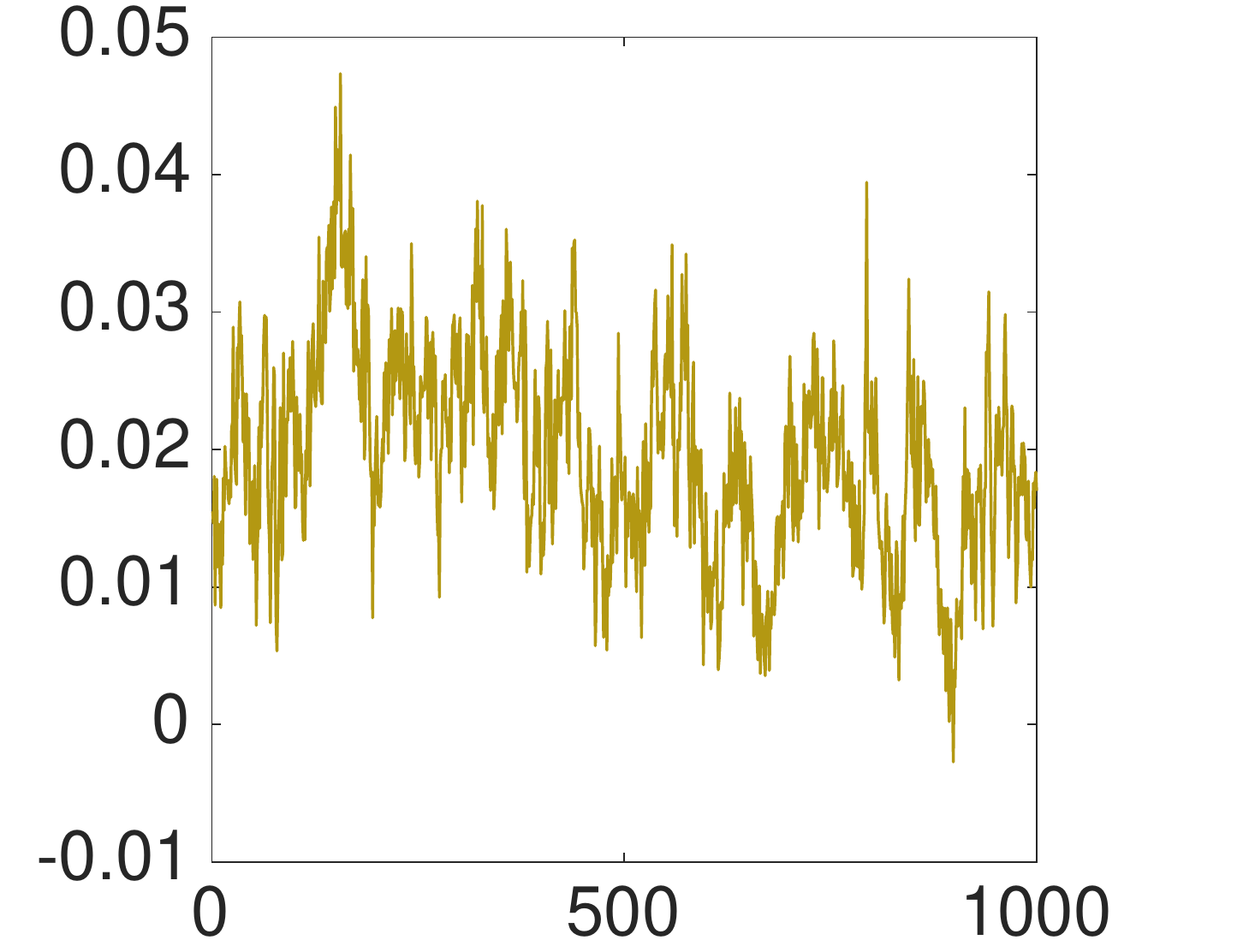}
            \includegraphics[width=\linewidth]{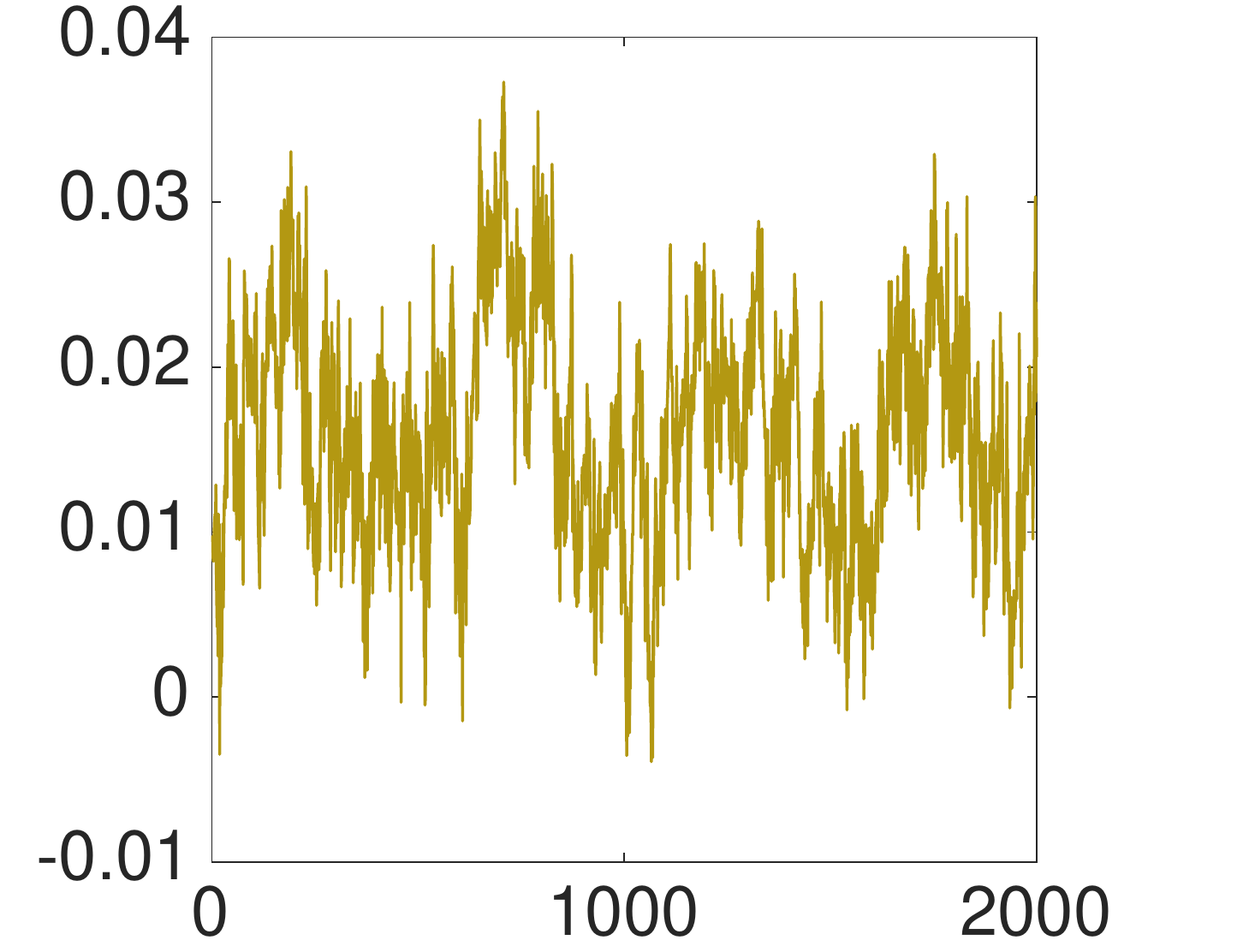}
            \caption{Iso 2nd }\end{subfigure}
        \begin{subfigure}[b]{\qhei}
            \includegraphics[width=\linewidth]{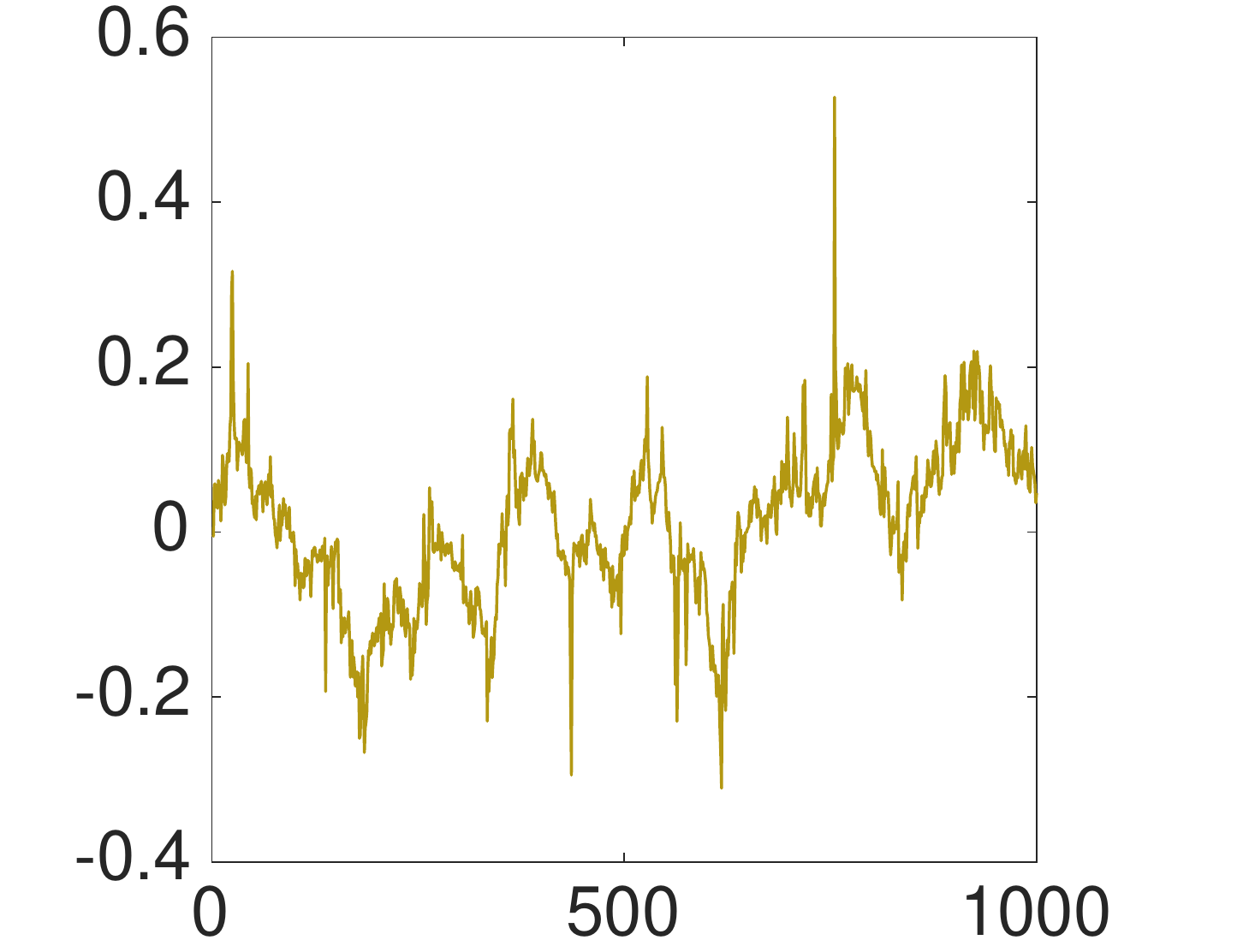}
            \includegraphics[width=\linewidth]{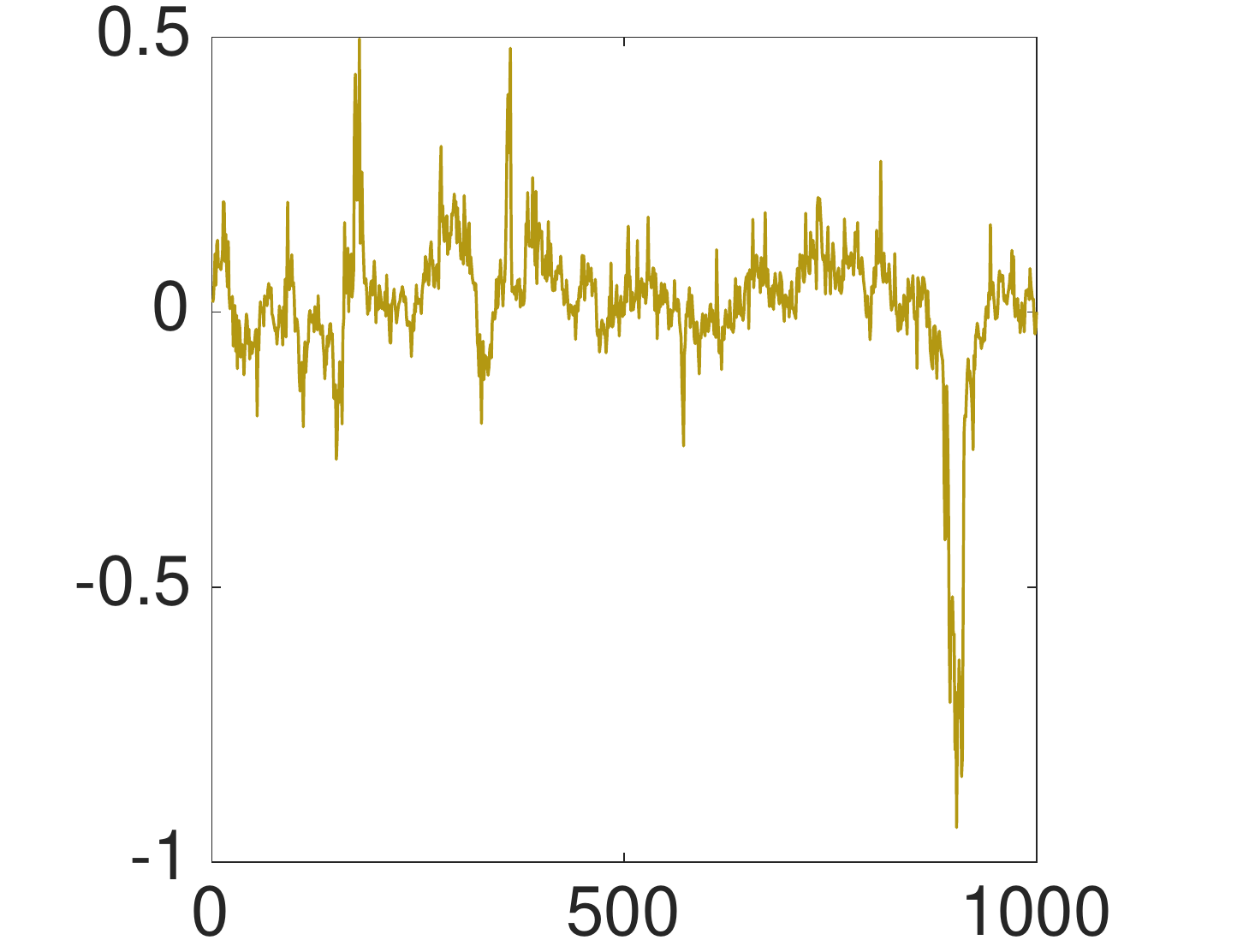}
            \includegraphics[width=\linewidth]{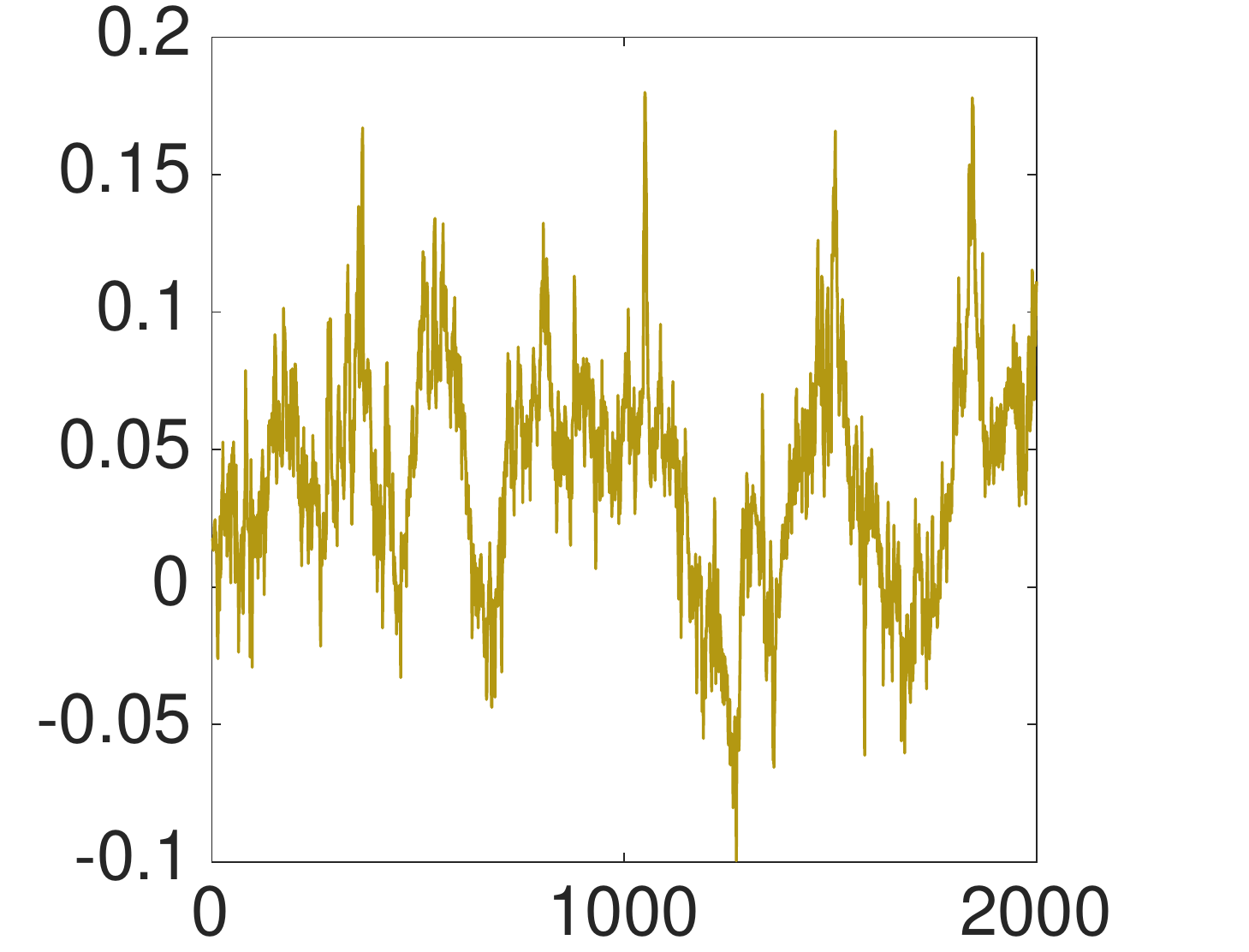}
            \caption{SPDE }\end{subfigure}
        \begin{subfigure}[b]{\qhei}
            \includegraphics[width=\linewidth]{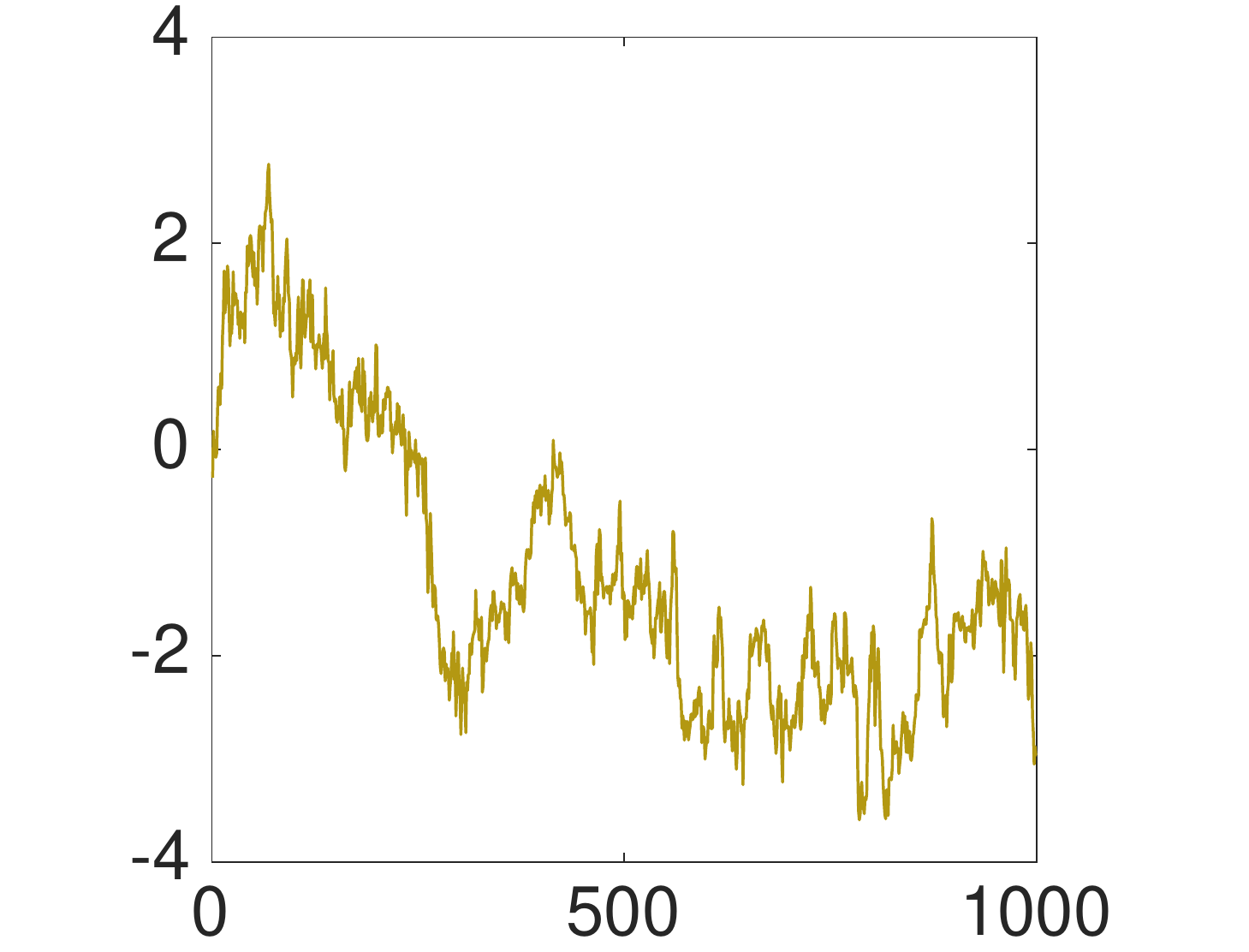}
            \includegraphics[width=\linewidth]{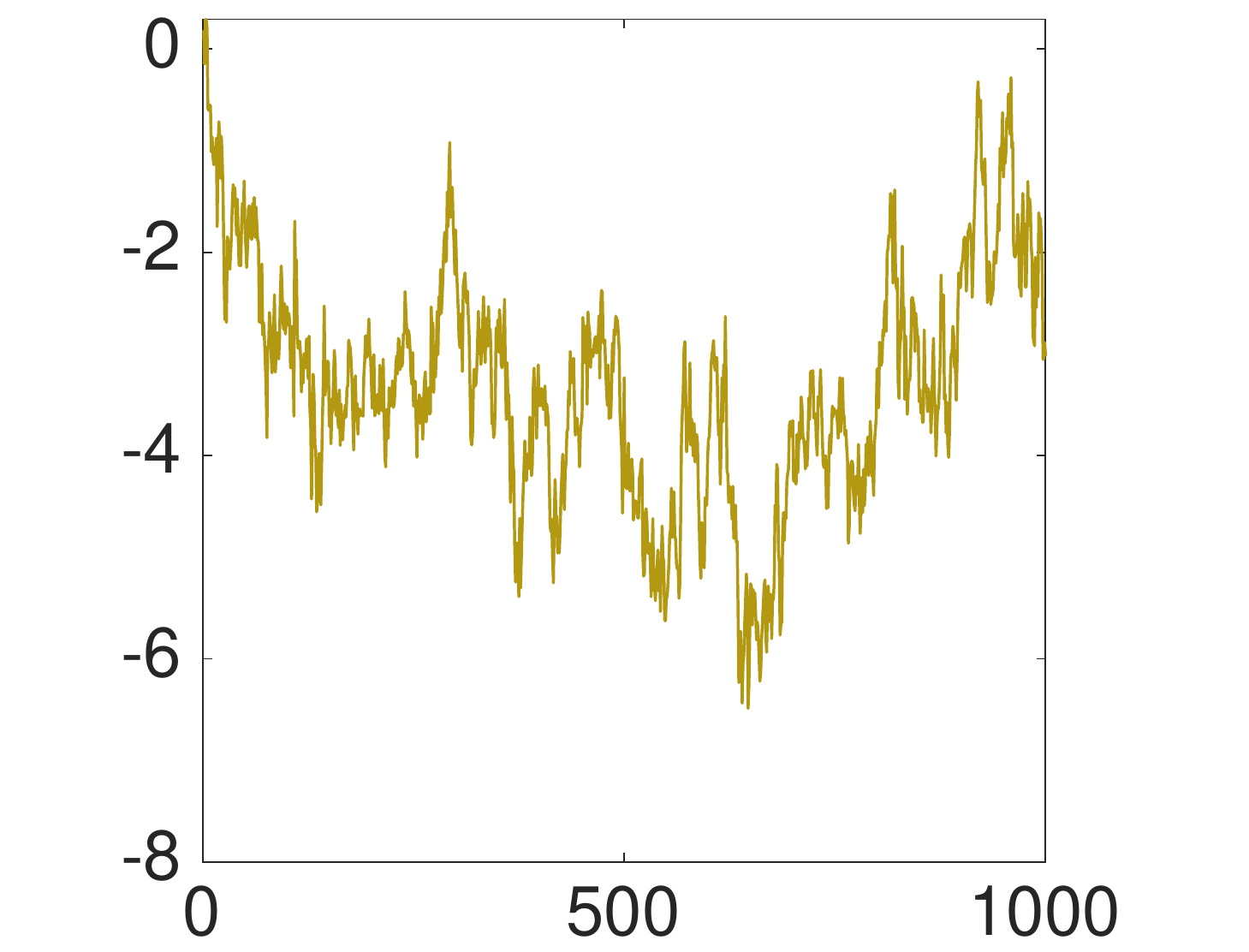}
            \includegraphics[width=\linewidth]{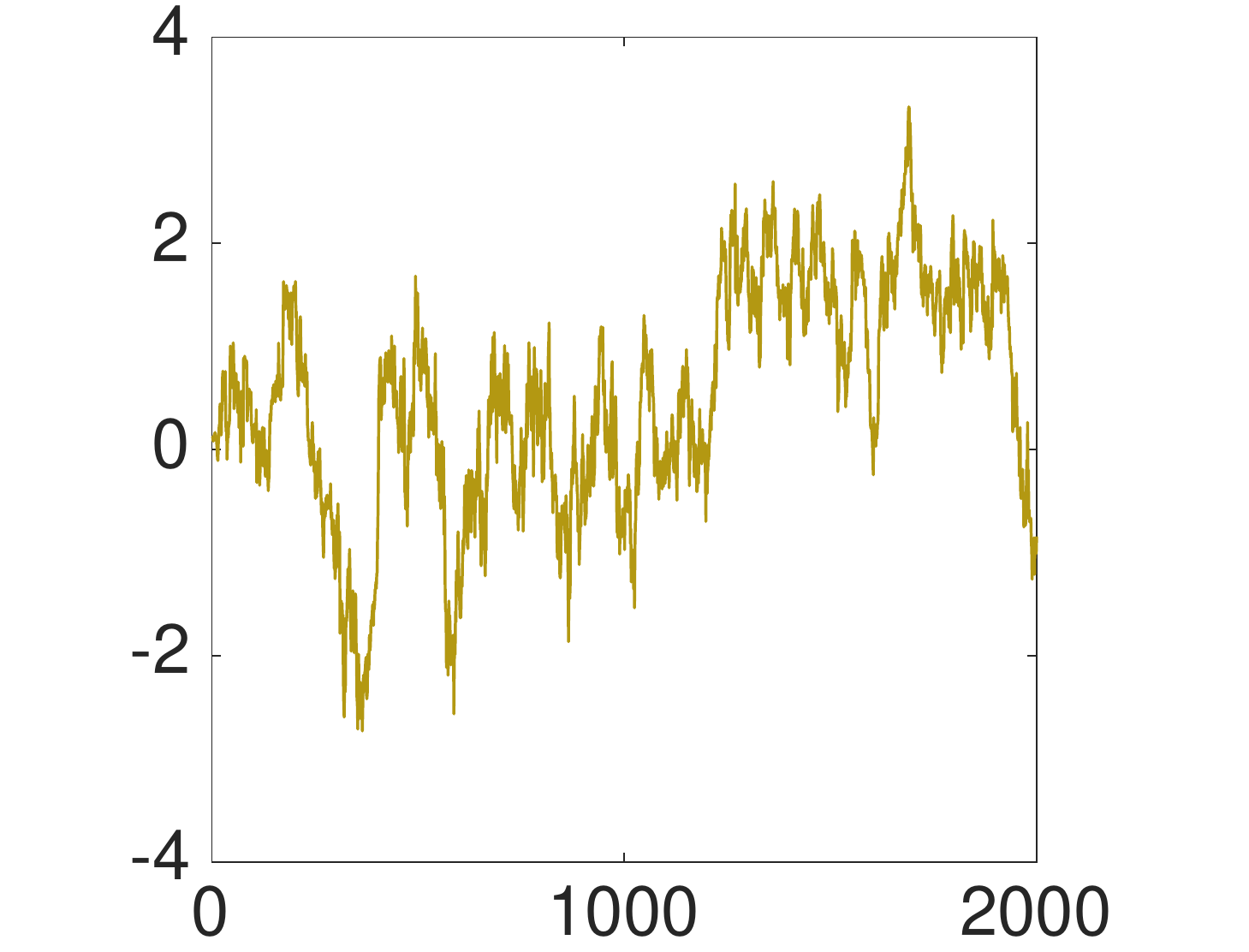}
            \caption{Sheet }\end{subfigure}
        
        \begin{subfigure}[b]{\qhei}
            \includegraphics[width=\linewidth]{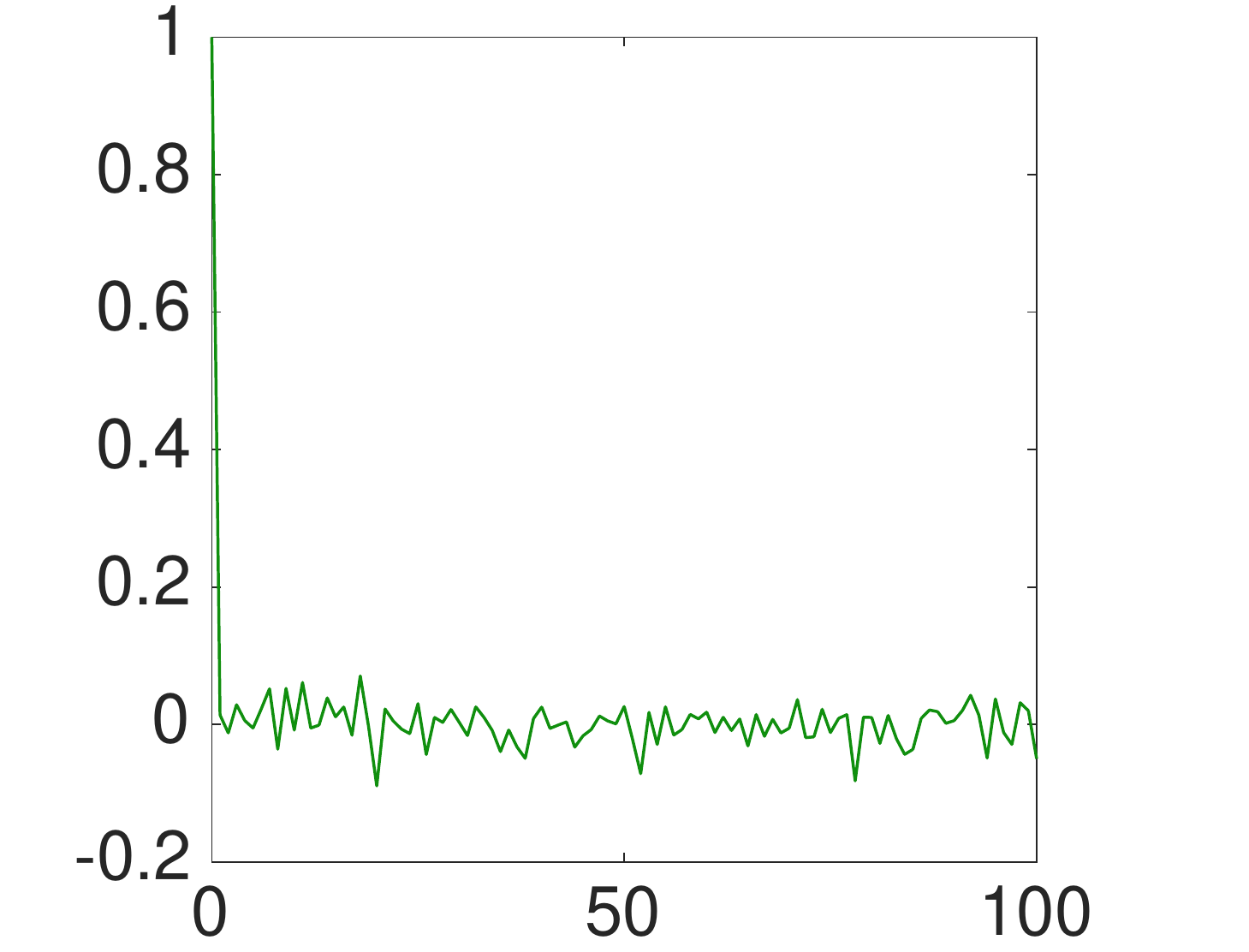}
            \includegraphics[width=\linewidth]{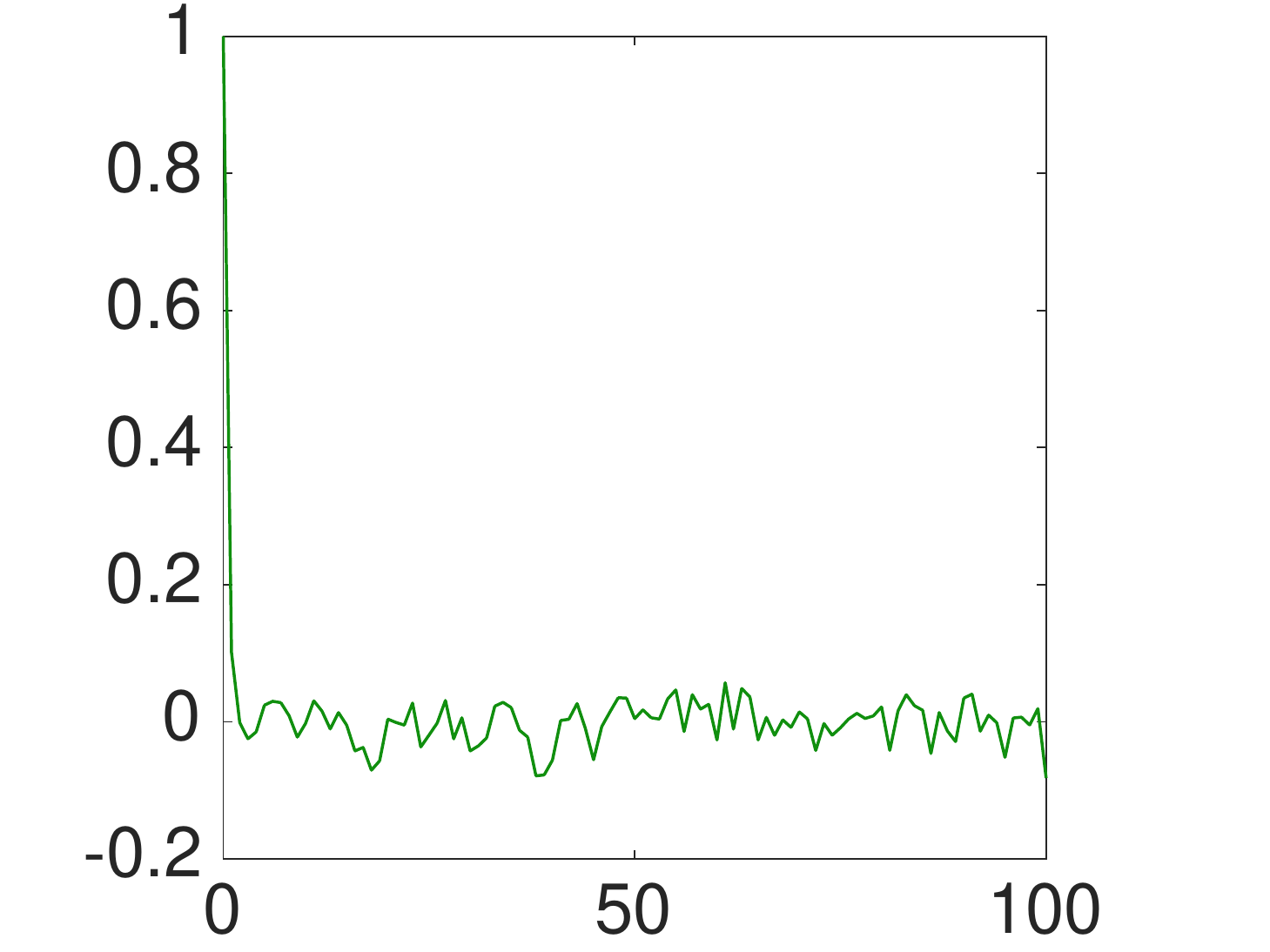}
            \includegraphics[width=\linewidth]{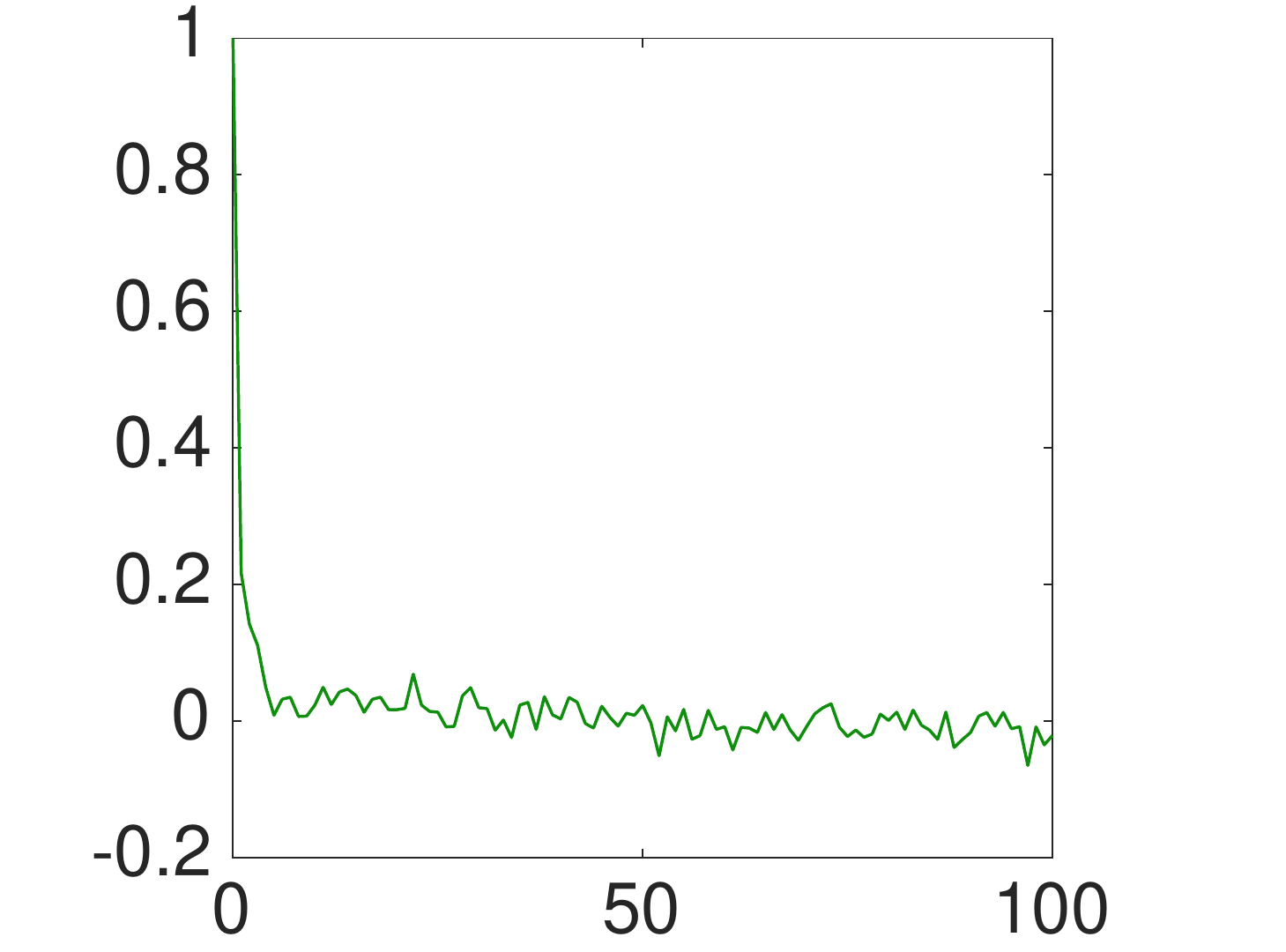}
            \caption{Aniso 1st }\end{subfigure}
        \begin{subfigure}[b]{\qhei}
            \includegraphics[width=\linewidth]{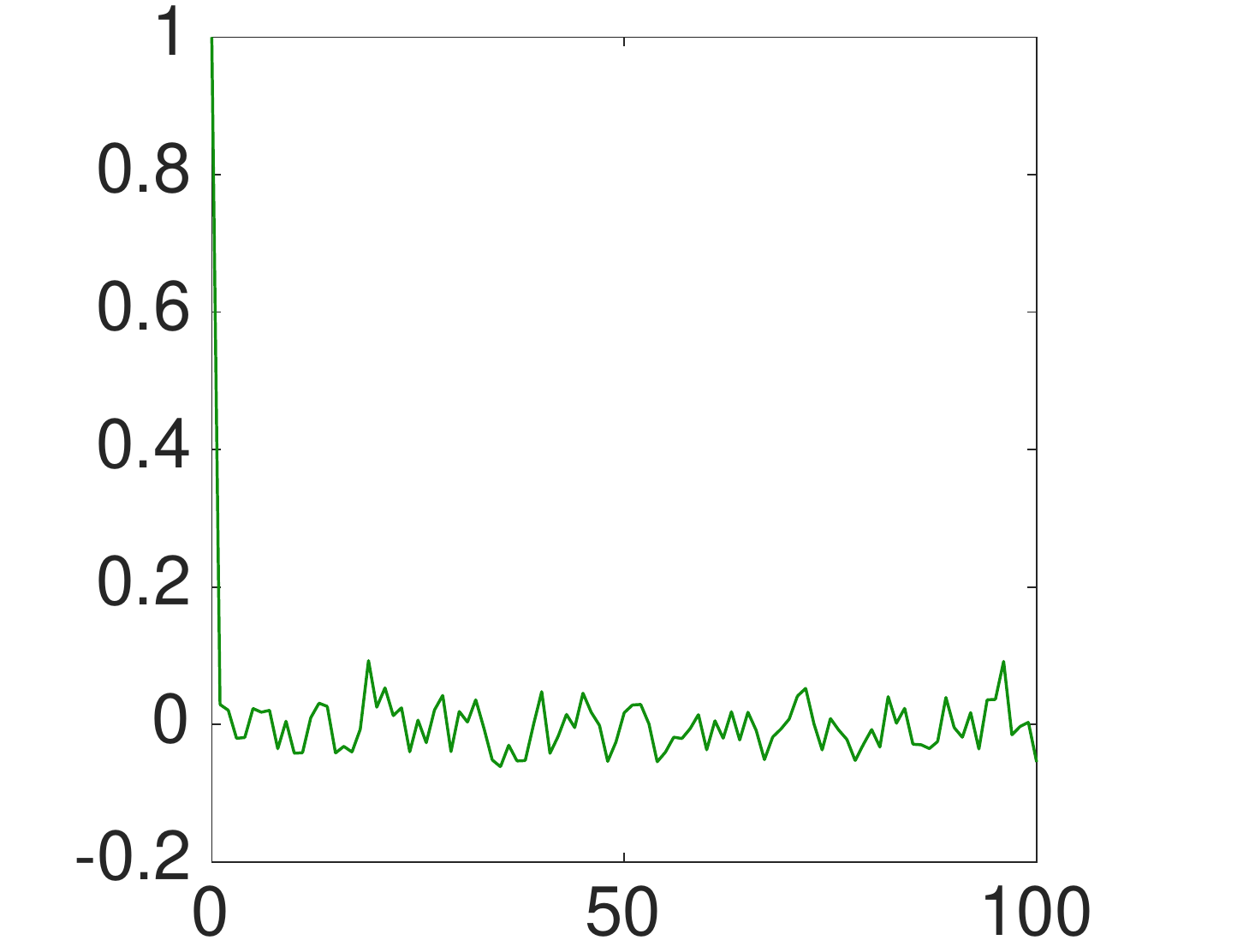}
            \includegraphics[width=\linewidth]{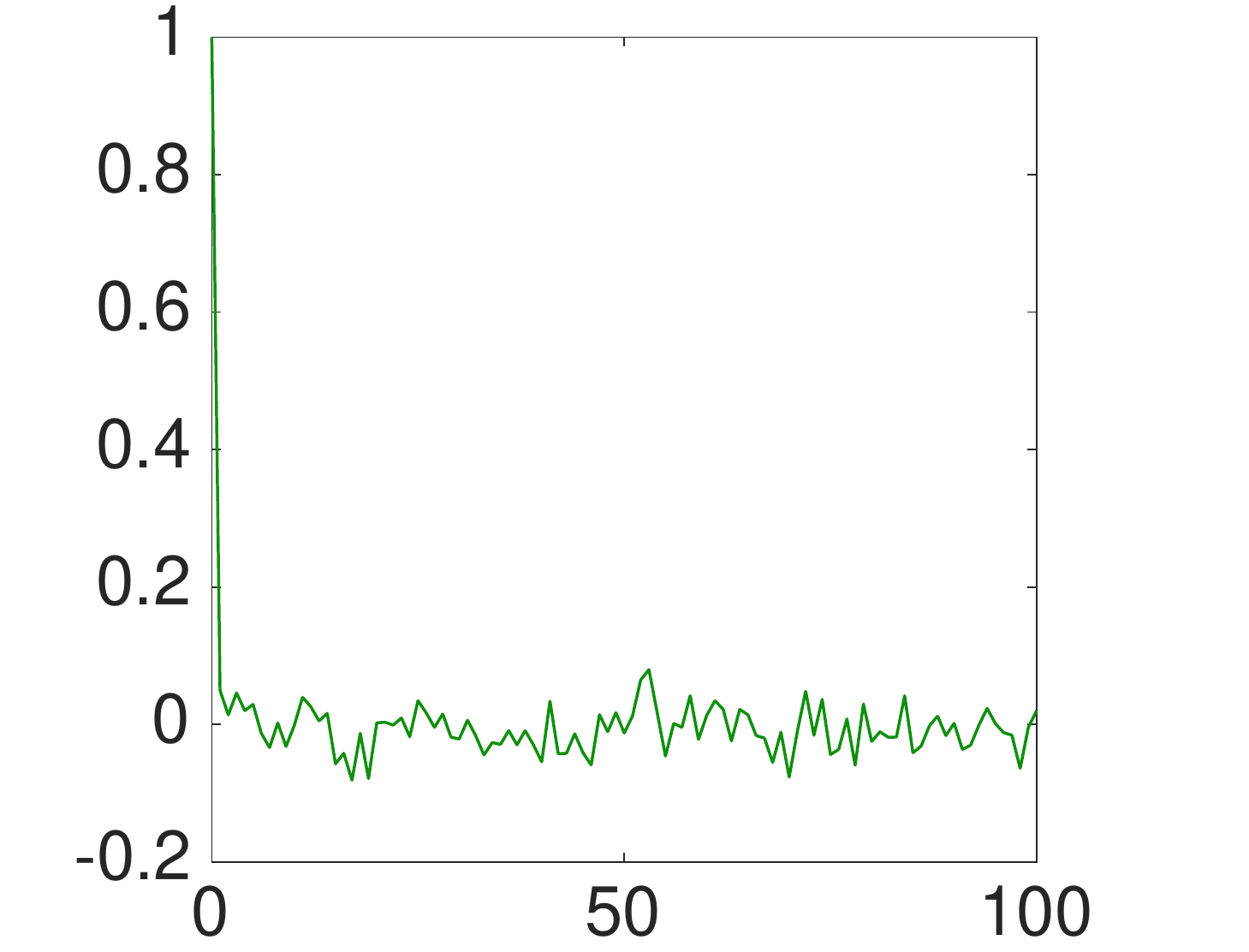}
            \includegraphics[width=\linewidth]{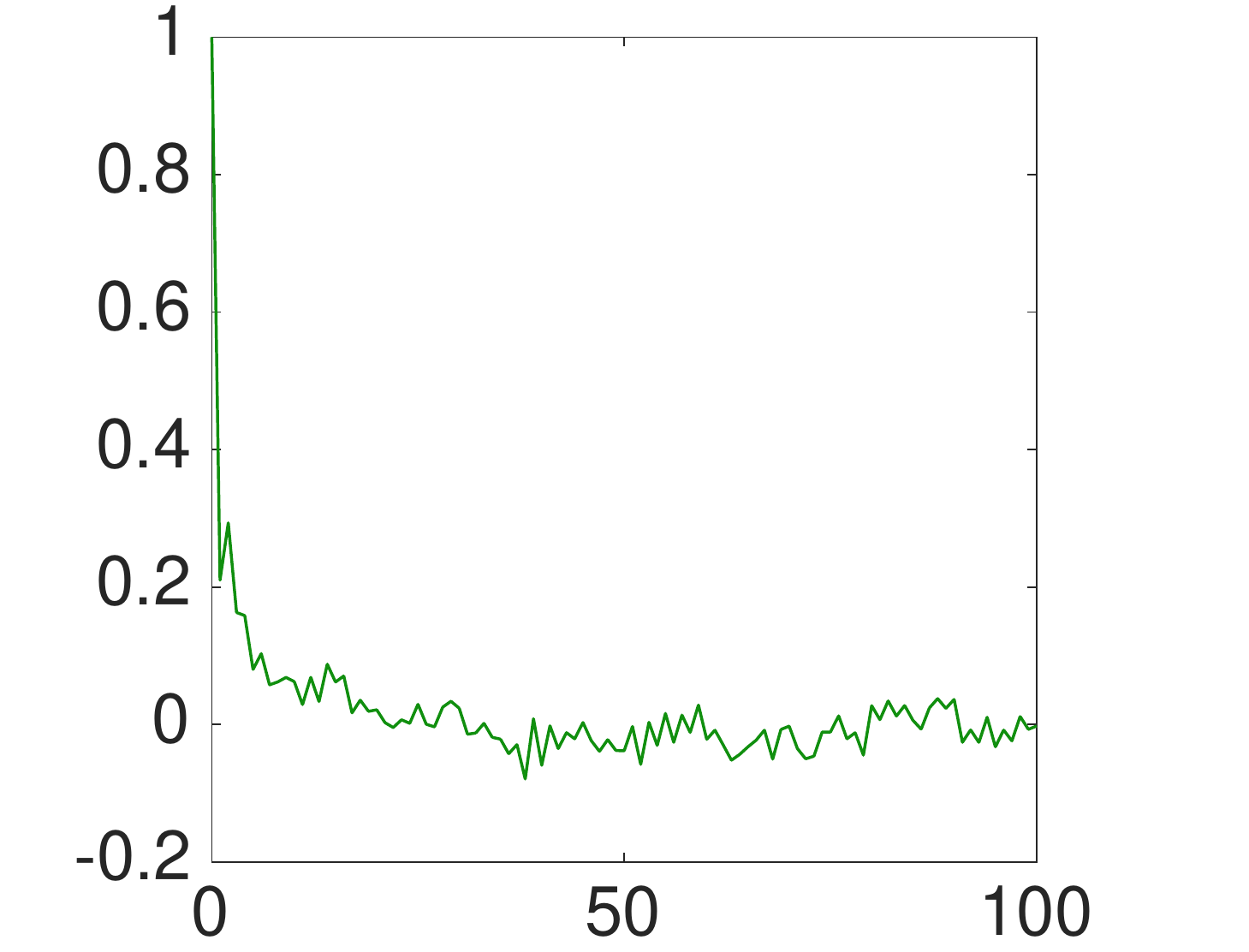}
            \caption{Iso 1st }\end{subfigure}
        \begin{subfigure}[b]{\qhei}
            \includegraphics[width=\linewidth]{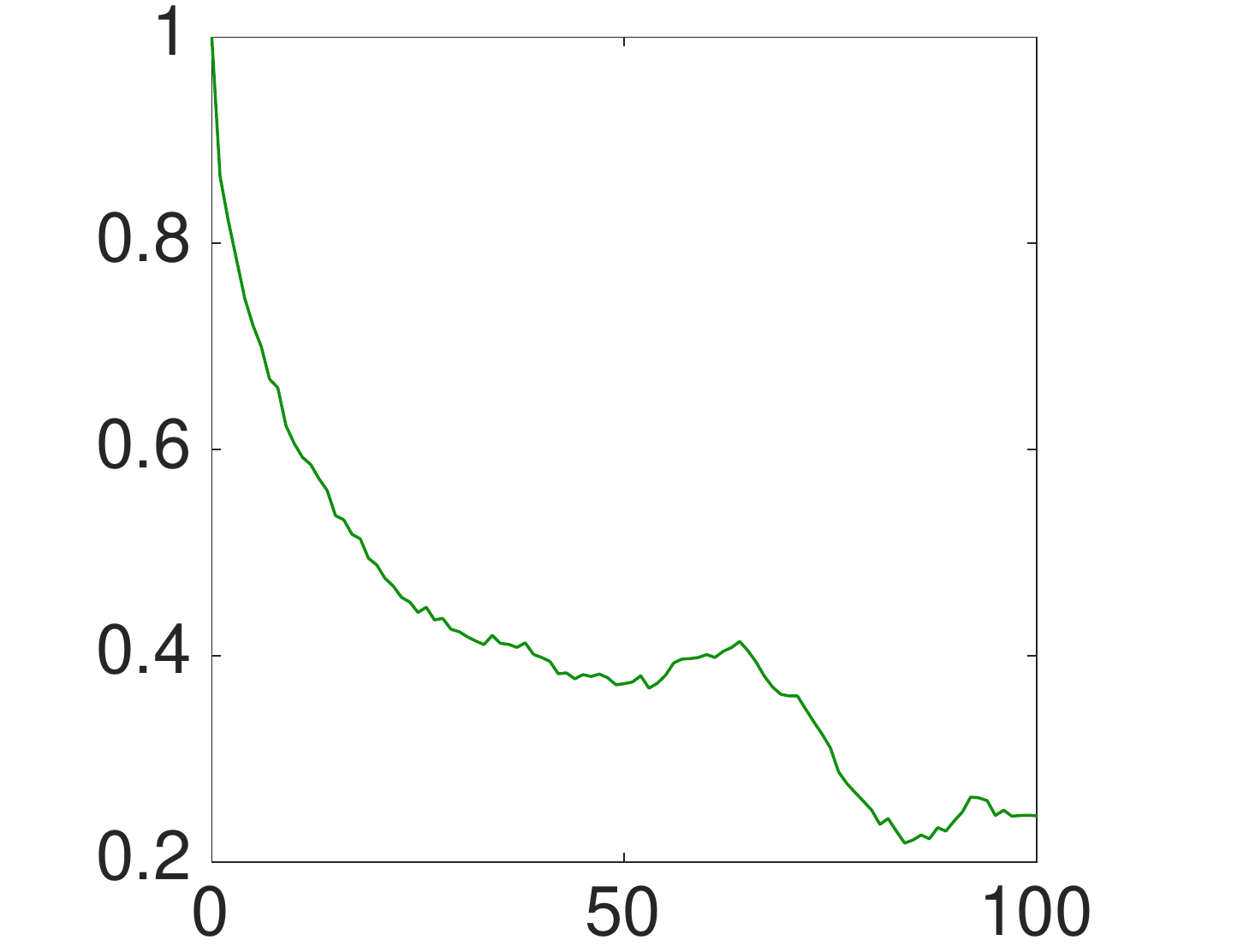}
            \includegraphics[width=\linewidth]{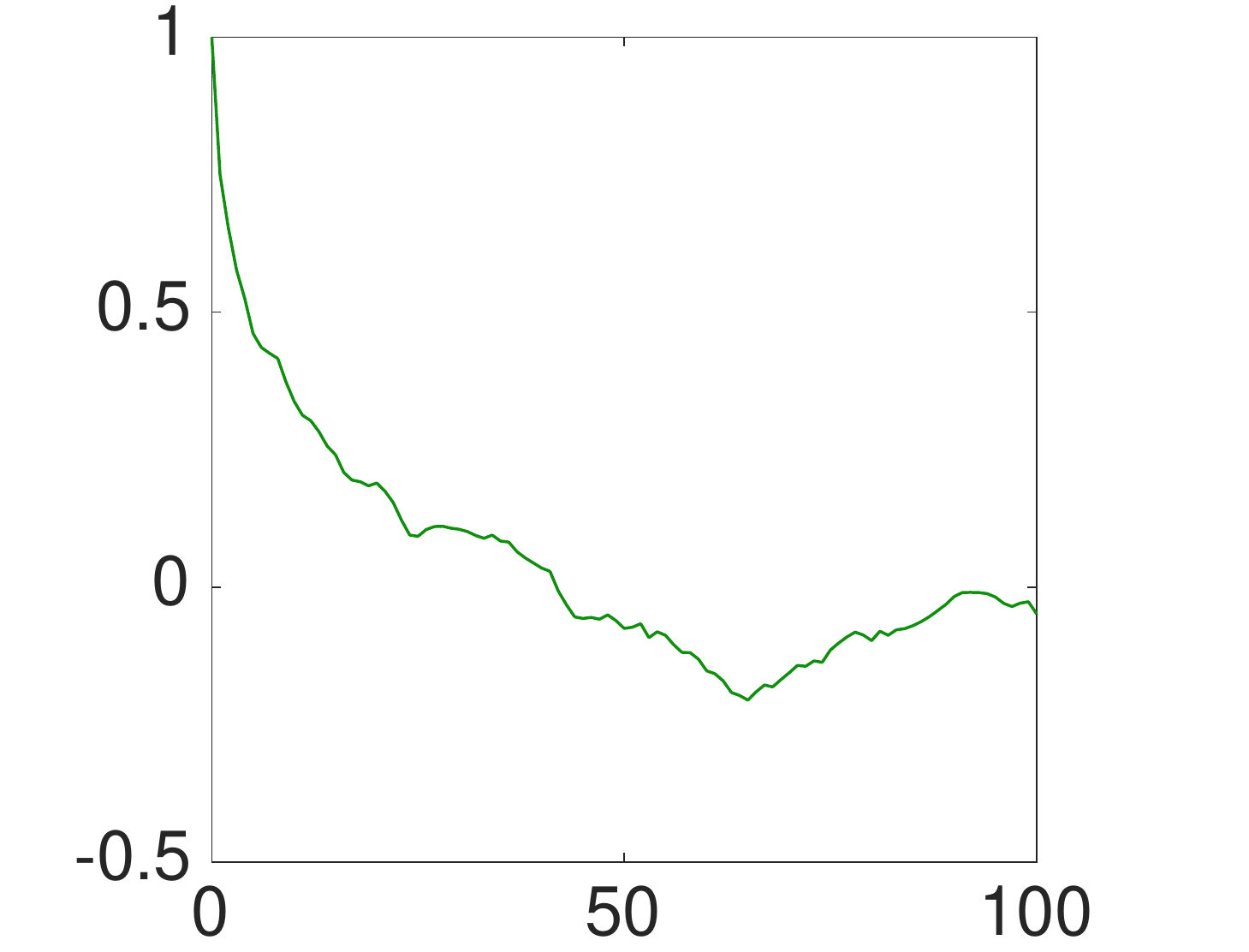}
            \includegraphics[width=\linewidth]{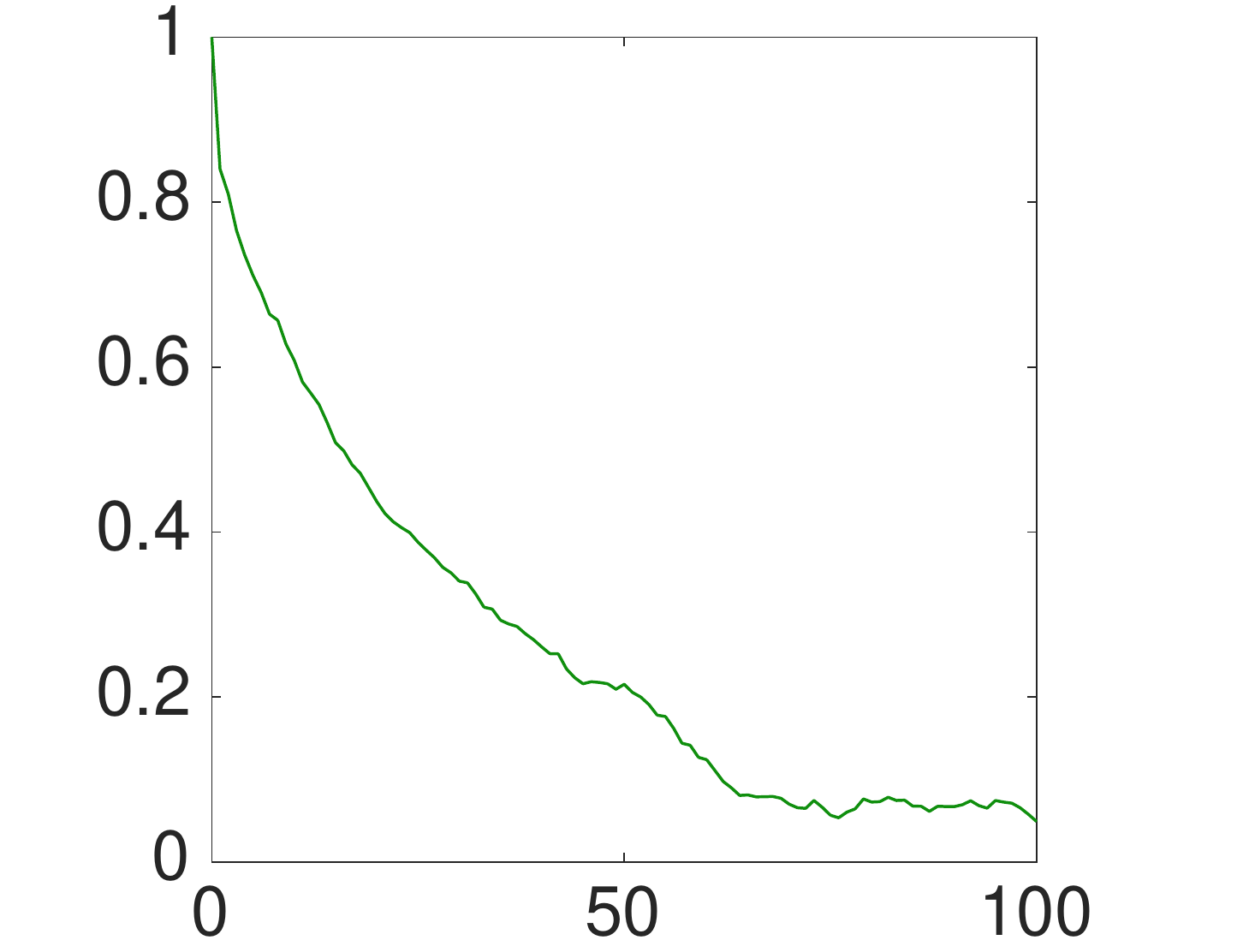}
            \caption{Aniso 2nd }\end{subfigure}
        \begin{subfigure}[b]{\qhei}
            \includegraphics[width=\linewidth]{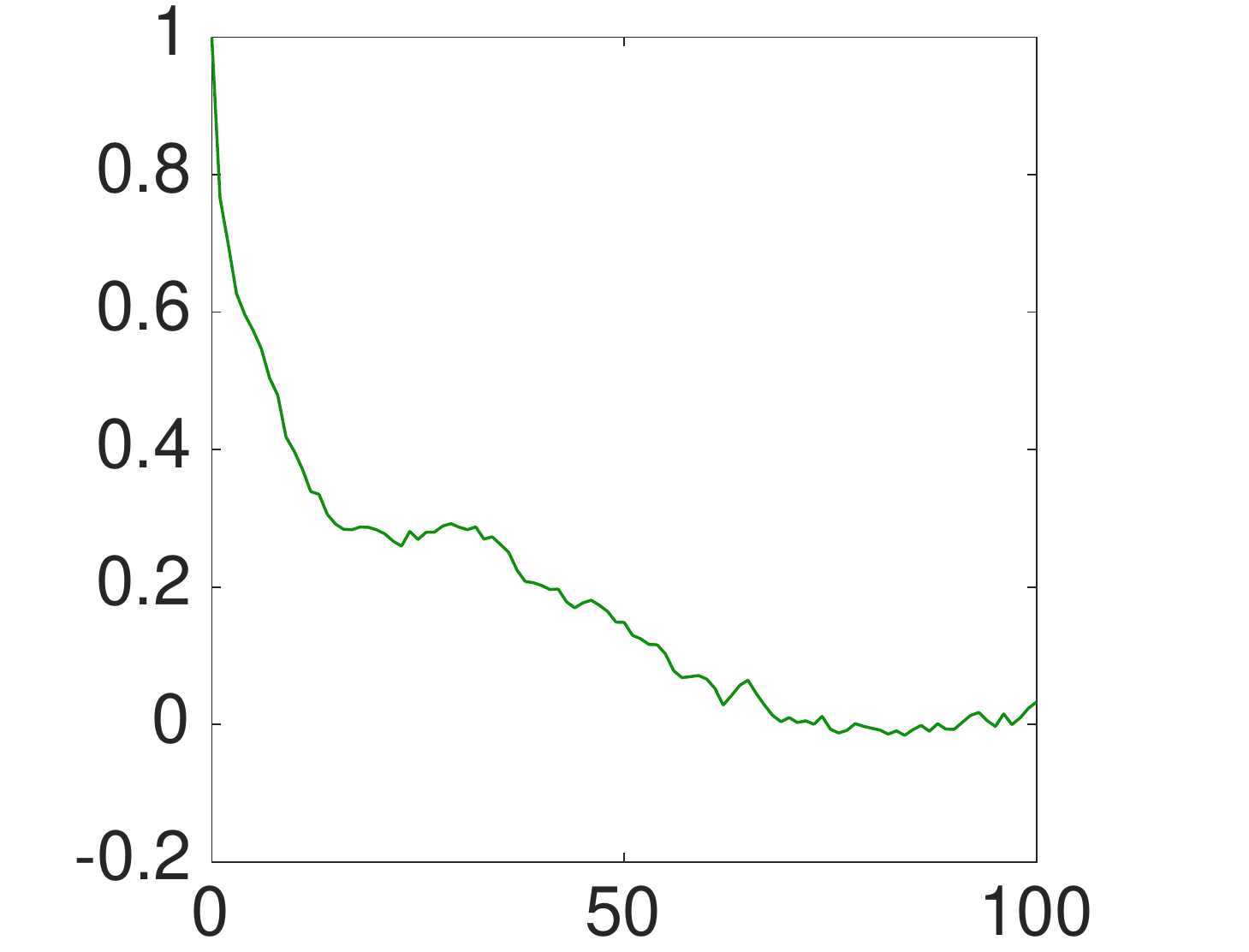}
            \includegraphics[width=\linewidth]{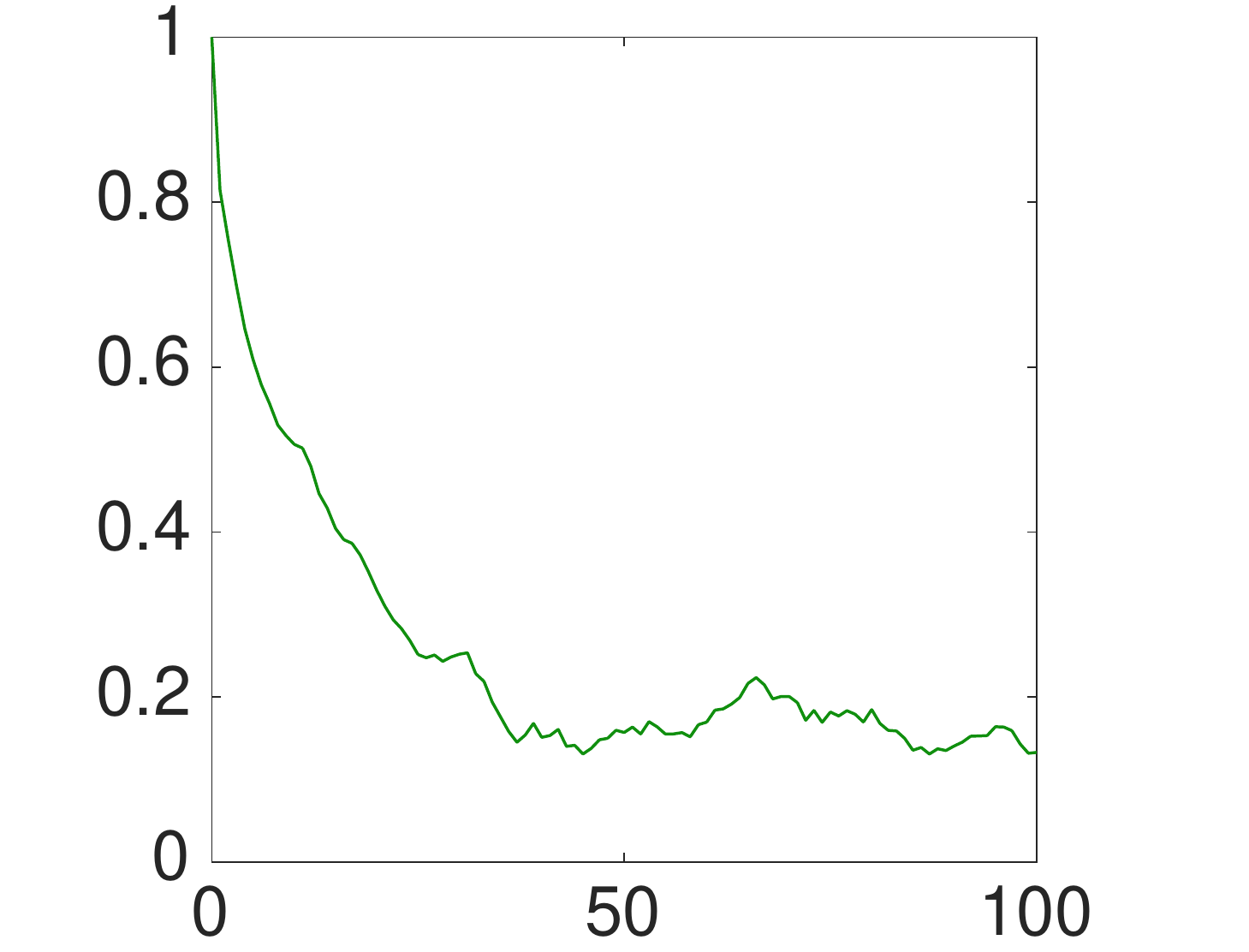}
            \includegraphics[width=\linewidth]{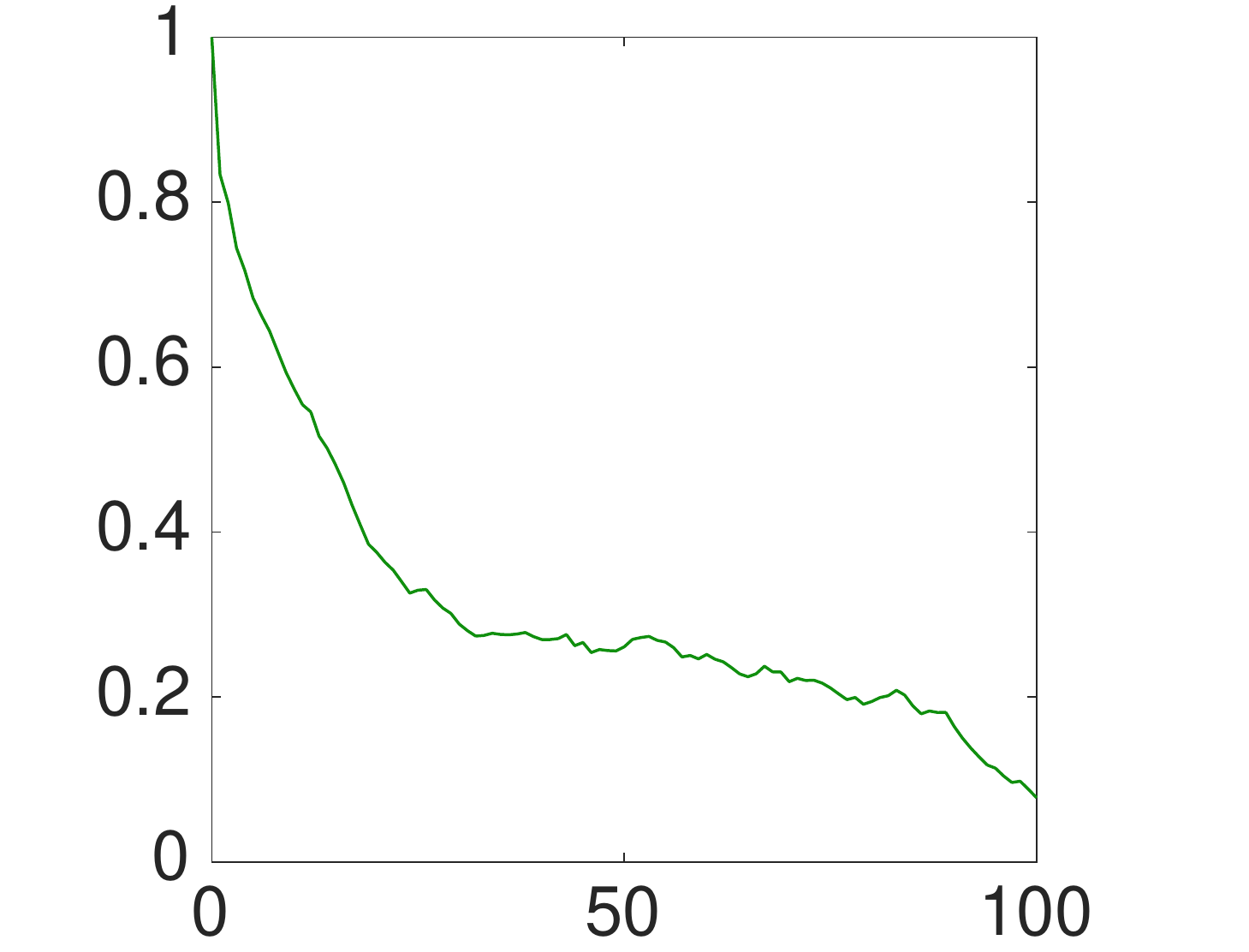}
            \caption{Iso 2nd }\end{subfigure}
        \begin{subfigure}[b]{\qhei}
            \includegraphics[width=\linewidth]{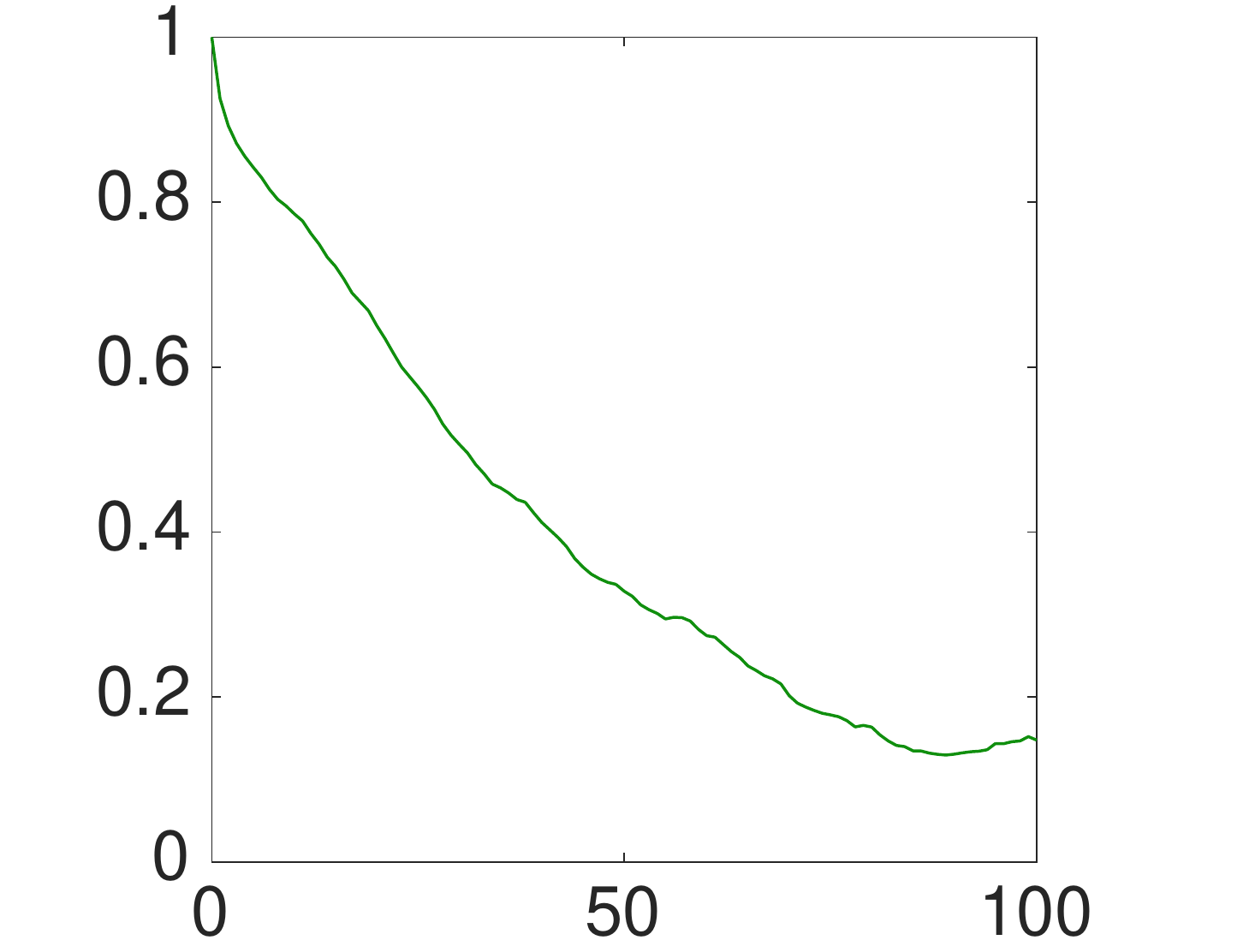}
            \includegraphics[width=\linewidth]{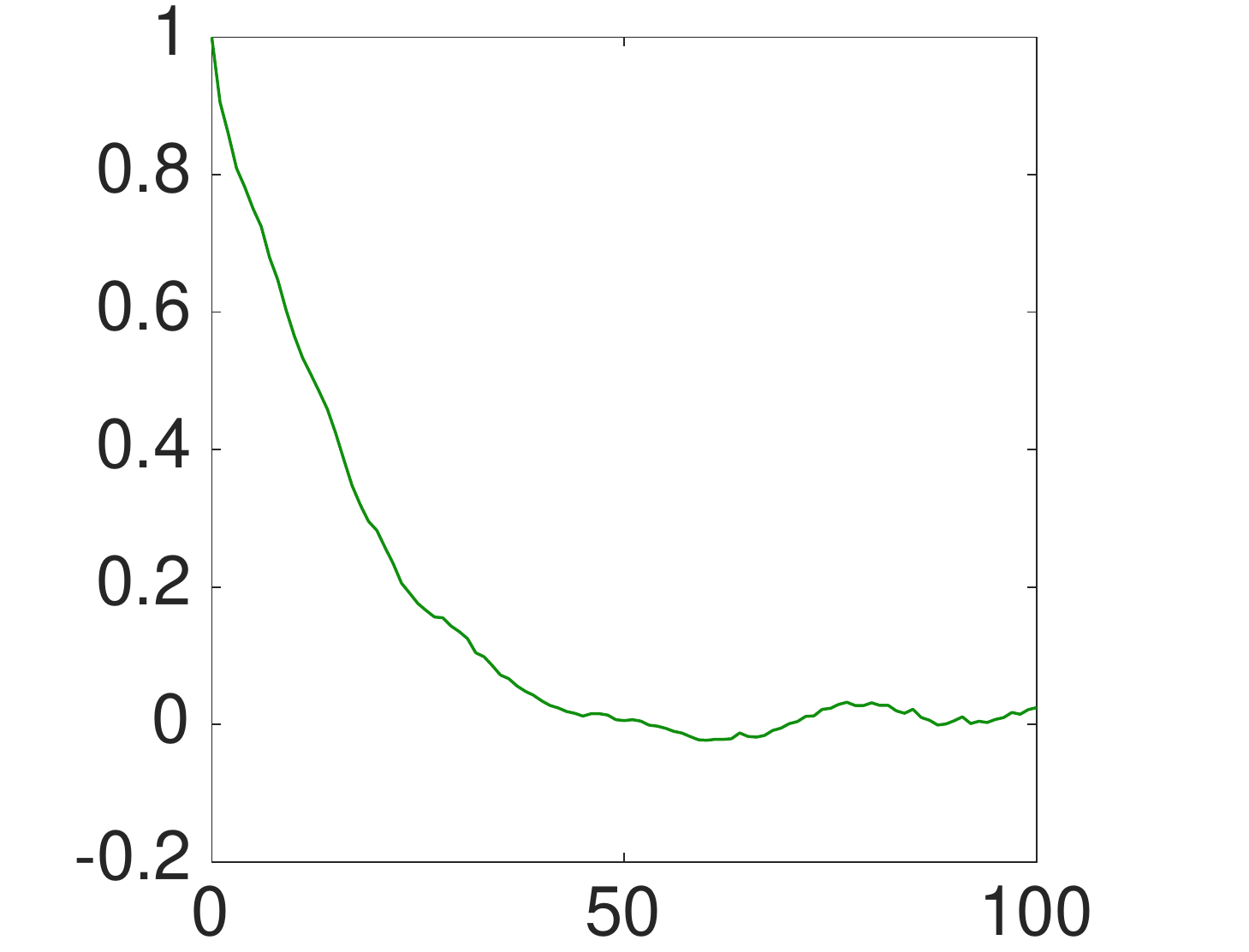}
            \includegraphics[width=\linewidth]{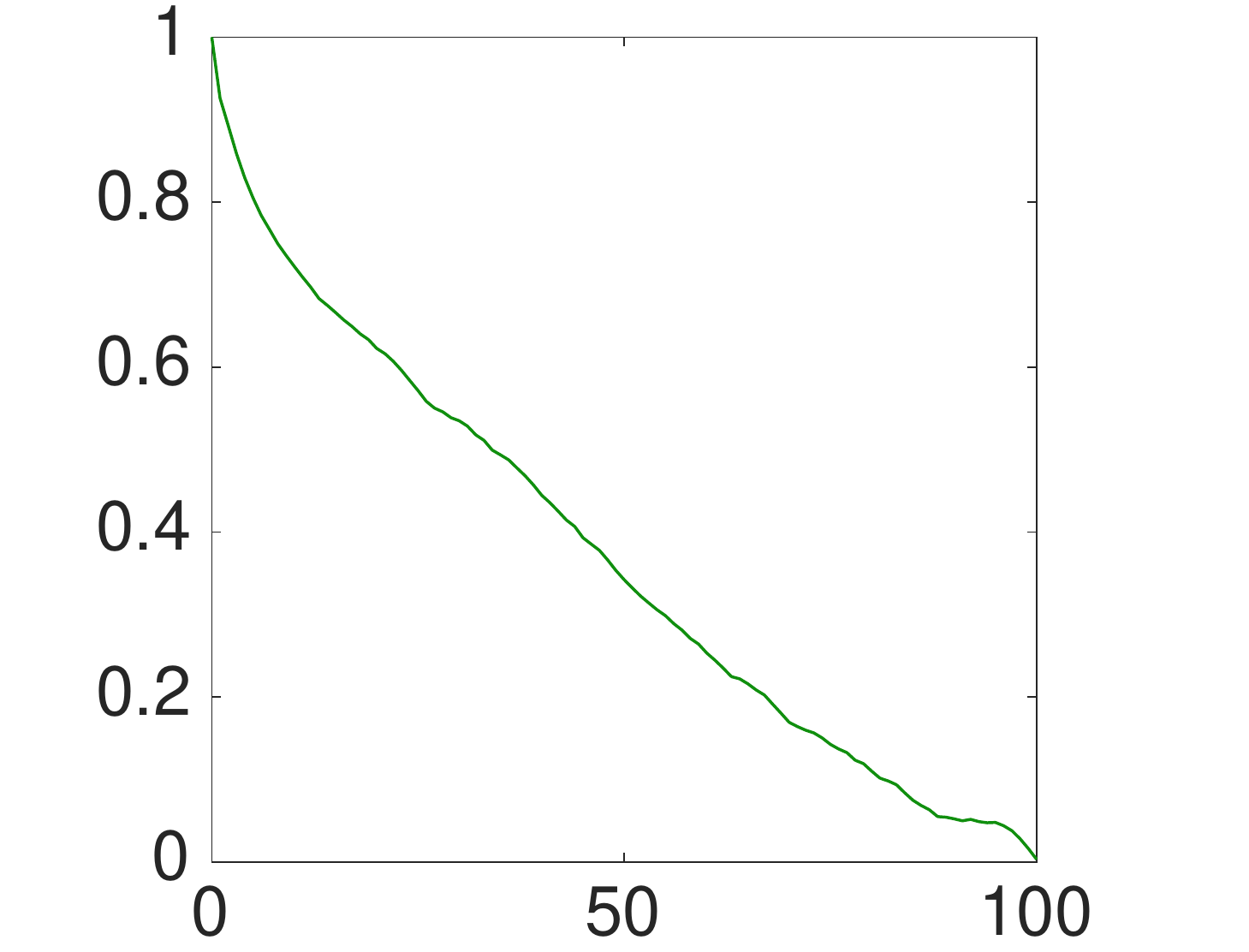}
            \caption{SPDE }\end{subfigure}
        \begin{subfigure}[b]{\qhei}
            \includegraphics[width=\linewidth]{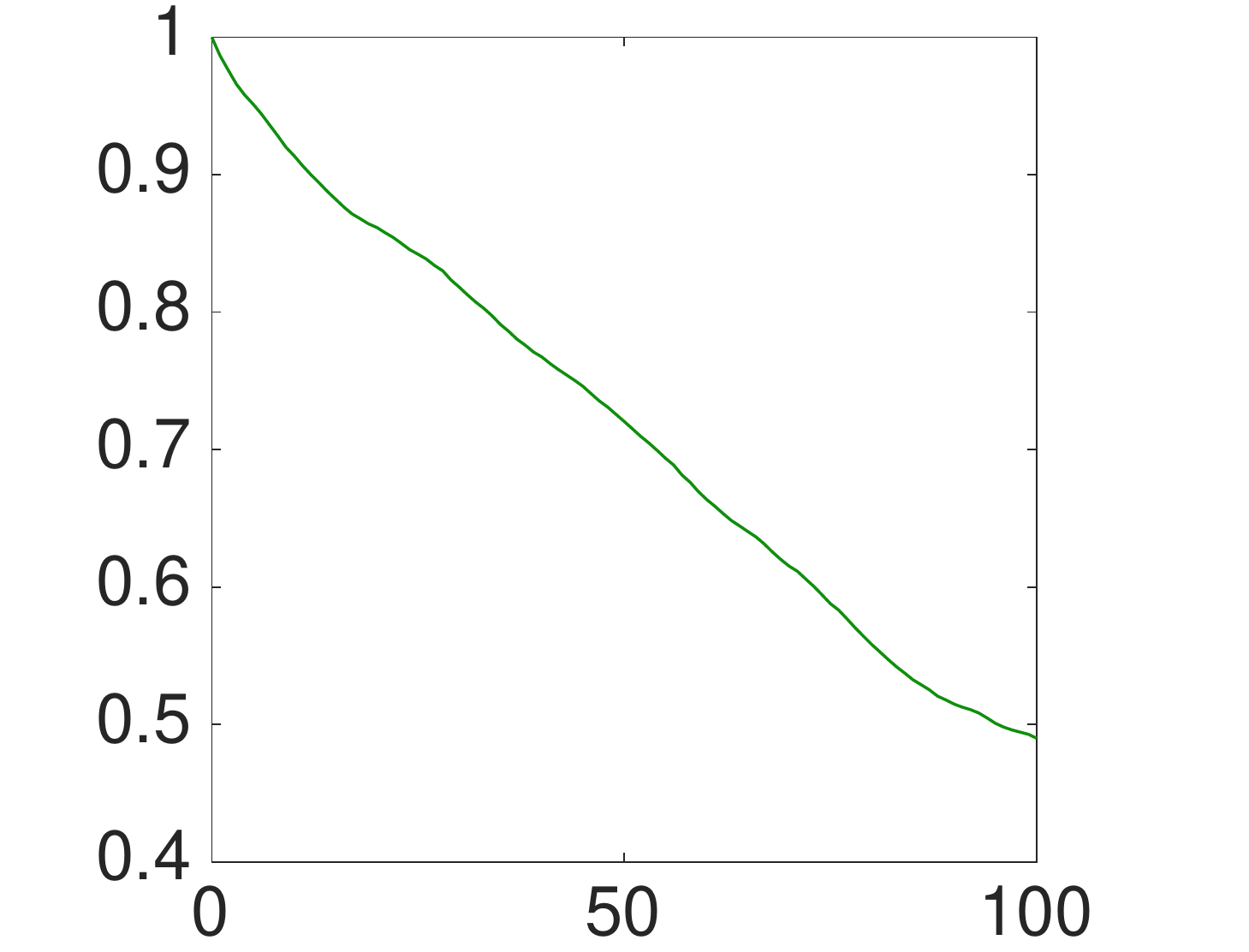}
            \includegraphics[width=\linewidth]{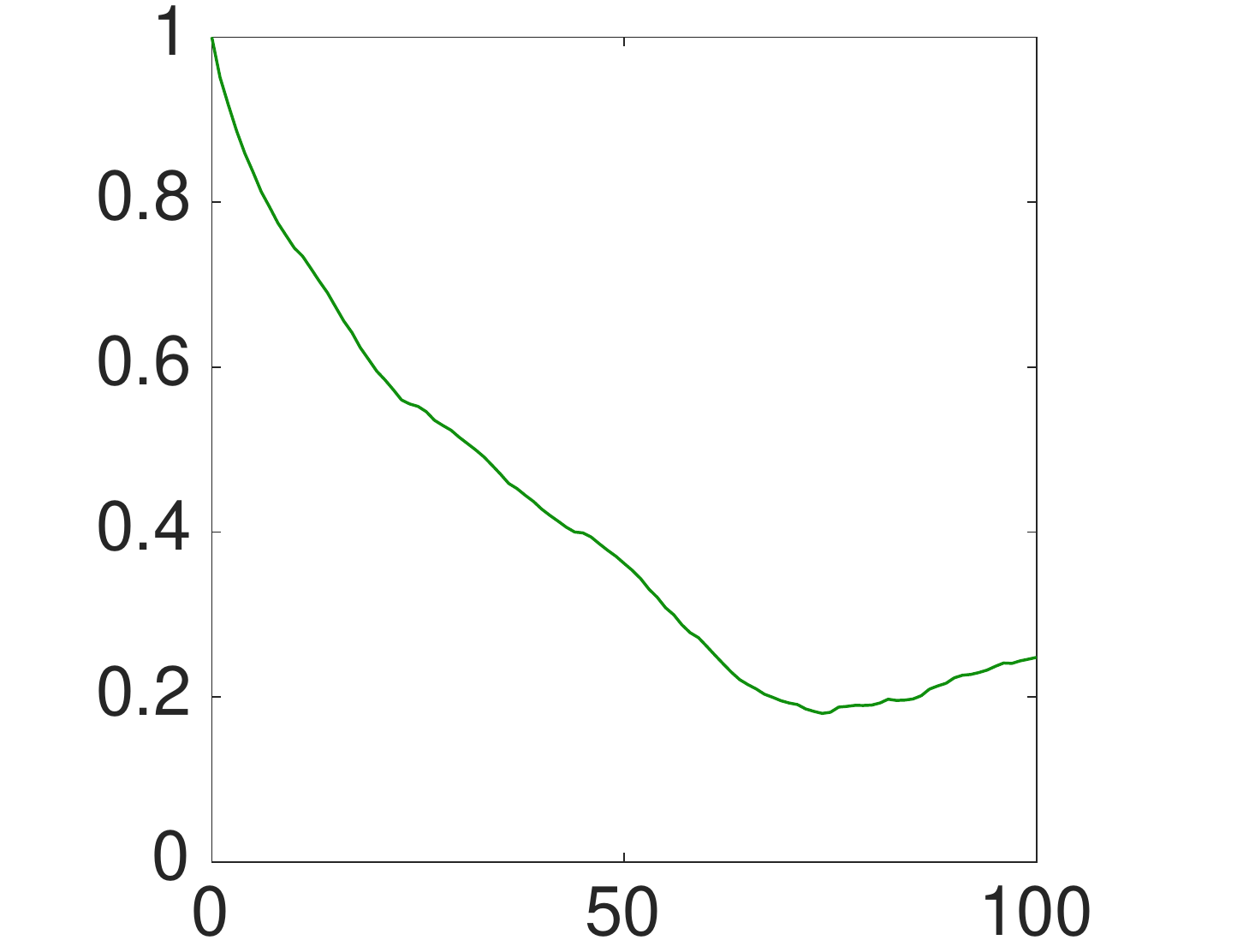}
            \includegraphics[width=\linewidth]{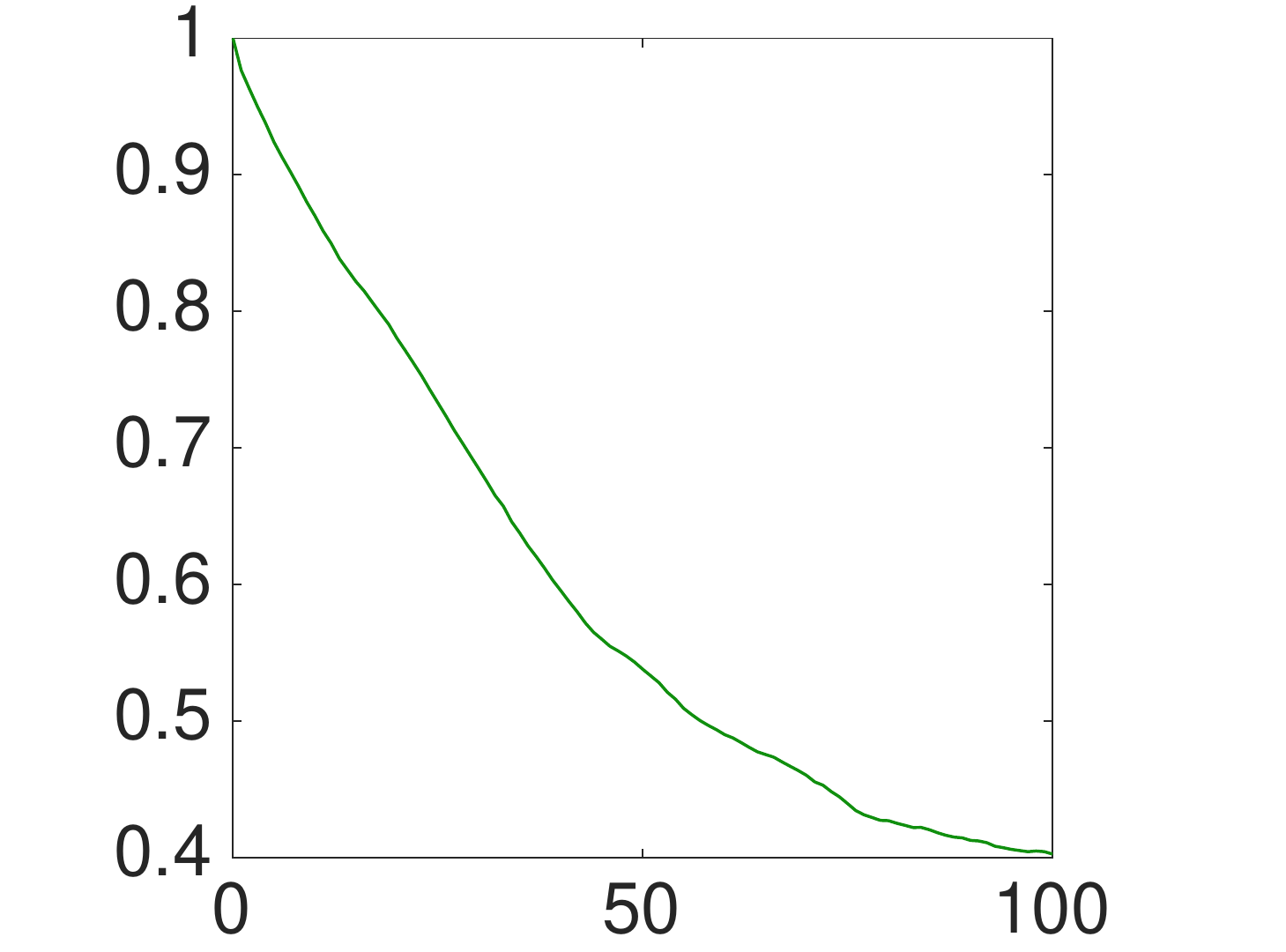}
            \caption{Sheet }\end{subfigure}
        
        \caption{  Trace plots for pixel $43\times77$.  Top rows: MwG. Middle rows: RAM. Bottom rows: HMC.}
        \label{t1}
    \end{figure}

\begin{figure}
        \centering
        \begin{subfigure}[b]{\qhei}
            \includegraphics[width=\linewidth]{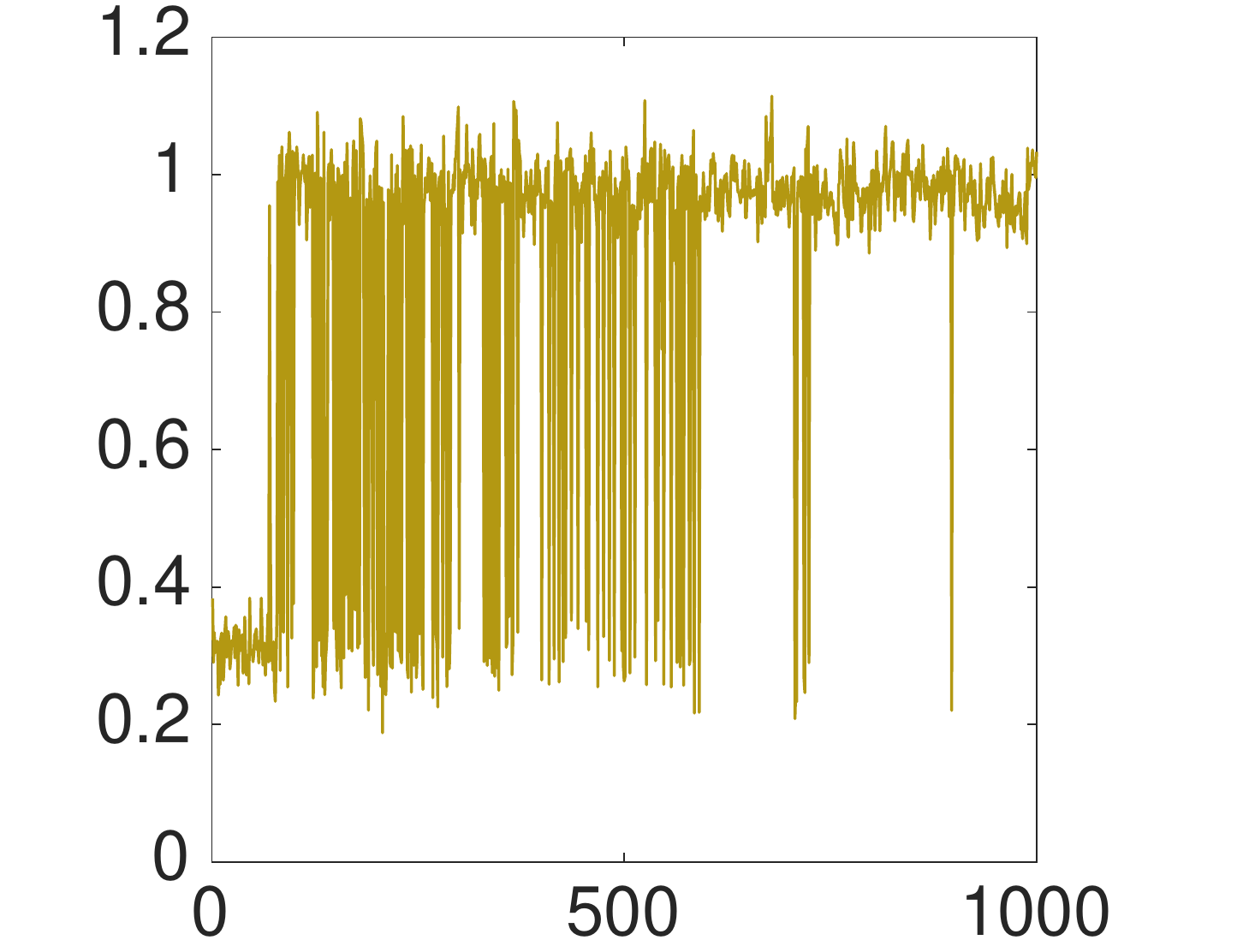}
            \includegraphics[width=\linewidth]{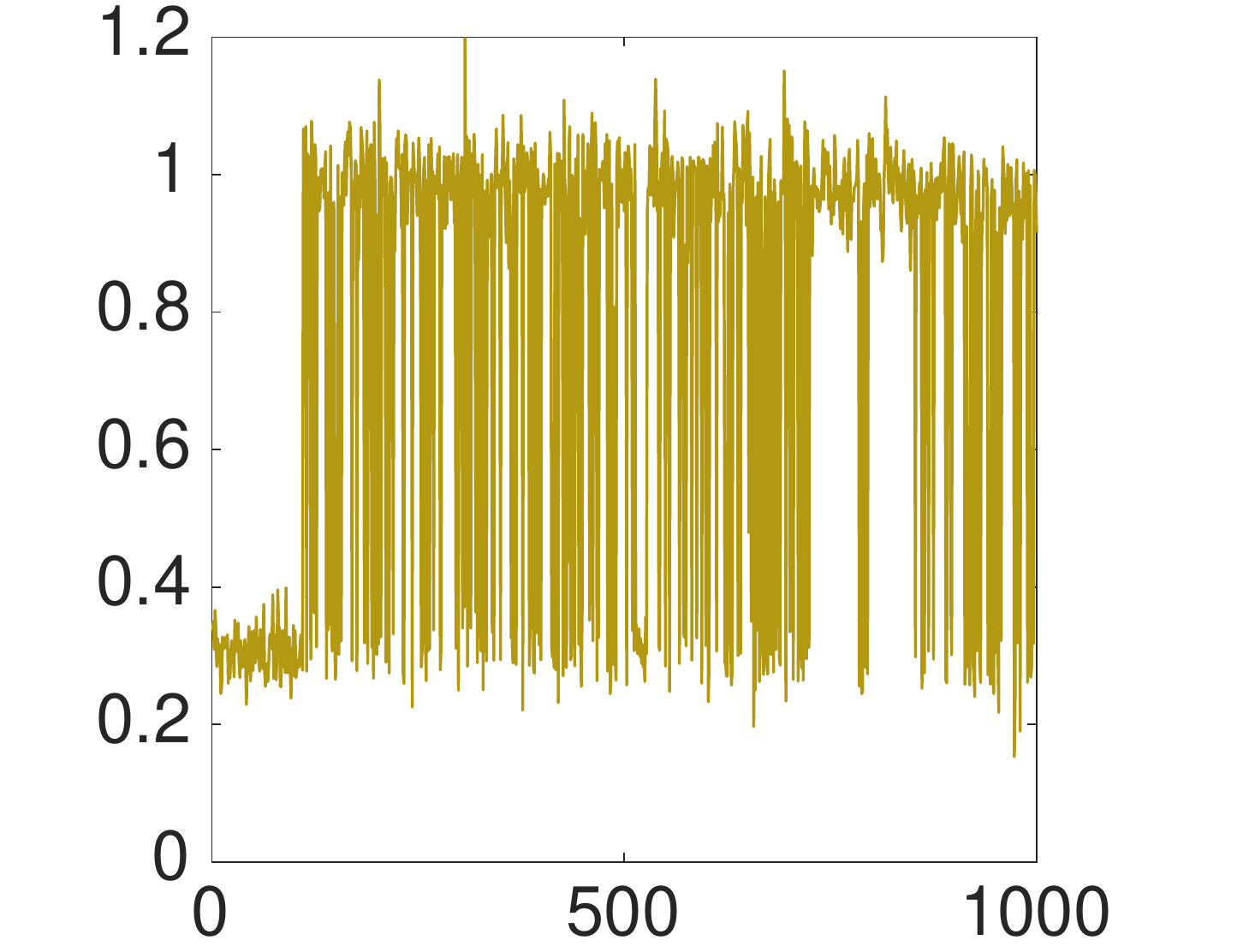}
            \includegraphics[width=\linewidth]{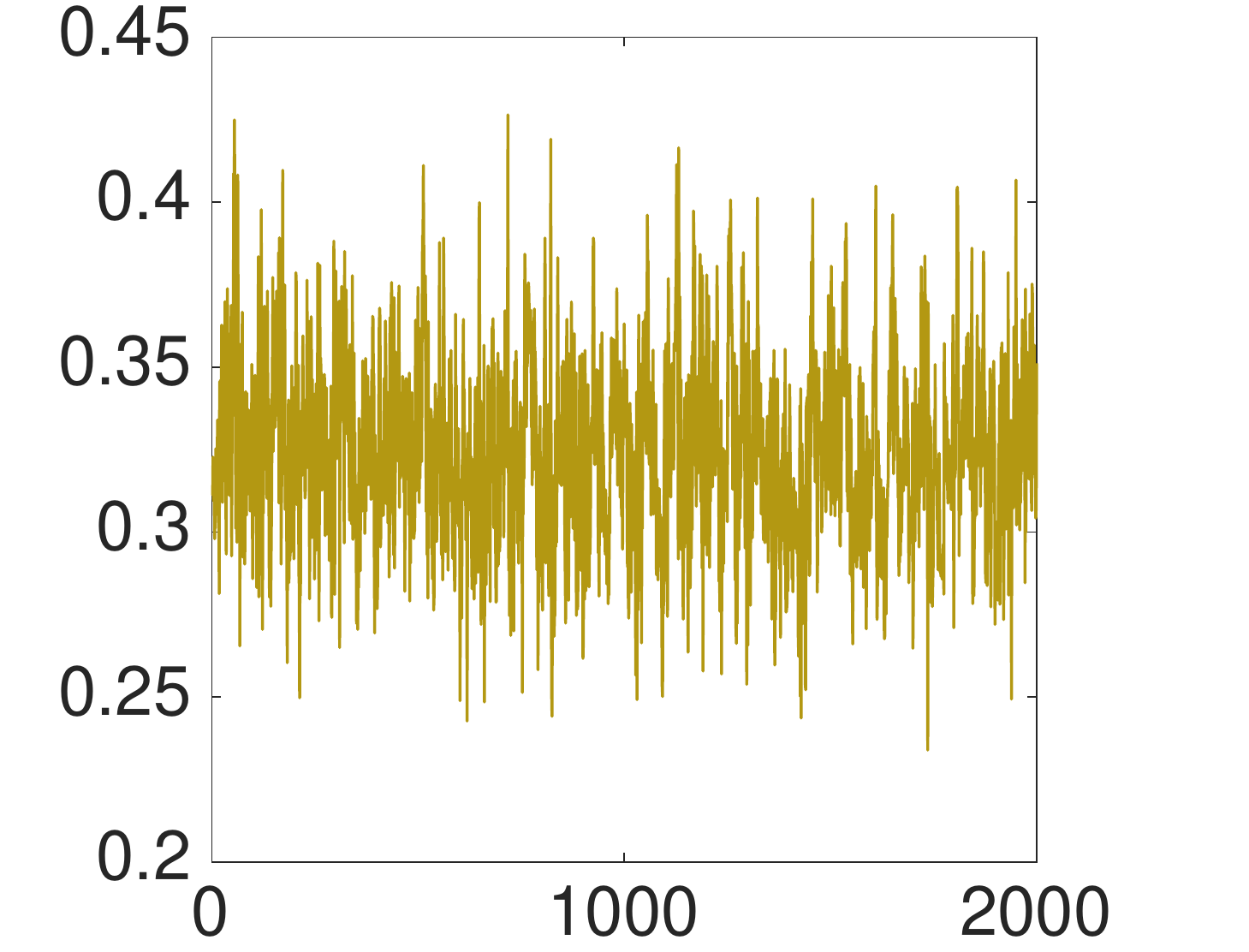}
            \caption{Aniso 1st }\end{subfigure}
        \begin{subfigure}[b]{\qhei}
            \includegraphics[width=\linewidth]{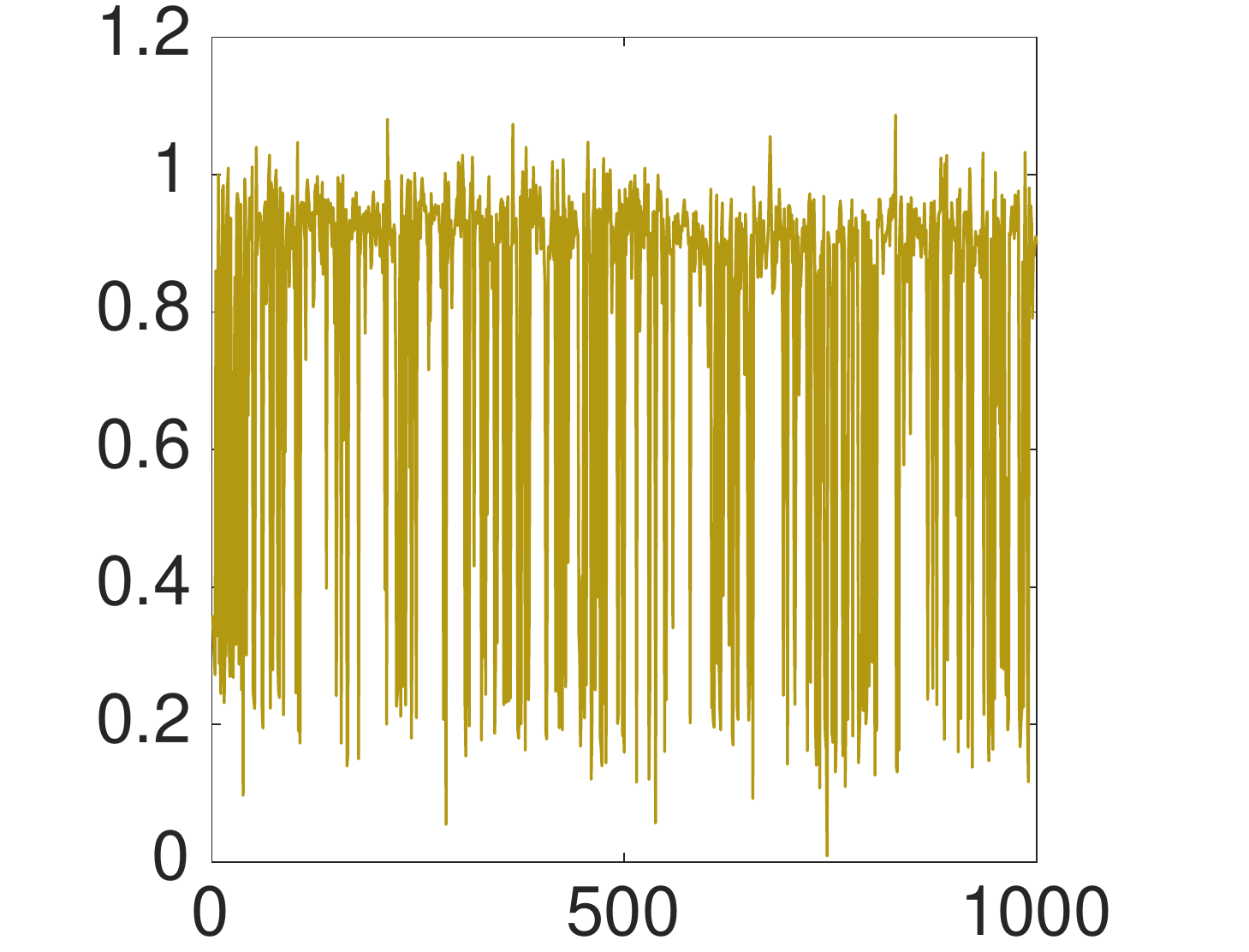}
            \includegraphics[width=\linewidth]{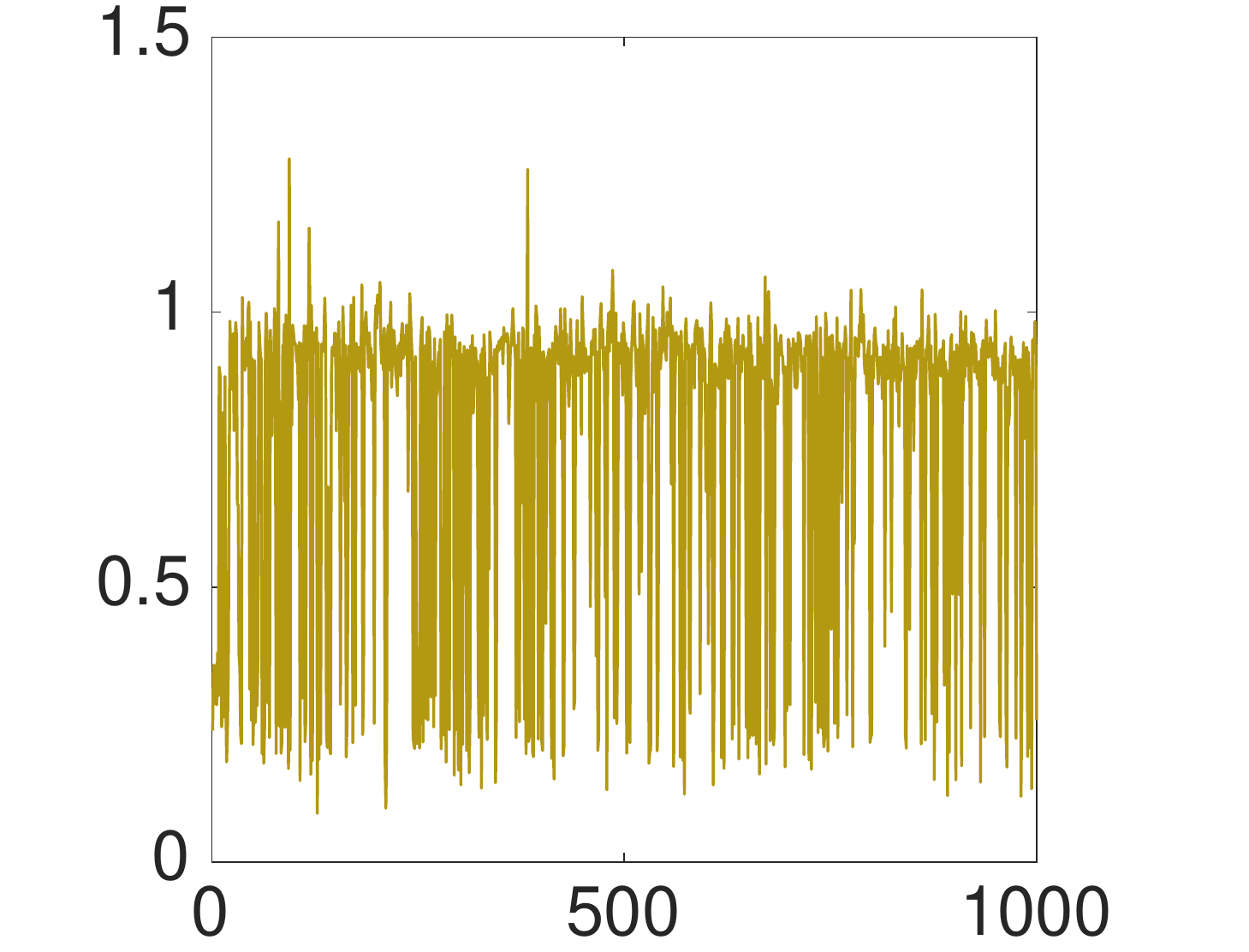}
            \includegraphics[width=\linewidth]{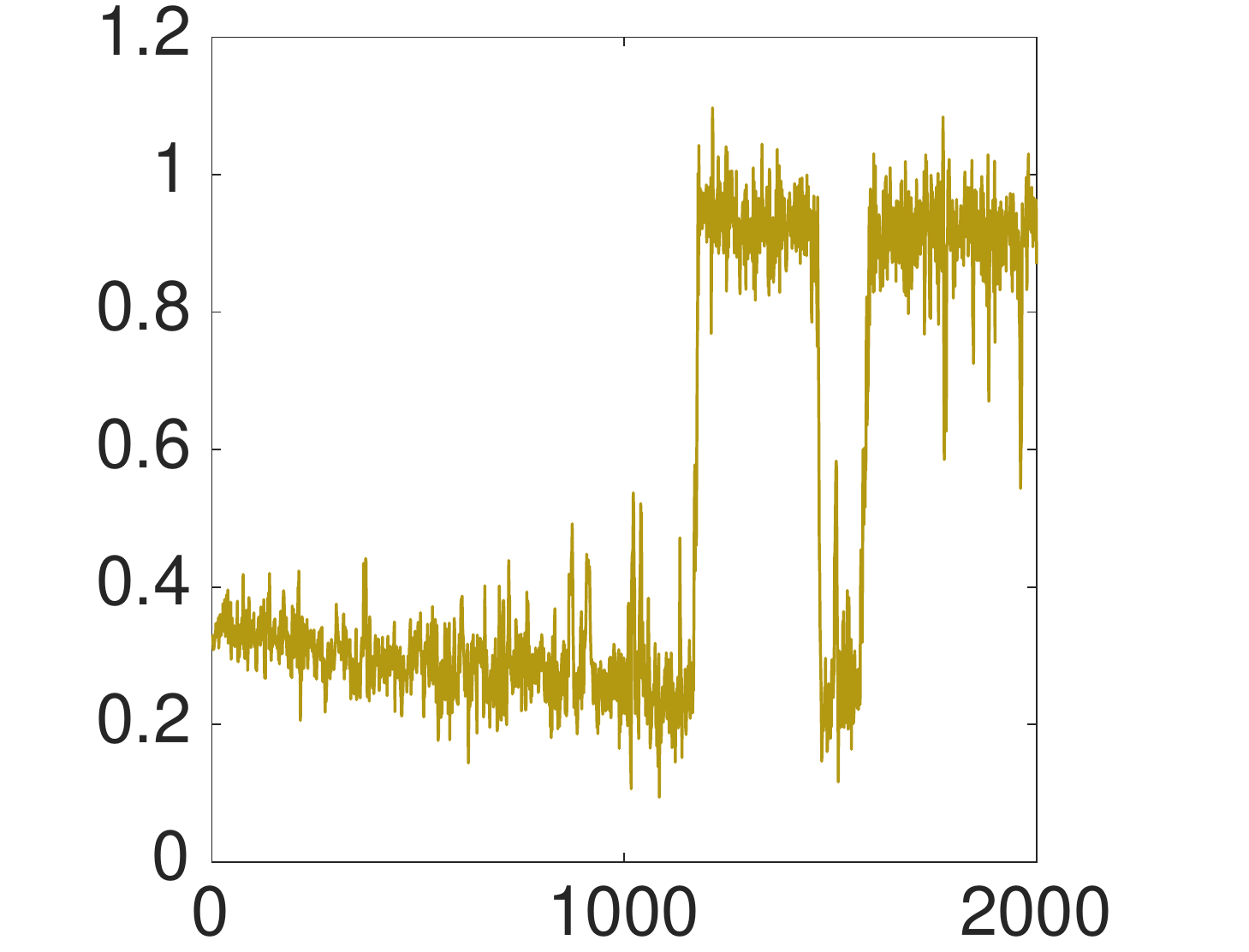}
            \caption{Iso 1st }\end{subfigure}
        \begin{subfigure}[b]{\qhei}
            \includegraphics[width=\linewidth]{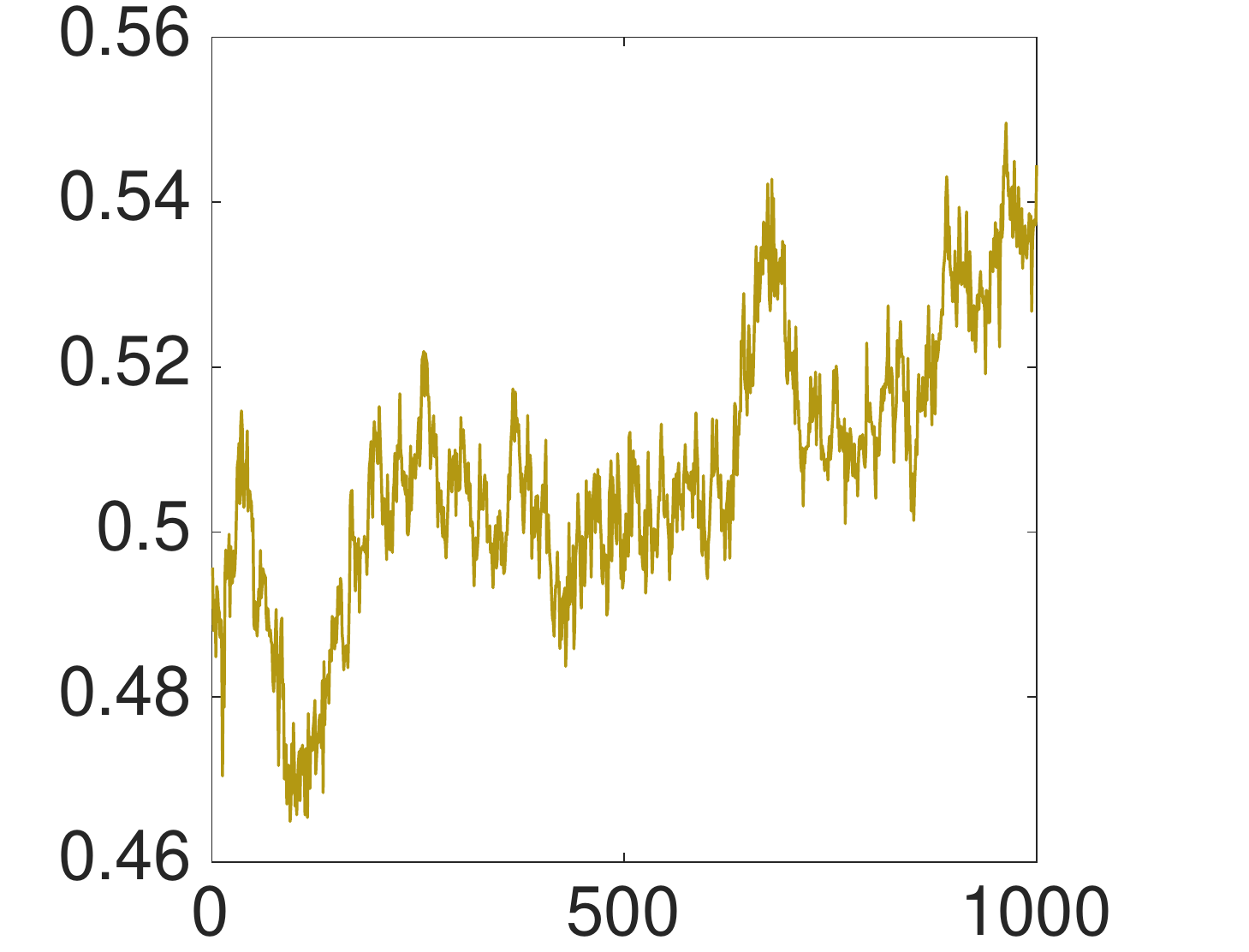}
            \includegraphics[width=\linewidth]{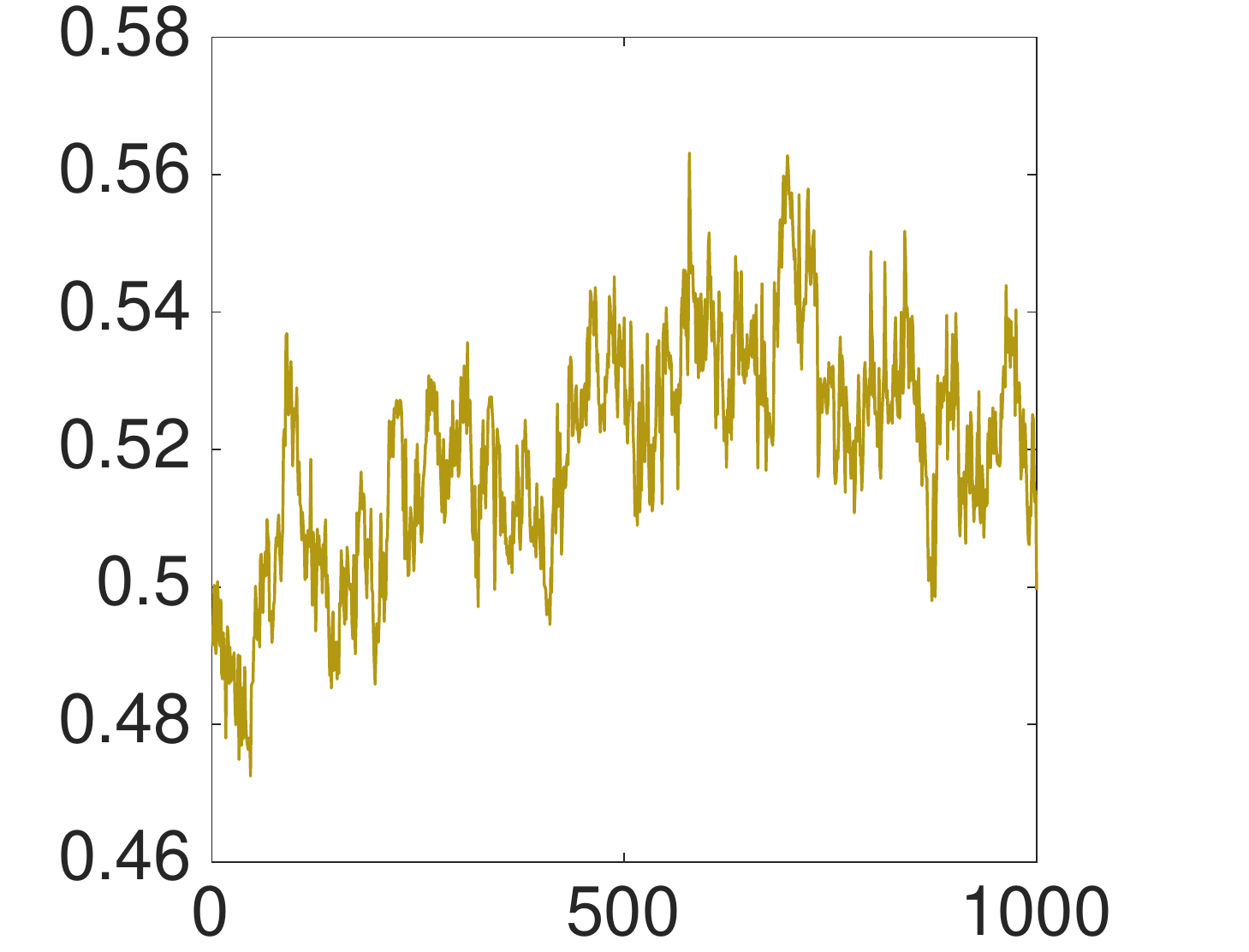}
            \includegraphics[width=\linewidth]{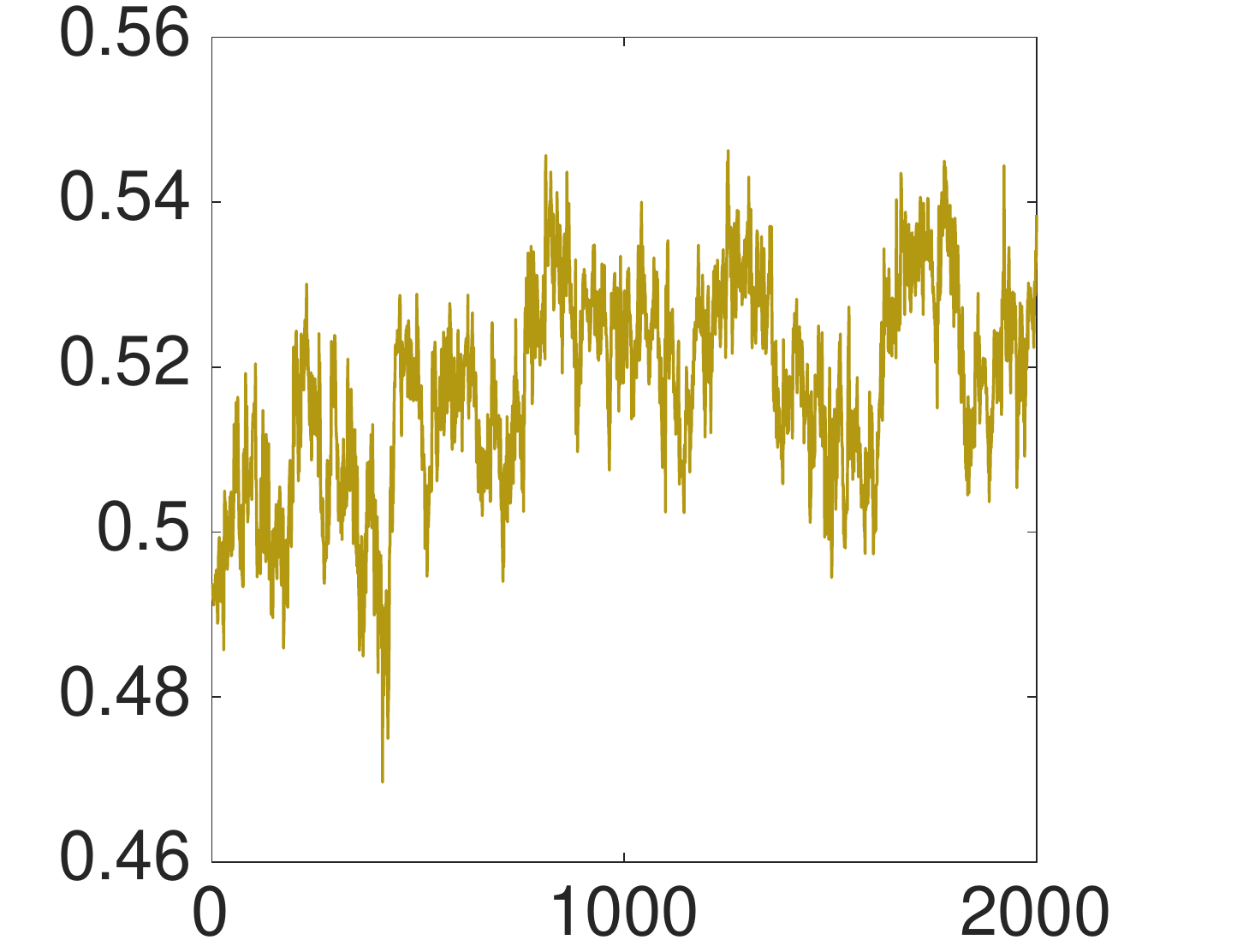}
            \caption{Aniso 2nd }\end{subfigure}
        \begin{subfigure}[b]{\qhei}
            \includegraphics[width=\linewidth]{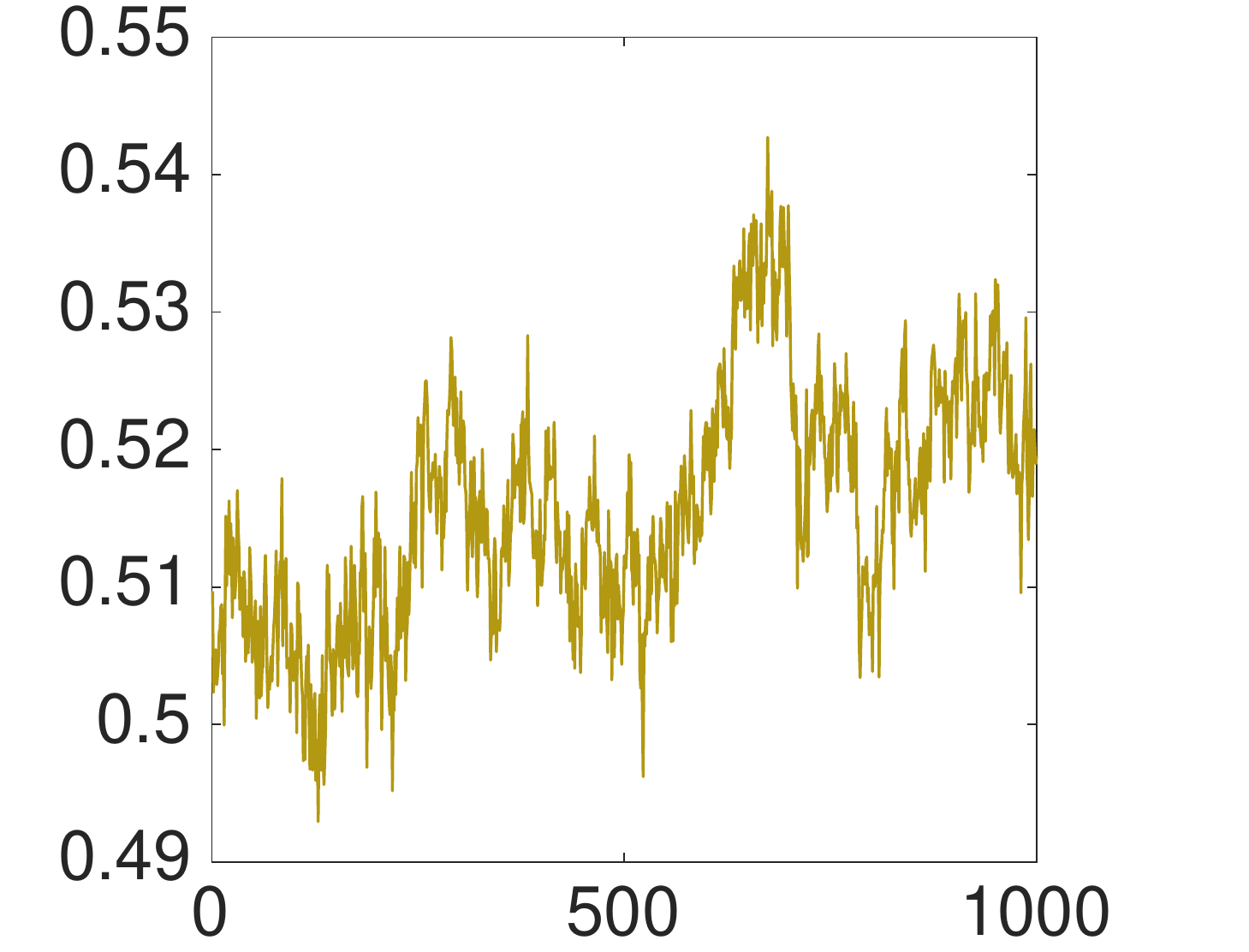}
            \includegraphics[width=\linewidth]{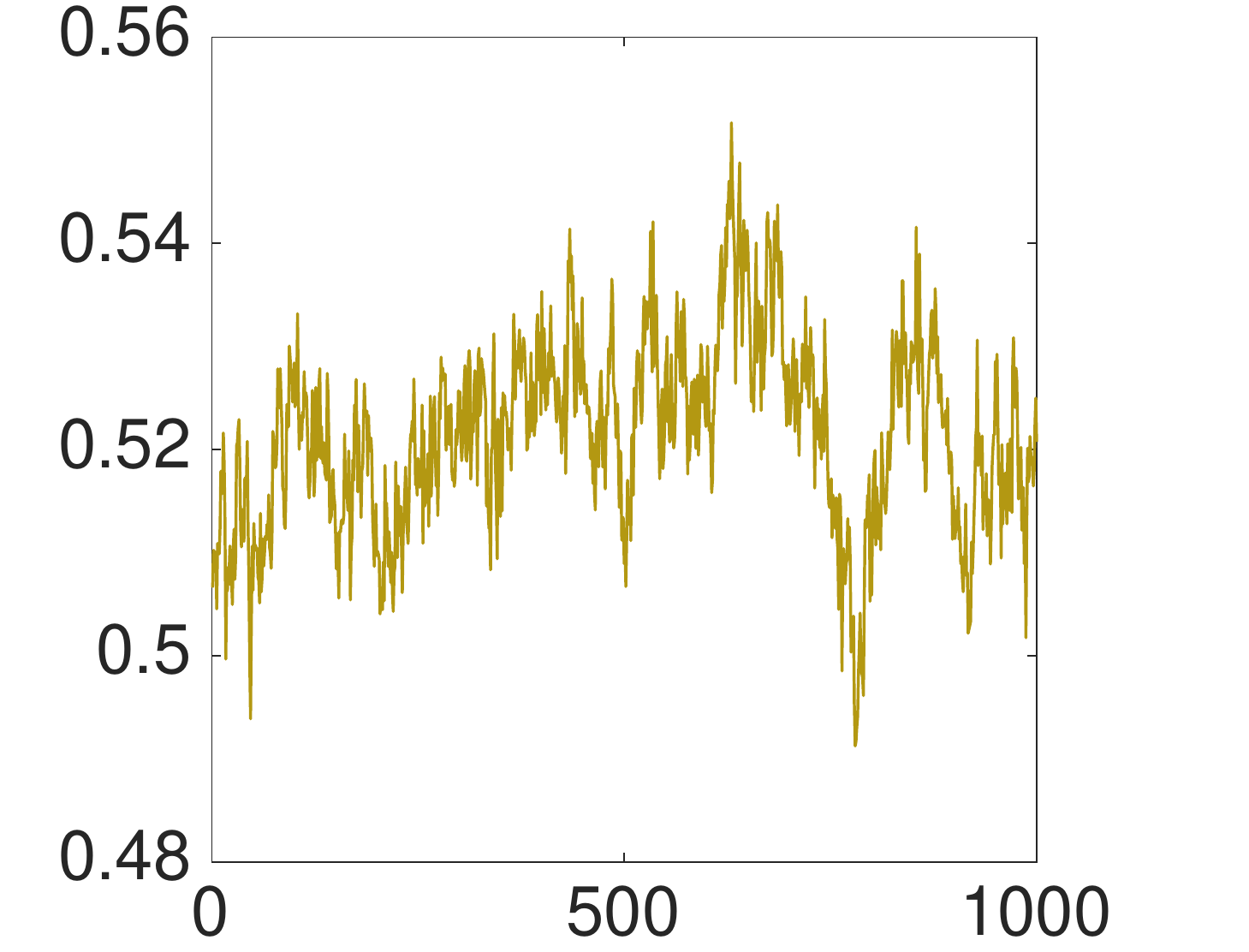}
            \includegraphics[width=\linewidth]{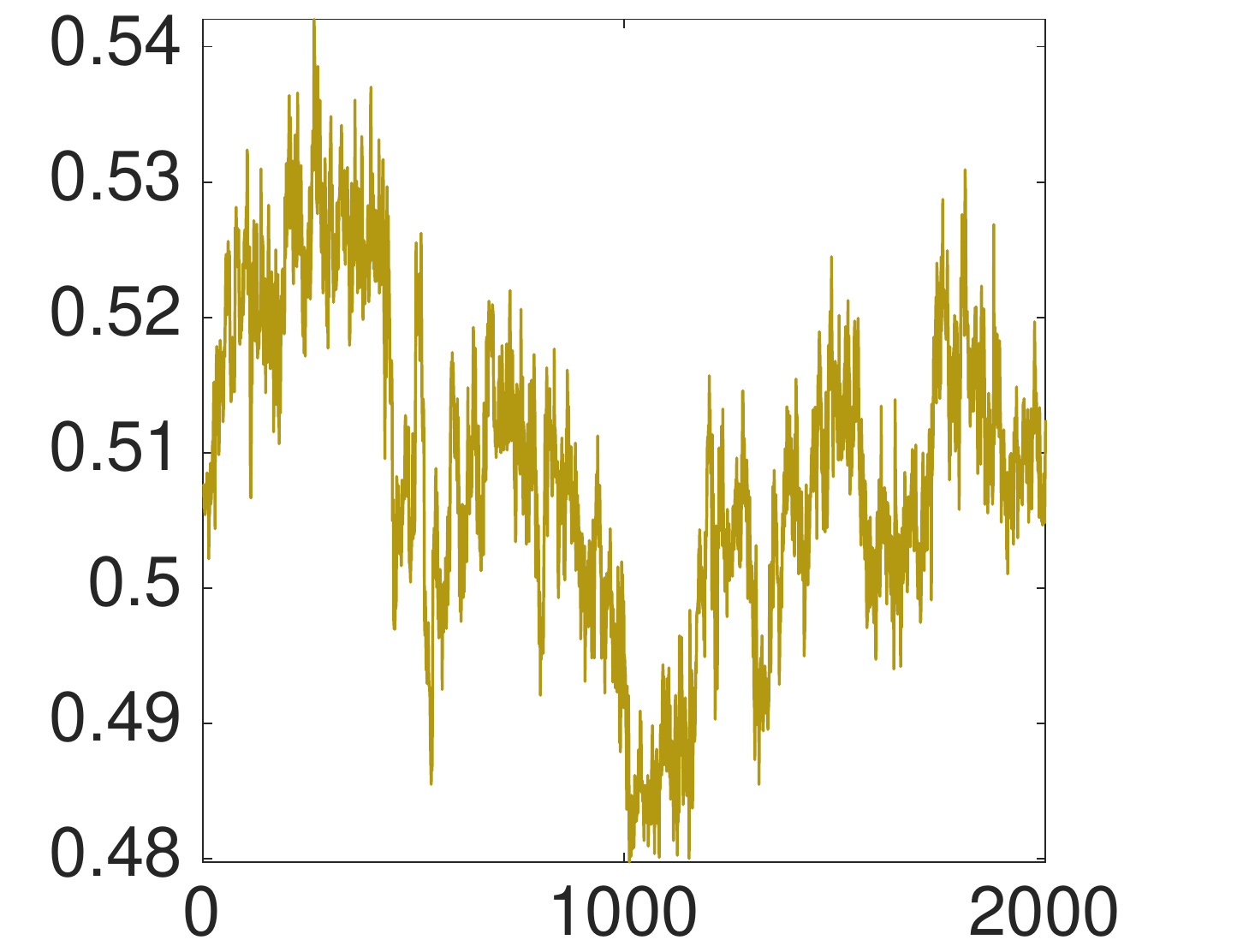}
            \caption{Iso 2nd }\end{subfigure}
        \begin{subfigure}[b]{\qhei}
            \includegraphics[width=\linewidth]{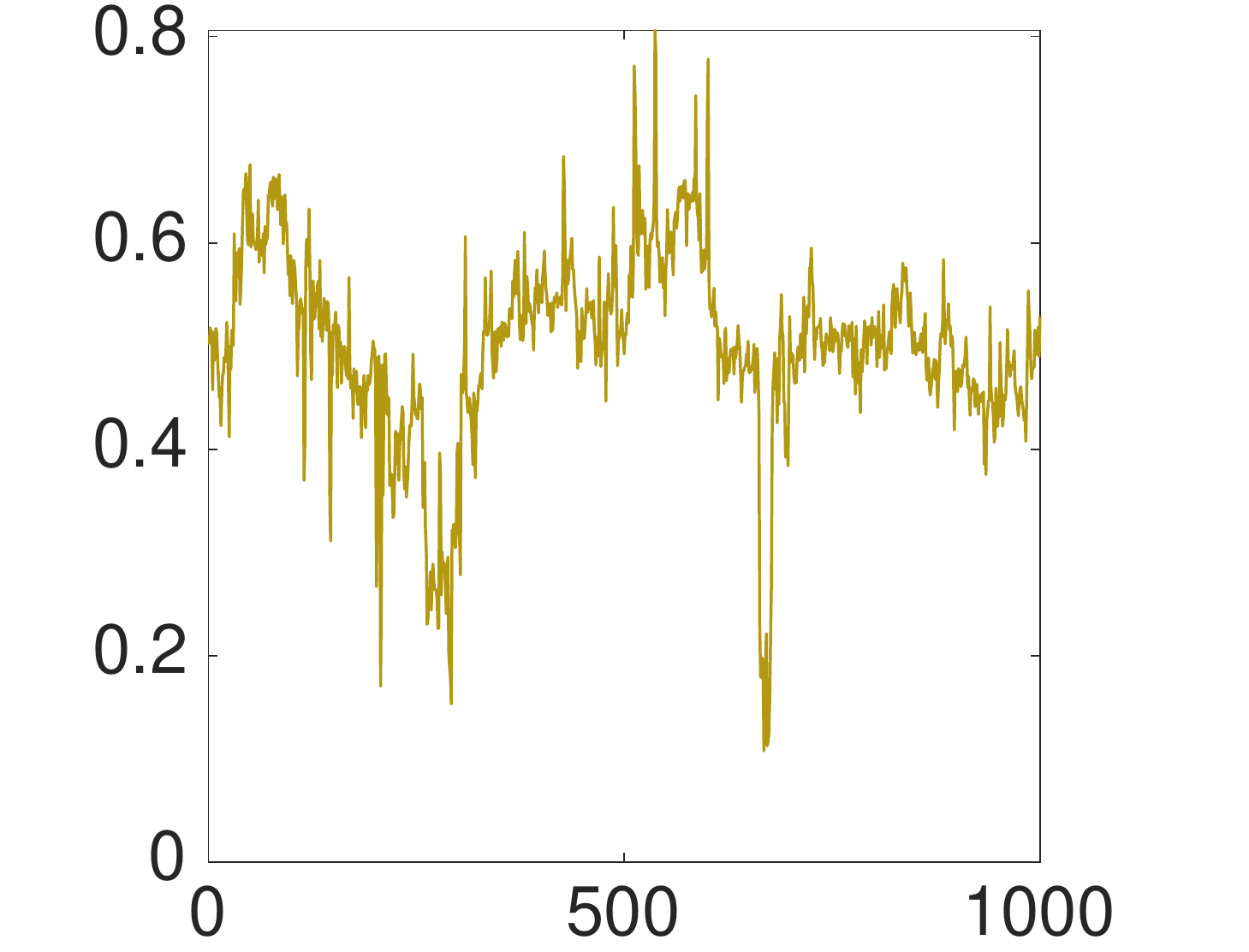}
            \includegraphics[width=\linewidth]{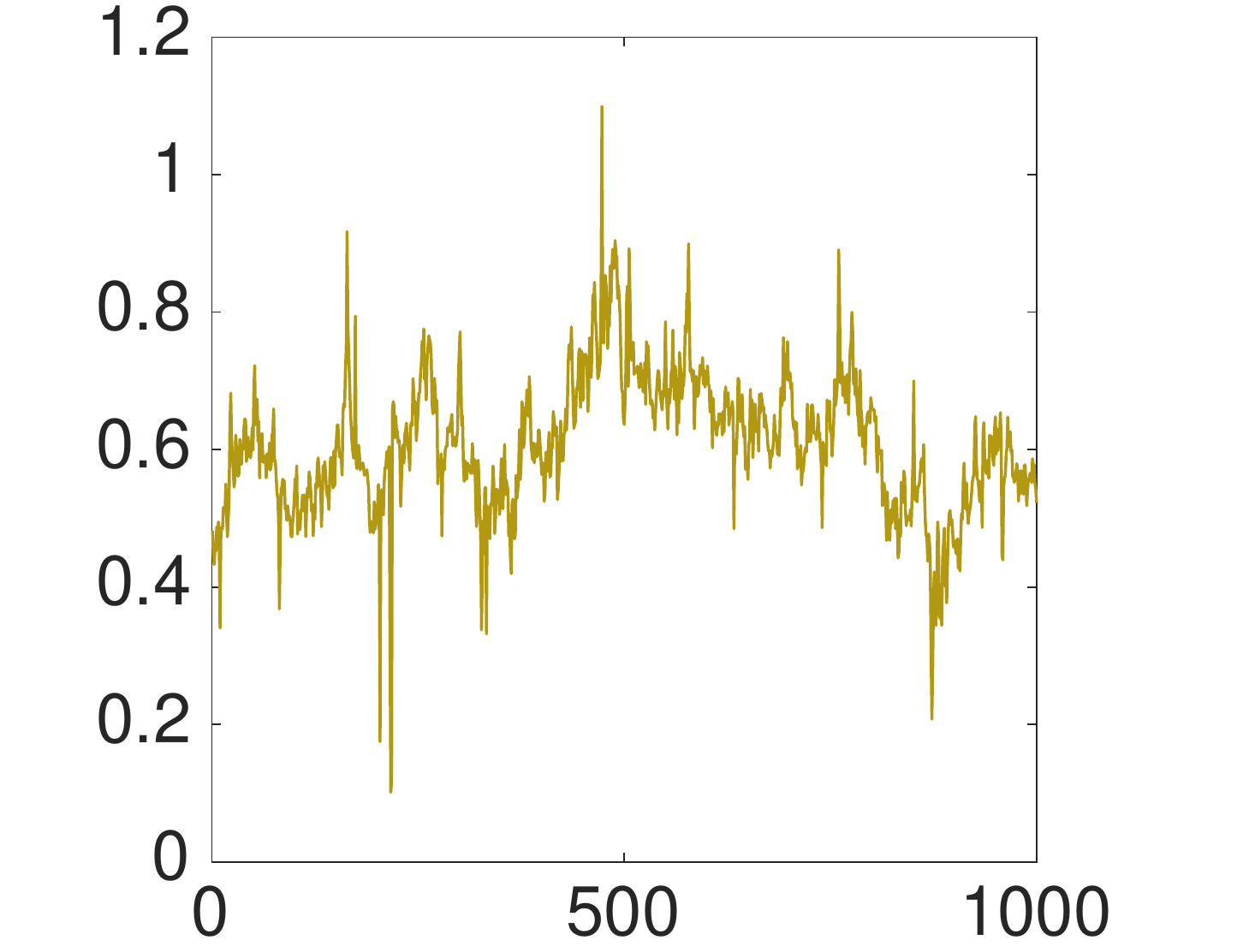}
            \includegraphics[width=\linewidth]{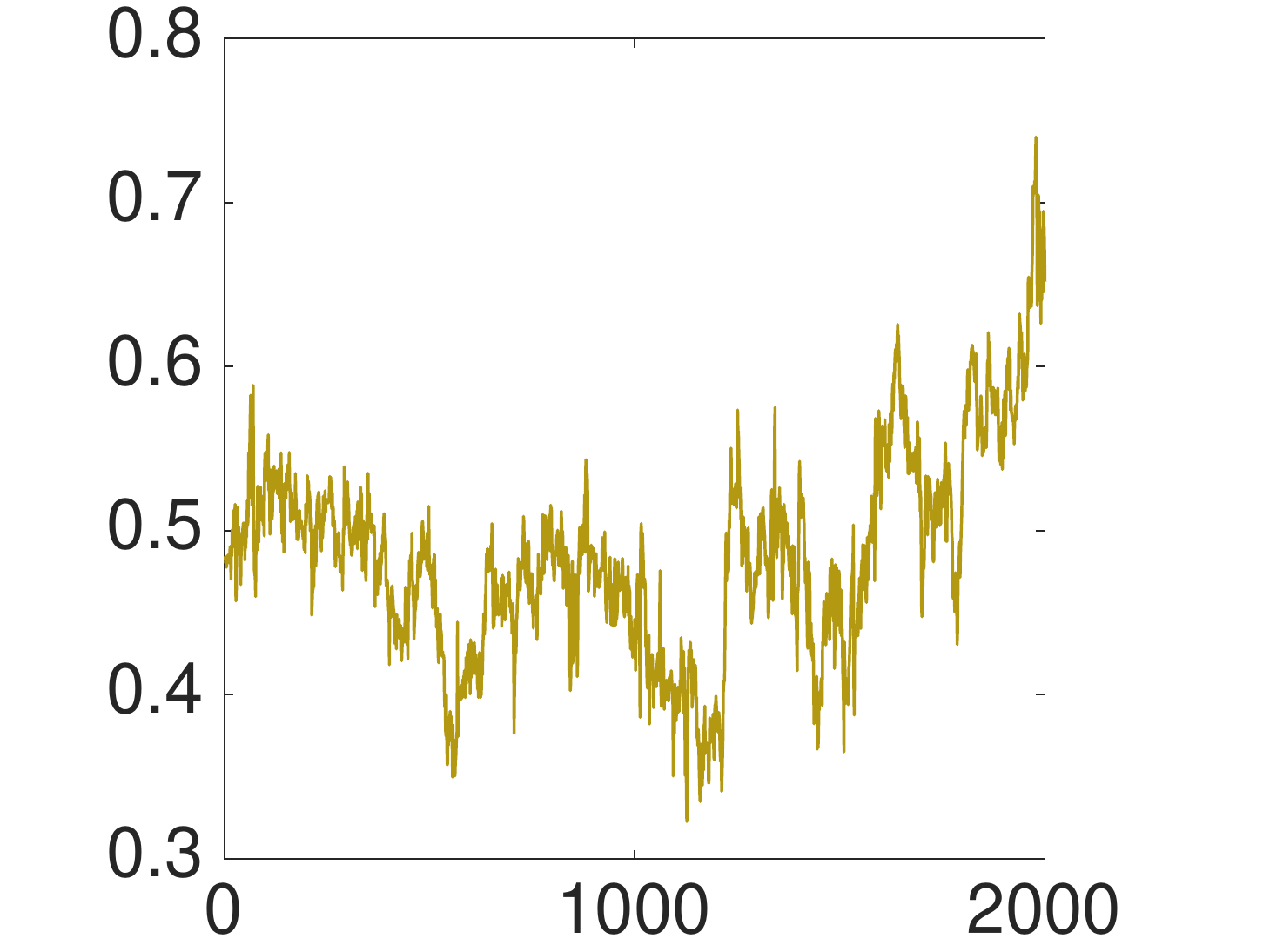}
            \caption{SPDE }\end{subfigure}
        \begin{subfigure}[b]{\qhei}
            \includegraphics[width=\linewidth]{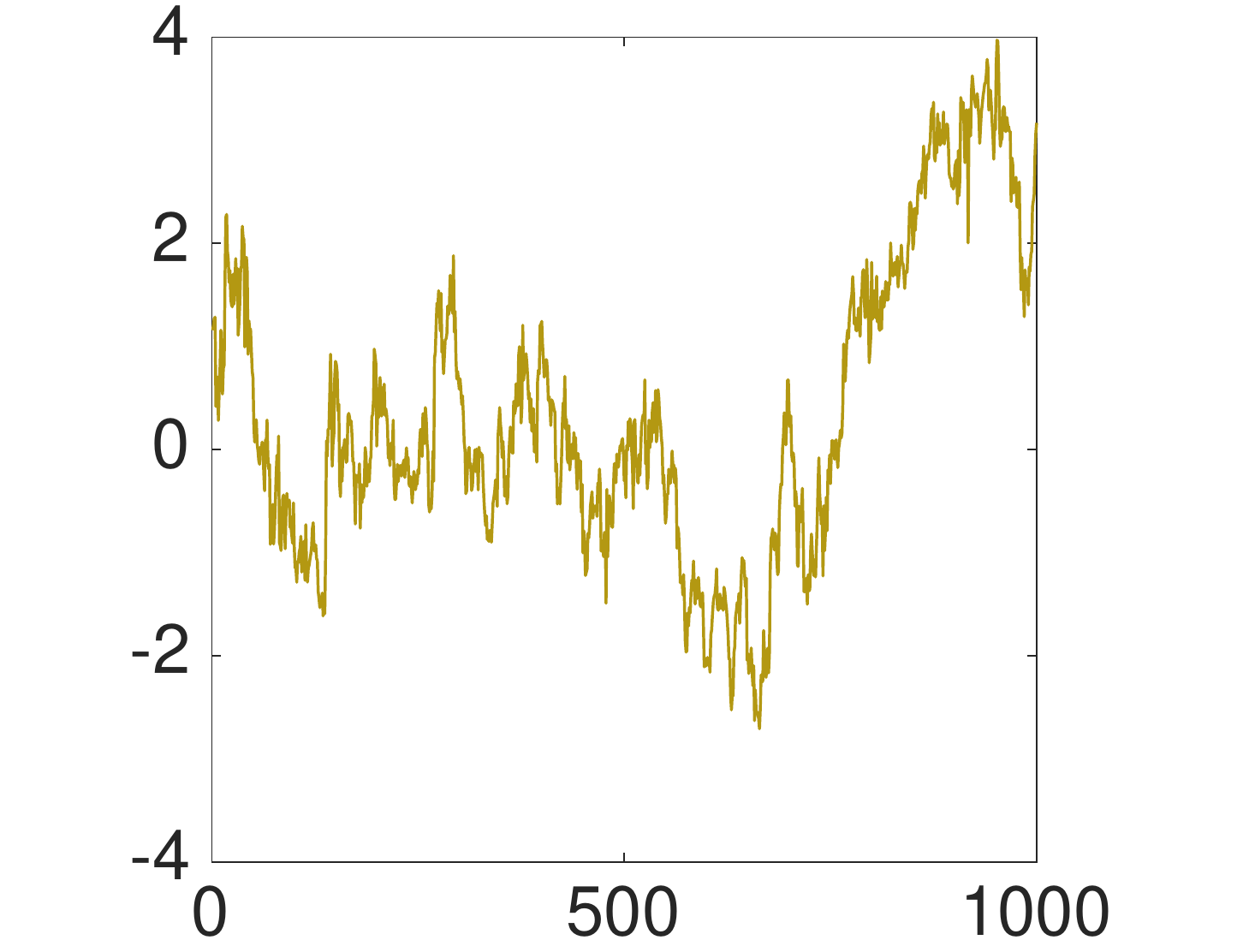}
            \includegraphics[width=\linewidth]{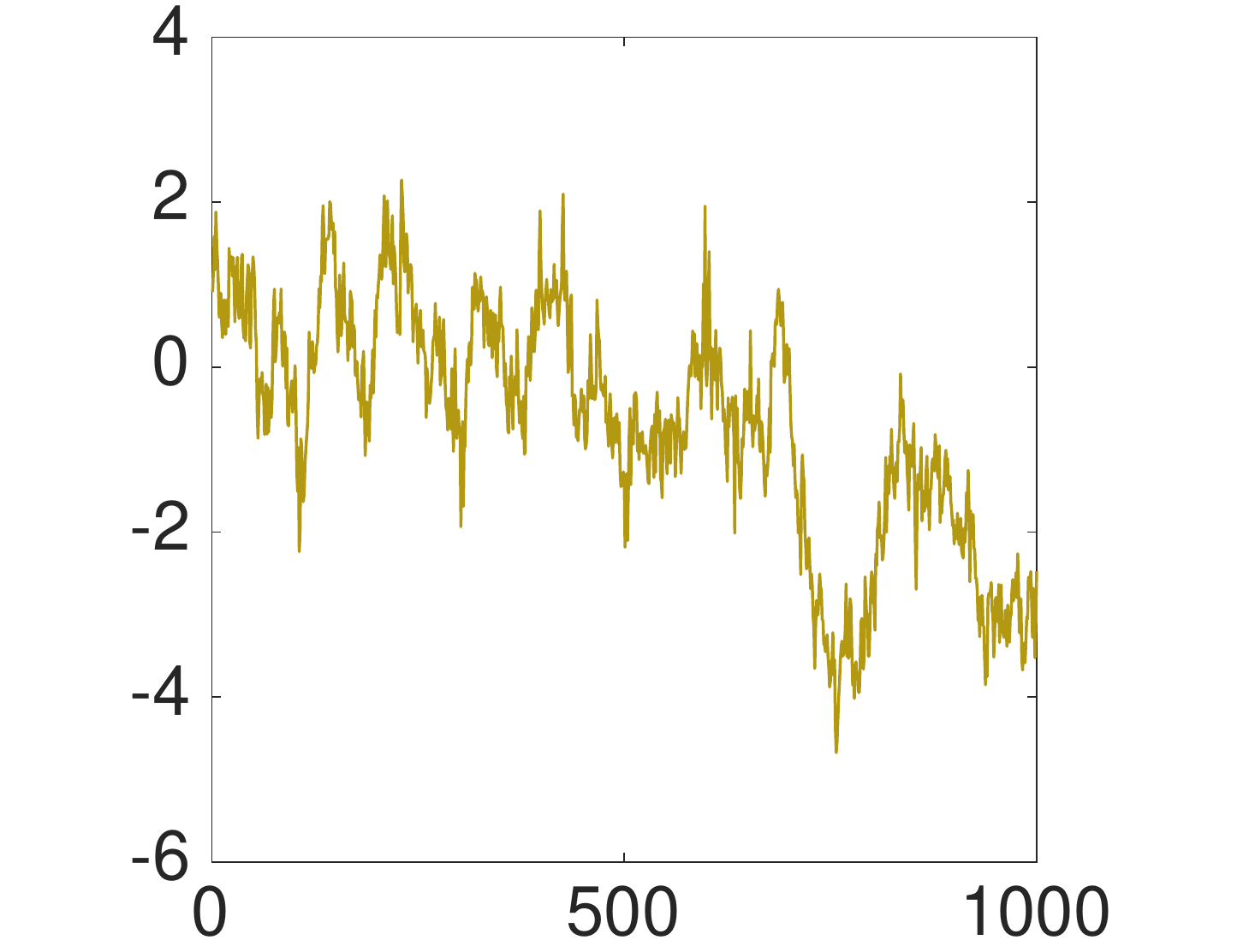}
            \includegraphics[width=\linewidth]{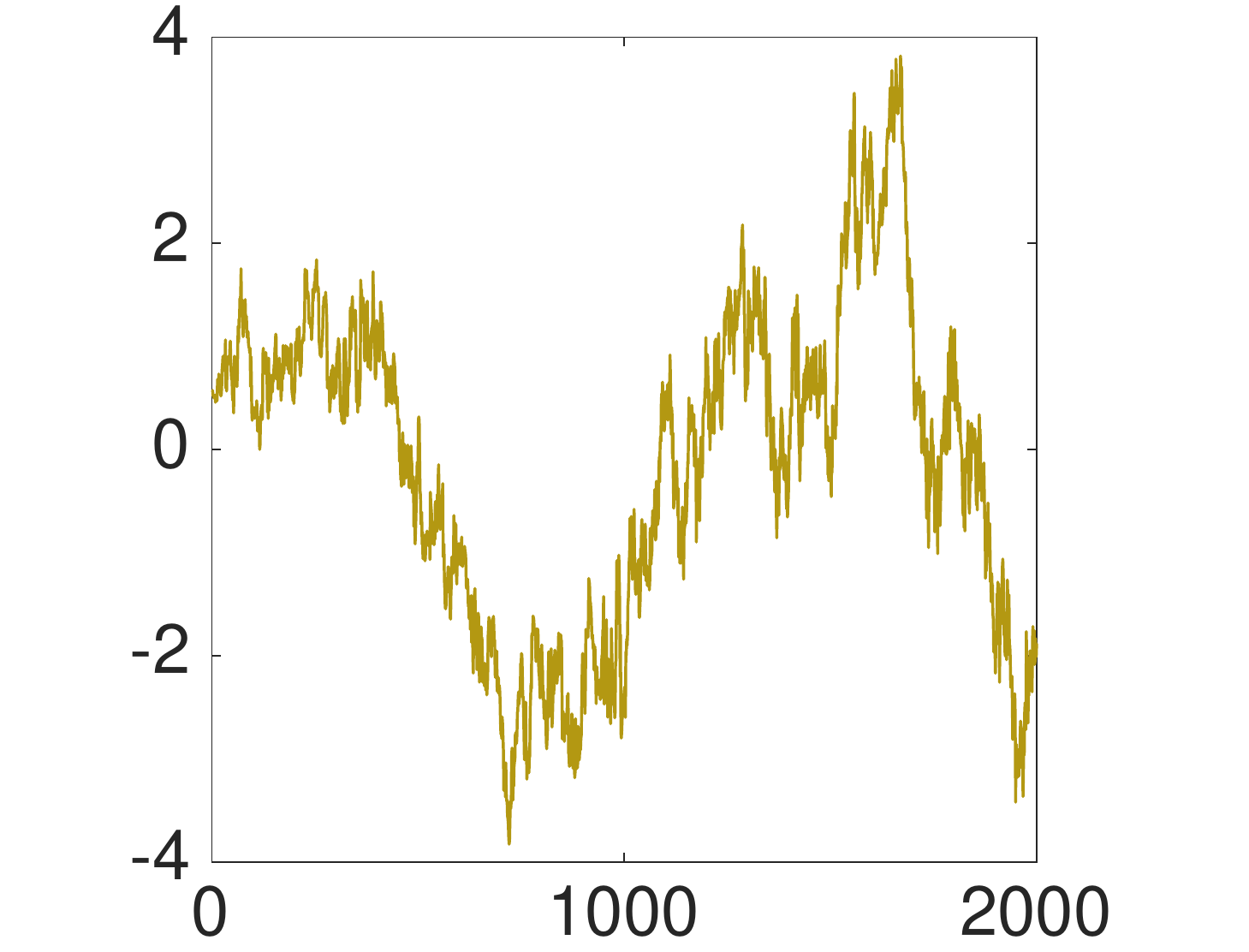}
            \caption{Sheet }\end{subfigure}
        
        \begin{subfigure}[b]{\qhei}
            \includegraphics[width=\linewidth]{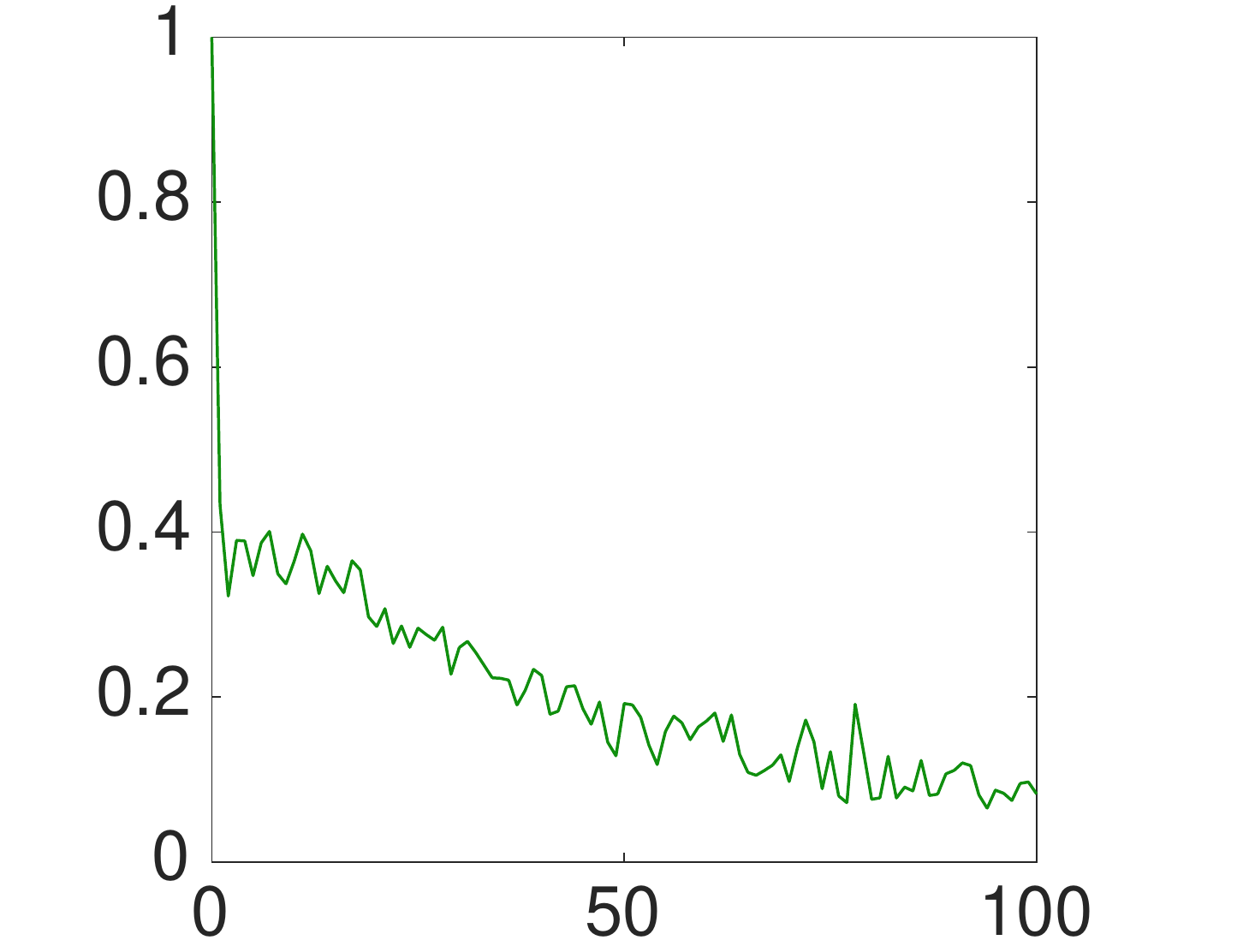}
            \includegraphics[width=\linewidth]{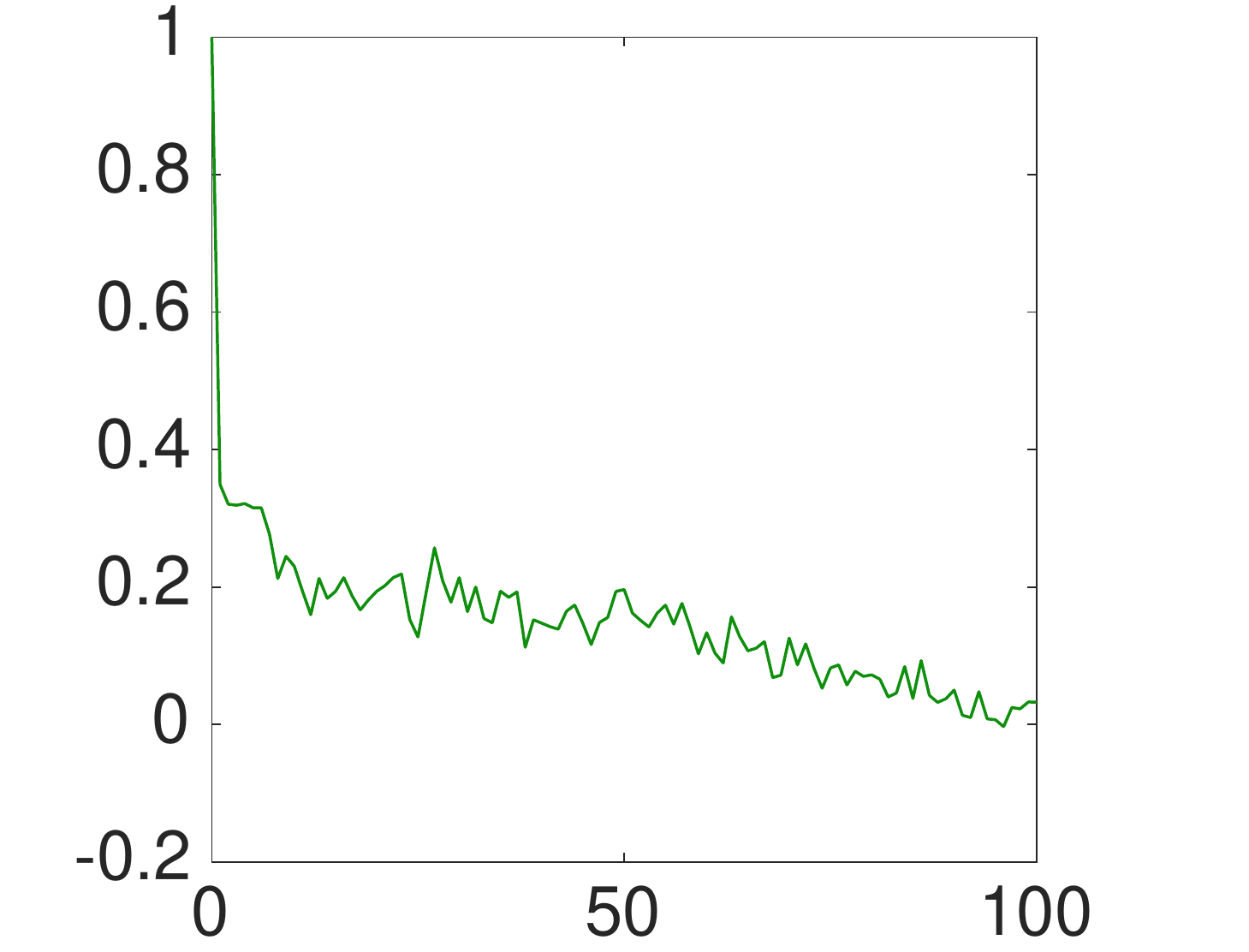}
            \includegraphics[width=\linewidth]{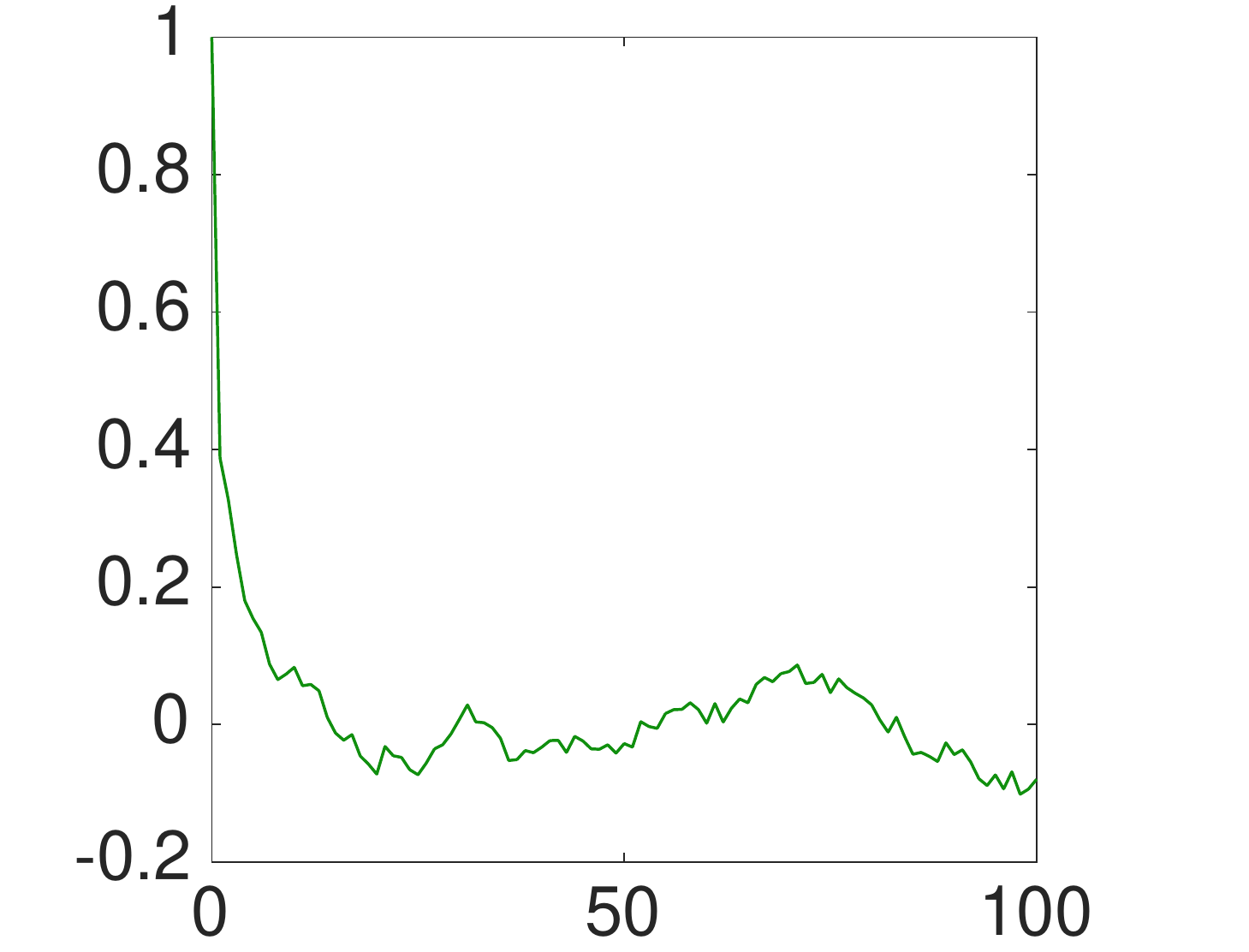}
            \caption{Aniso 1st }\end{subfigure}
        \begin{subfigure}[b]{\qhei}
            \includegraphics[width=\linewidth]{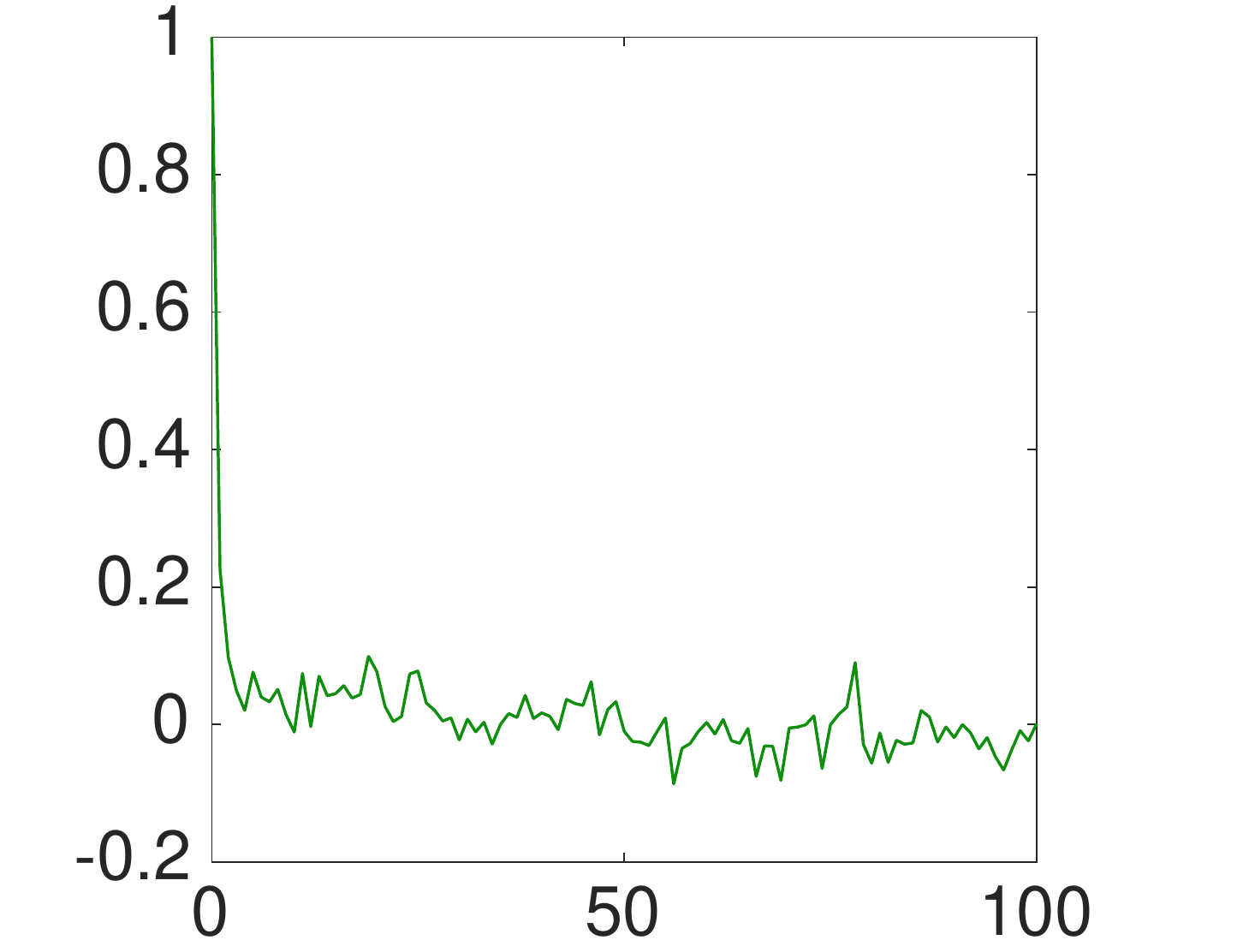}
            \includegraphics[width=\linewidth]{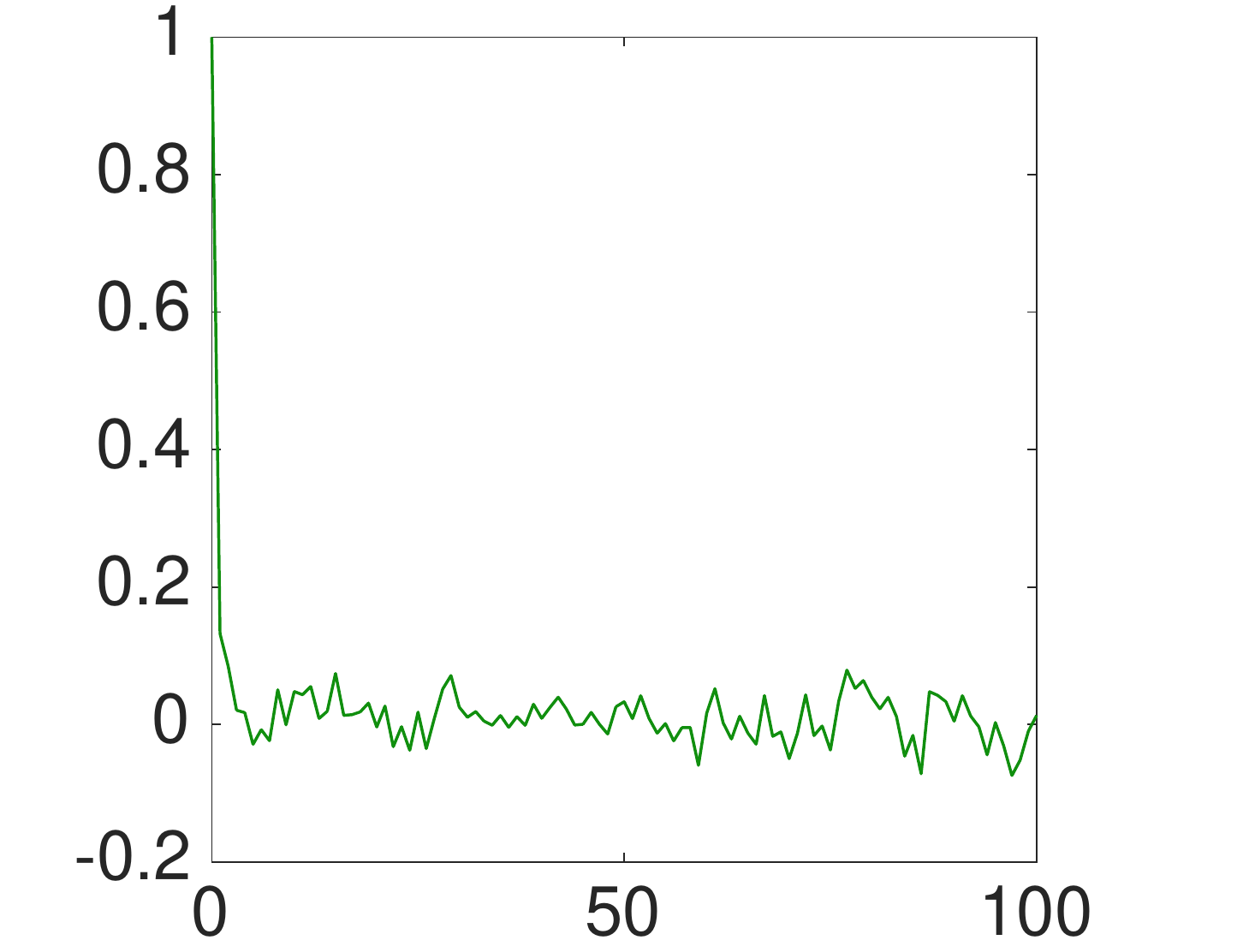}
            \includegraphics[width=\linewidth]{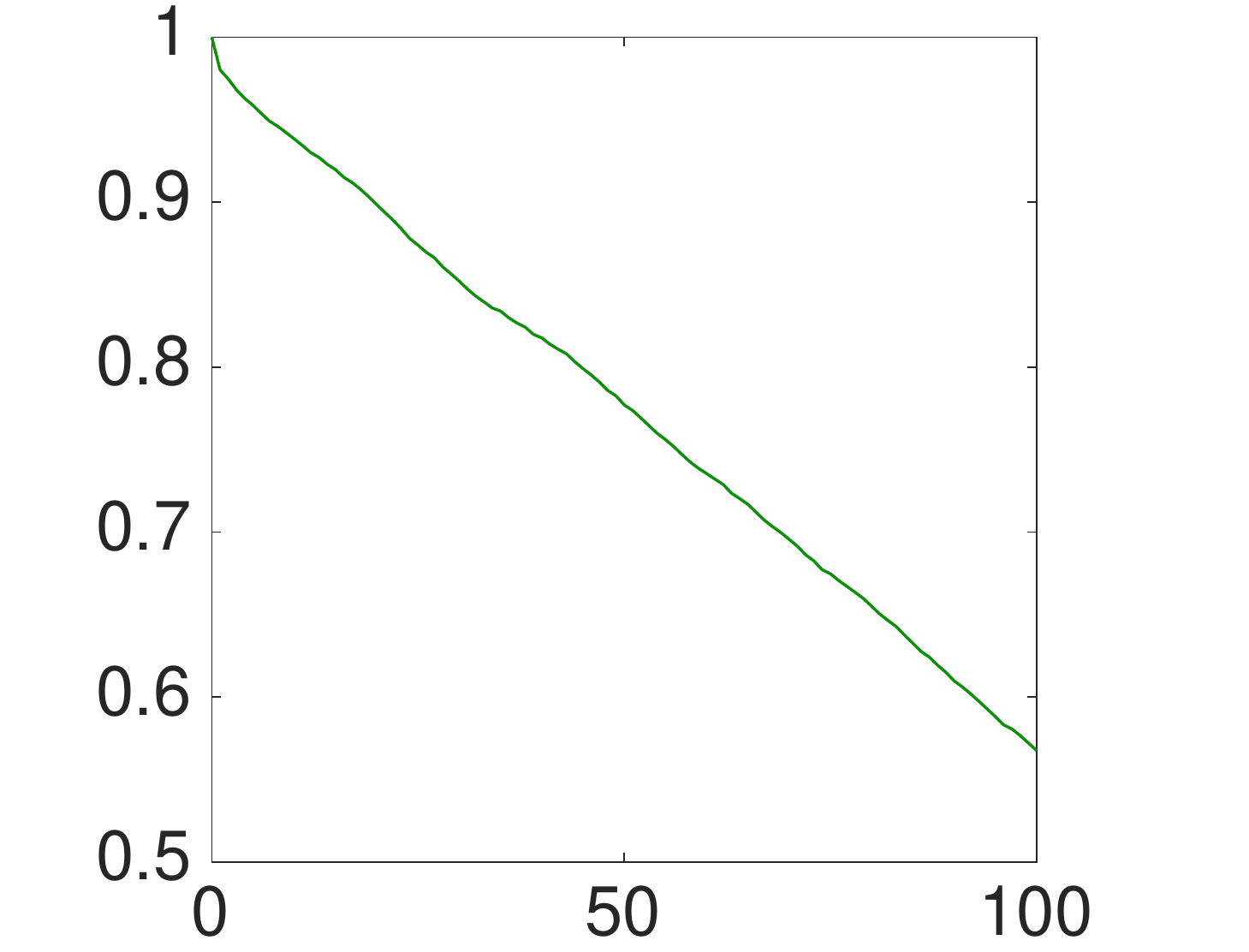}
            \caption{Iso 1st }\end{subfigure}
        \begin{subfigure}[b]{\qhei}
            \includegraphics[width=\linewidth]{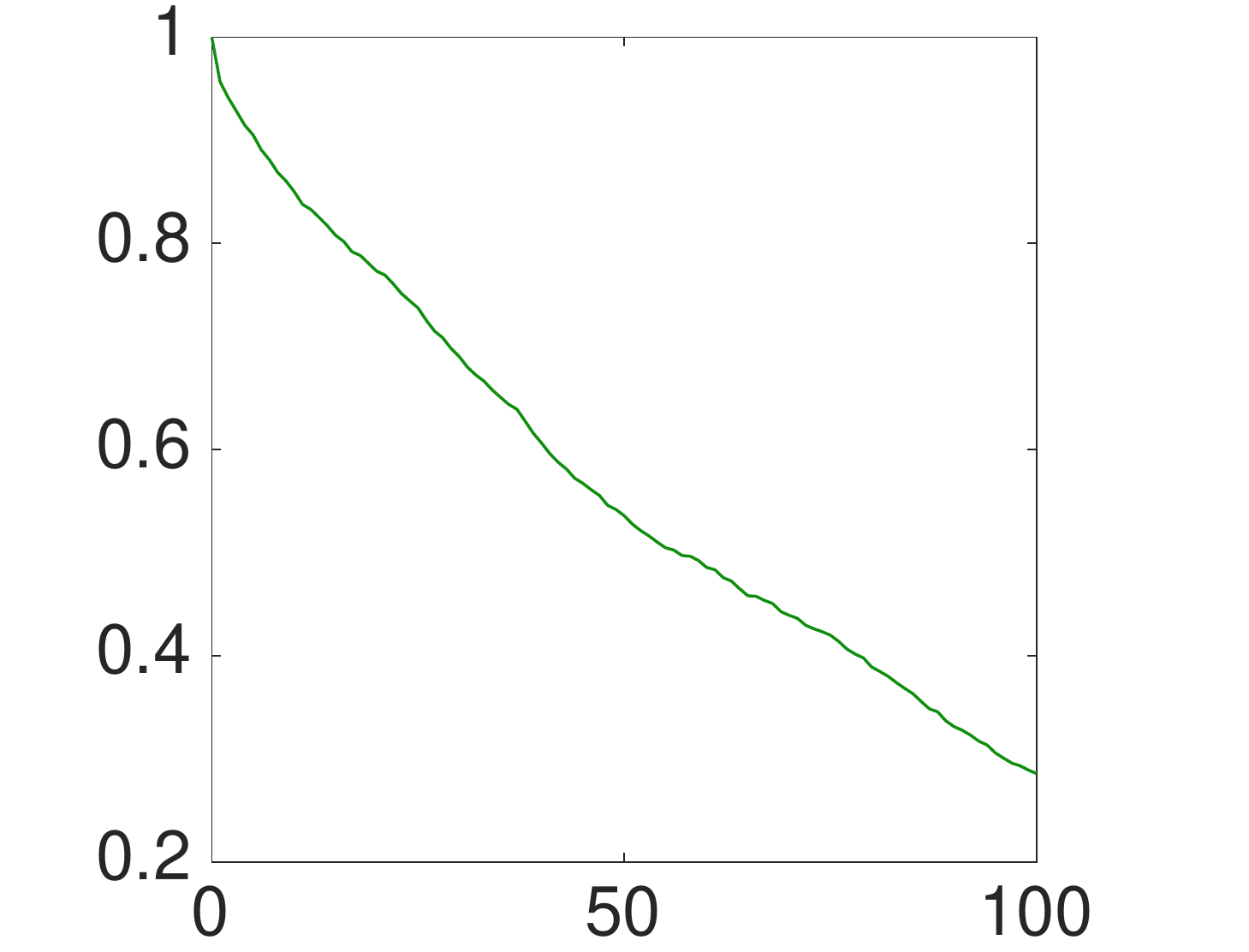}
            \includegraphics[width=\linewidth]{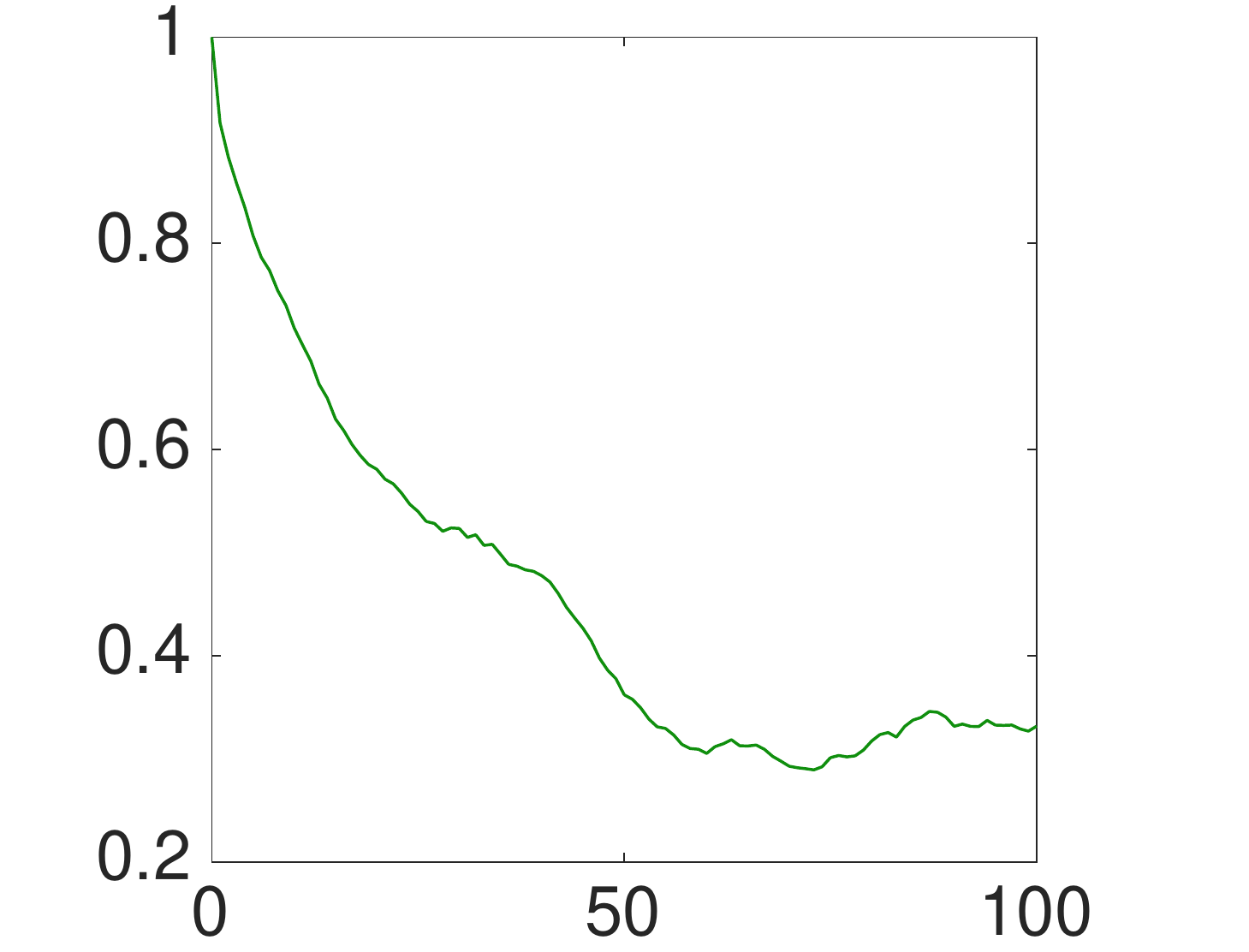}
            \includegraphics[width=\linewidth]{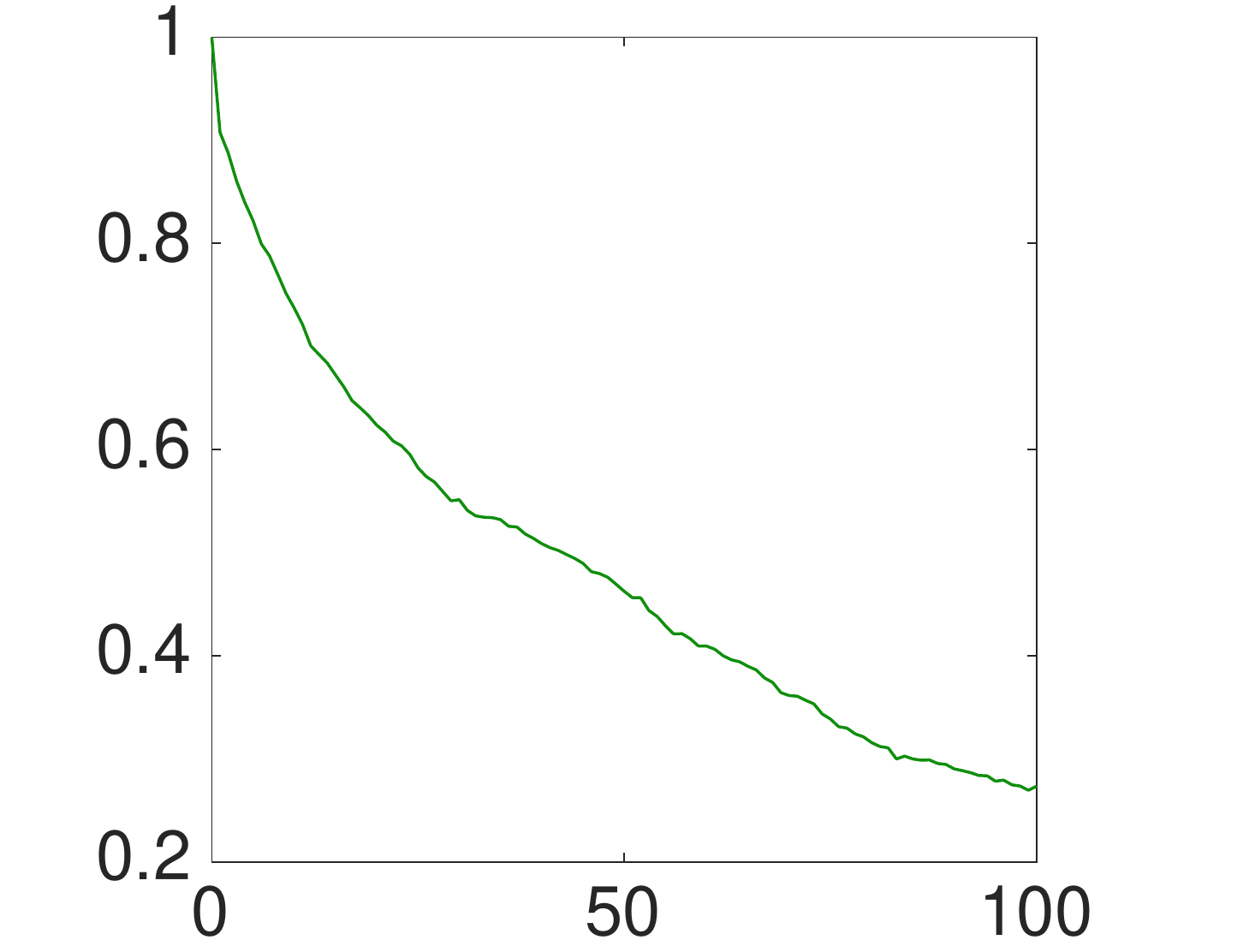}
            \caption{Aniso 2nd }\end{subfigure}
        \begin{subfigure}[b]{\qhei}
            \includegraphics[width=\linewidth]{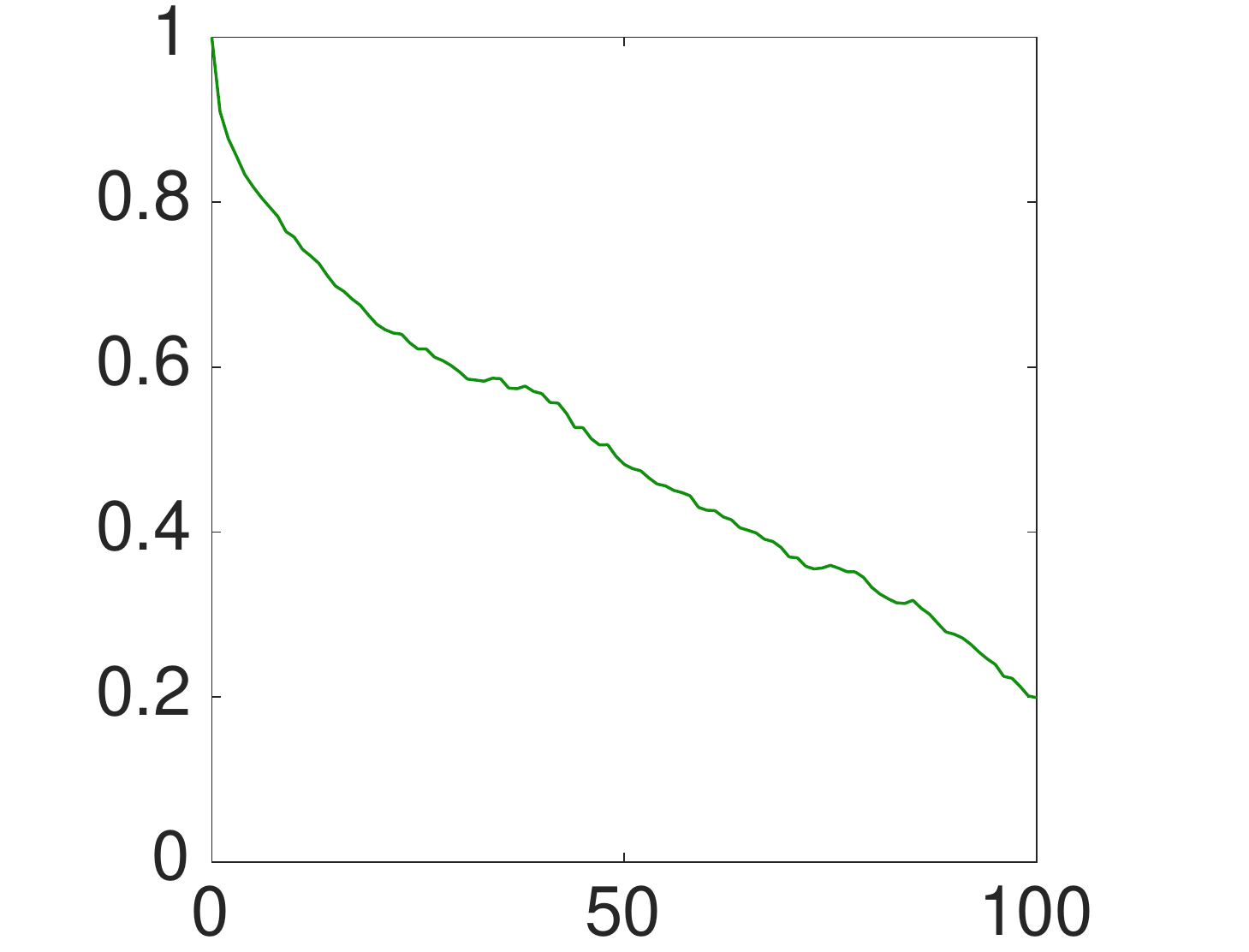}
            \includegraphics[width=\linewidth]{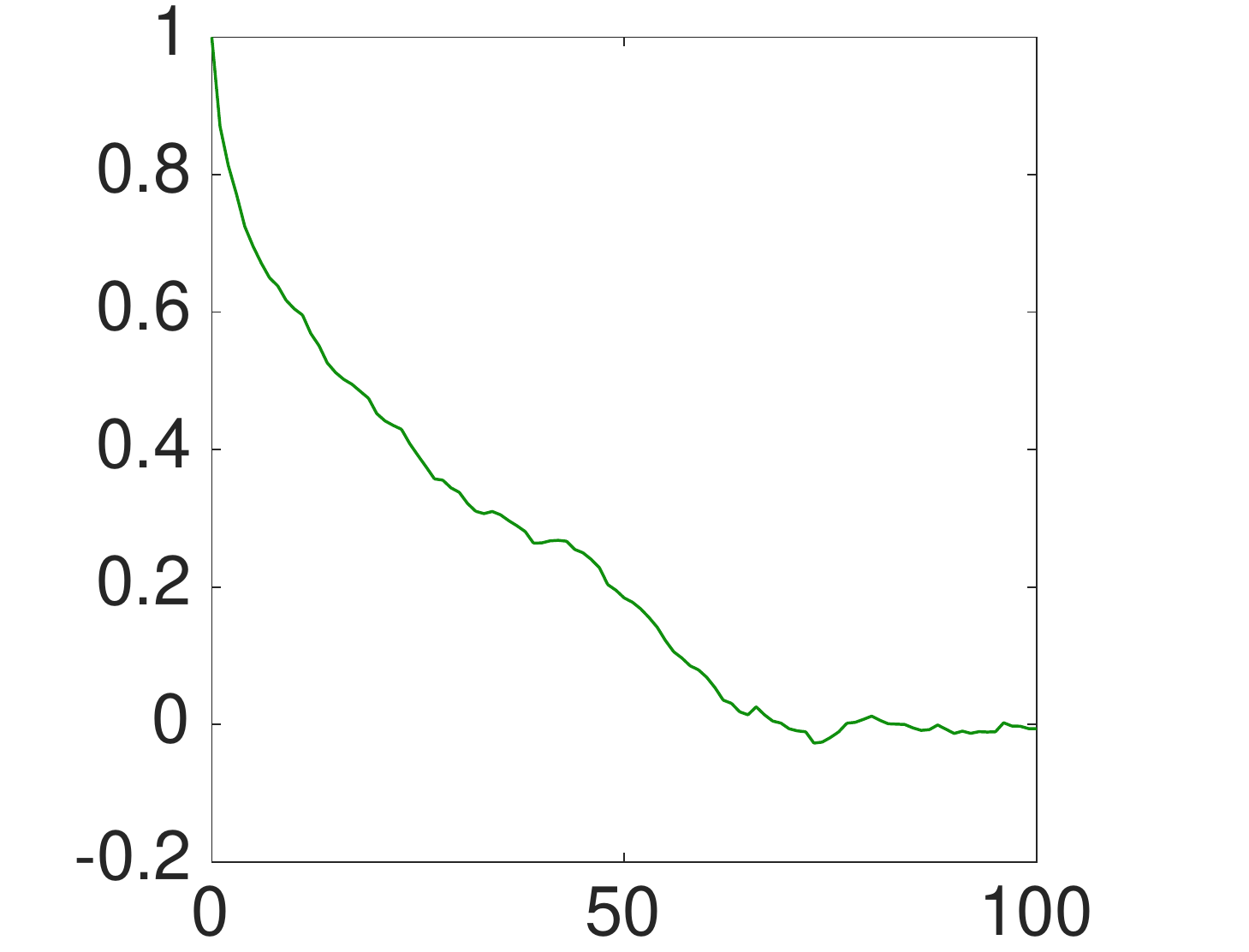}
            \includegraphics[width=\linewidth]{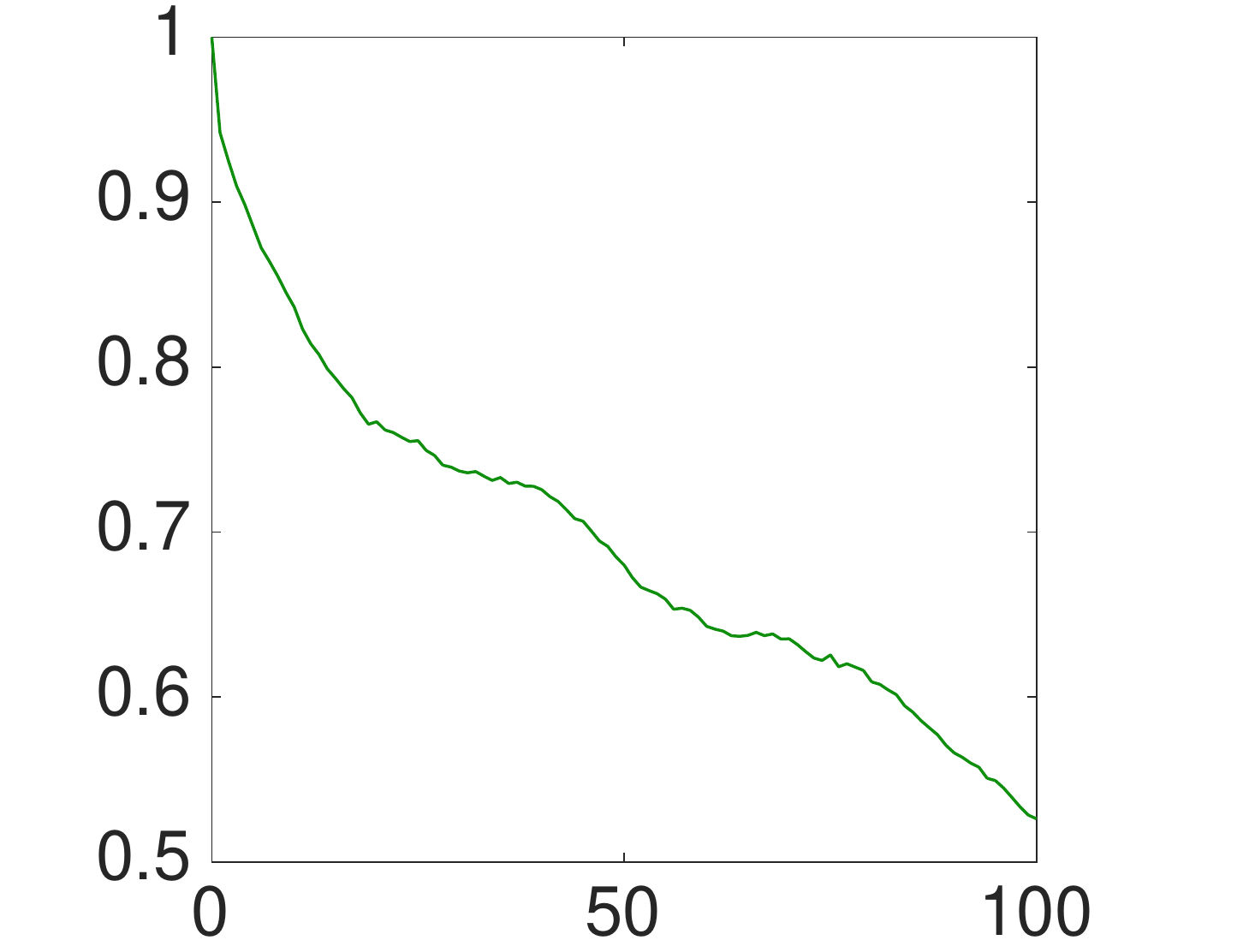}
            \caption{Iso 2nd }\end{subfigure}
        \begin{subfigure}[b]{\qhei}
            \includegraphics[width=\linewidth]{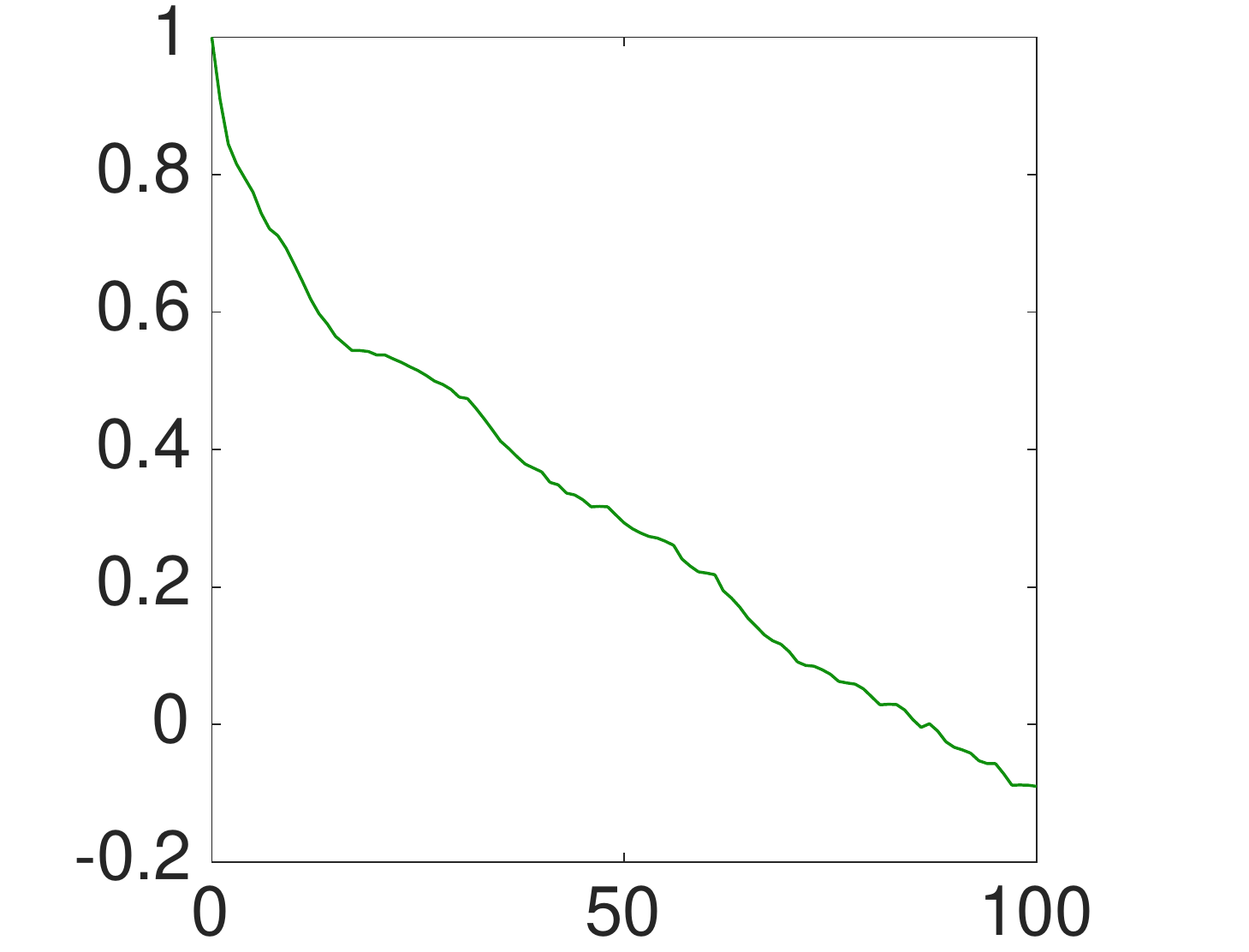}
            \includegraphics[width=\linewidth]{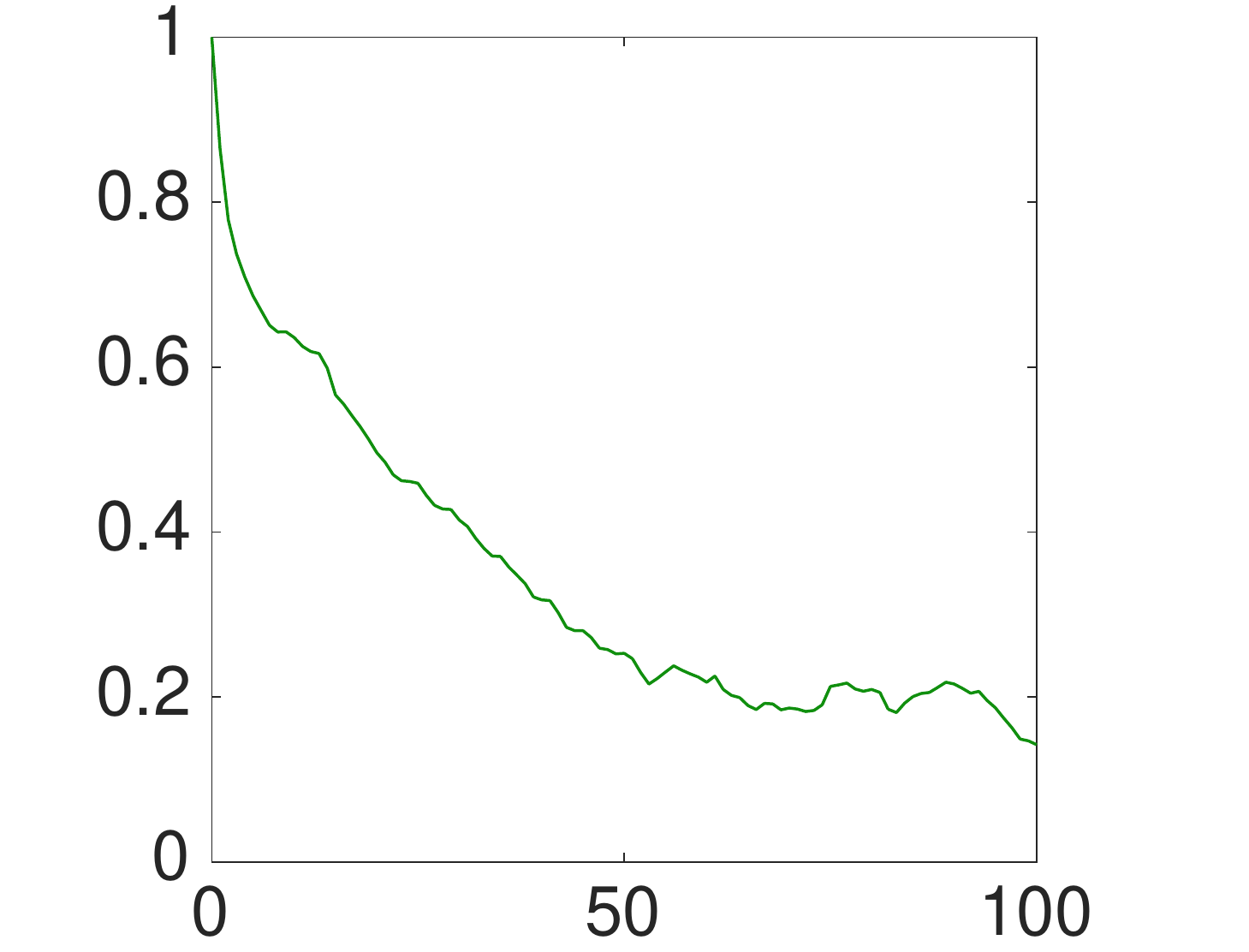}
            \includegraphics[width=\linewidth]{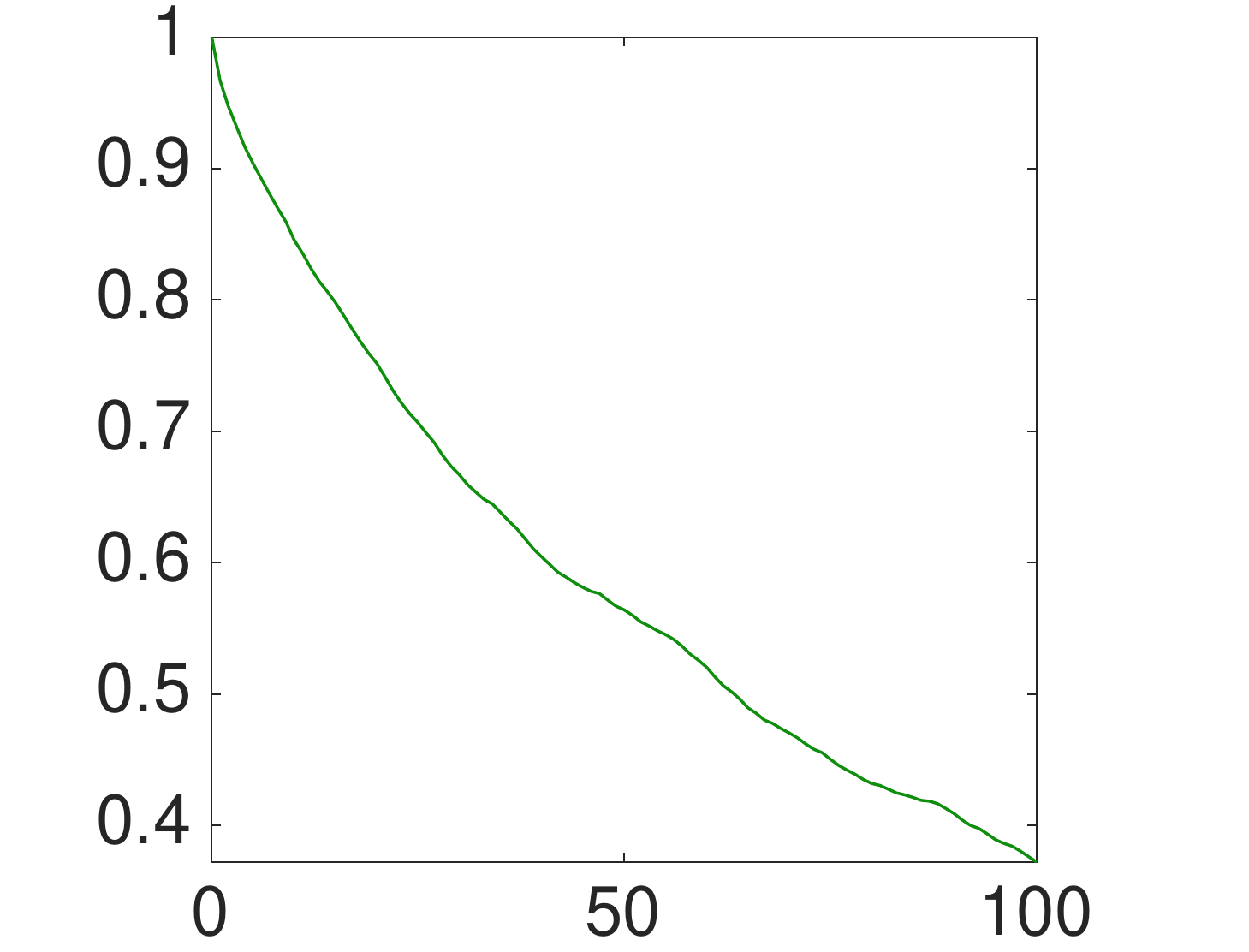}
            \caption{SPDE }\end{subfigure}
        \begin{subfigure}[b]{\qhei}
            \includegraphics[width=\linewidth]{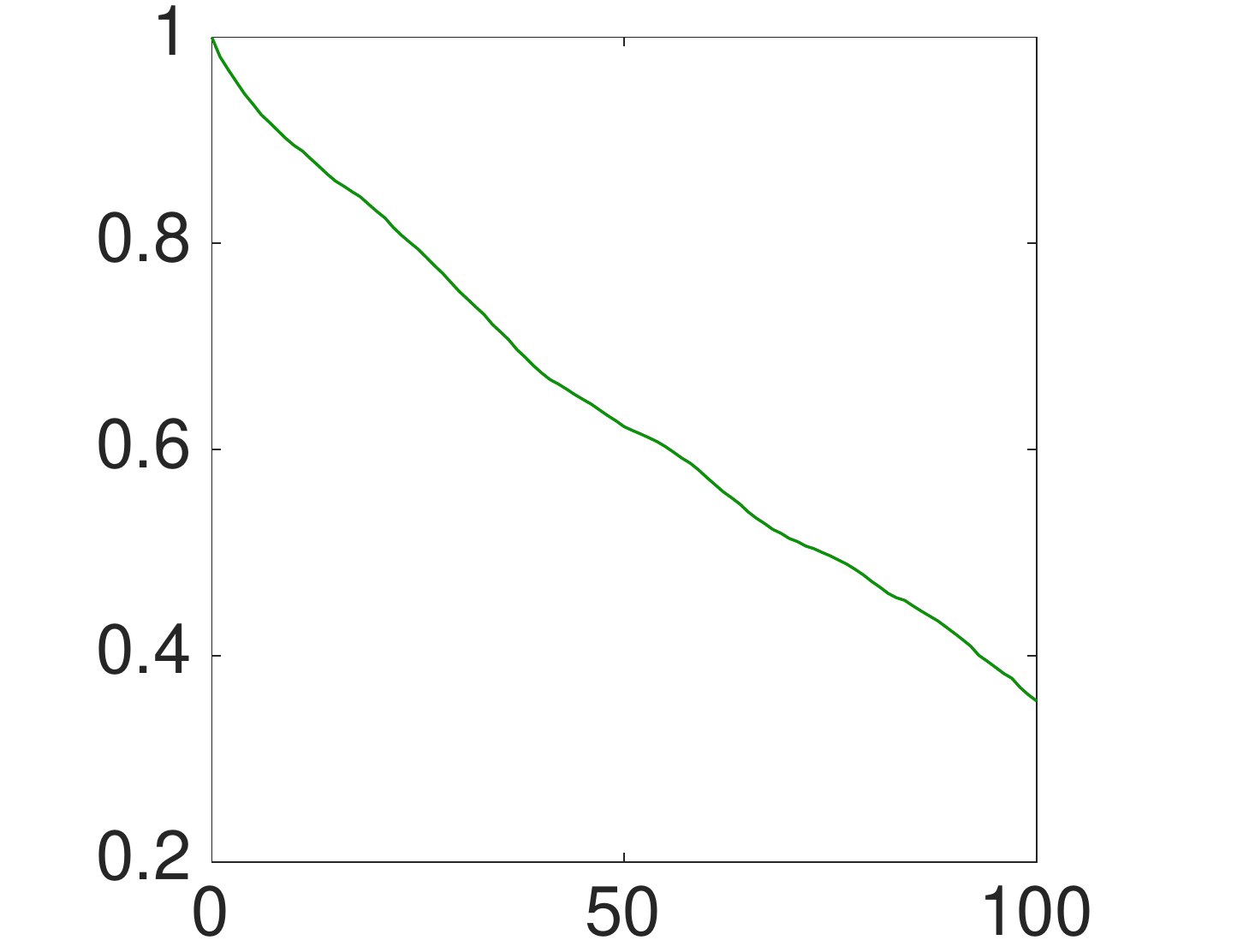}
            \includegraphics[width=\linewidth]{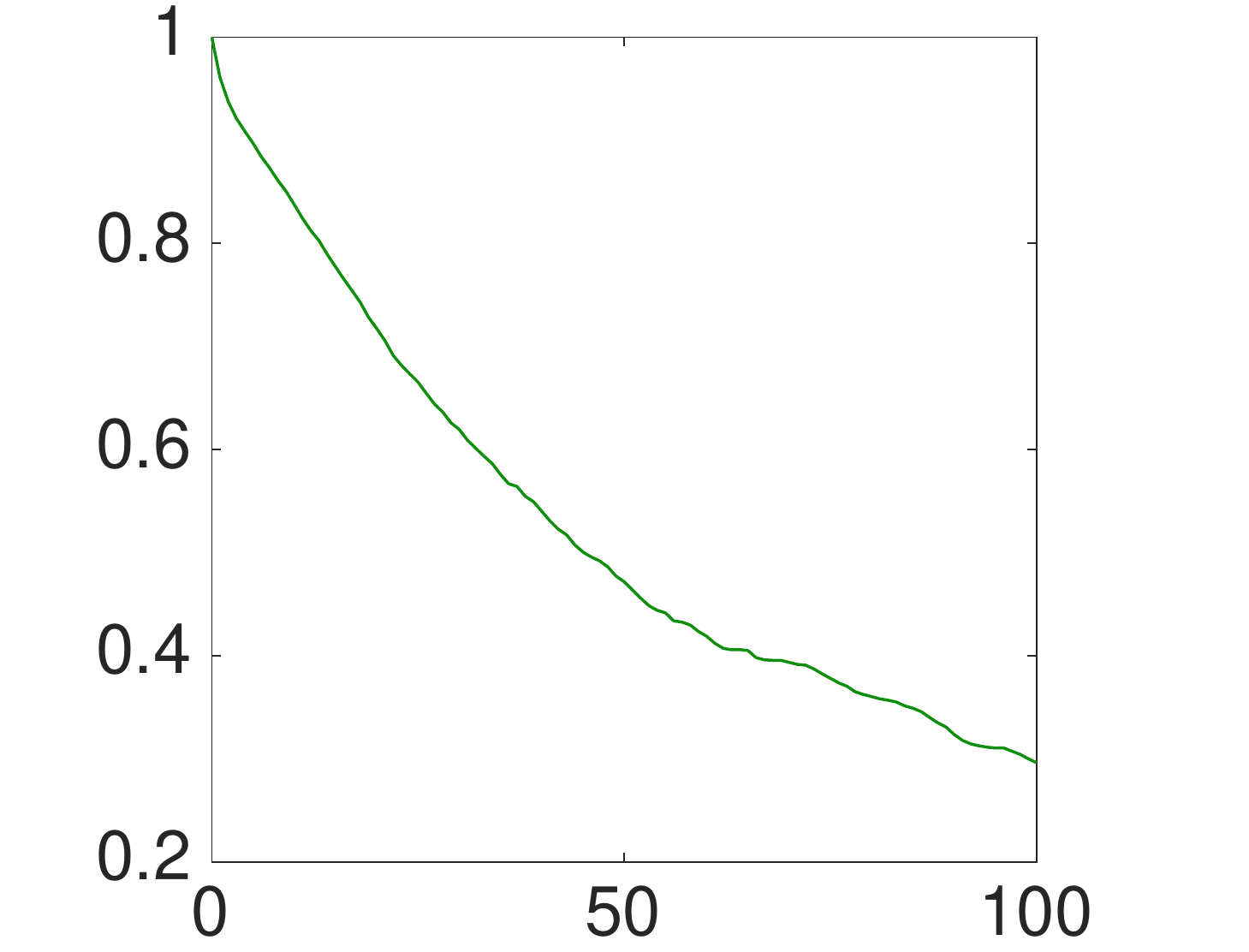}
            \includegraphics[width=\linewidth]{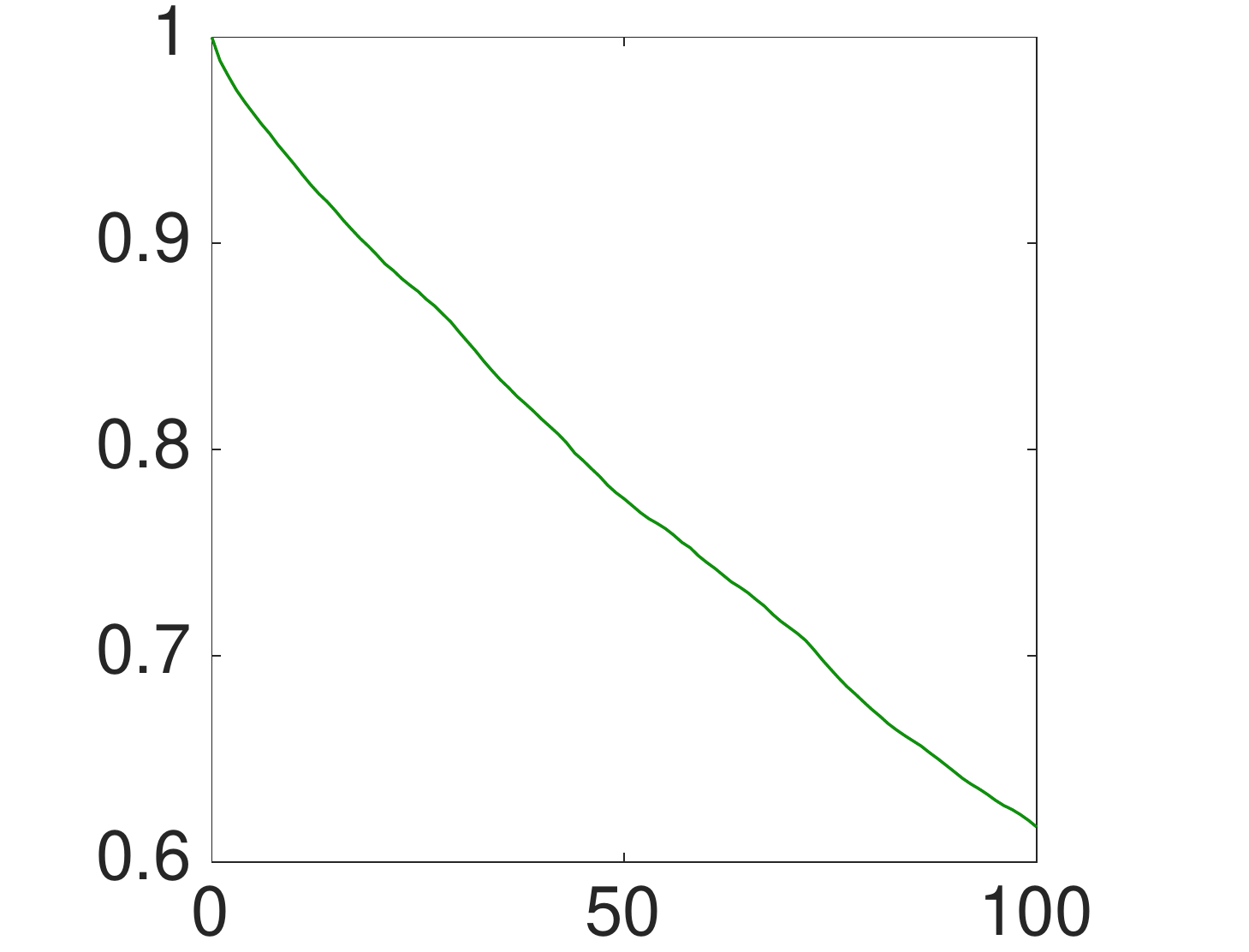}
            \caption{Sheet }\end{subfigure}
        
        \caption{  Trace plots and autocorrelation functions for pixel $39\times129$.  Top rows: MwG. Middle rows: RAM. Bottom rows: HMC. }
        \label{t2}
    \end{figure}

    The performance difference of the three  MCMC algorithms in sampling the posteriors seems not to be major. However, the PSRF values of the MCMC runs suggest that the RAM is a bit more robust \textcolor{black}{than} other two methods in \textcolor{black}{the} sampling. The PSRF difference is seen  best from the trace and PSRF plots of the first order Cauchy priors. If a marginal distribution is highly multimodal as some of the marginals of the posteriors with the first order difference prior (\Cref{t2}) are, NUTS struggles to sample the modes compared to RAM and MwG. On the other hand, the marginal chains of unimodal nodes (1st order difference priors in \Cref{t1}) reveal no great differences between the methods. In terms of PSRF, the geometry of the first order isotropic difference priors  seems slightly more beneficial compared to the anisotropic ones  due to the additional multimodality likely induced by the anisotropicity, the same conclusion \textcolor{black}{cannot} be said for posteriors with second order difference, as seen in \Cref{andiff2d}.
    
\begin{figure}
        \centering
        \begin{subfigure}[b]{\qhei}
            \includegraphics[width=\linewidth]{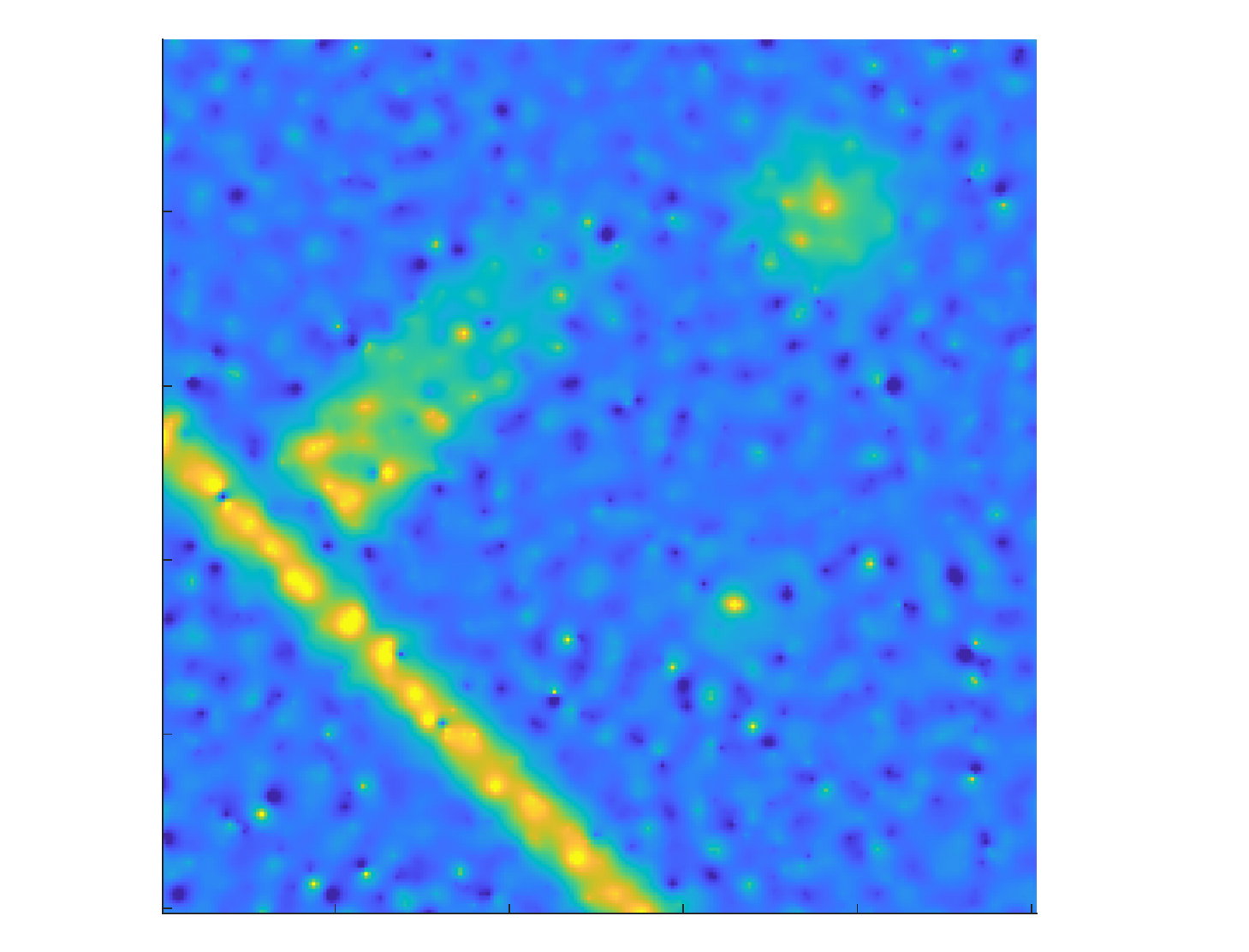}
            \includegraphics[width=\linewidth]{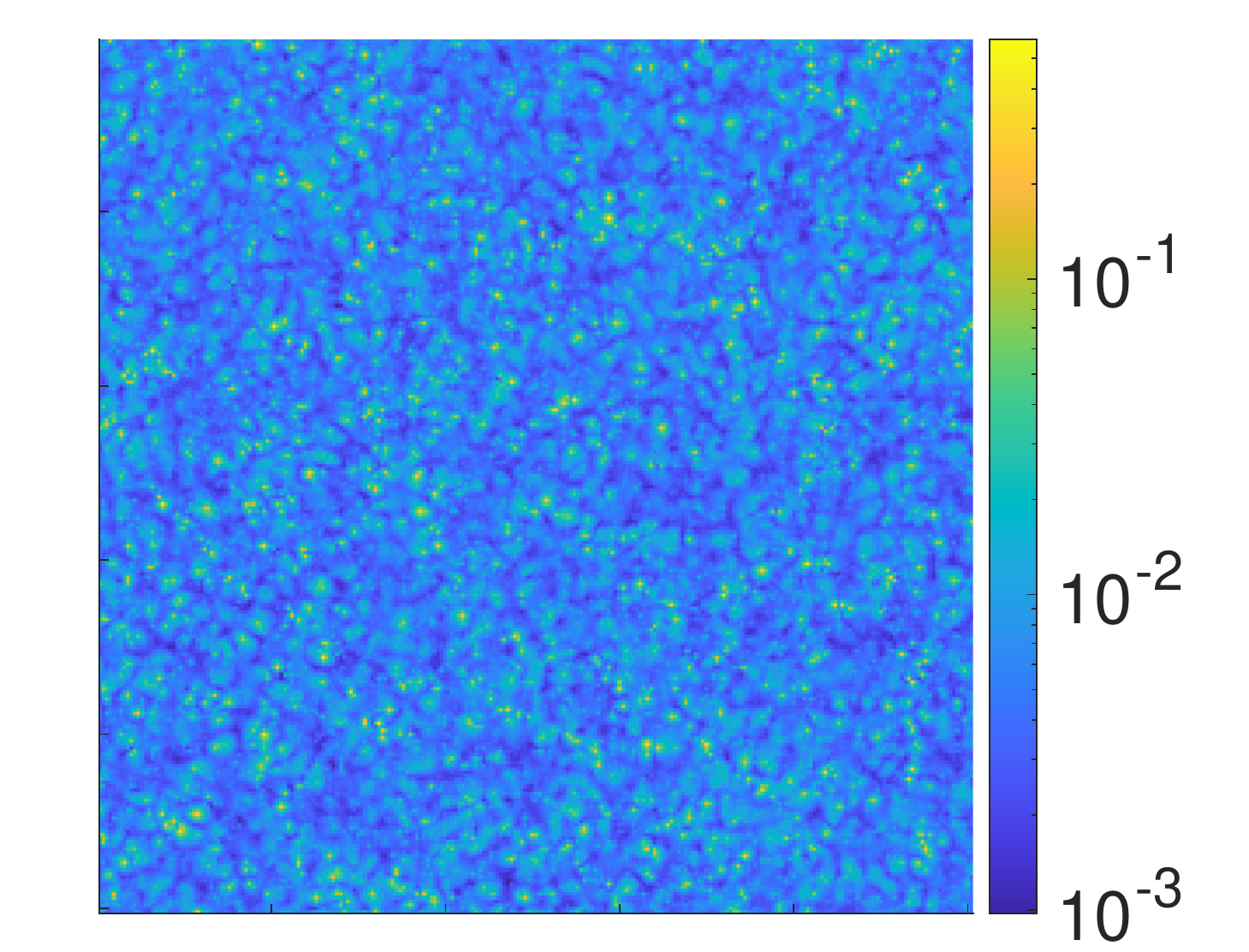}
            \includegraphics[width=\linewidth]{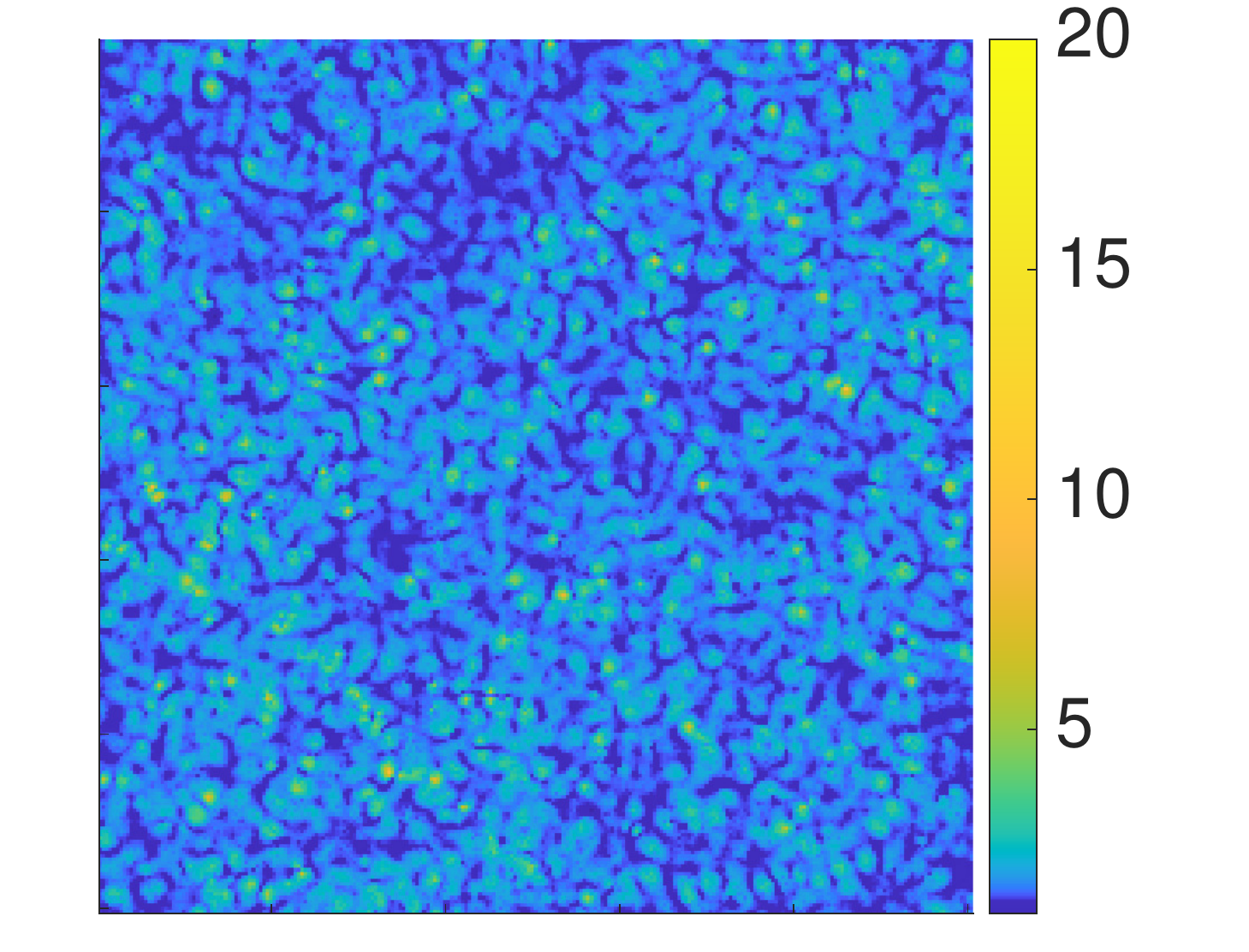}
            \caption{SPDE, MwG }\label{spde_mwg_mean}\end{subfigure}
        \begin{subfigure}[b]{\qhei}
            \includegraphics[width=\linewidth]{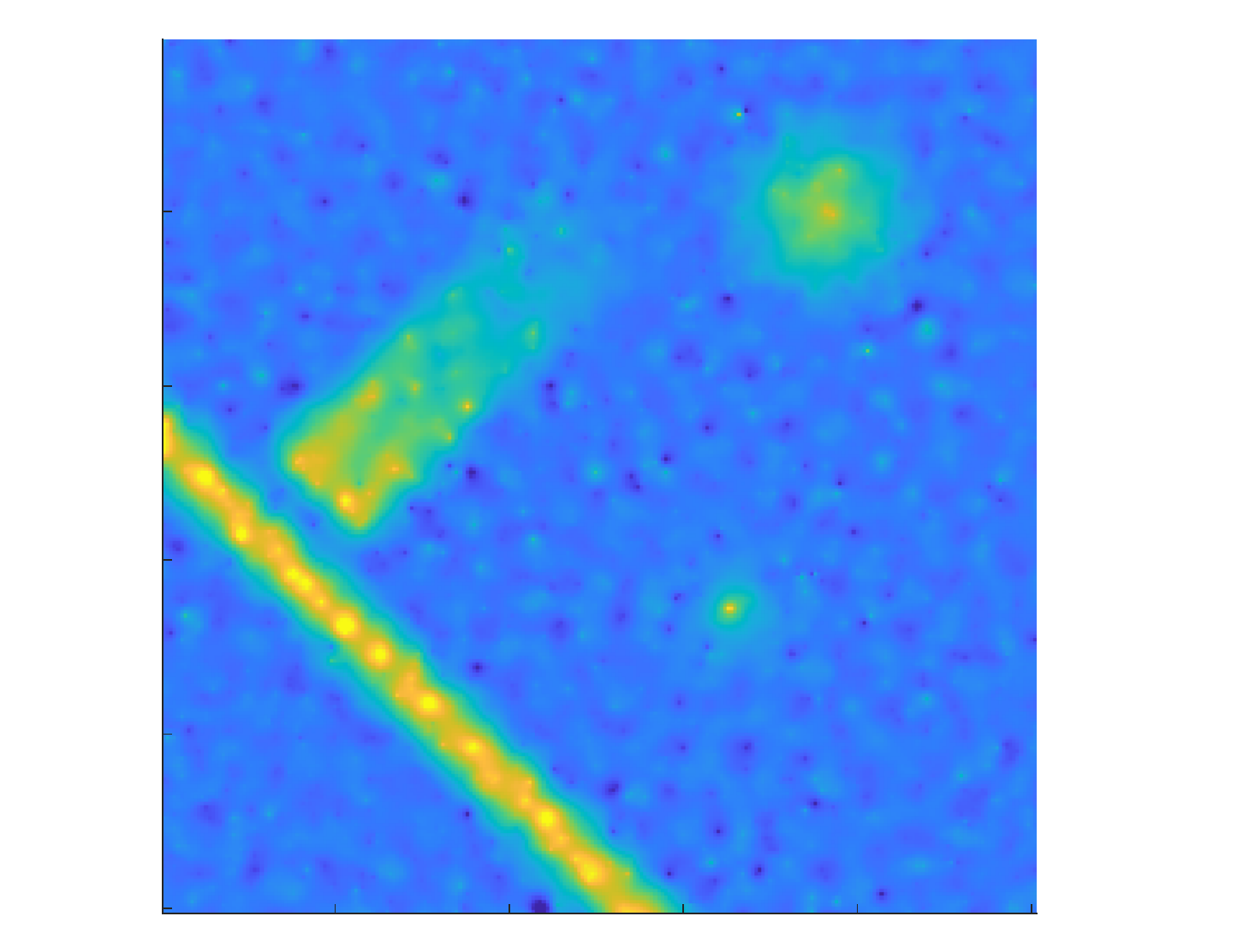}
            \includegraphics[width=\linewidth]{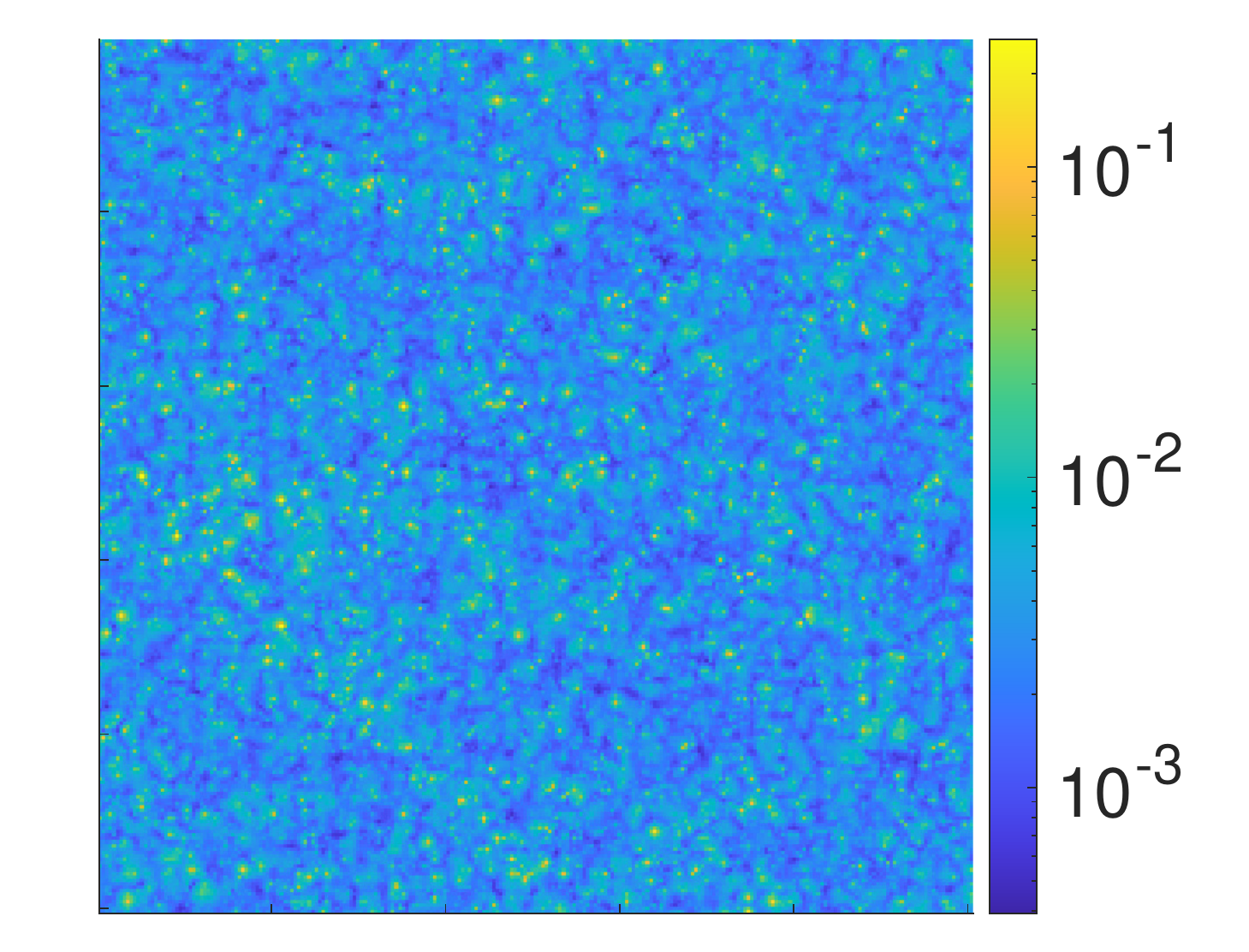}
            \includegraphics[width=\linewidth]{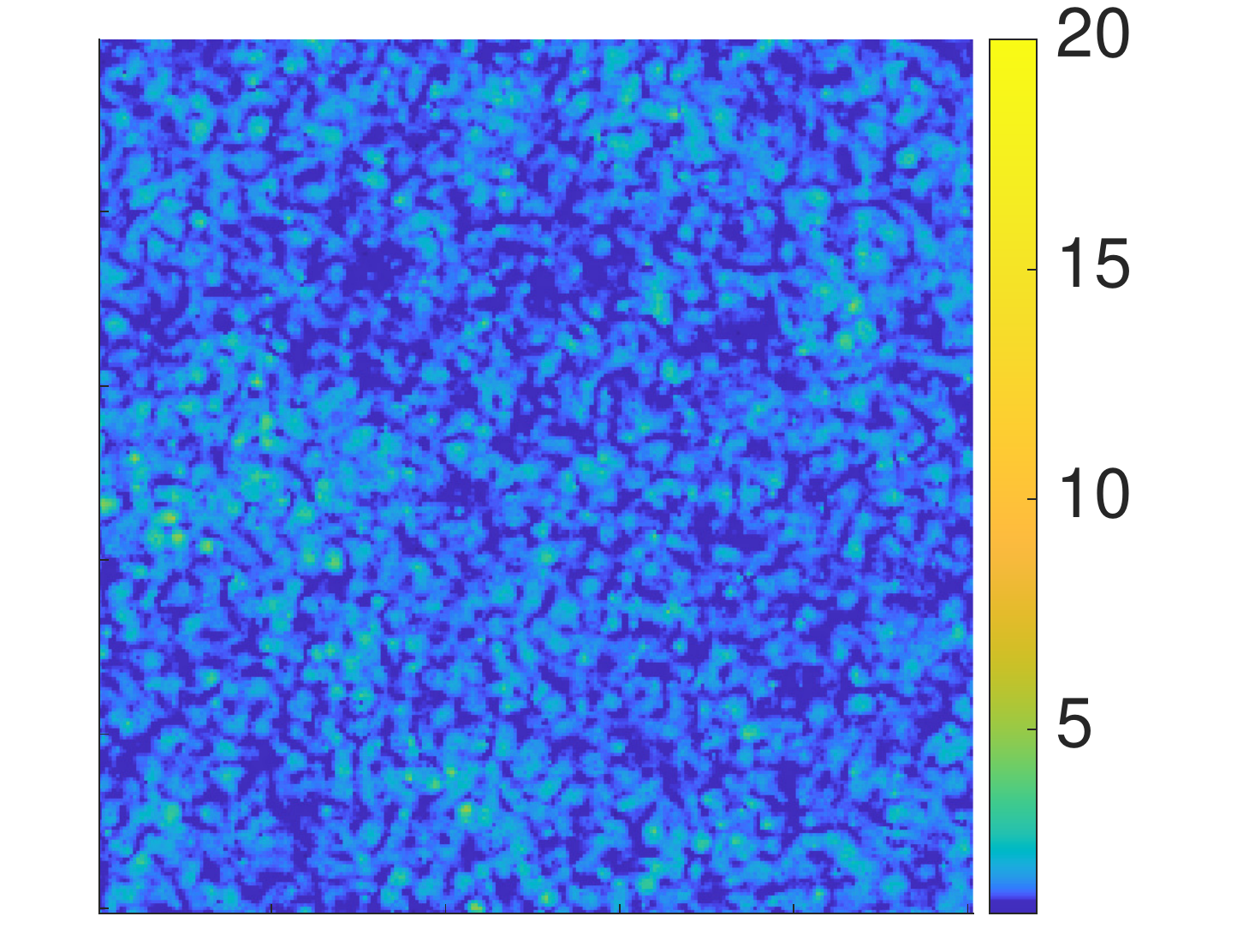}
            \caption{SPDE, NUTS }\label{spde_nuts_mean}\end{subfigure}
        \begin{subfigure}[b]{\qhei}
            \includegraphics[width=\linewidth]{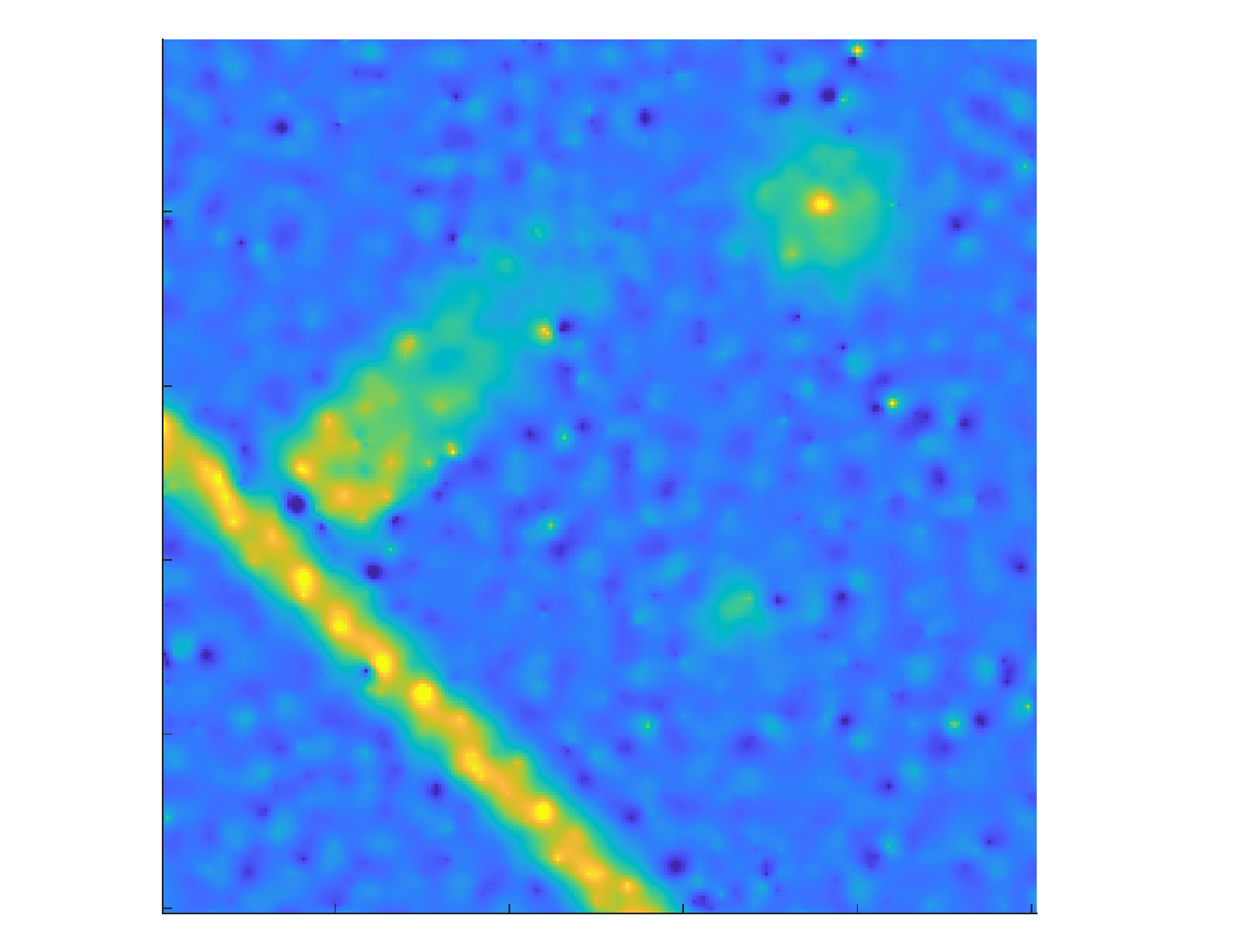}
            \includegraphics[width=\linewidth]{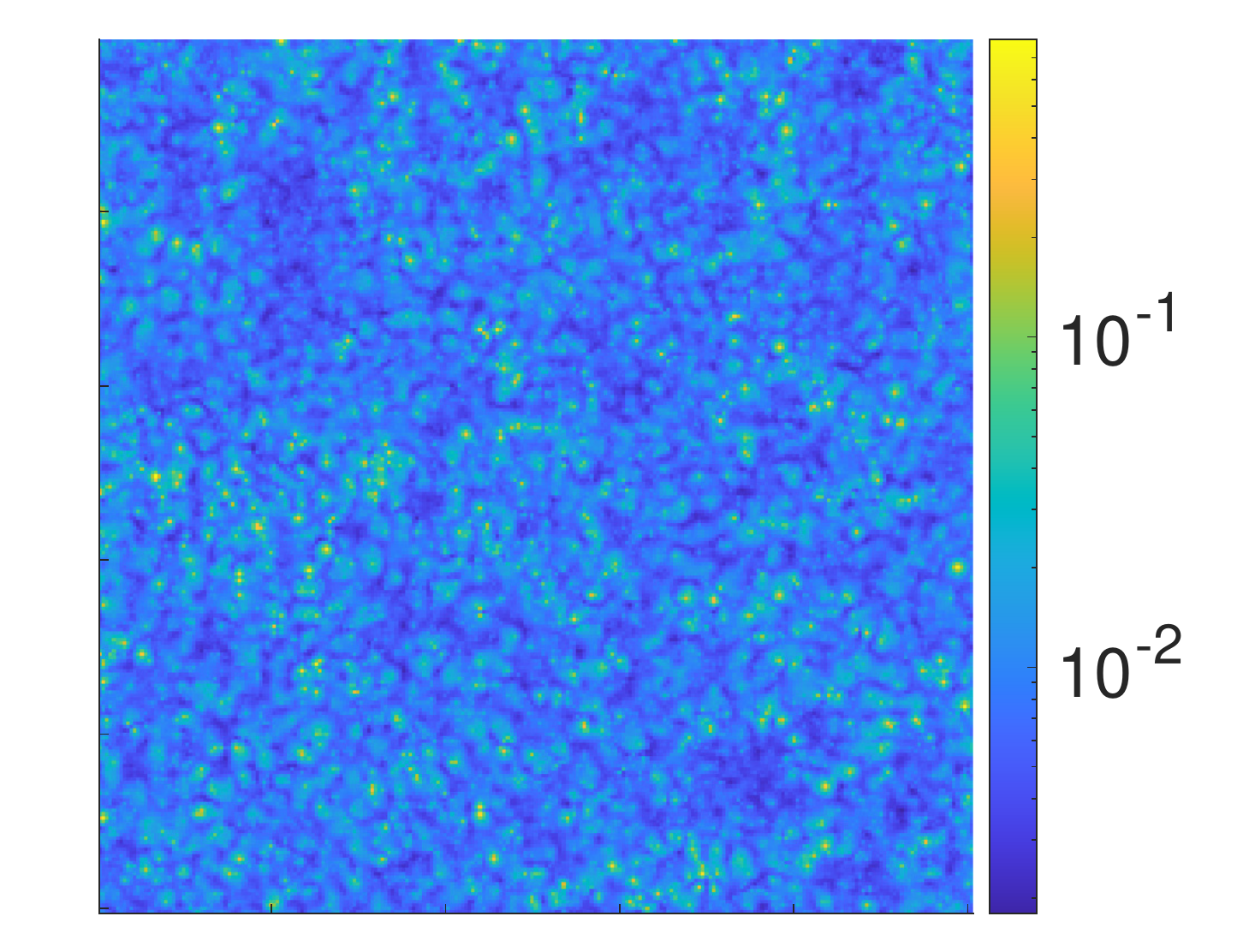}
            \includegraphics[width=\linewidth]{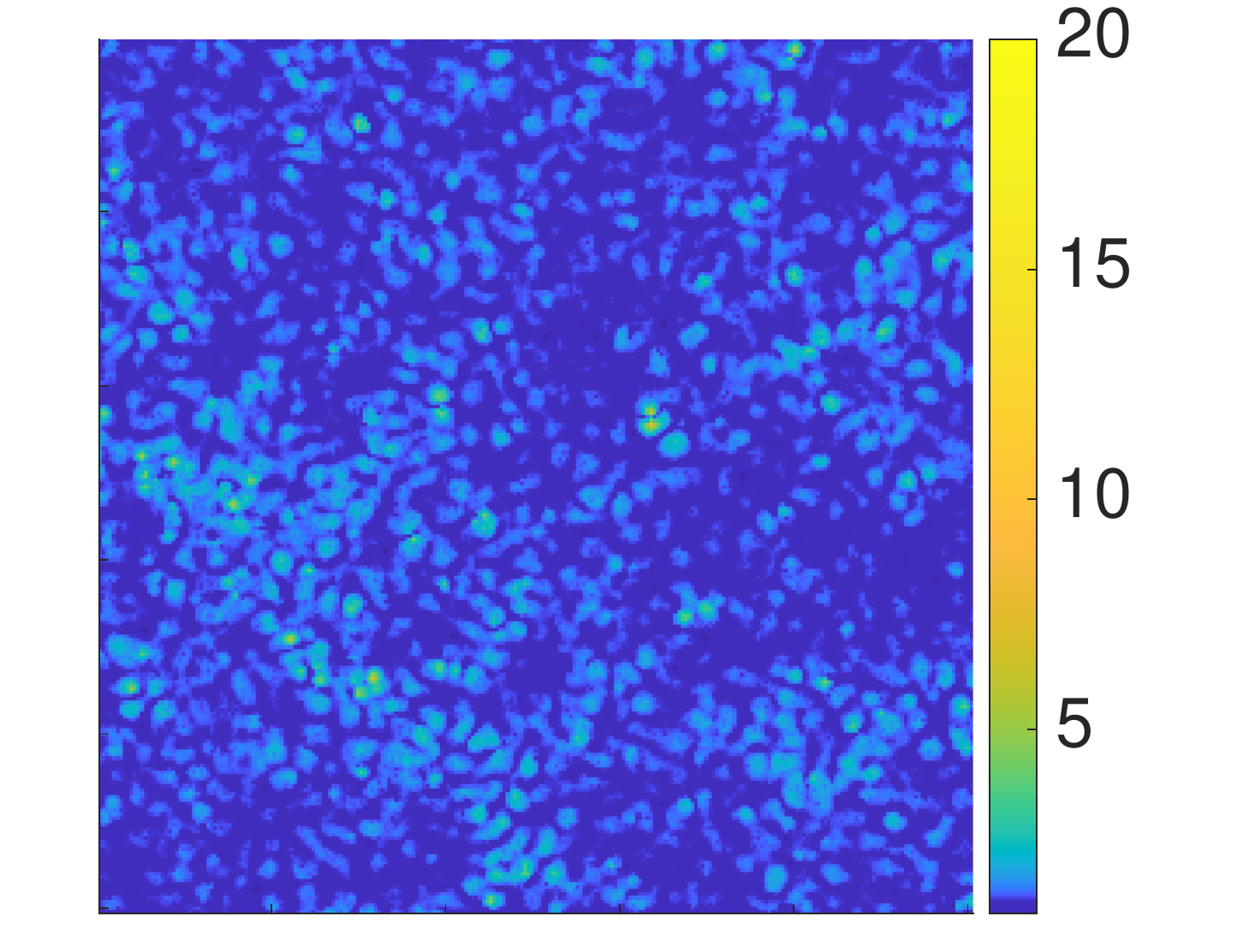}
            \caption{SPDE, RAM }\label{spde_ram_mean}\end{subfigure}
        \begin{subfigure}[b]{\qhei}
            \includegraphics[width=\linewidth]{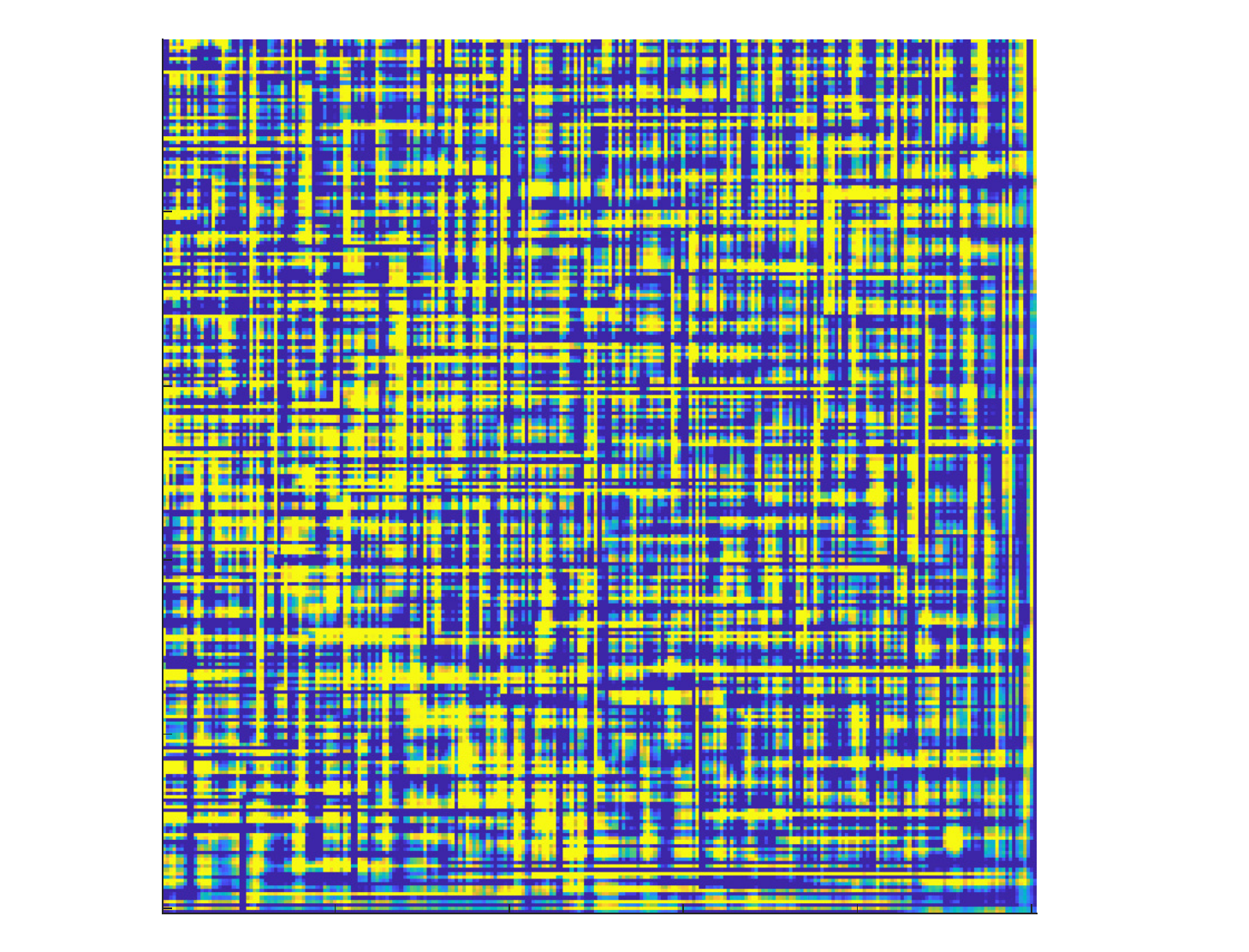}
            \includegraphics[width=\linewidth]{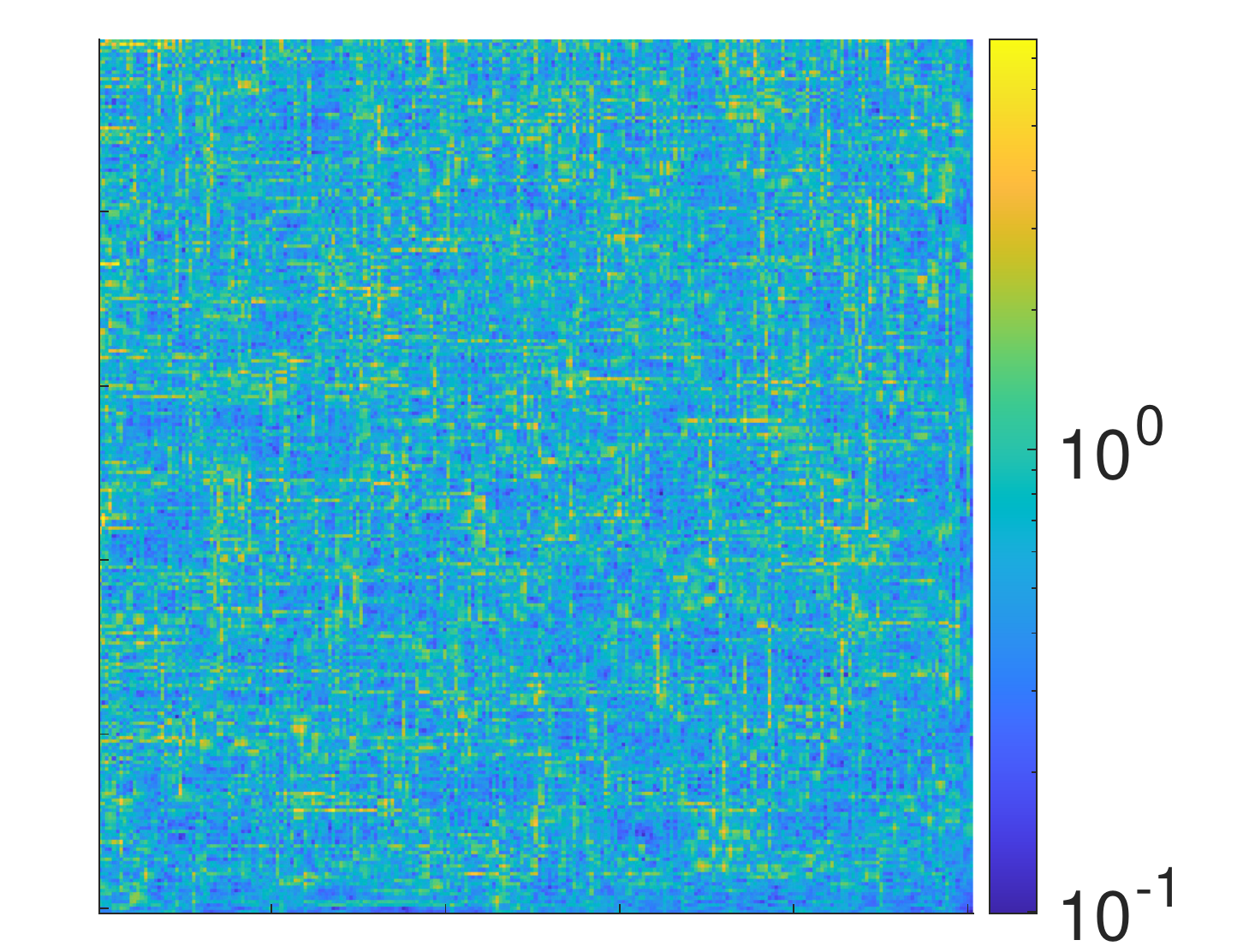}
            \includegraphics[width=\linewidth]{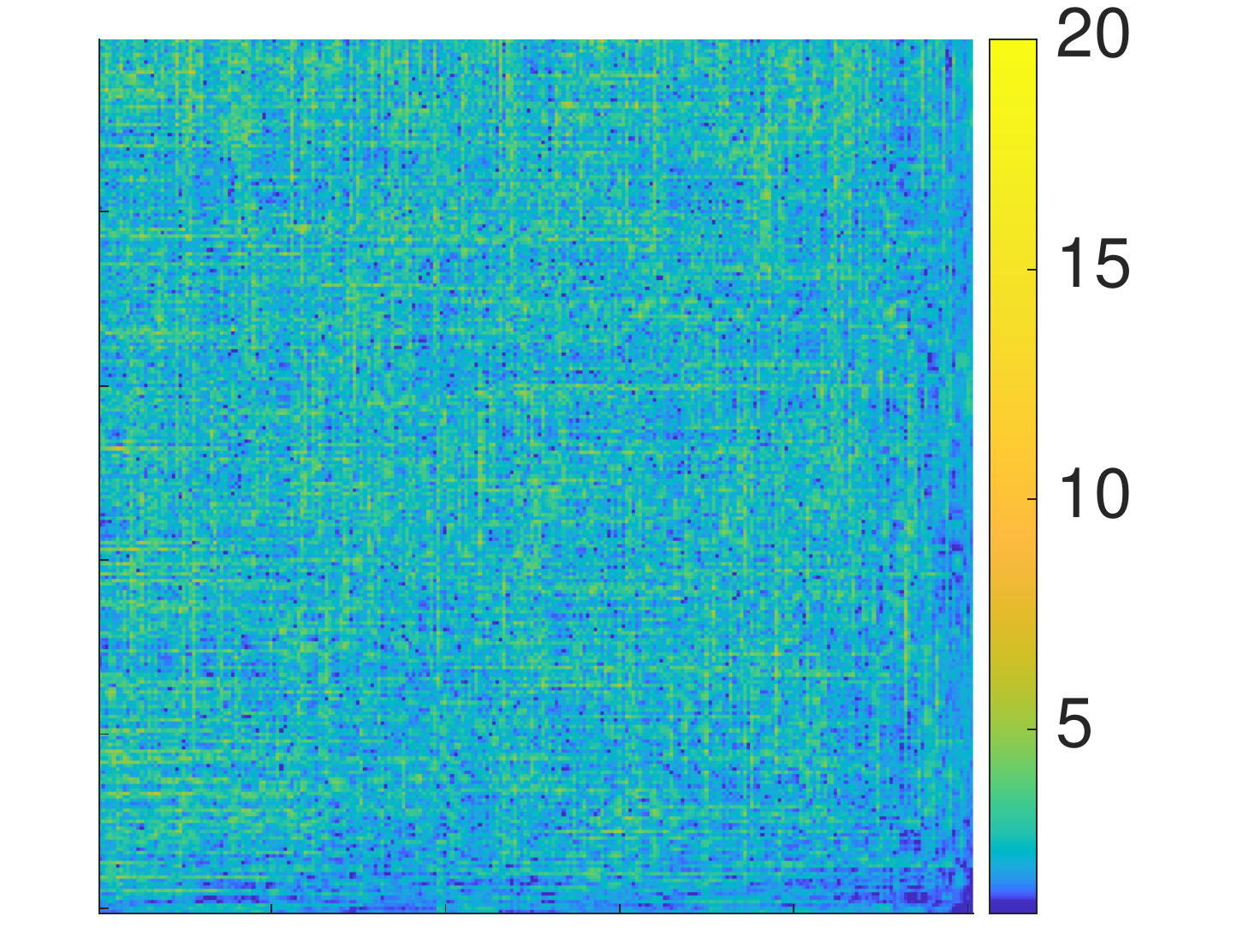}
            \caption{Sheet, MwG }\label{sheet_mwg_mean}\end{subfigure}
        \begin{subfigure}[b]{\qhei}
            \includegraphics[width=\linewidth]{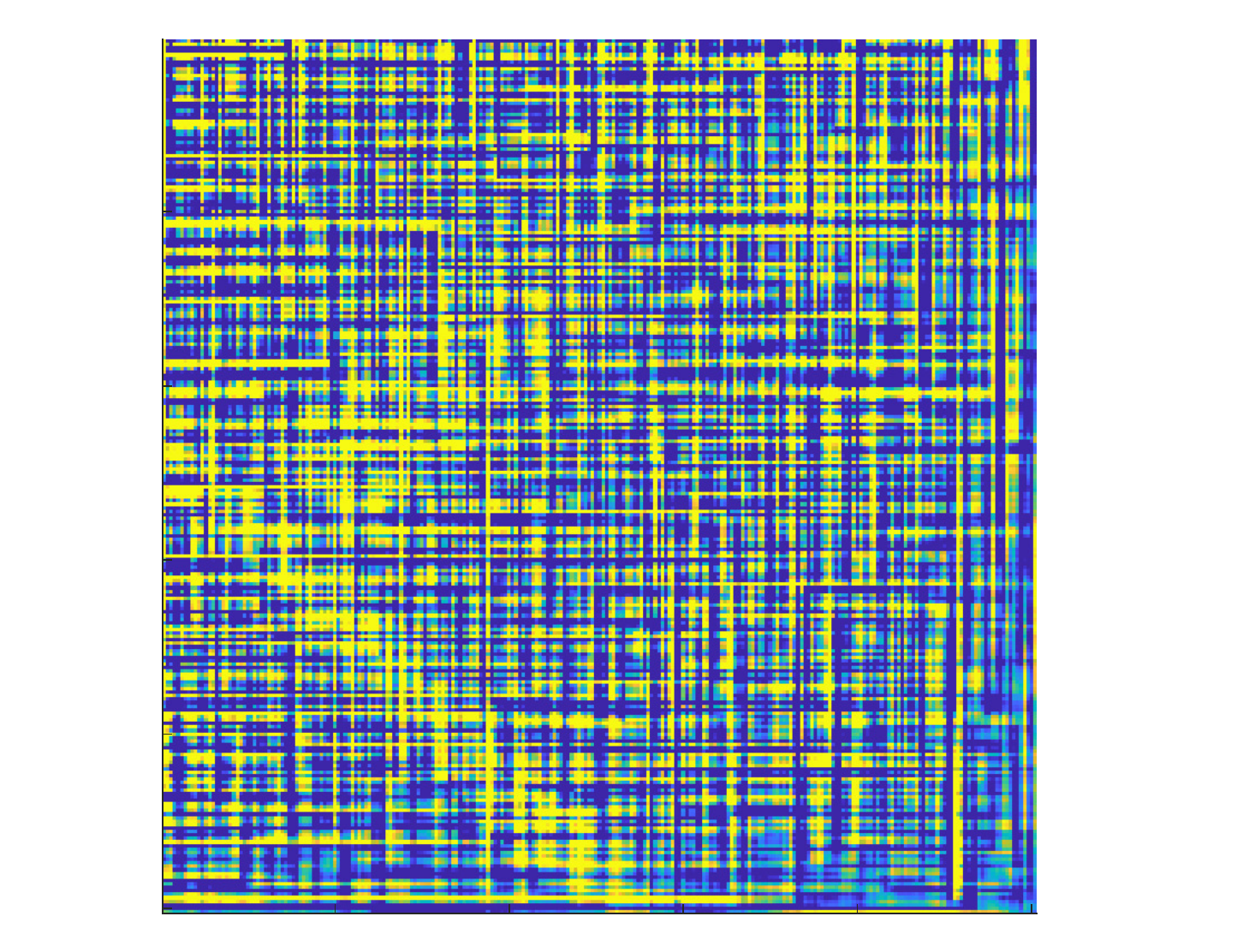}
            \includegraphics[width=\linewidth]{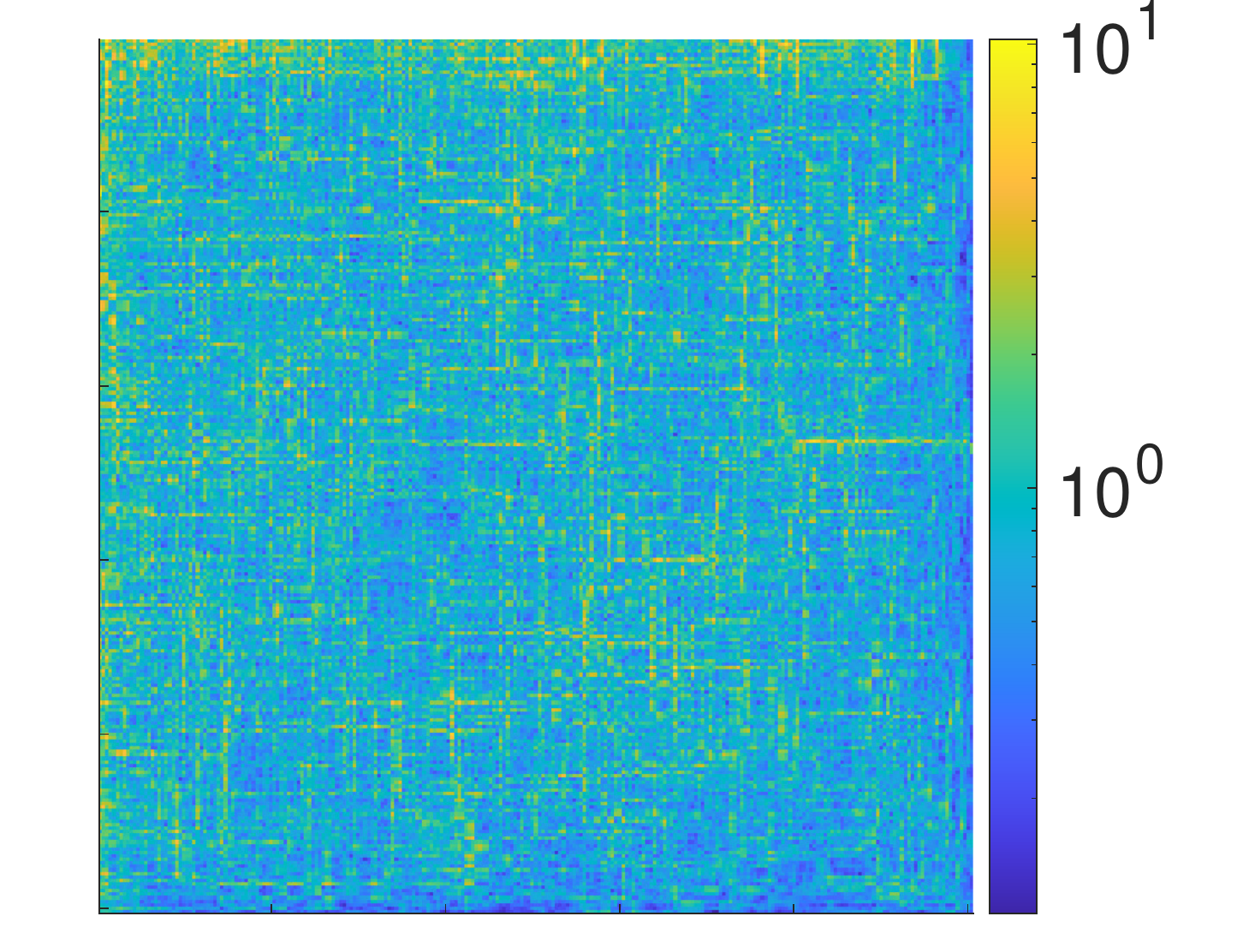}
            \includegraphics[width=\linewidth]{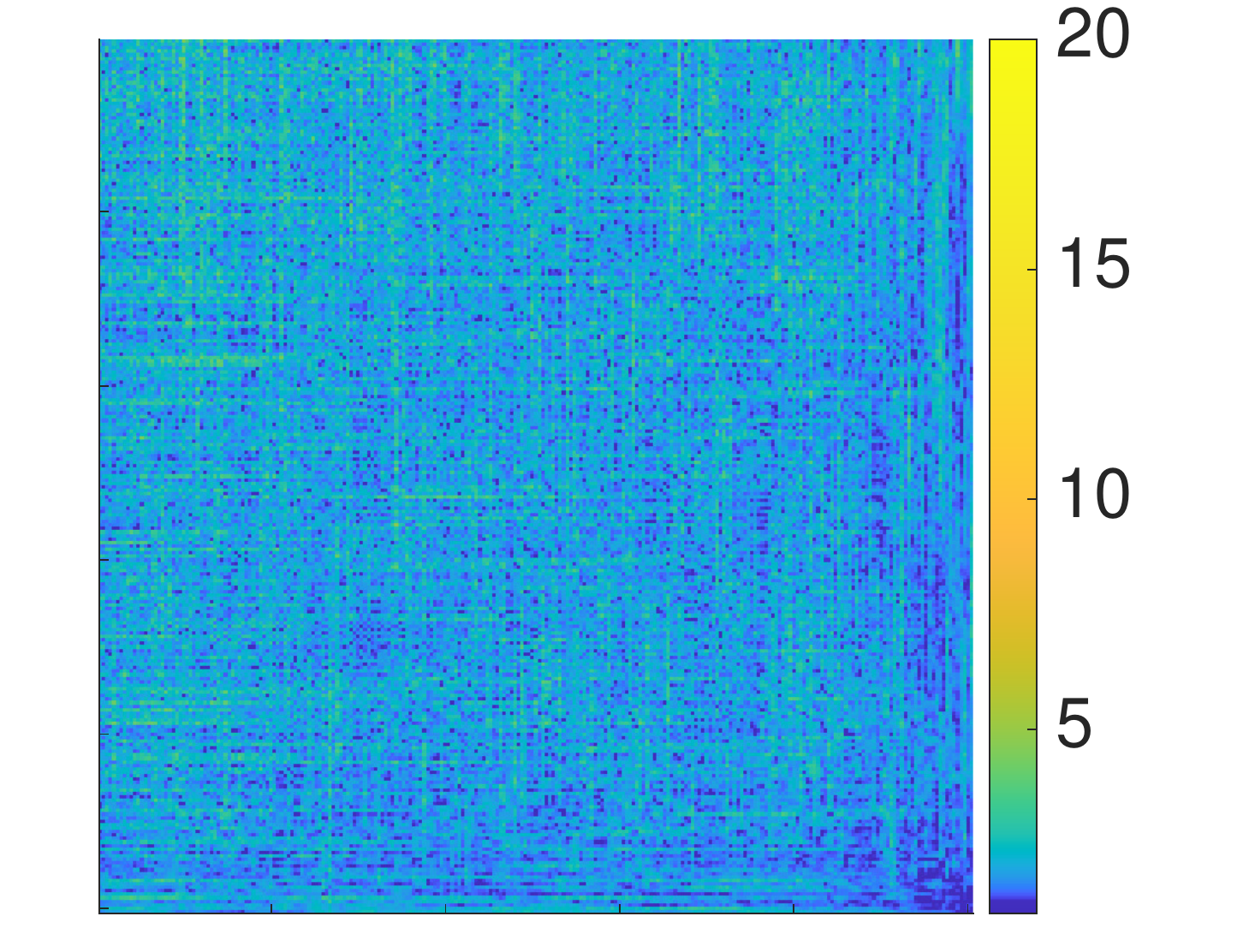}
            \caption{Sheet, NUTS }\label{sheet_nuts_mean}\end{subfigure}
        \begin{subfigure}[b]{\qhei}
            \includegraphics[width=\linewidth]{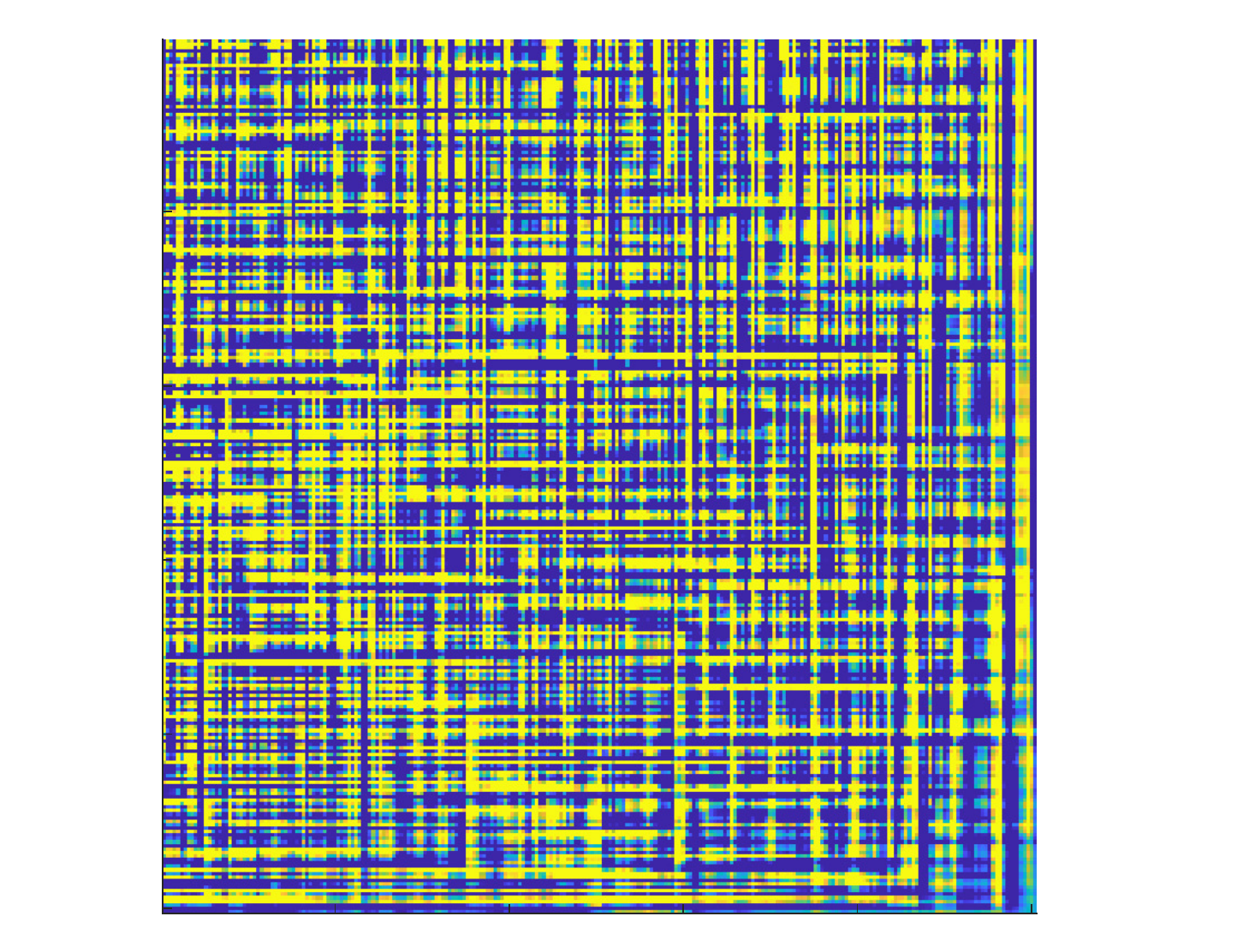}
            \includegraphics[width=\linewidth]{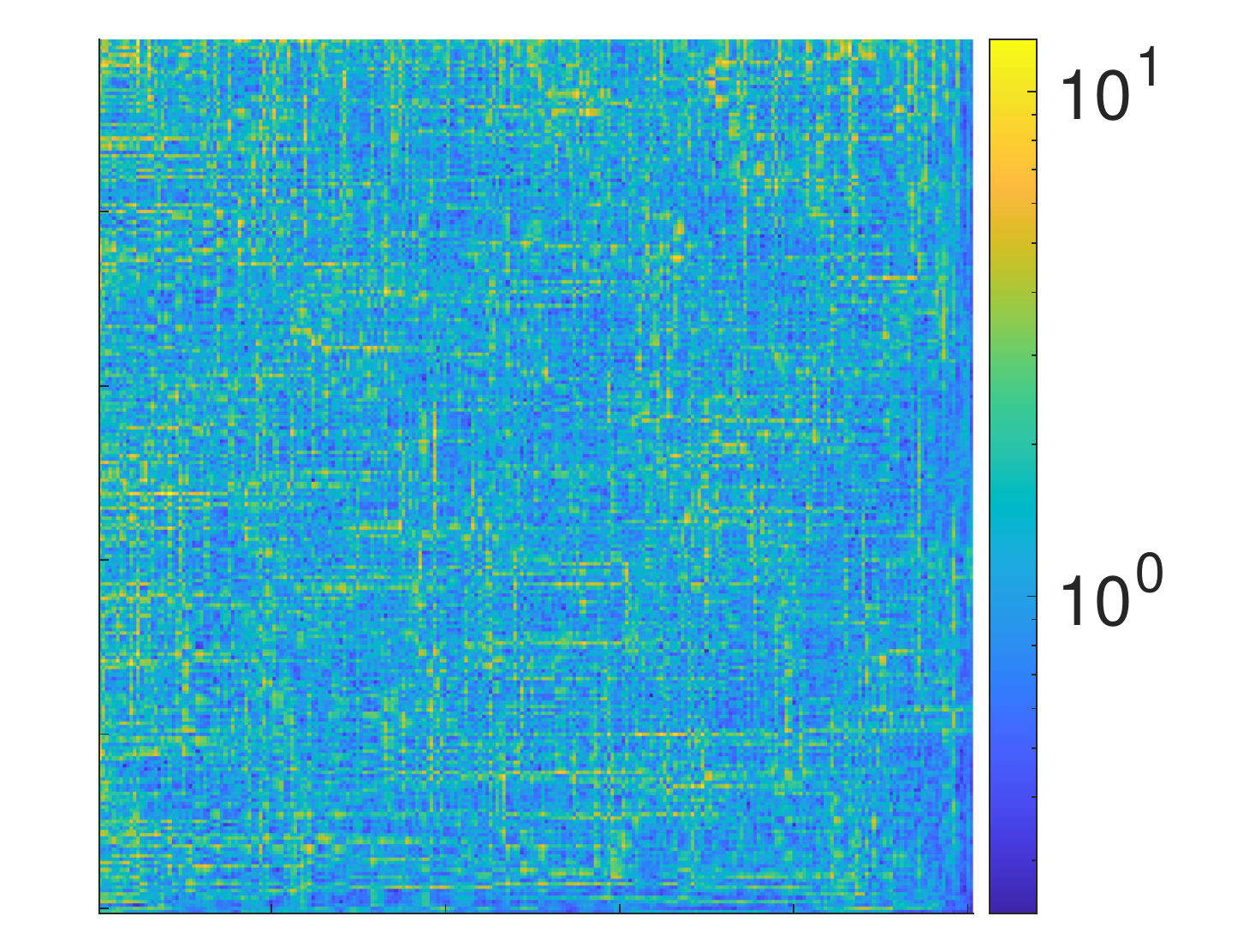}
            \includegraphics[width=\linewidth]{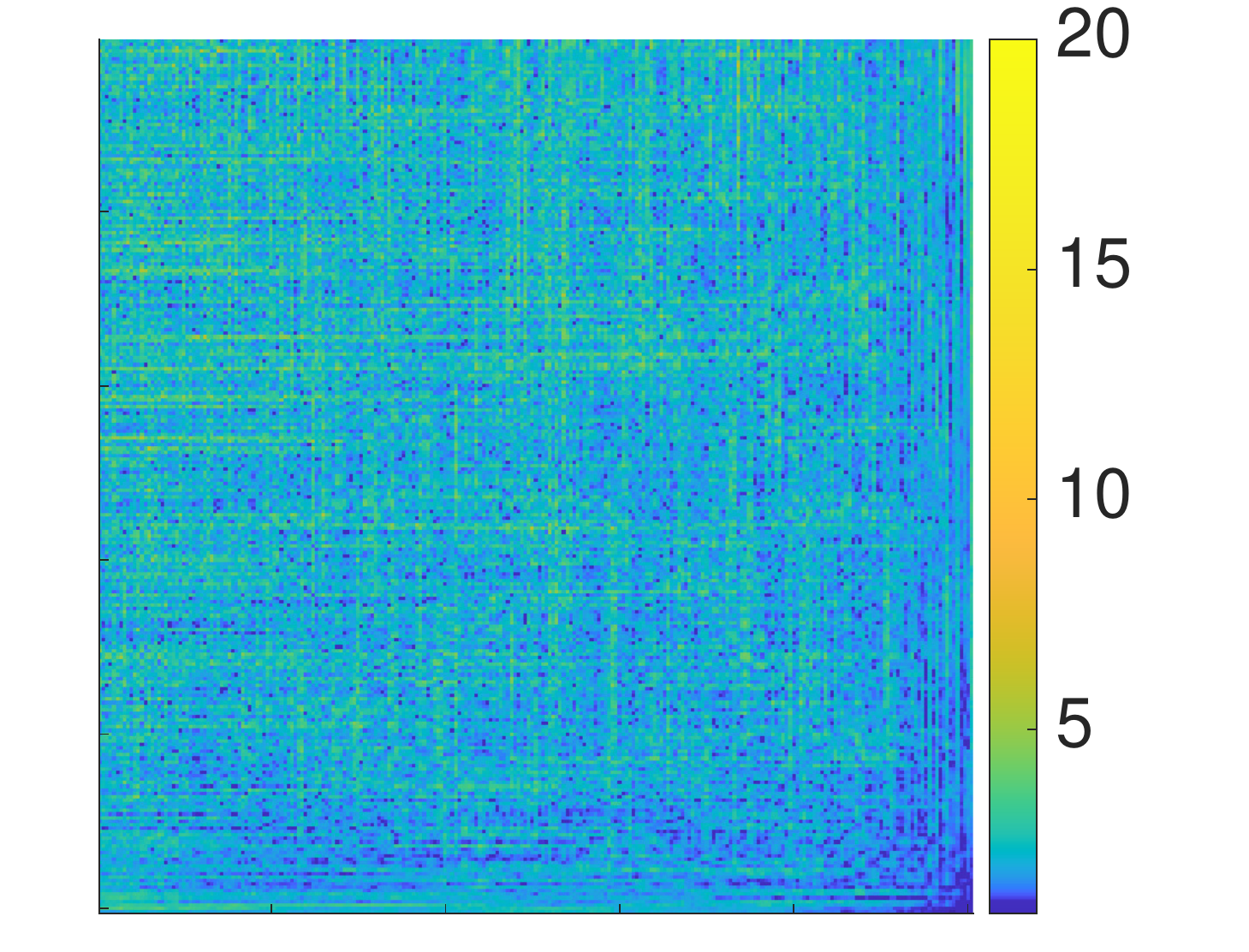}
            \caption{Sheet, RAM }\label{sheet_ram_mean}\end{subfigure}
        
        \caption{  SPDE and  Cauchy sheet priors -- Top row:  mean estimates. Middle row: Variance estimates. Bottom row: PSRF convergence  diagnostics. }
        \label{spde}
    \end{figure}

    Great differences between  the MCMC methods cannot be declared, but again, the RAM seems to have edge over the two other methods. In fact, all the methods exhibit equally faint performance when sampling the posteriors with the second order difference, the SPDE and the sheet priors. We speculate the inefficiency is caused by higher-order non-Gaussian correlations between the nodes. The joint distribution is too heavy-tailed, which renders the marginal chains to slowly drift and oscillate close to their mean values.
    The posteriors with the SPDE and the Cauchy sheet prior seem to be extremely challenging to sample effectively with the current MCMC methods. The PSRF diagnostics of the posteriors of SPDE prior exhibit large values throughout in the domain, what is once again explained by the heavy-tailedness and multimodality of the conditional distributions and make the MCMC converge extremely slowly. As a result, the conditional mean estimators are filled with sponge-kind of peaks and valleys, that can be seen from \Cref{spde}. The sheet prior and the selection of its parameter are definitely  worth of further inspecting, but at least these results are not  promising for applications.

    \begin{figure}
        \begin{subfigure}[b]{\qhei}
            \includegraphics[width=\linewidth]{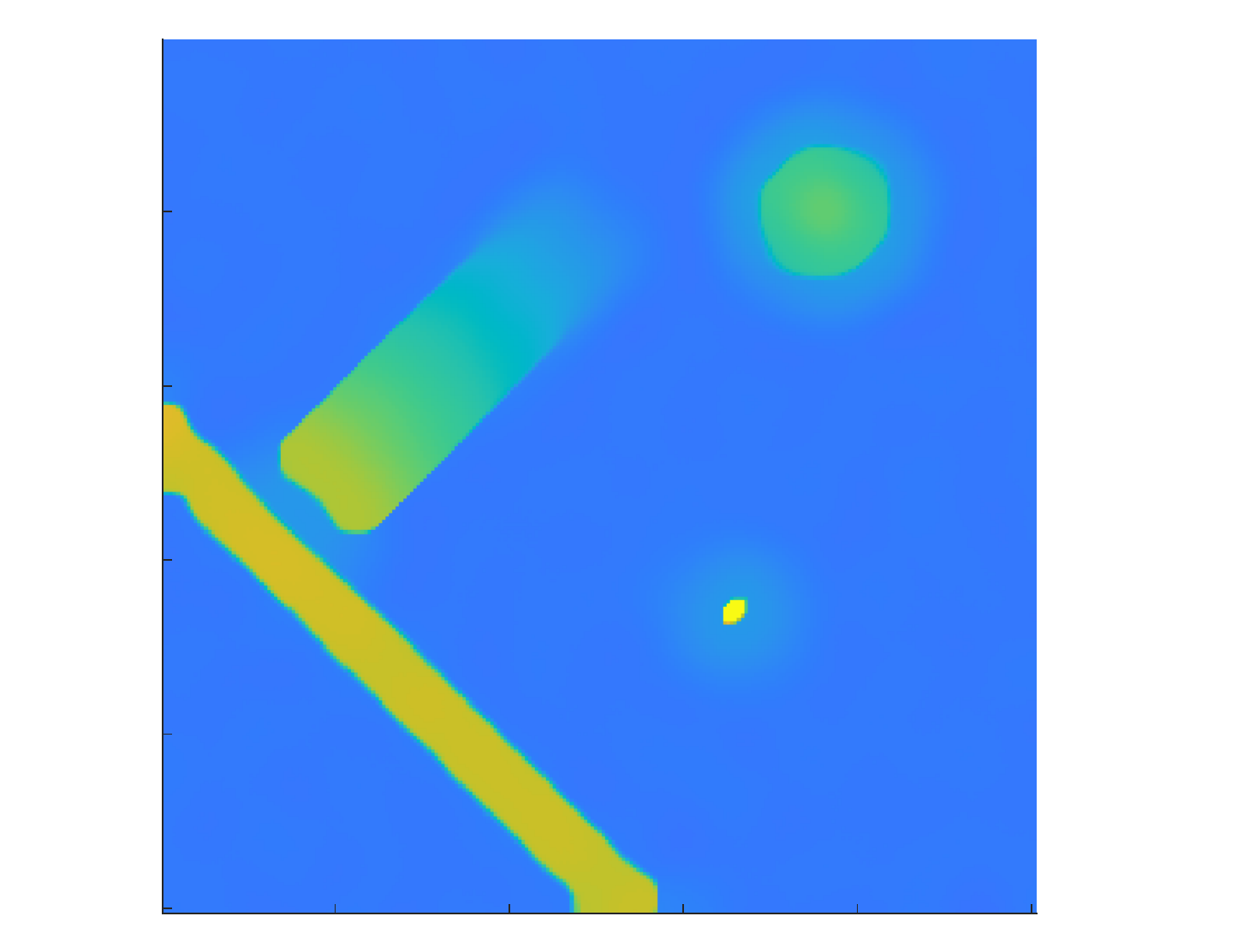} 
            \includegraphics[width=\linewidth]{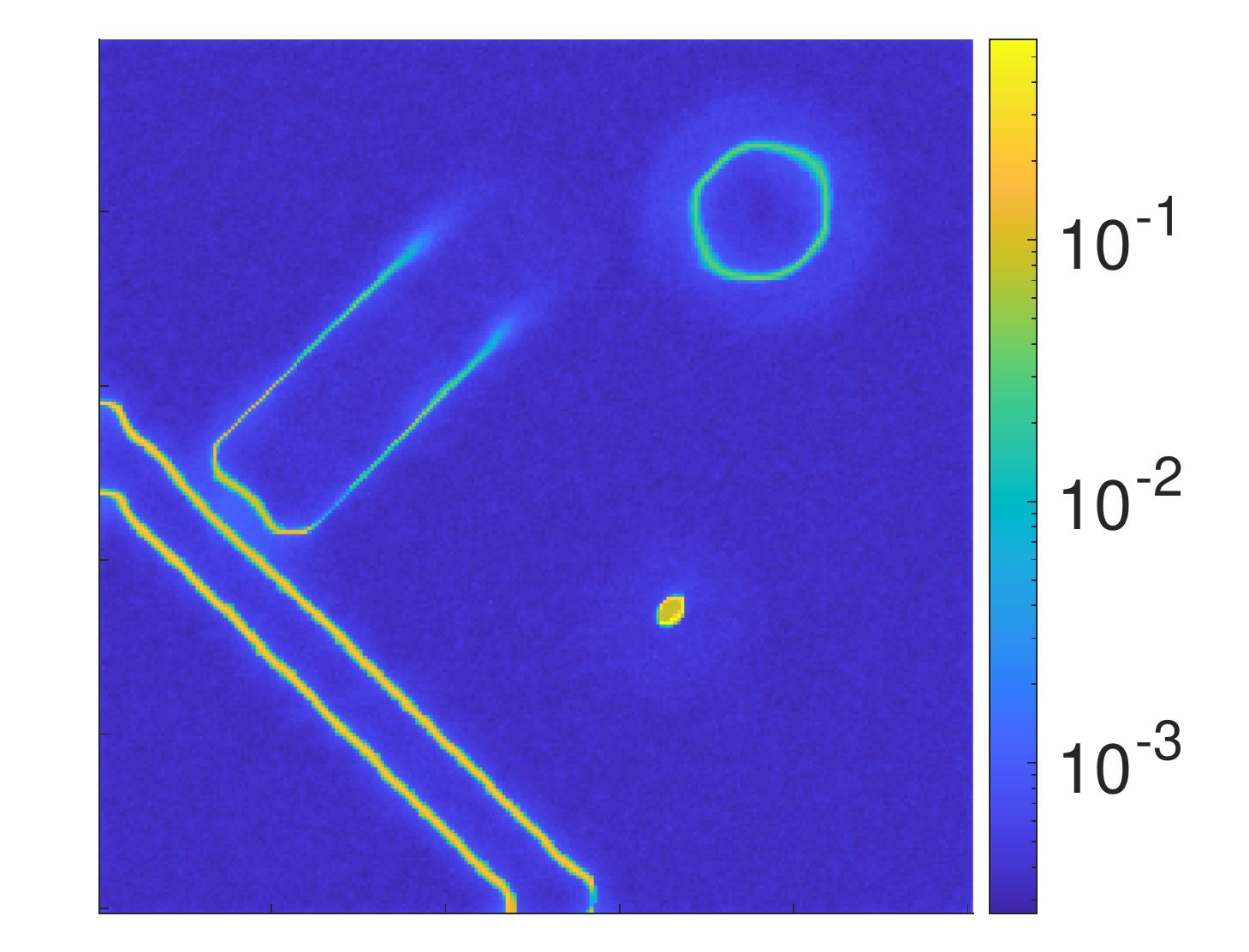}
            \includegraphics[width=\linewidth]{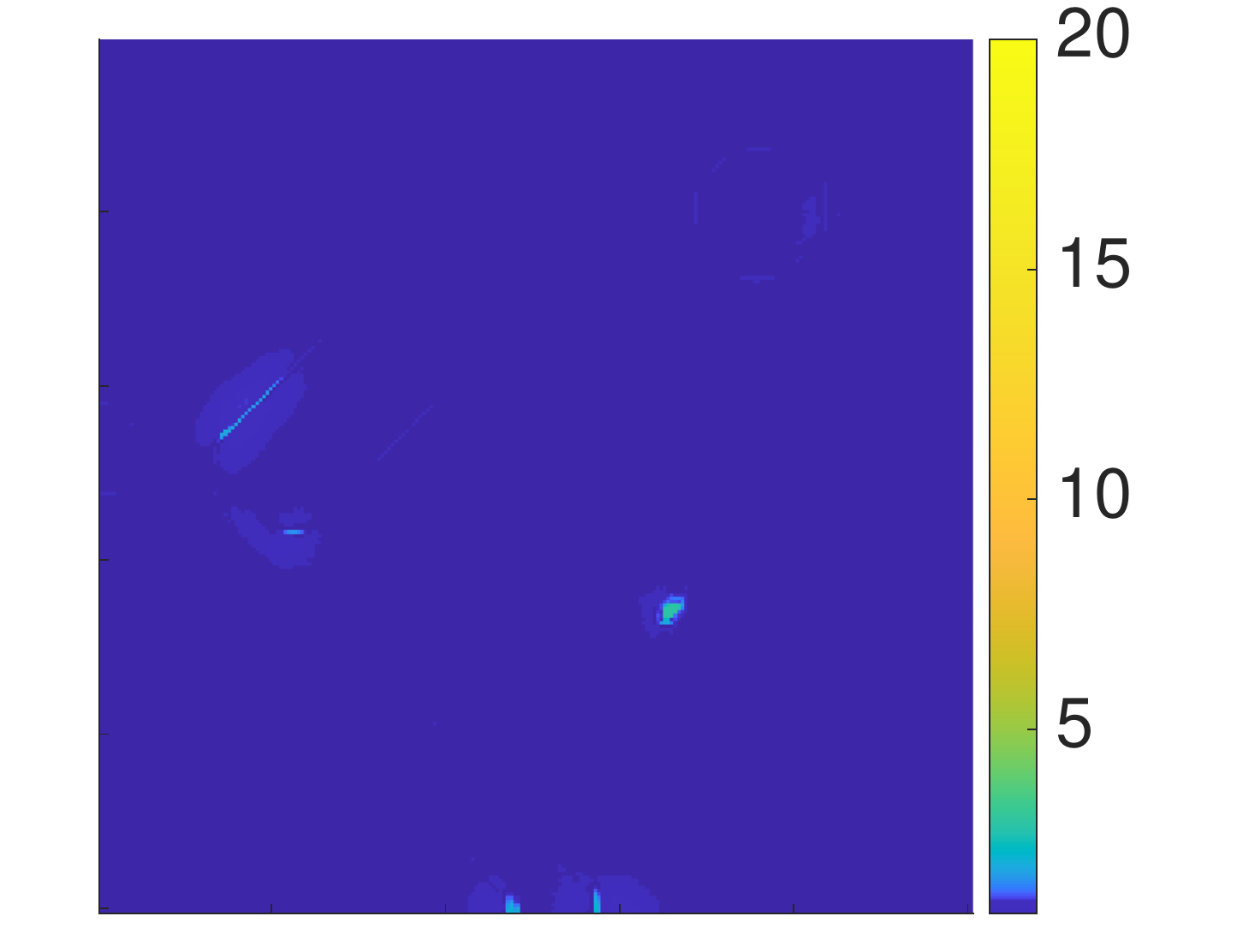}
            \caption{Isotropic, MwG }\label{isod1_mwg_mean}\end{subfigure}
        \begin{subfigure}[b]{\qhei}
            \includegraphics[width=\linewidth]{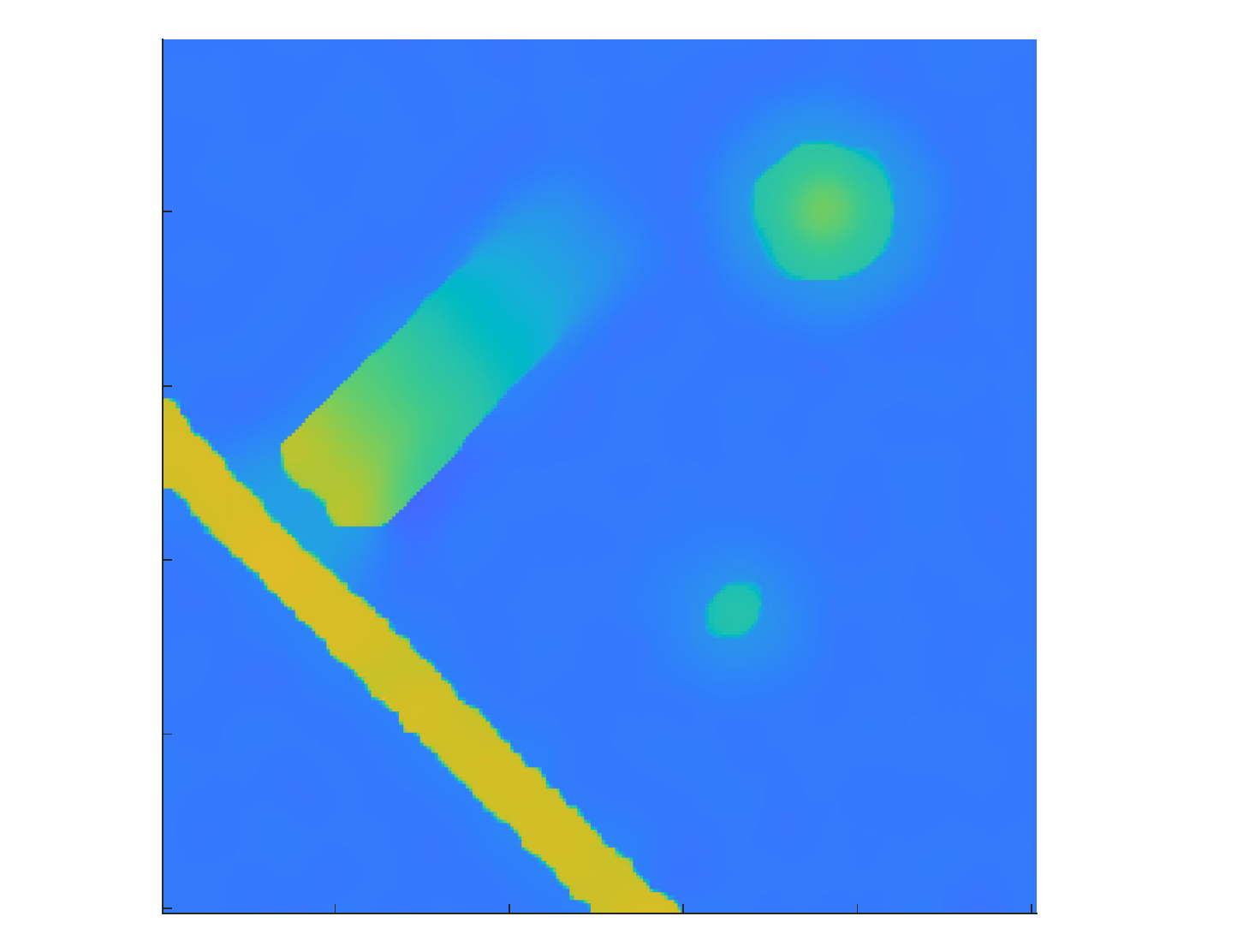}
            \includegraphics[width=\linewidth]{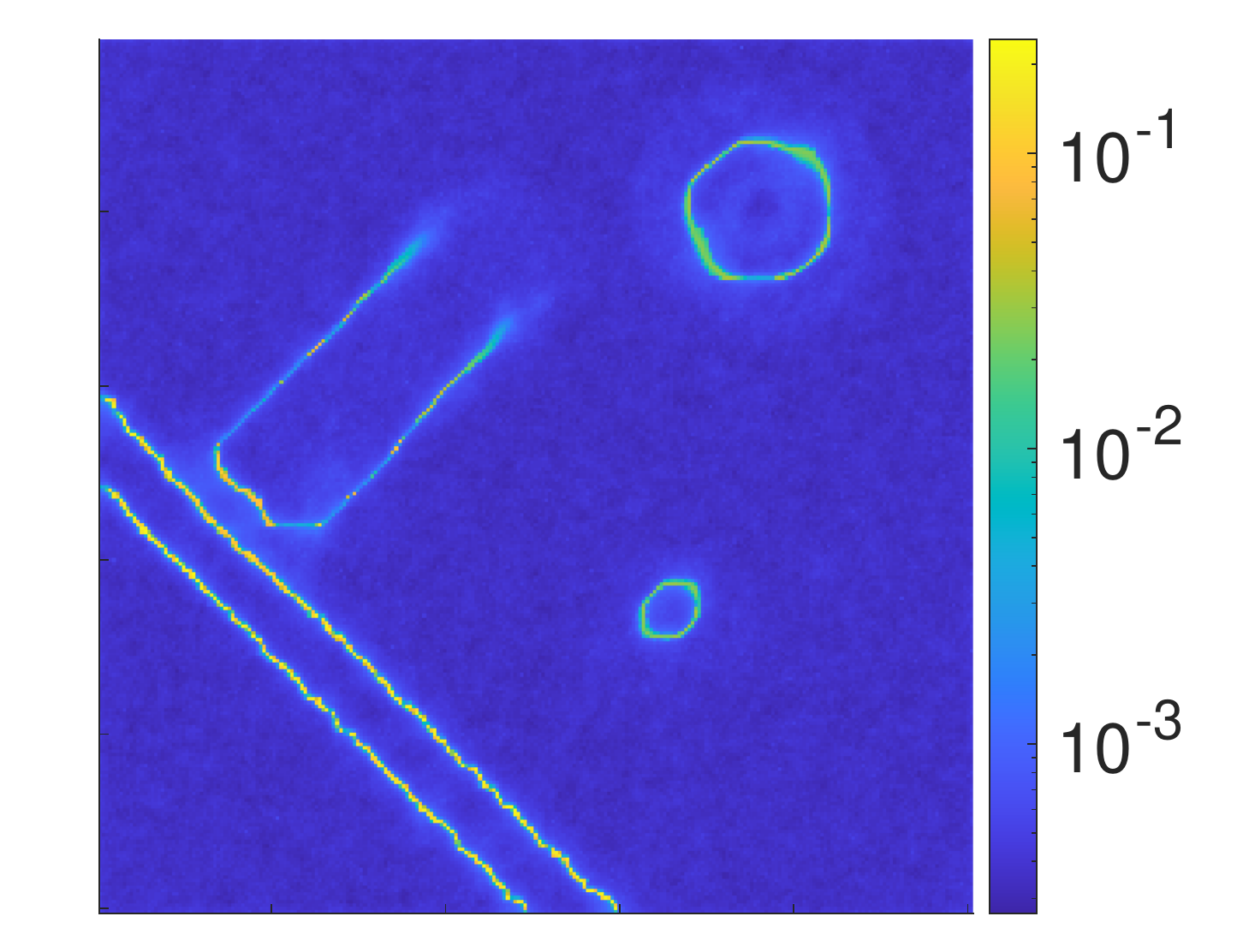}
            \includegraphics[width=\linewidth]{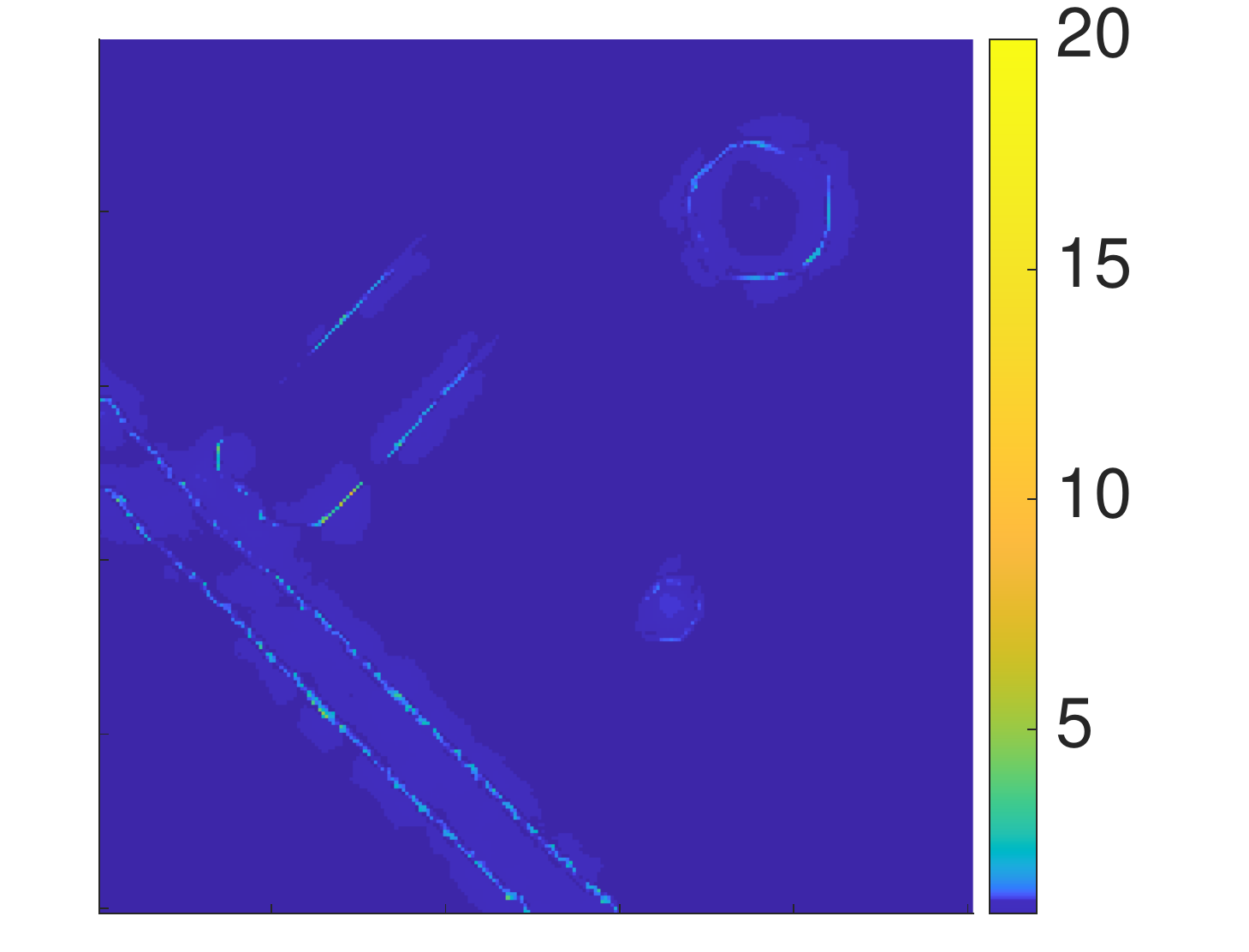}
            \caption{Isotropic, NUTS }\label{isod1_nuts_mean}\end{subfigure}
        \begin{subfigure}[b]{\qhei}
            \includegraphics[width=\linewidth]{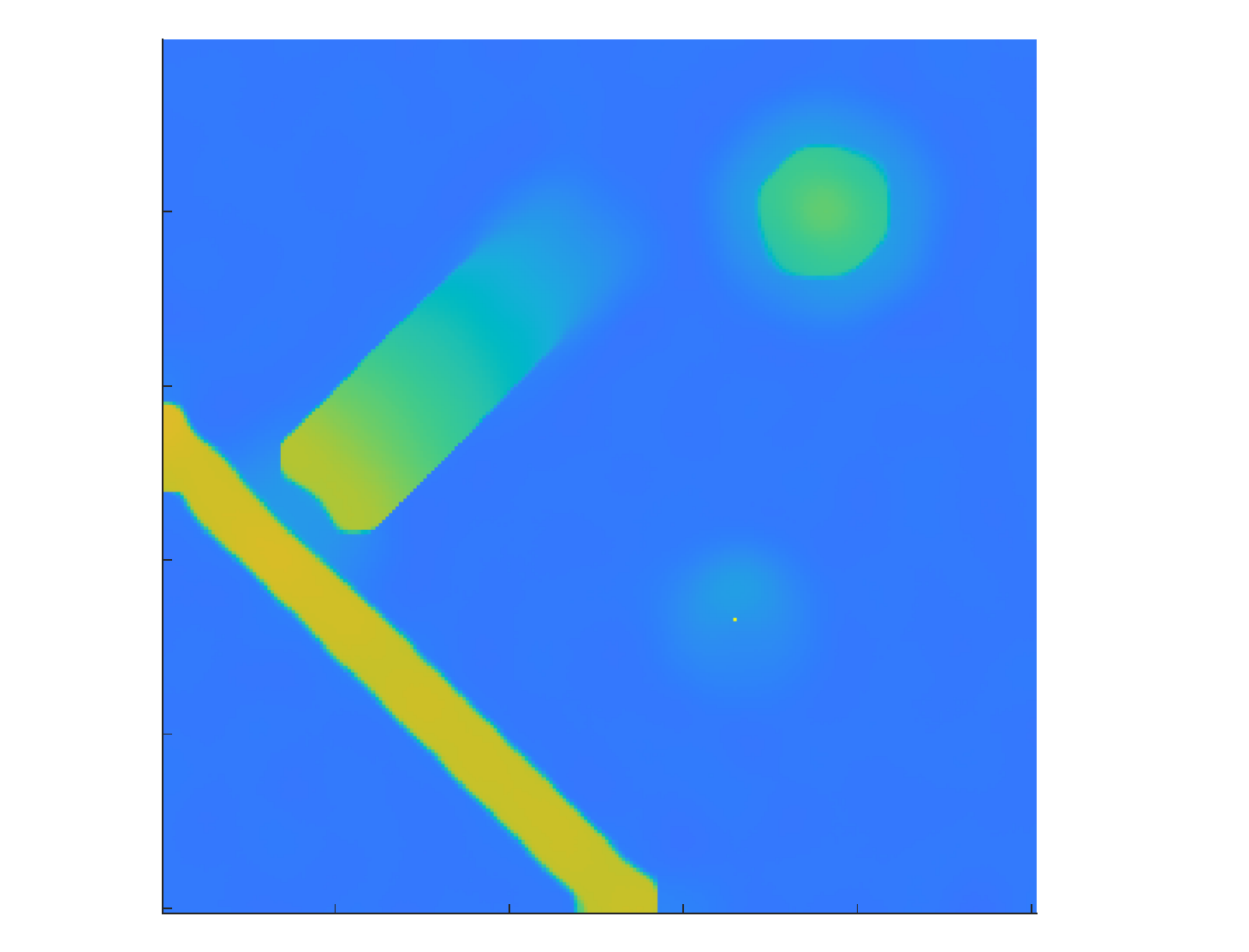}
            \includegraphics[width=\linewidth]{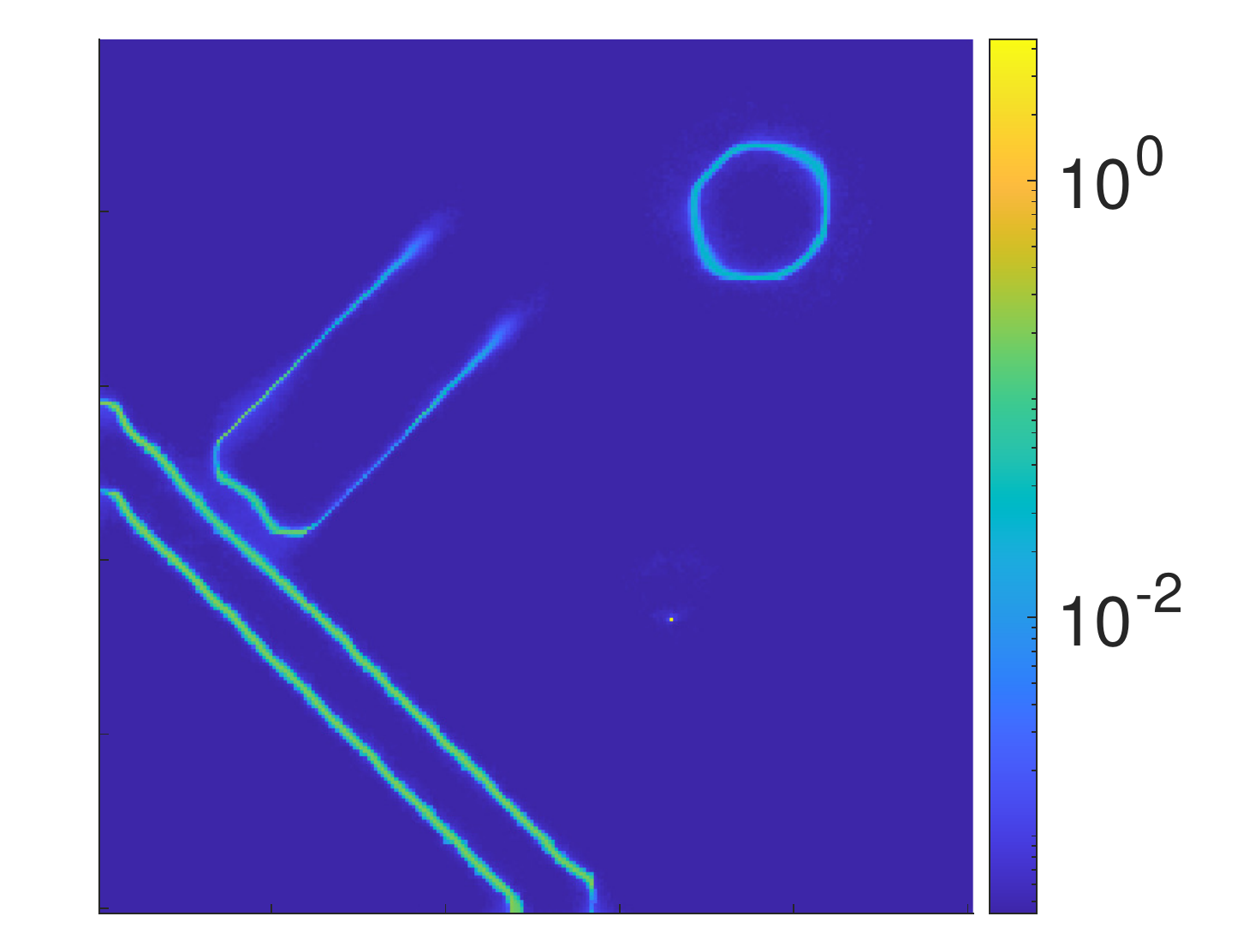}
            \includegraphics[width=\linewidth]{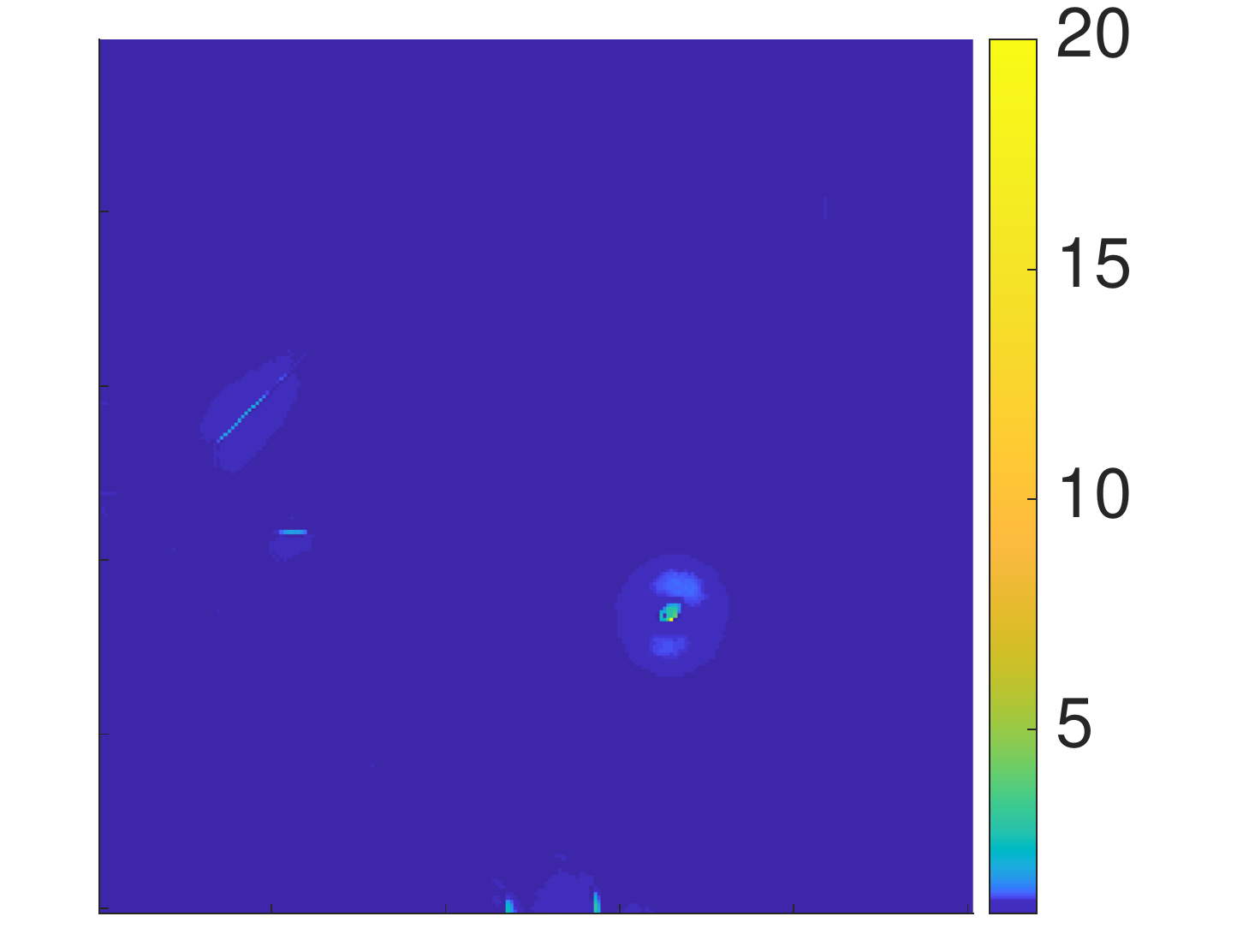}
            \caption{Isotropic, RAM }\label{isod1_ram_mean}\end{subfigure}
        \begin{subfigure}[b]{\qhei}
            \includegraphics[width=\linewidth]{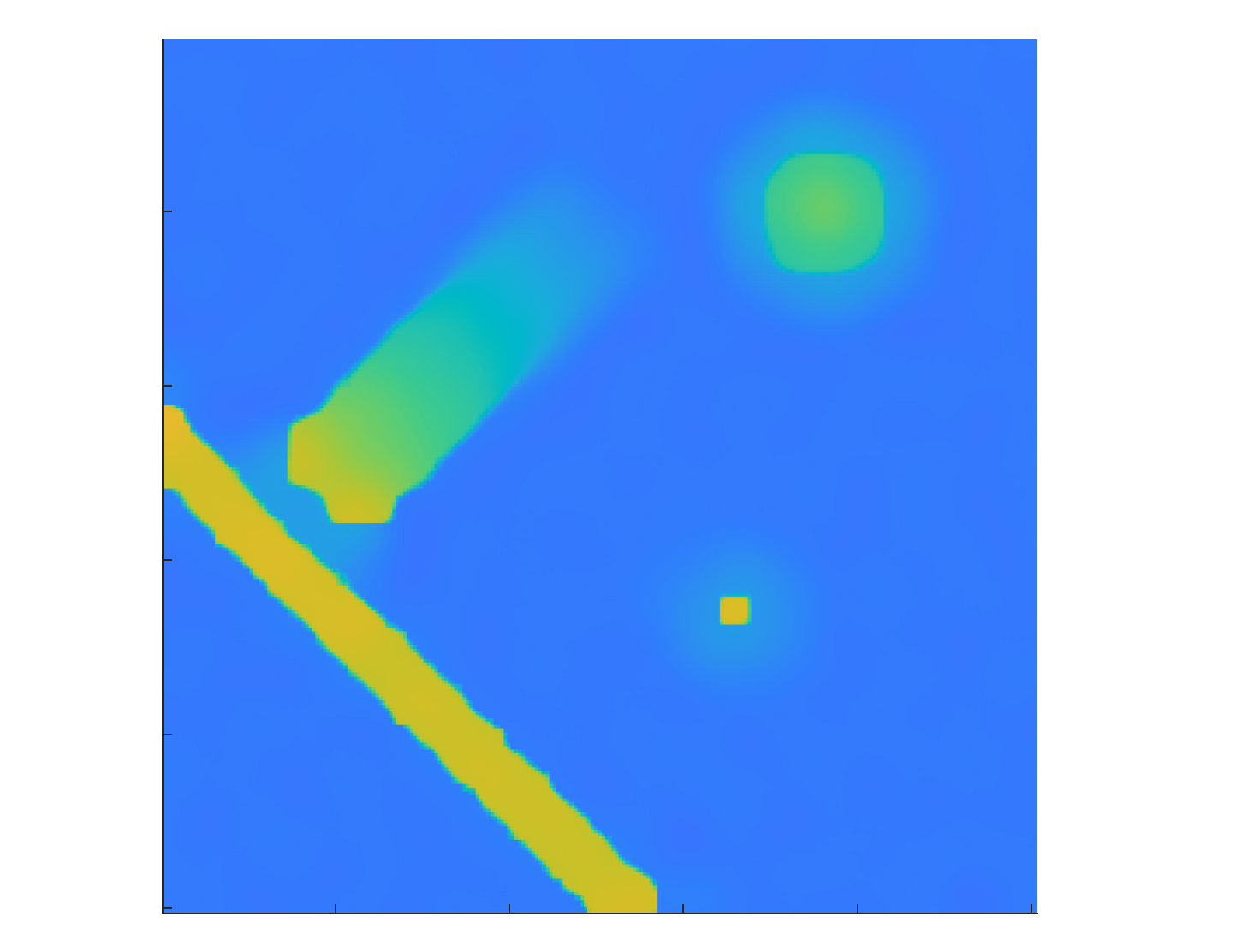}
            \includegraphics[width=\linewidth]{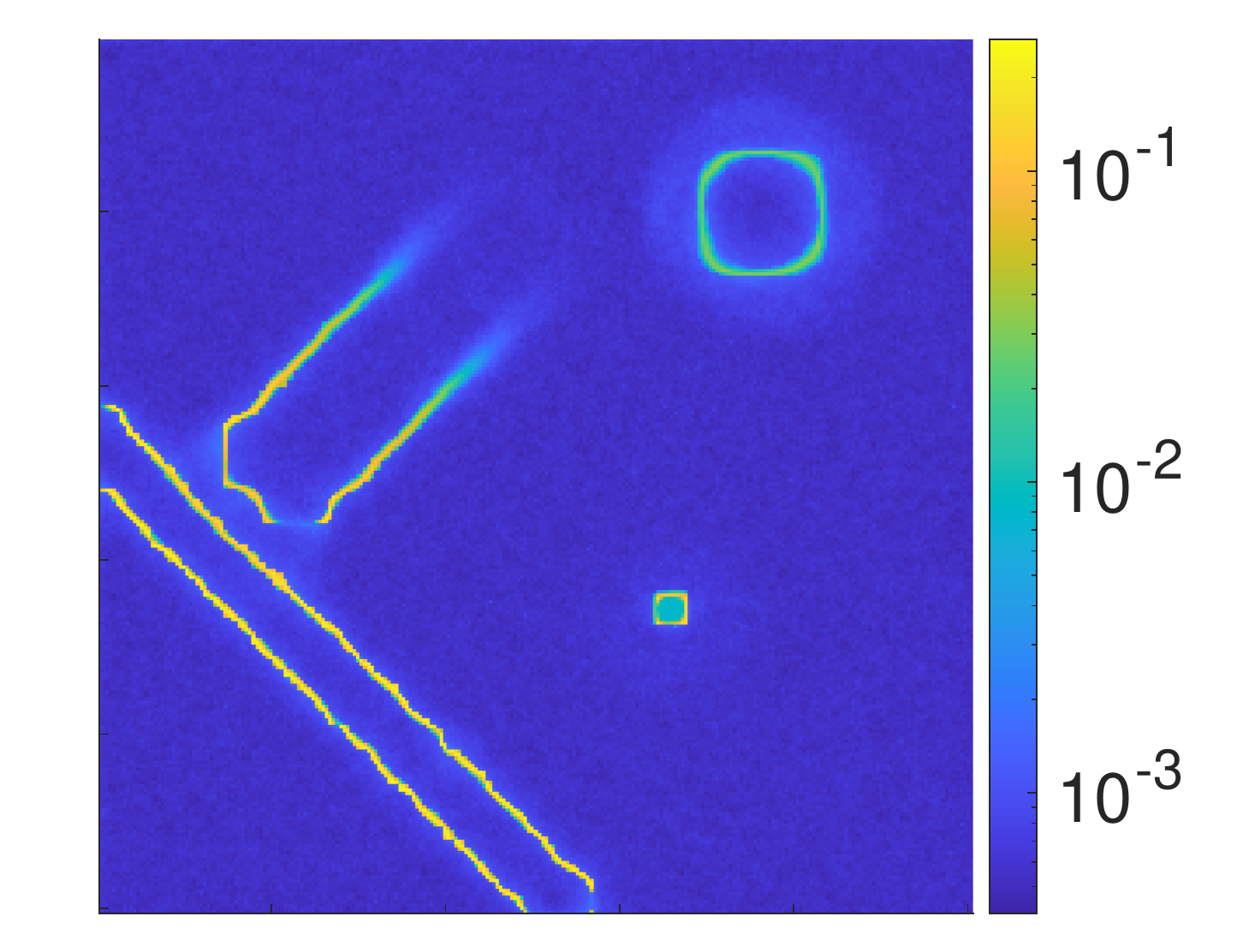}
            \includegraphics[width=\linewidth]{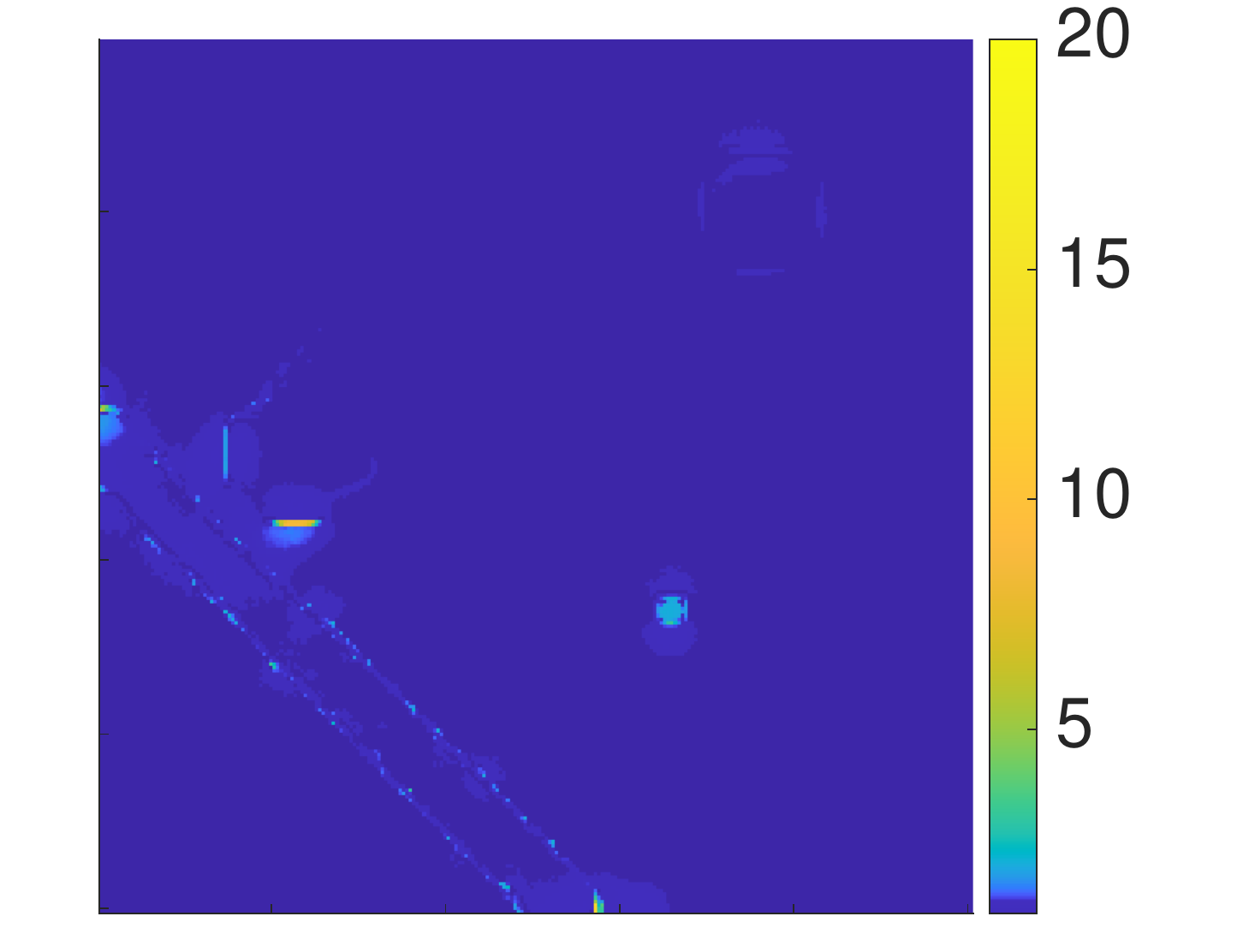}
            \caption{Anisotropic, MwG }\label{anisod1_mwg_mean}\end{subfigure}
        \begin{subfigure}[b]{\qhei}
            \includegraphics[width=\linewidth]{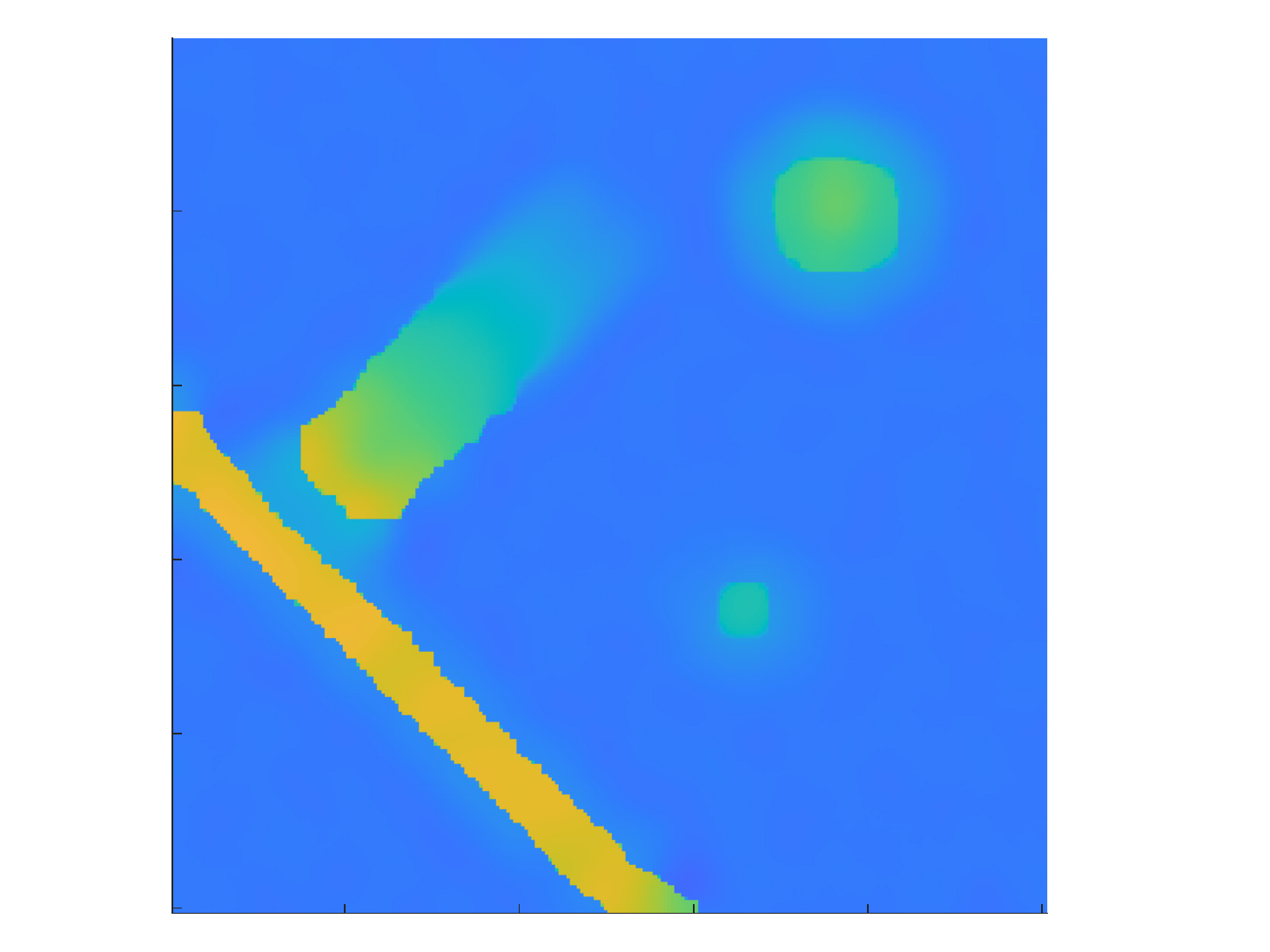}
            \includegraphics[width=\linewidth]{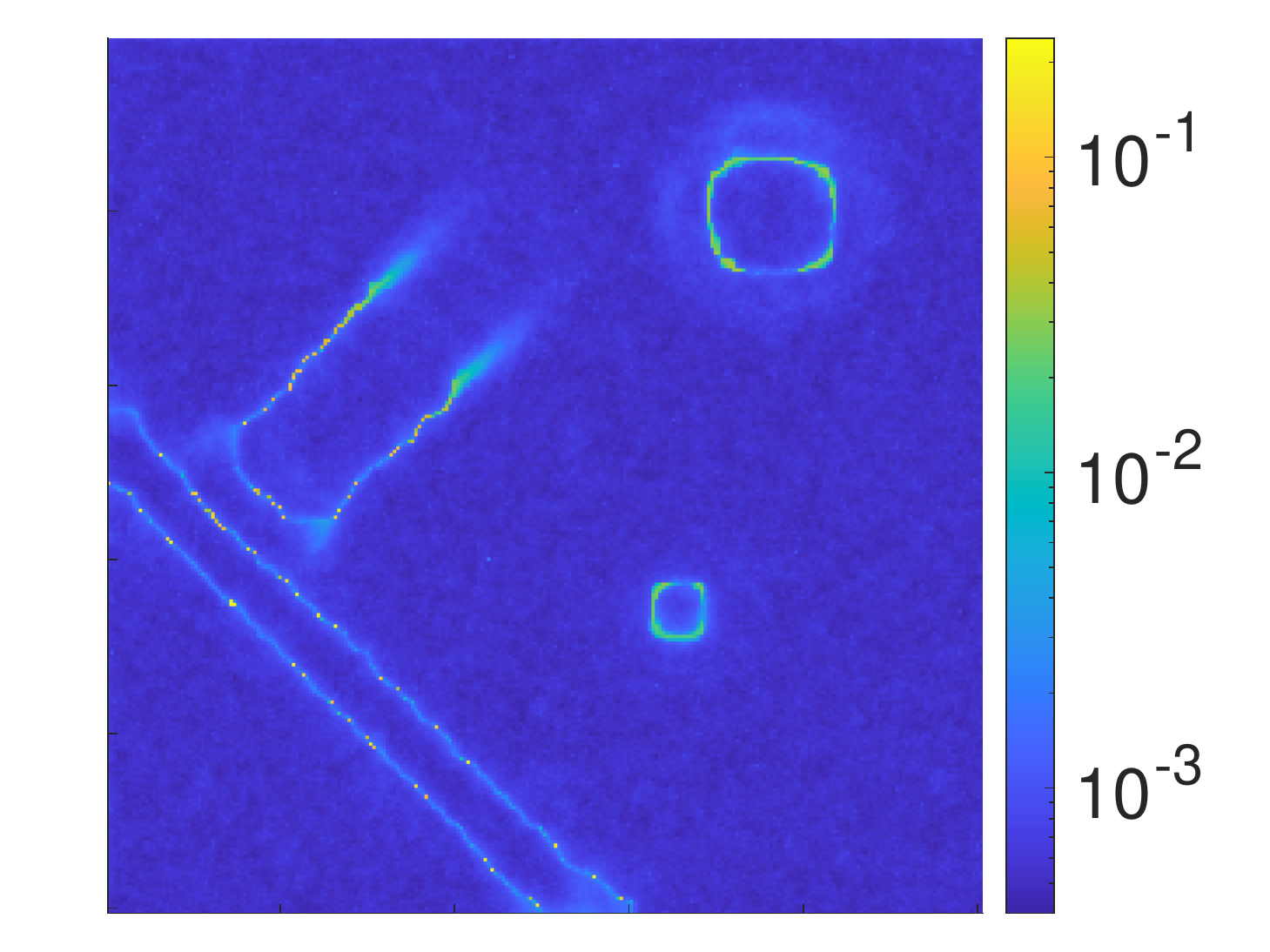}
            \includegraphics[width=\linewidth]{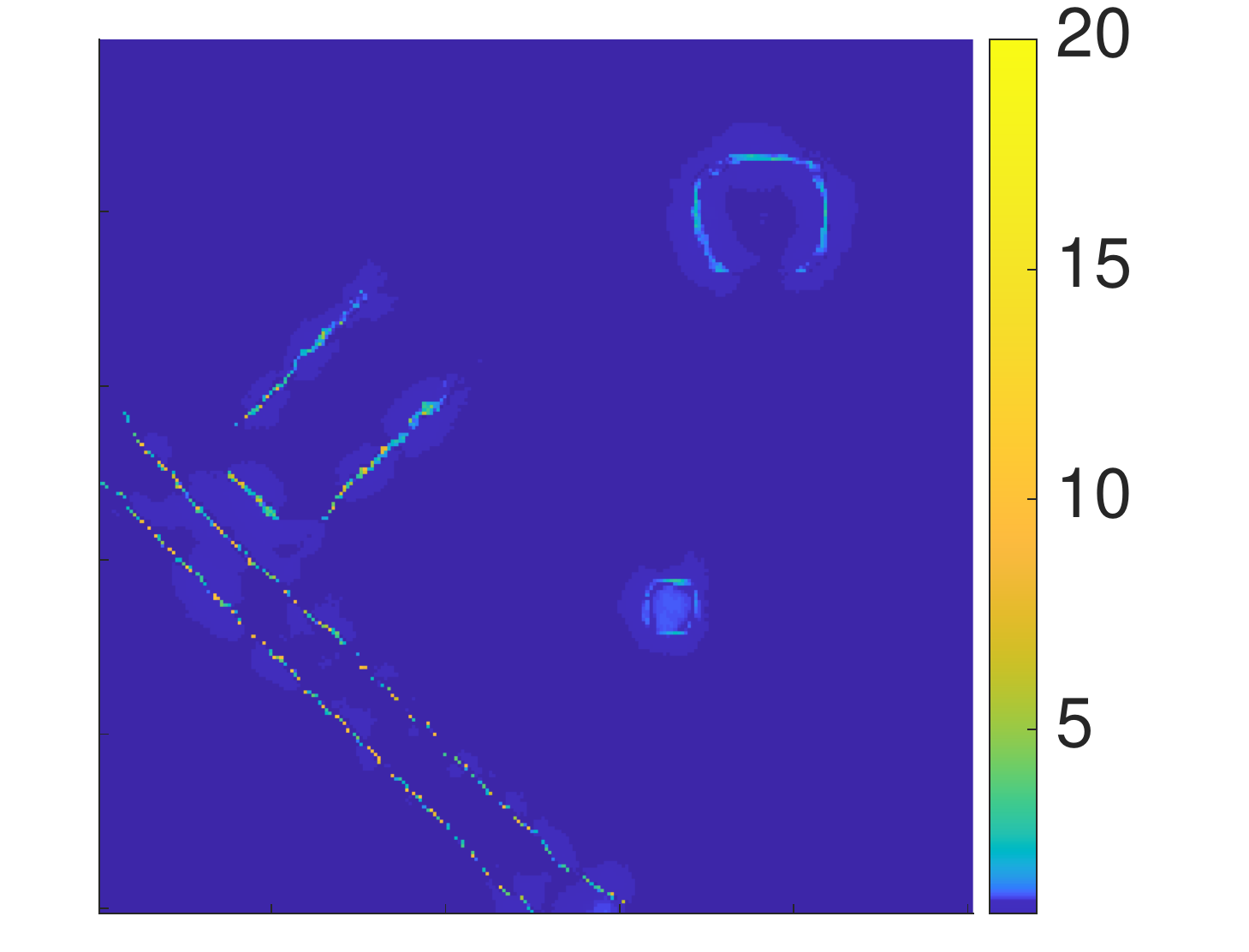}
            \caption{Anisotropic, NUTS }\label{anisod1_nuts_mean}\end{subfigure}
        \begin{subfigure}[b]{\qhei}
            \includegraphics[width=\linewidth]{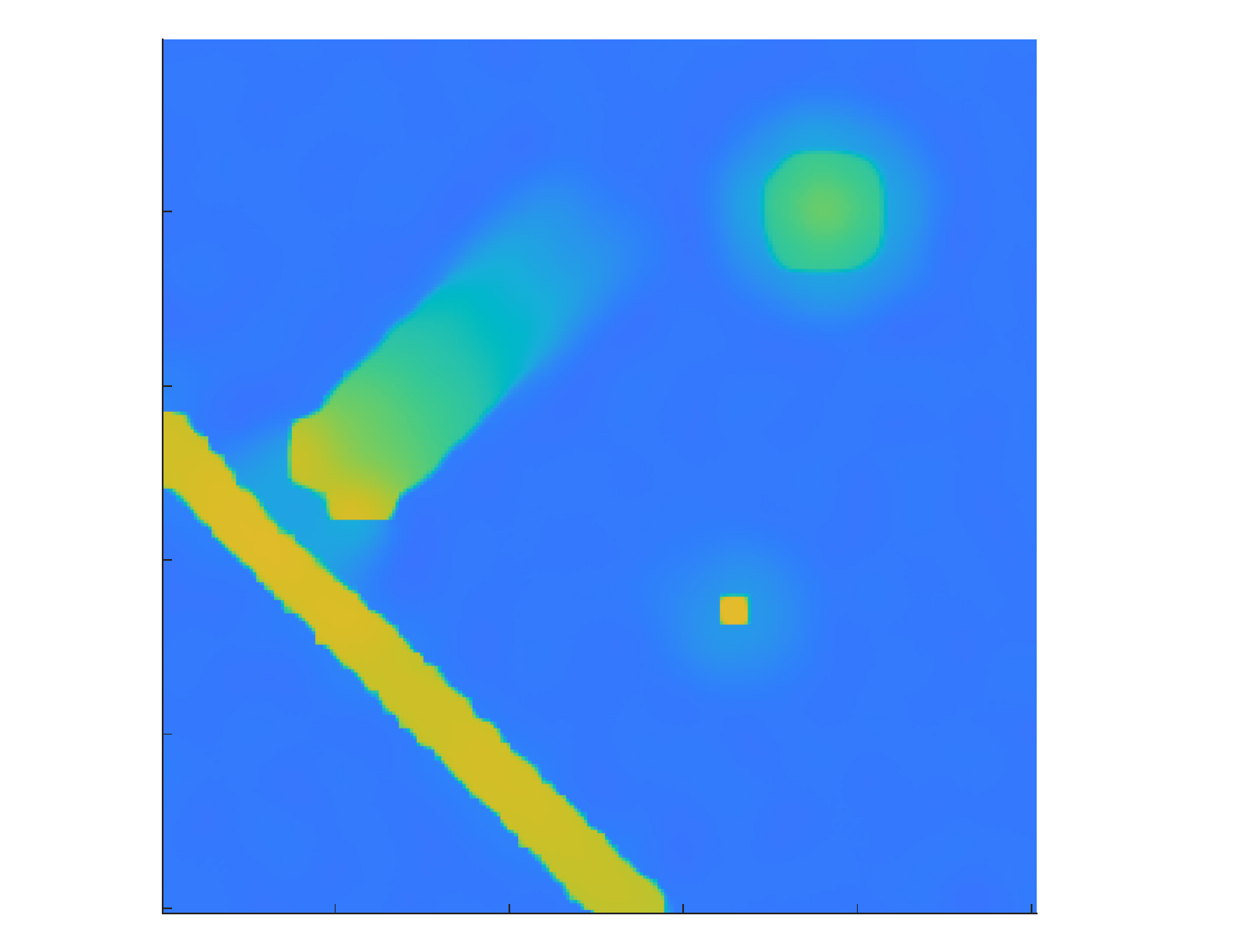}
            \includegraphics[width=\linewidth]{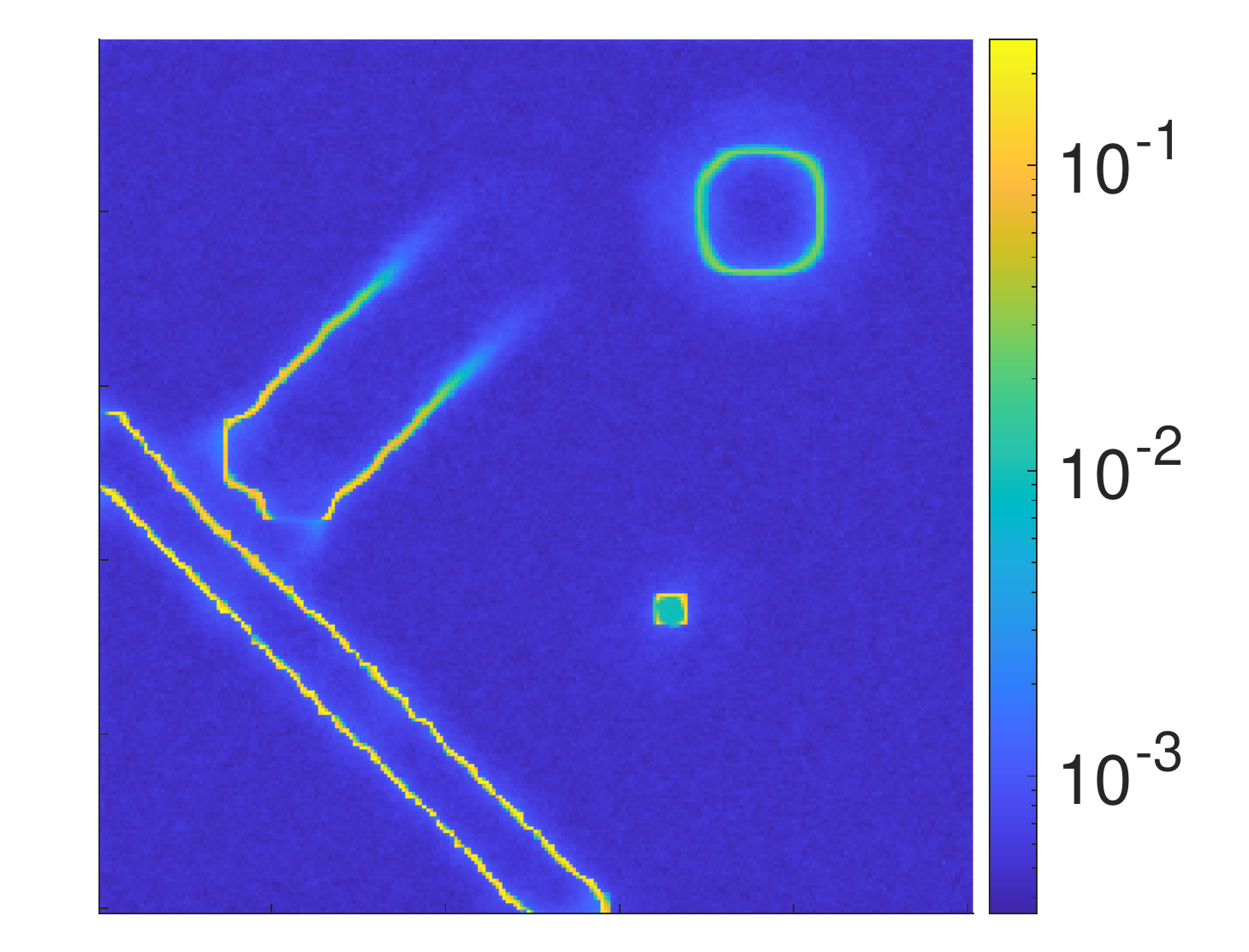}
            \includegraphics[width=\linewidth]{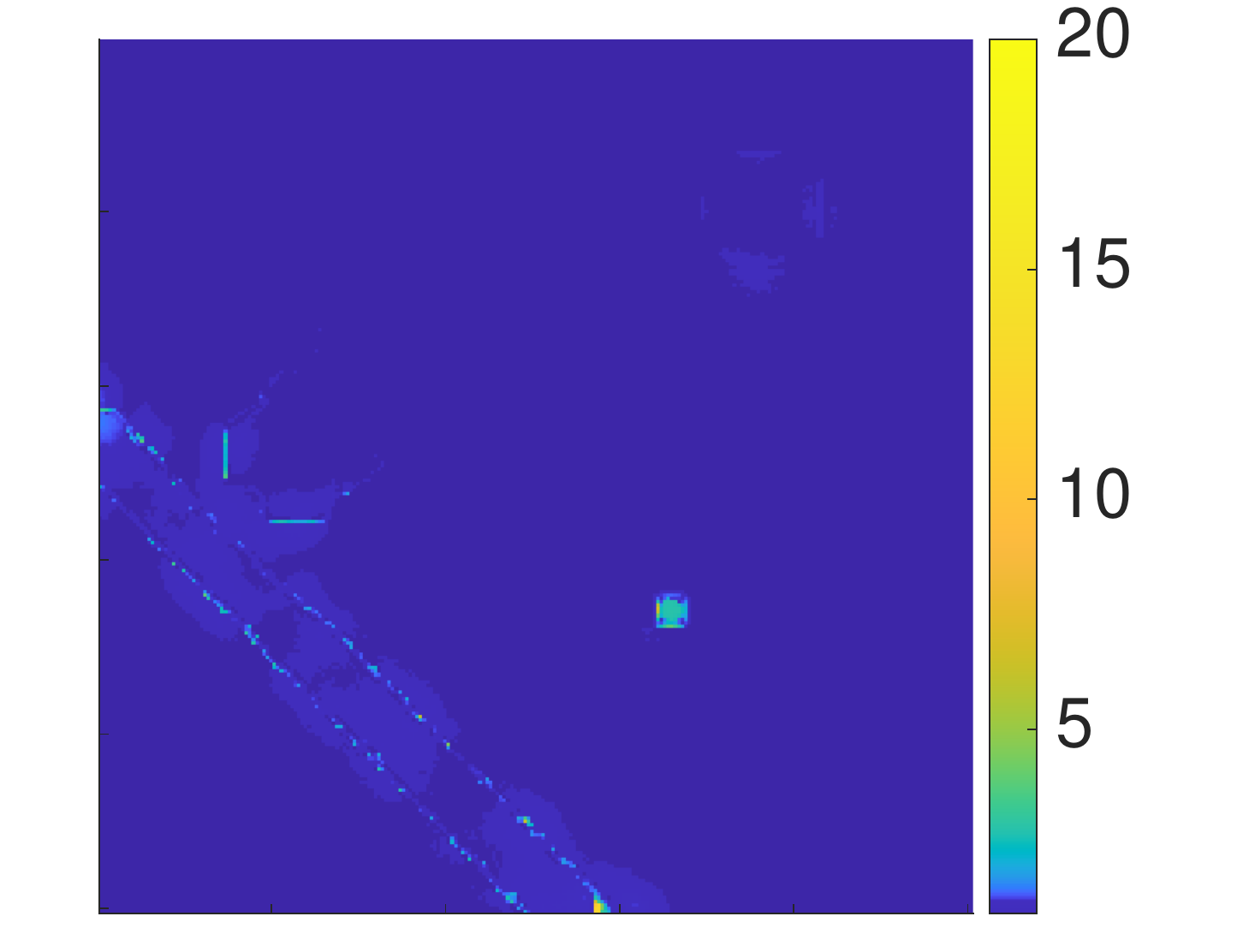}
            \caption{Anisotropic, RAM }\label{anisod1_ram_mean}\end{subfigure}
        \caption{ First order Cauchy priors -- Top row:  mean estimates. Middle row: Variance estimates. Bottom row: PSRF convergence  diagnostics. }       
        \label{andiff1d}
    \end{figure}

        \subsection{Comparison to Gaussian and total variation priors}
    \textcolor{black}{
    As the final numerical experiment, we compare a selection of the Cauchy priors to common Gaussian random field priors and isotropic total variation priors in order to illustrate the  unique features of the priors in Bayesian inversion.  In addition to the isotropic first and second order Cauchy difference prior and Cauchy SPDE prior, we employ the first and second order isotropic total variation (TV) prior \citep{GKS17}, and the first and second order Gaussian difference priors, as well as Gaussian Mat\'ern random field prior. We employ the smooth Charbonnier approximation \citep{CBAB97} of the isotropic total variation priors, which renders the log posteriors differentiable and hence easier for L-BFGS to deal with. The first order isotropic total variation prior in a equispaced lattice is of form
     \begin{equation}
        \label{eq:tv}
        \pi(\mathbf{u}) \propto 
        \pi_{\partial \Omega}(\mathbf{u})
        \prod_{i=1}^{N-1}  \prod_{j=1}^{N-1} \exp \left(-\zeta \sqrt { (u_{i+1,j} - u_{i,j})^2 + (u_{i,j+1} - u_{i,j})^2 + \delta^2 }\right),
        \end{equation} 
        with 
        \begin{equation*}
        \pi_{\partial \Omega}(\mathbf{u}) = \prod_{(i,j) \in \partial \Omega} \exp \left (-\zeta' \sqrt{{u_i,j}^2 + \delta} \right ).
        \end{equation*}
        Similarly, the second order isotropic total variation prior can be expressed as
         \begin{equation}
        \label{eq:tv2}
        \pi(\mathbf{u}) \propto 
        \pi_{\partial \Omega}(\mathbf{u})
        \prod_{i=2}^{N-1}  \prod_{2=1}^{N-1} \exp \left(-\zeta \sqrt { (u_{i+1,j} - 2u_{i,j} + u_{i-1,j})^2 + (u_{i,j+1} - 2u_{i,j} - u_{i,j-1})^2 + \delta^2 }\right), 
        \end{equation} 
        with
        \begin{equation*}
       \pi_{\partial \Omega}(\mathbf{u}) = \prod_{(i,j) \in \partial \Omega} \exp \left (-\psi \sqrt{u_{i,j}^2 + \delta^2} -\zeta' \sqrt{(u_{i,j} - u_{i-\cdot,j-\cdot})^2 + \delta^2}  \right ).
        \end{equation*}
    Additionally, the first order Gaussian difference prior is 
     \begin{equation}
        \label{eq:g1}
        \pi(\mathbf{u}) \propto 
        \pi_{\partial \Omega}(\mathbf{u})
        \prod_{i=1}^{N-1}  \prod_{j=1}^{N-1} \exp \left(-\frac{1}{2 \sigma_1^2}  \left( (u_{i+1,j} - u_{i,j})^2 + (u_{i,j+1} - u_{i,j})^2  \right)\right),
        \end{equation} 
        where
        \begin{equation*}
        \pi_{\partial \Omega}(\mathbf{u}) = \prod_{(i,j) \in \partial \Omega} \exp \left (-\frac{1}{2 \sigma_0^2} u_{i,j}^2 \right ),
        \end{equation*}
        and the second order Gaussian difference prior is of form
         \begin{equation}
        \label{eq:g2}
        \pi(\mathbf{u}) \propto 
        \pi_{\partial \Omega}(\mathbf{u})
        \prod_{i=2}^{N-1}  \prod_{2=1}^{N-1} \exp \left( -\frac{1}{2 \sigma_2^2} \left(  (u_{i+1,j} - 2u_{i,j} + u_{i-1,j})^2 + (u_{i,j+1} - 2u_{i,j} - u_{i,j-1})^2 \right) \right), 
        \end{equation} 
        where
        \begin{equation*}
       \pi_{\partial \Omega}(\mathbf{u}) = \prod_{(i,j) \in \partial \Omega} \exp \left (-\frac{1}{2 \sigma_0^2} u_{i,j}^2  -\frac{1}{2 \sigma_1^2} \left(u_{i,j} - u_{i-\cdot,j-\cdot}\right)^2  \right ).
        \end{equation*}}

 \textcolor{black}{Like before,  by $u_{i-\cdot,j-\cdot}$ we denote the closest lattice point inside the domain $\Omega$ of the pixel $(i,j)$, and the prior parameters $\zeta, \zeta',\psi, \psi', \sigma_0, \sigma_1, \sigma_2 >0$. In both the first and the second order TV priors, we set $\delta^2$ to $10^{-5}$. The construction of the Gaussian priors is similar to the Cauchy and total variation priors.  In the Gaussian difference priors, we discard the square roots in Equations (\ref{eq:tv}) and (\ref{eq:tv2}) as well as the Charbonnier approximation term $\delta$. The Mat\'ern SPDE prior with Gaussian noise is constructed with the sample discretized operator as in \ref{eq:spde}, but we set the auxiliary terms $p_{i,j}$ to follow Gaussian distribution:
        \begin{equation}
          \pi_\textrm{p}(\mathbf{u}) \propto \prod_{(i,j)}  \exp \left ( -\frac{1}{2\sigma_w^2} p_{i,j}^2\right).   
        \end{equation}
Similar to the previous two numerical examples, we employ the priors for Bayesian deconvolution. We discretize the reconstruction grid in $256\times256$ pixels, and we simulated the noisy convolutions in a grid of $160\times160$ pixels by a help of a larger grid of the phantom of $300\times300$ pixels and the corresponding convolution matrix. The phantom was hand-crafted and thus it does not have any analytical expression. We compute the MAP estimates using the L-BFGS method. The MAP estimates are plotted in \ref{others}. The logarithms of the gradient norms as a function of the L-BFGS iteration number are presented in \ref{others-stats}. We set  $\lambda$ of the first order Cauchy difference prior to 0.1 and $\gamma$ to 1.0; $\lambda$ of the second order Cauchy difference prior to 0.001 and $\lambda'$ and  $\gamma$ to 1.0; the $\ell$ parameter of the Cauchy SPDE prior to $0.01$ and the scale of the Cauchy noise to $3.0$. Likewise, we set $\sigma_1$ of the Gaussian first order difference prior to 0.1 and $\sigma_0$ to 10;  $\sigma_2$ of the second order difference prior to 0.0005,  $\sigma_1 $  and $\sigma_0$ to 10.0; the $\ell$ parameter of the Gaussian SPDE prior to 0.005 and the variance of the Gaussian noise ($\sigma_w$) to $3.0^2$. Finally, we set  $\zeta$ of the first order total variation prior to 5.0 and $\zeta'$ to 0.001; $\zeta$ of the second order total variation prior to 1000.0 and $\zeta'$ and $\psi$ to 0.001.}

\textcolor{black}{The MAP estimate with first order Cauchy difference prior favors both abrupt discontinuities and very gradually decreasing or increasing smooth features. On the other hand, the Gaussian first order difference prior is unable to model discontinuities, which appears as smooth reconstructions.  As the total variation priors have finite variance, the MAP estimates with first order TV priors do not contain as sharp-edged features as the first order Cauchy prior does. The MAP estimate with the first order TV prior has pictorial features, since it does not contain very steep discontinuities, but also seems to compose of a discrete number of levels.  The second order priors are more difficult to describe, but the infinite variance of the Cauchy distribution allows existence of features that have piecewise constant first order partial derivatives. The second order Gaussian difference prior smooths out all the features.  The second order total variation prior cannot  model vigorously enough first order  partial derivatives that are either almost constant, which renders the MAP estimates to have gradual transitions for the first order partial derivatives like in the  pyramid-shaped part of the test function in Figure \ref{others}.}

\textcolor{black}{Finally, the motivation for Mat\'ern SPDE with Cauchy noise is justified if the prior should seek for exponential peaks and valleys. The Mat\'ern SPDE prior with Gaussian noise is unable to do so without sacrificing the capability to discard noise and irrelevant features or without making the prior less informative and hence more prone to overfitting. Thanks to the infinite variance of the Cauchy distribution, the peaks can be detected, but the drawback is that the prior favors such peaks elsewhere, too.}

\textcolor{black}{In the light of the gradient norm of the log posteriors in L-BFGS iteration,  the difficulties MCMC algorithms face when sampling posteriors with Cauchy priors, are more understandable. That is, even L-BFGS struggles with the heavy-tailed posterior distributions that are possibly multimodal. Nevertheless, the behavior is not surprising, as L-BFGS approximates the Hessian of the target functional with a low rank matrix that is known to perform well with functionals like Gaussian log densities. The issue is the worst with the posteriors having the Cauchy noise SPDE prior or the second order Cauchy difference prior.}
    

\newcommand{\stathei}{4.2cm}
\newcommand{\maphei}{3.9cm}

    \begin{figure}
        \centering
         \begin{subfigure}[b]{\maphei}
         \includegraphics[width=\linewidth]{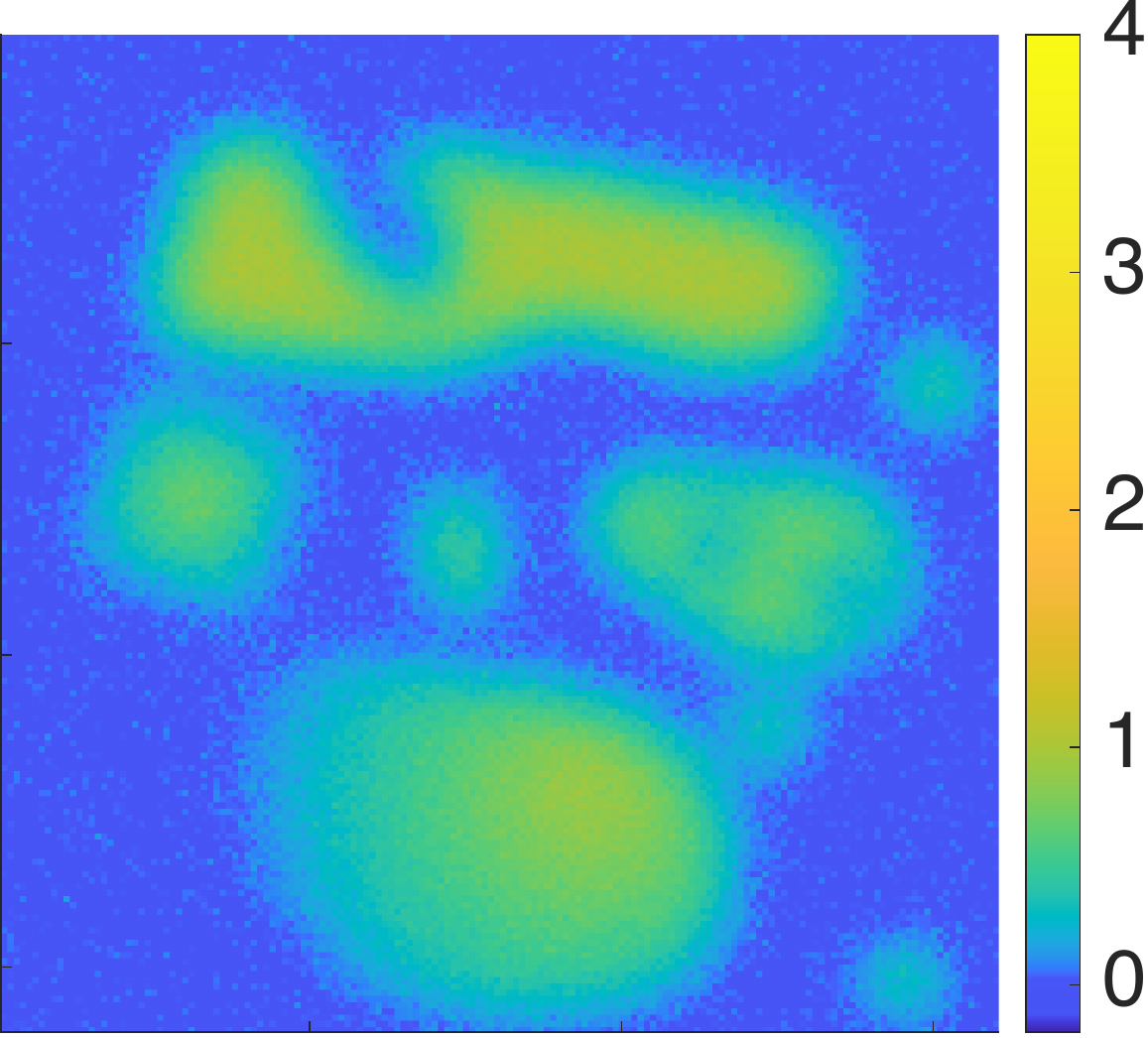}
             \includegraphics[width=\linewidth]{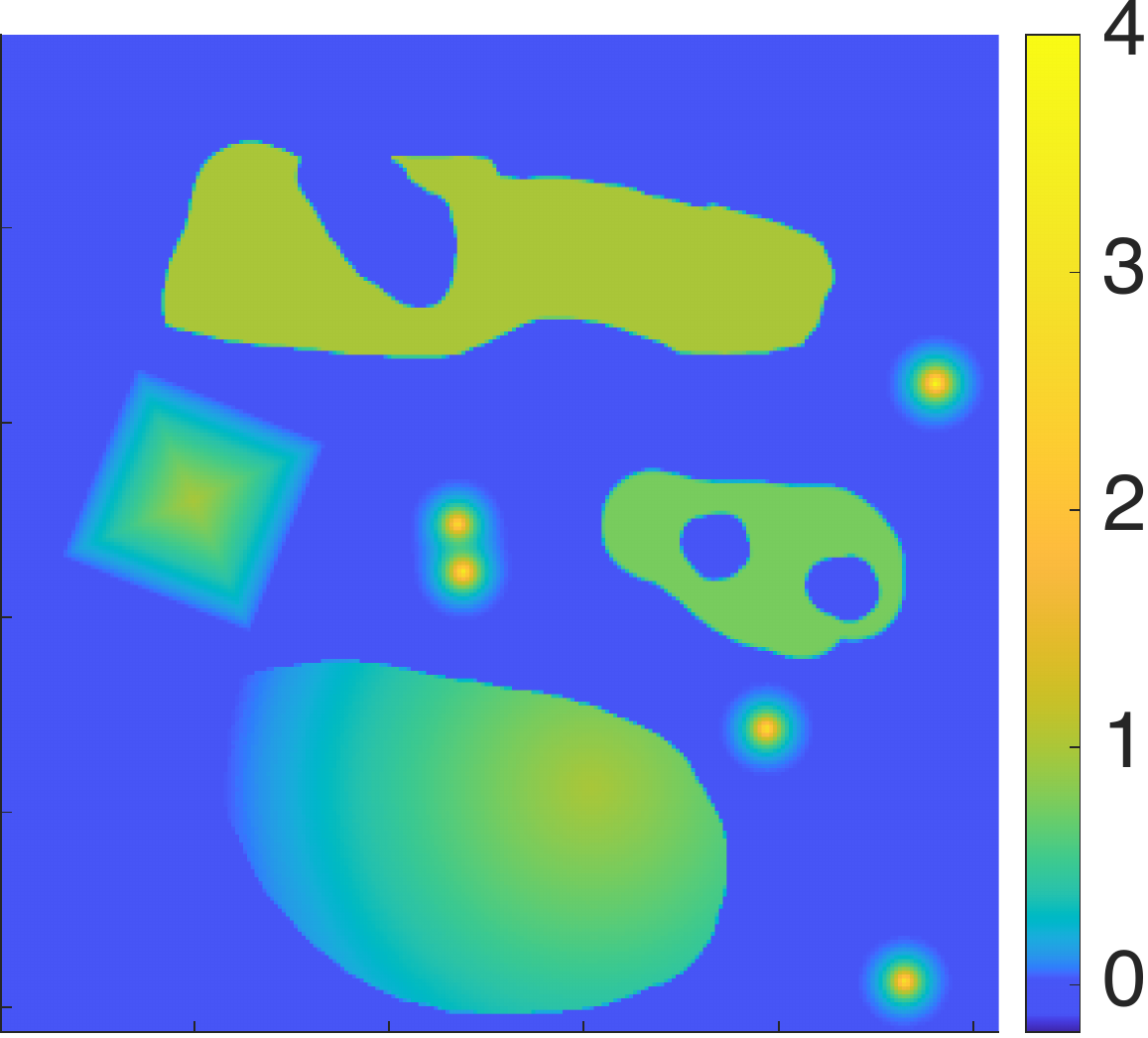}
            \caption{Meas. and truth}\end{subfigure}
        \begin{subfigure}[b]{\maphei}
        \includegraphics[width=\linewidth]{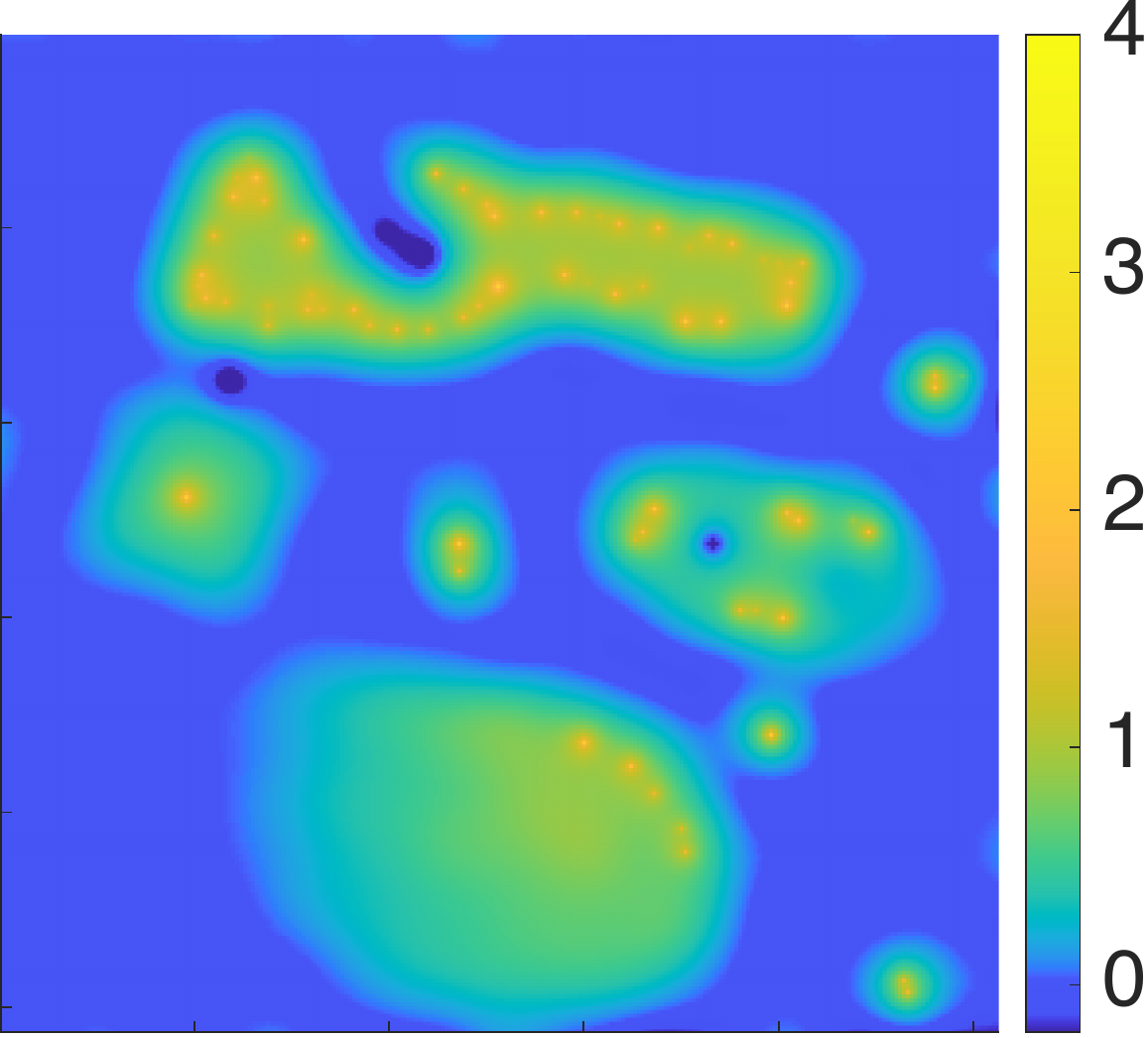}
            \includegraphics[width=\linewidth]{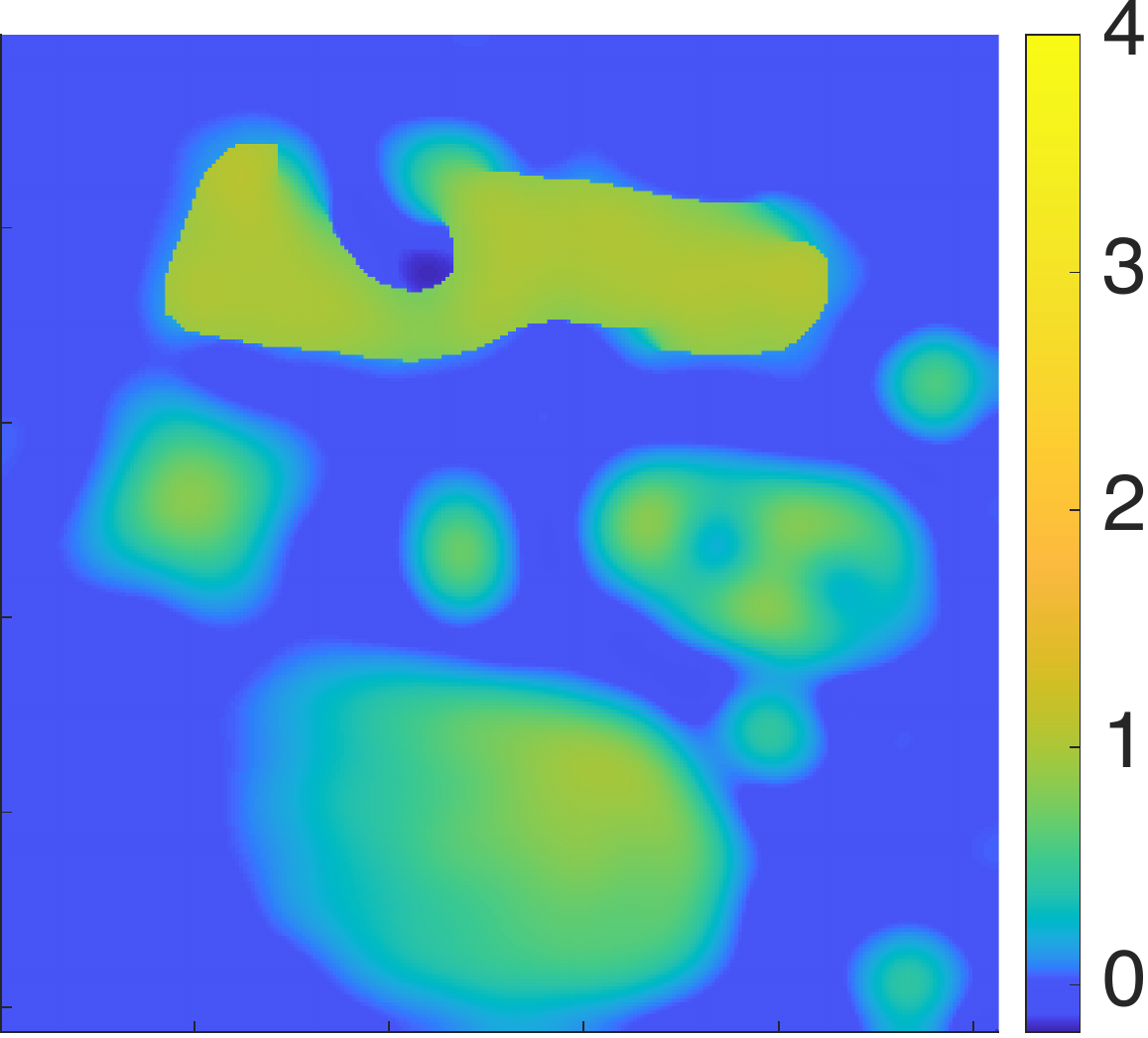}
            \includegraphics[width=\linewidth]{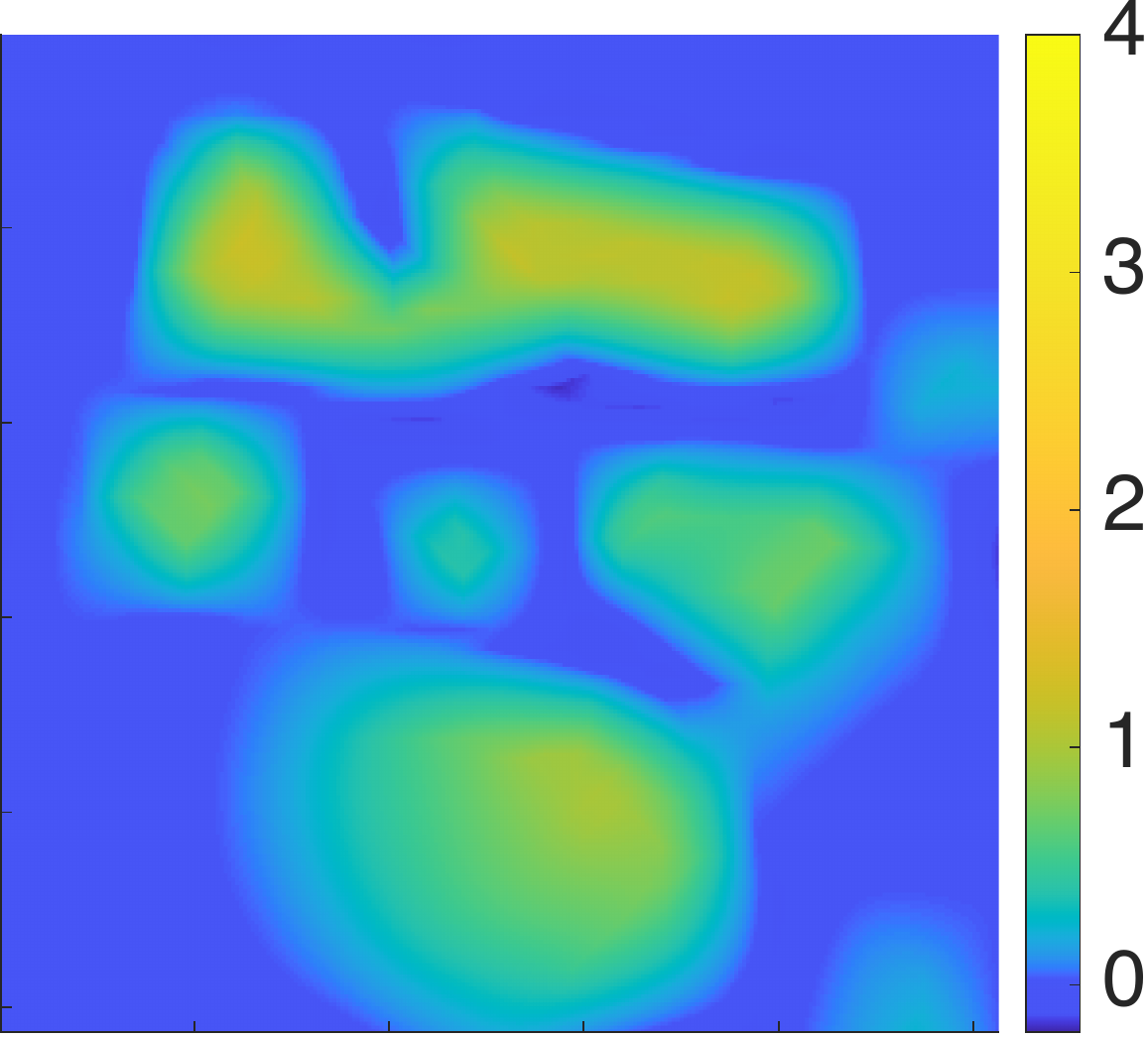}
            \caption{Cauchy}\end{subfigure}
        \begin{subfigure}[b]{\maphei}
        \includegraphics[width=\linewidth]{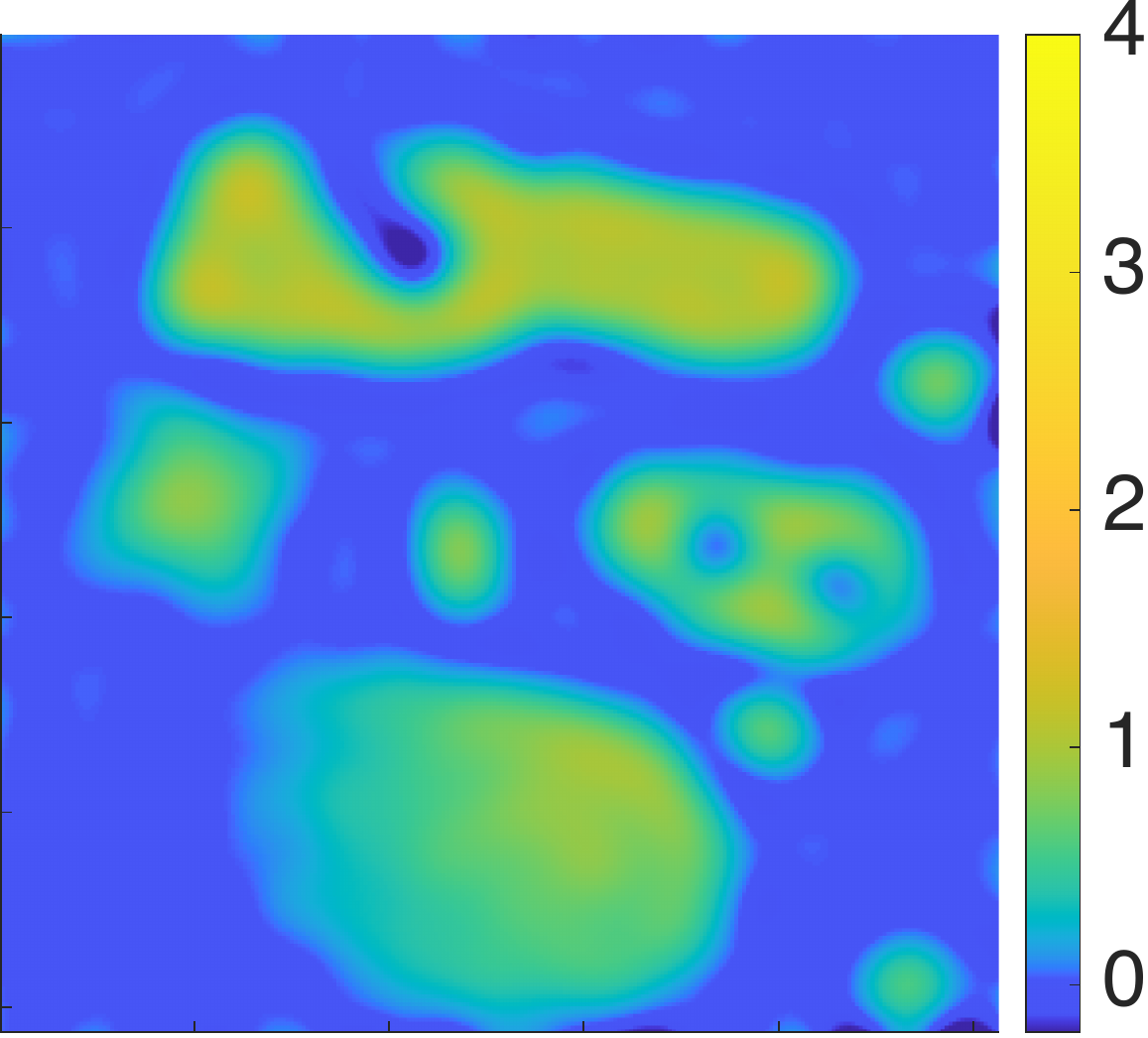}
            \includegraphics[width=\linewidth]{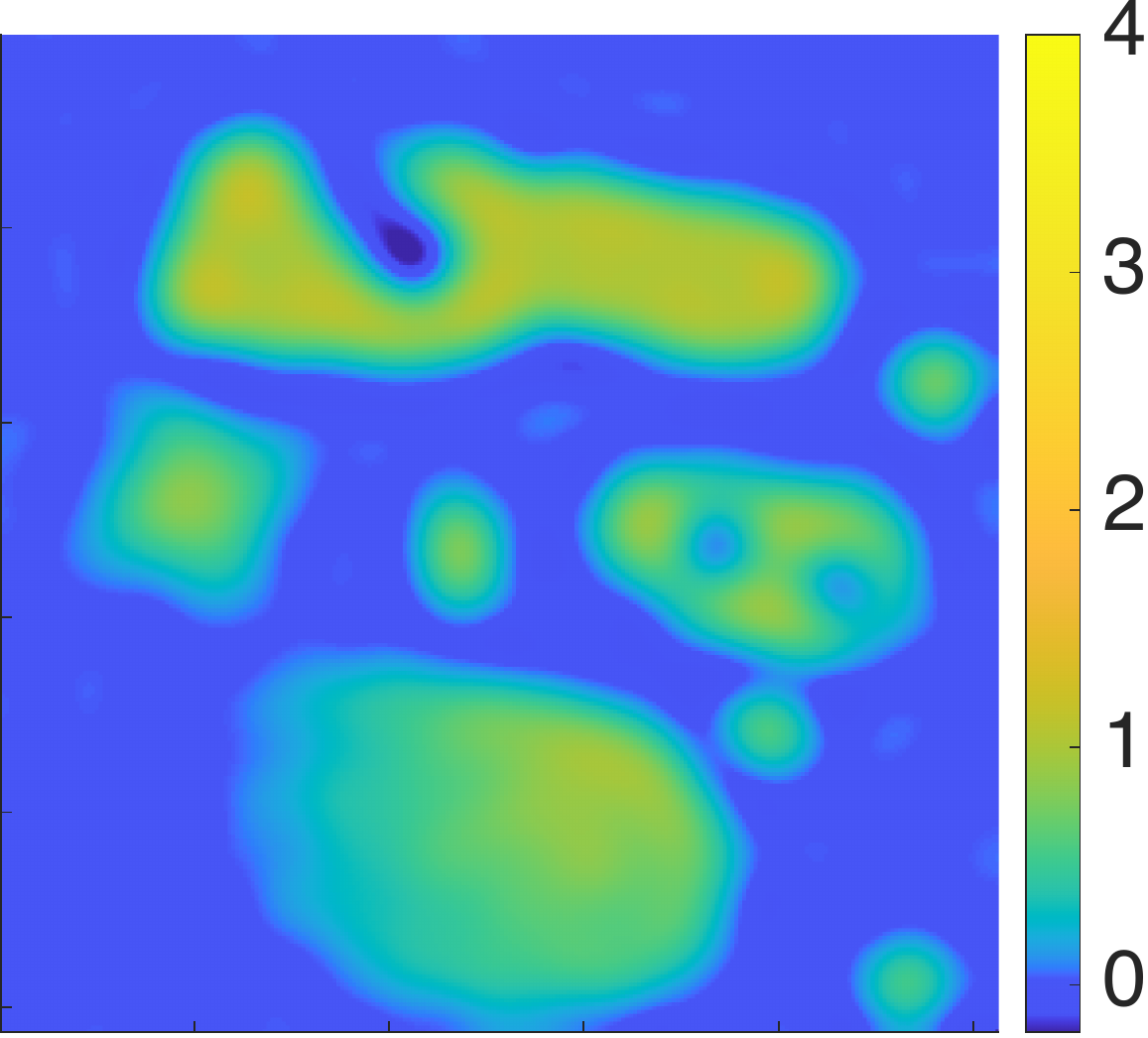}
            \includegraphics[width=\linewidth]{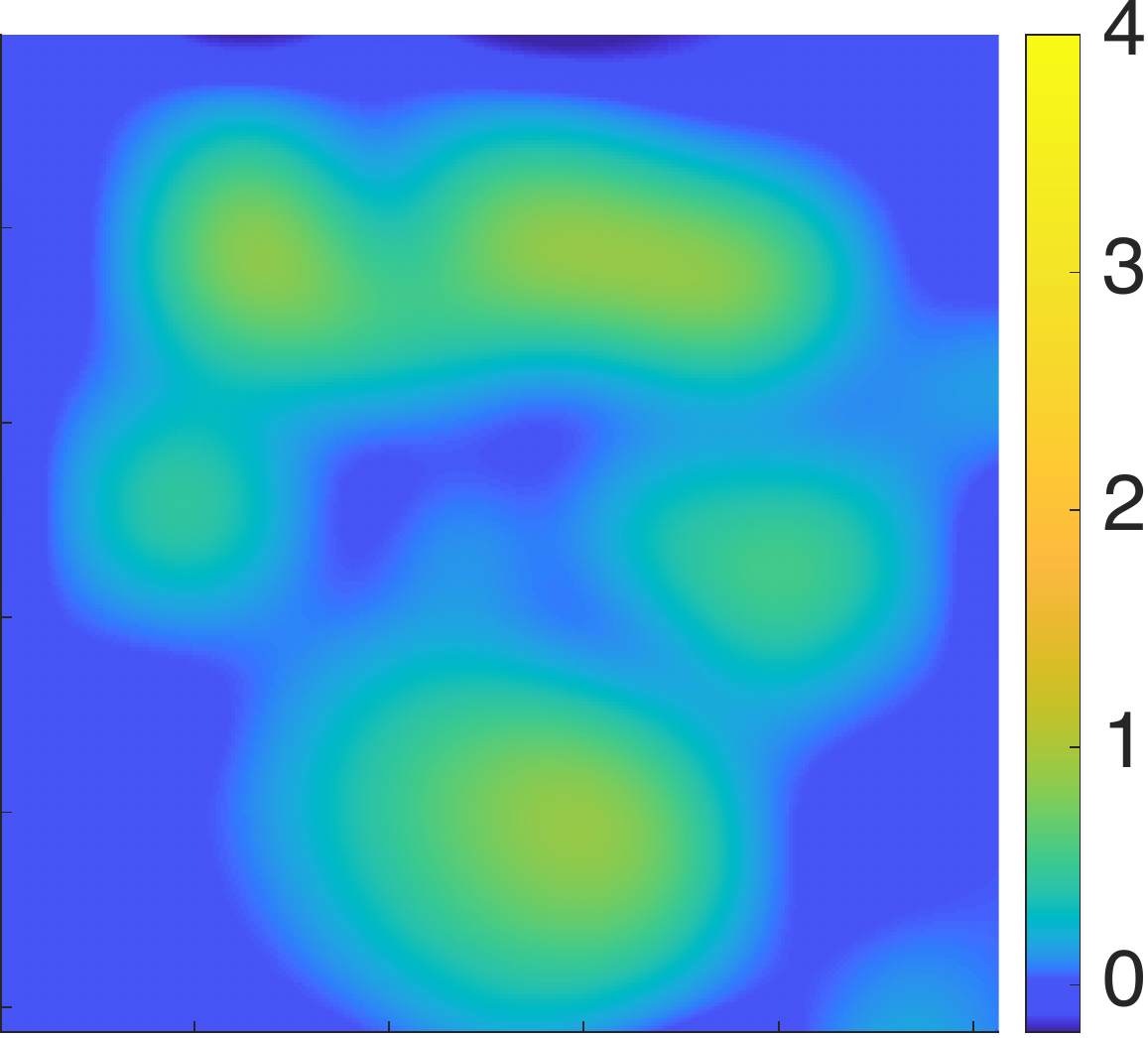}
            \caption{Gaussian }\end{subfigure}
        \begin{subfigure}[b]{\maphei}
             \includegraphics[width=\linewidth]{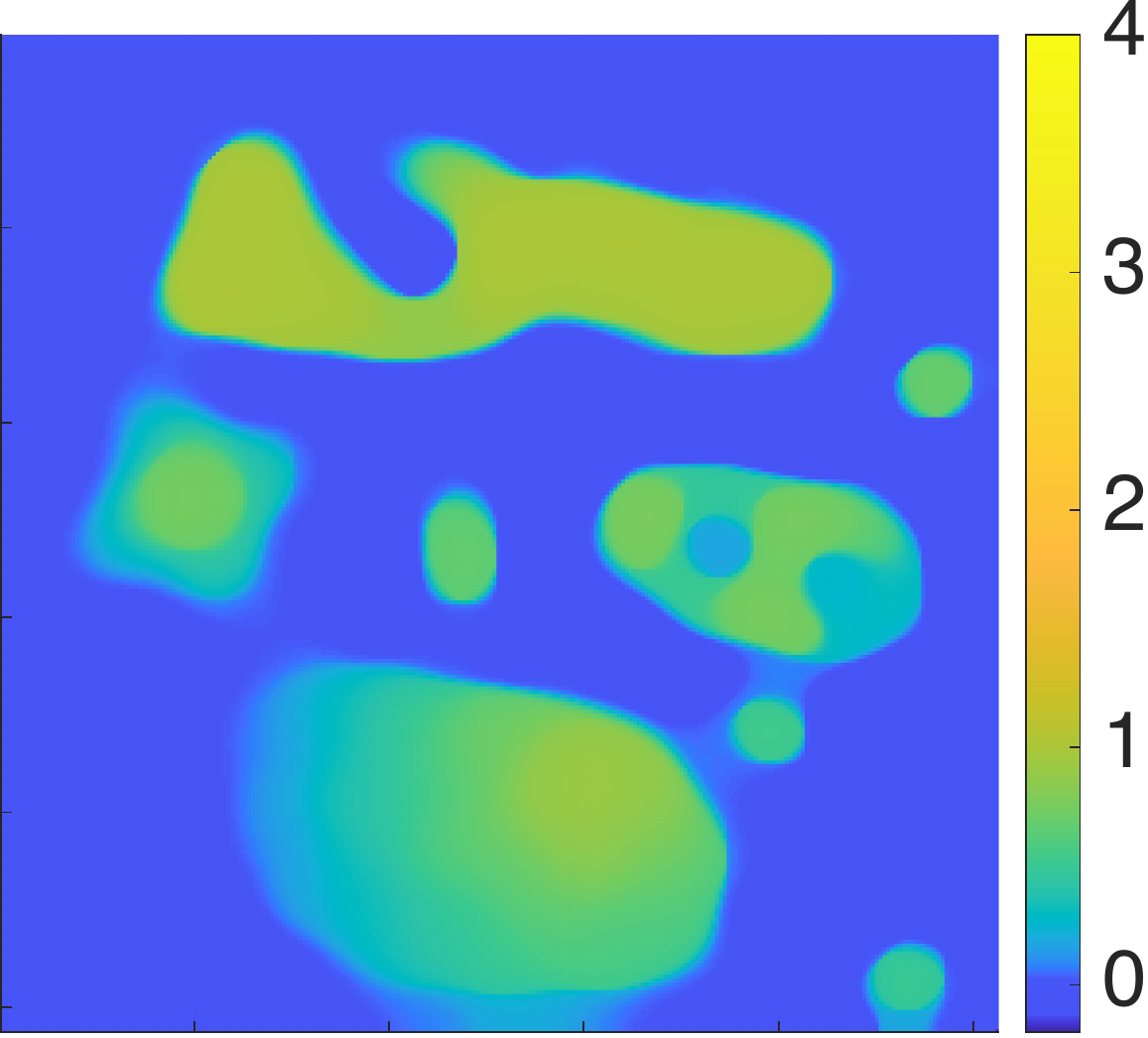}
            \includegraphics[width=\linewidth]{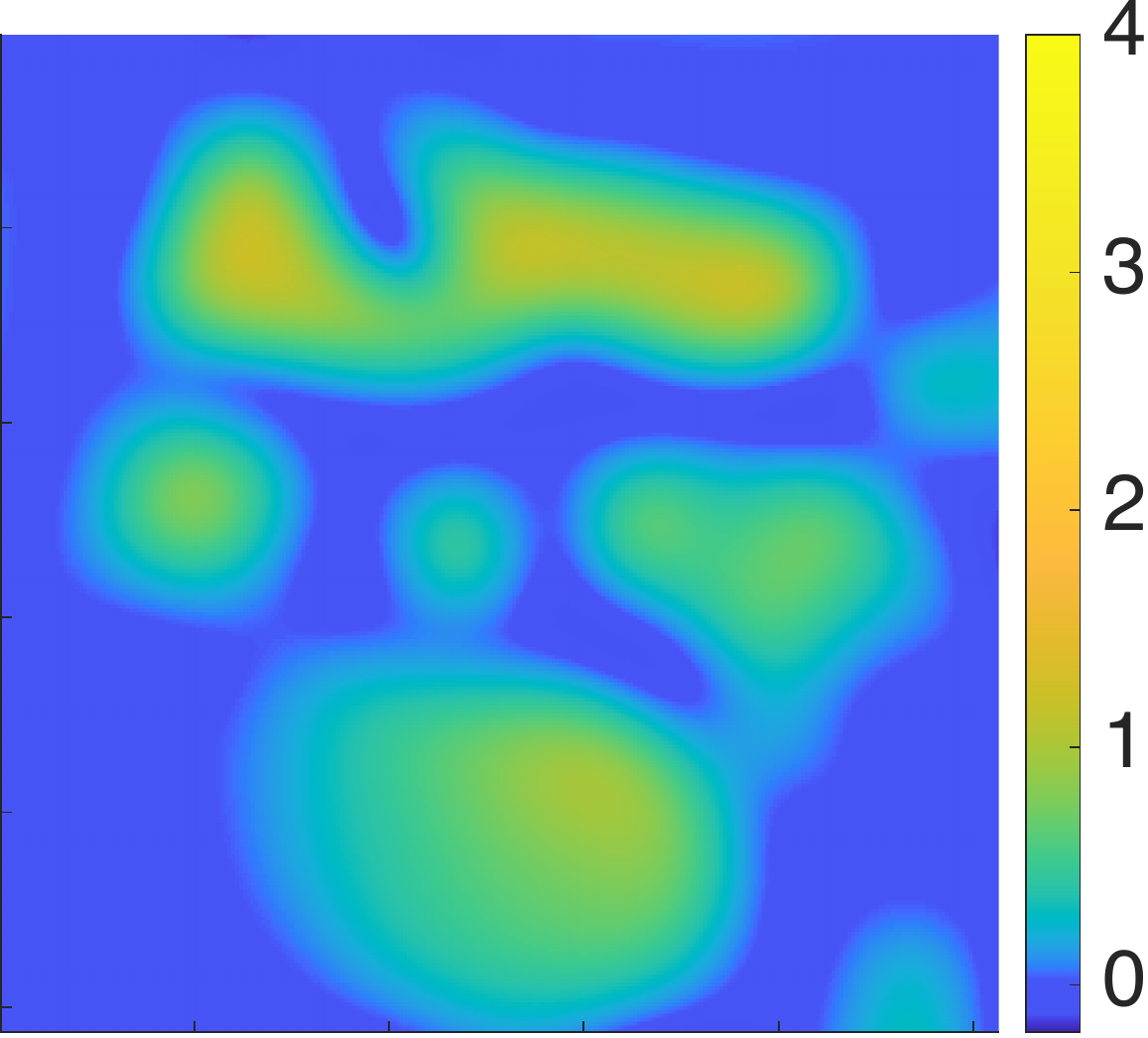}
            \caption{TV}\end{subfigure}
            \caption{MAP estimates with Cauchy, Gaussian and total variation priors and the ground truth and measurement data. Top row: SPDE based priors. Middle row: the simulated measurement data and the first order difference priors. Bottom row: ground truth and the second order difference priors.}
        \label{others}
    \end{figure}
    
        \begin{figure}
        \centering
        \begin{subfigure}[b]{\stathei}
        \includegraphics[width=\linewidth]{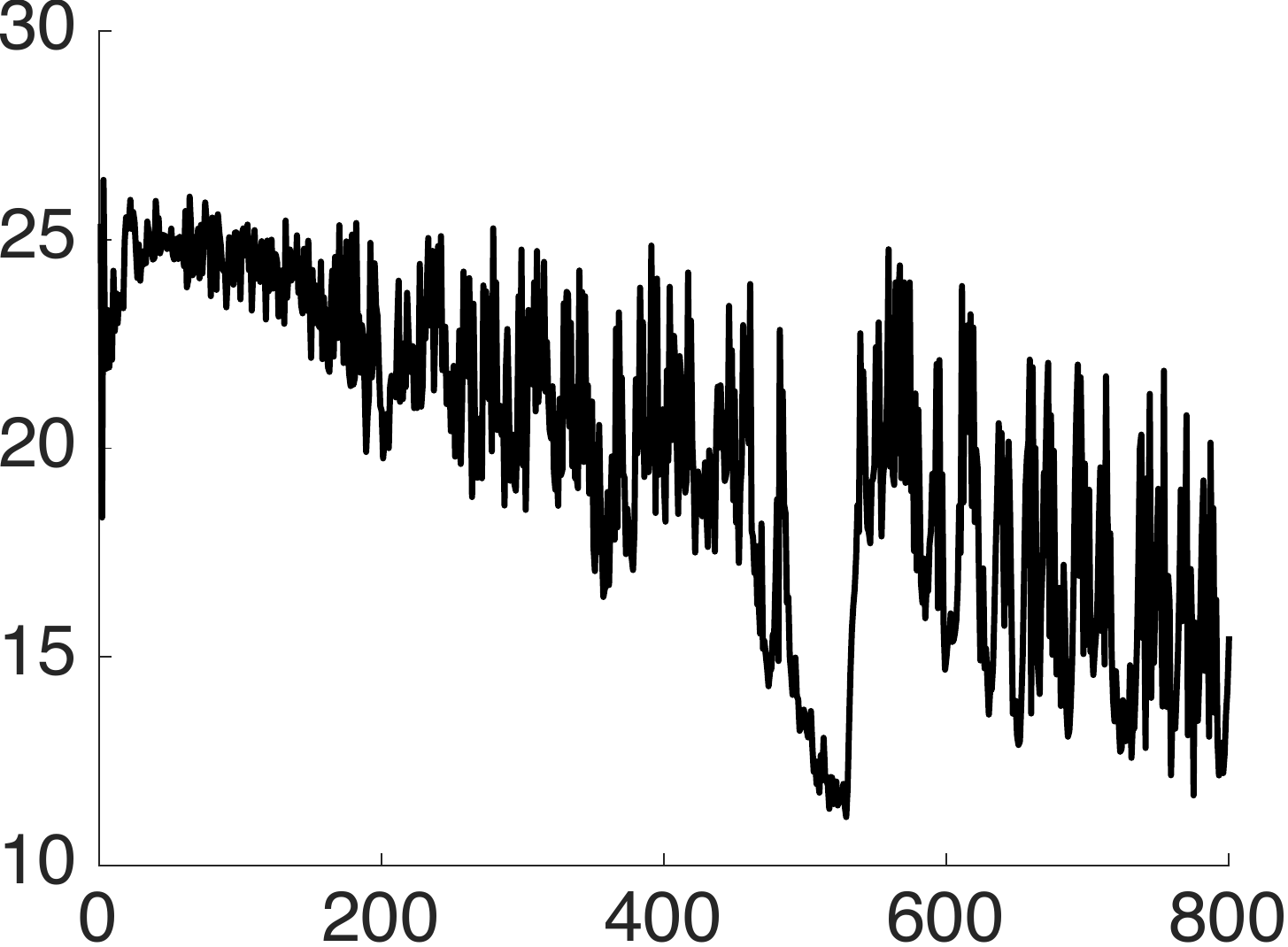}
            \includegraphics[width=\linewidth]{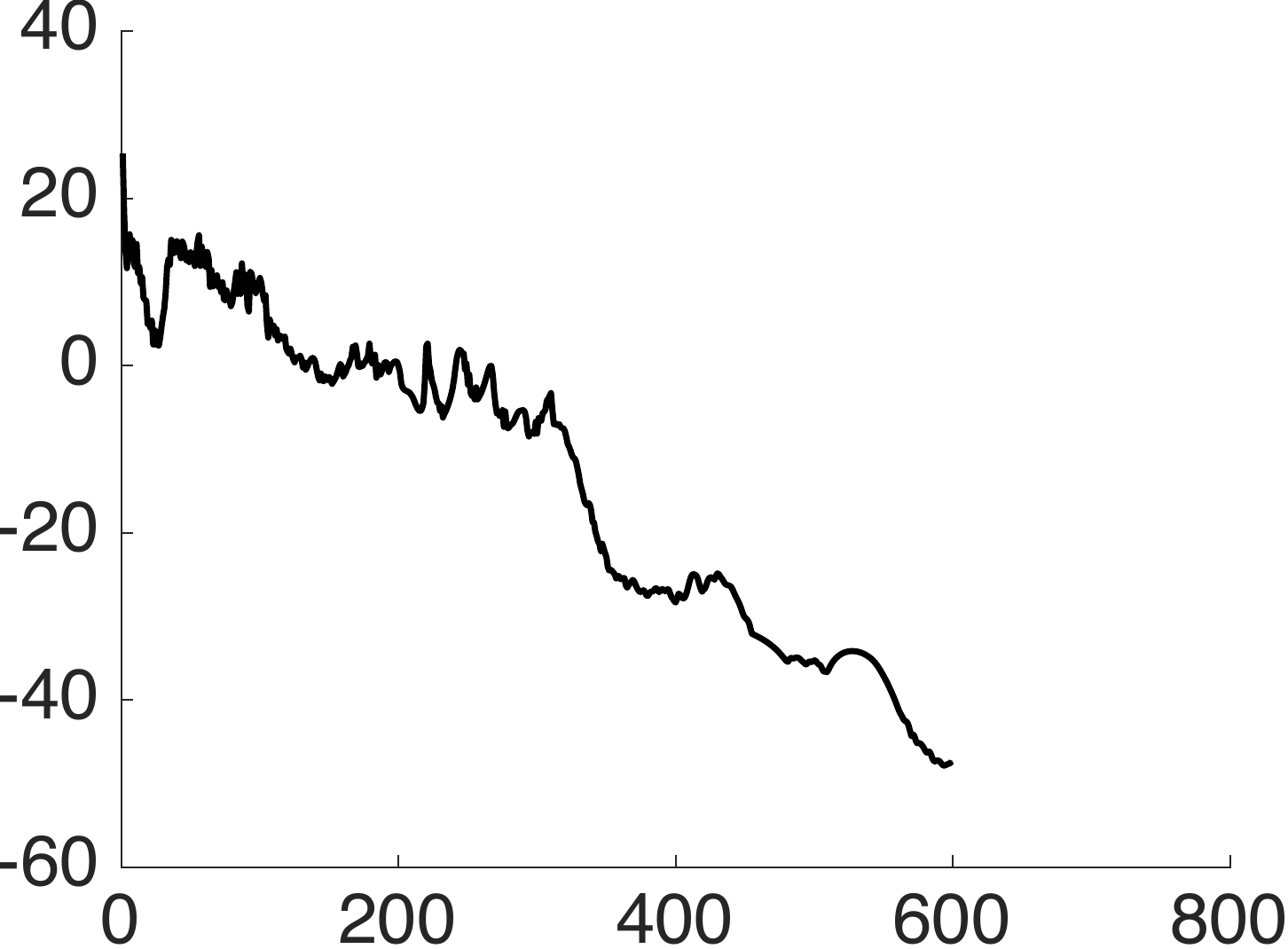}
            \includegraphics[width=\linewidth]{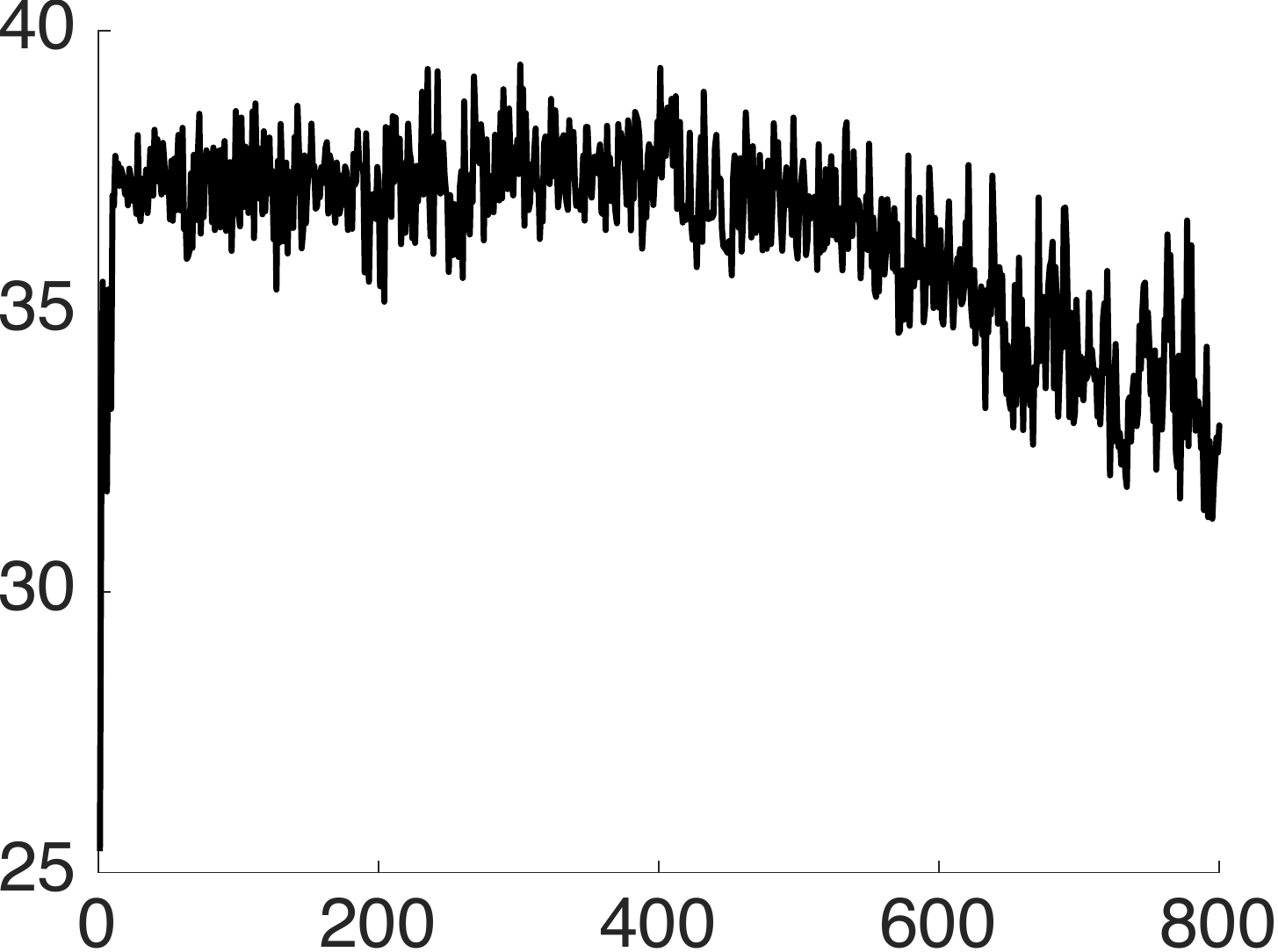}
            \caption{Cauchy}\end{subfigure}
        \begin{subfigure}[b]{\stathei}
        \includegraphics[width=\linewidth]{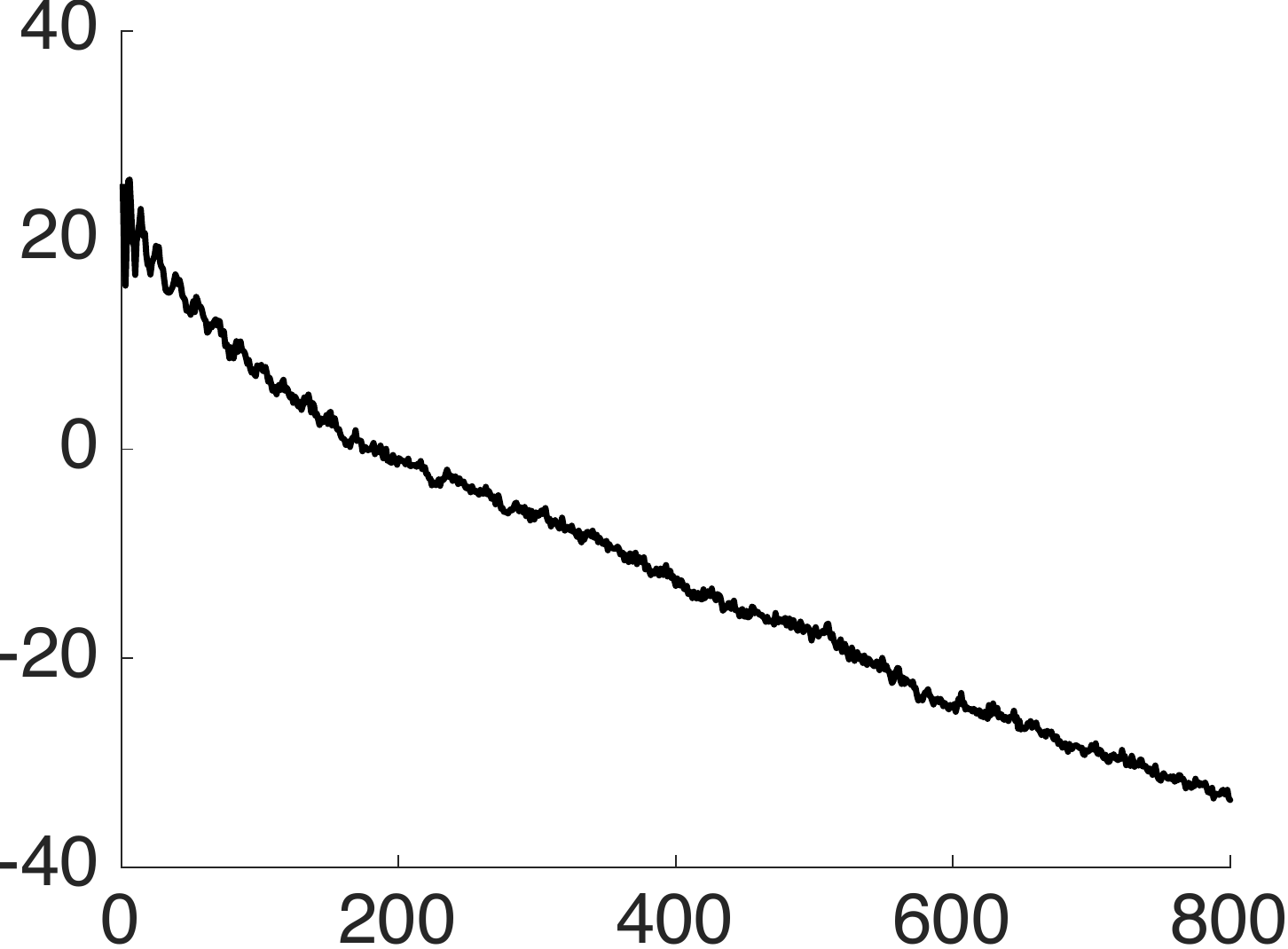}
            \includegraphics[width=\linewidth]{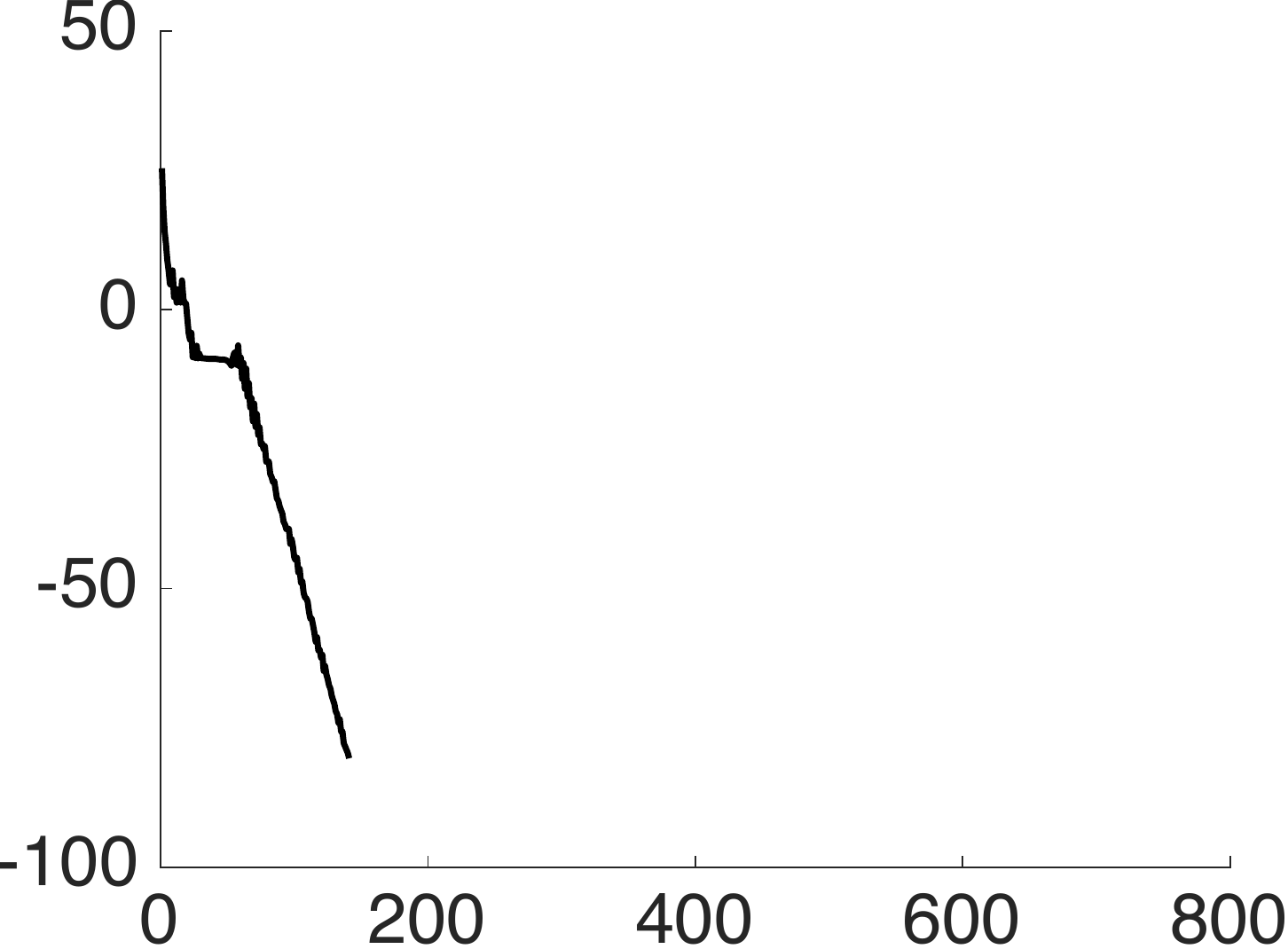}
            \includegraphics[width=\linewidth]{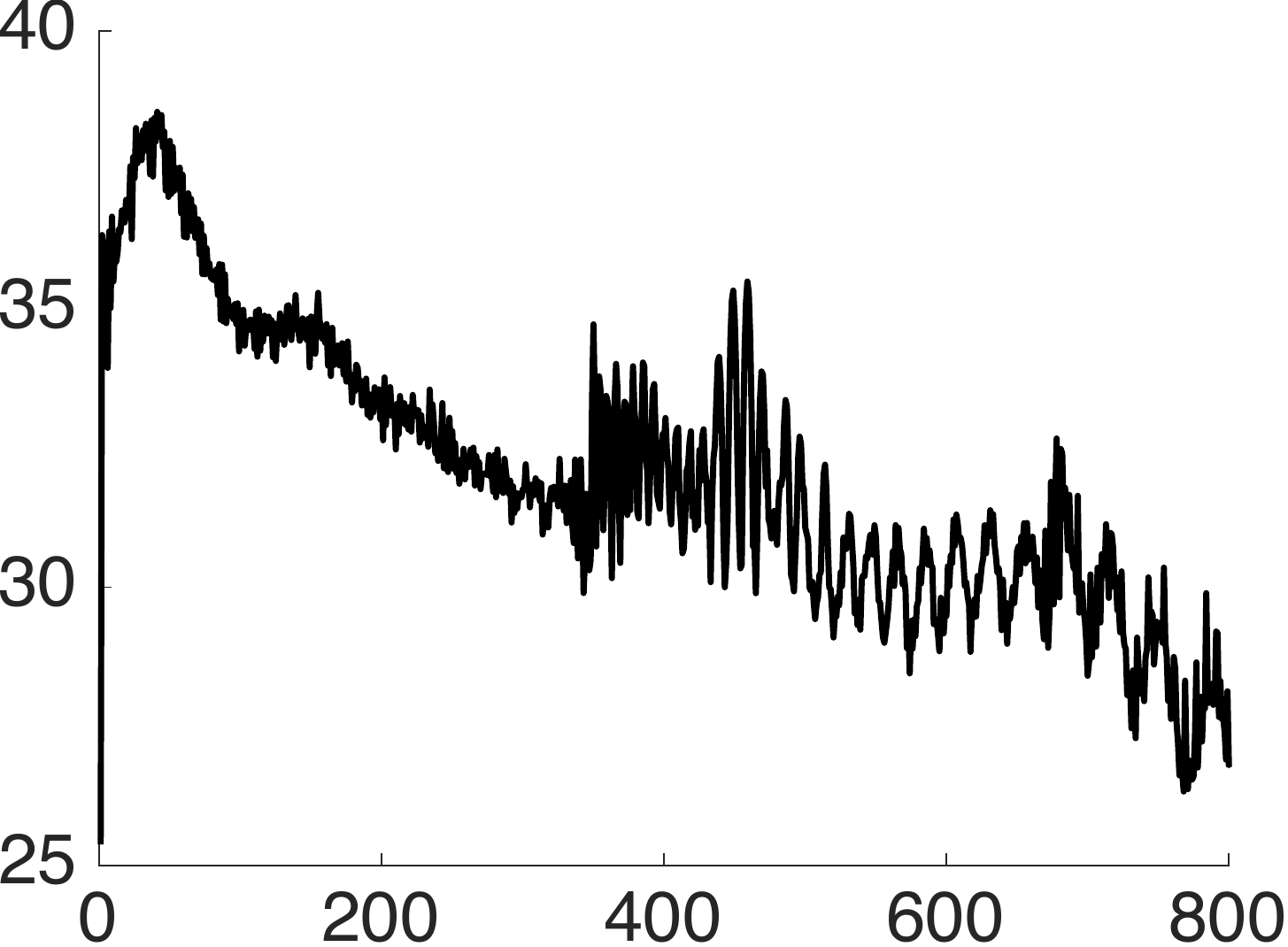}
            \caption{Gaussian }\end{subfigure}
        \begin{subfigure}[b]{\stathei}
             \includegraphics[width=\linewidth]{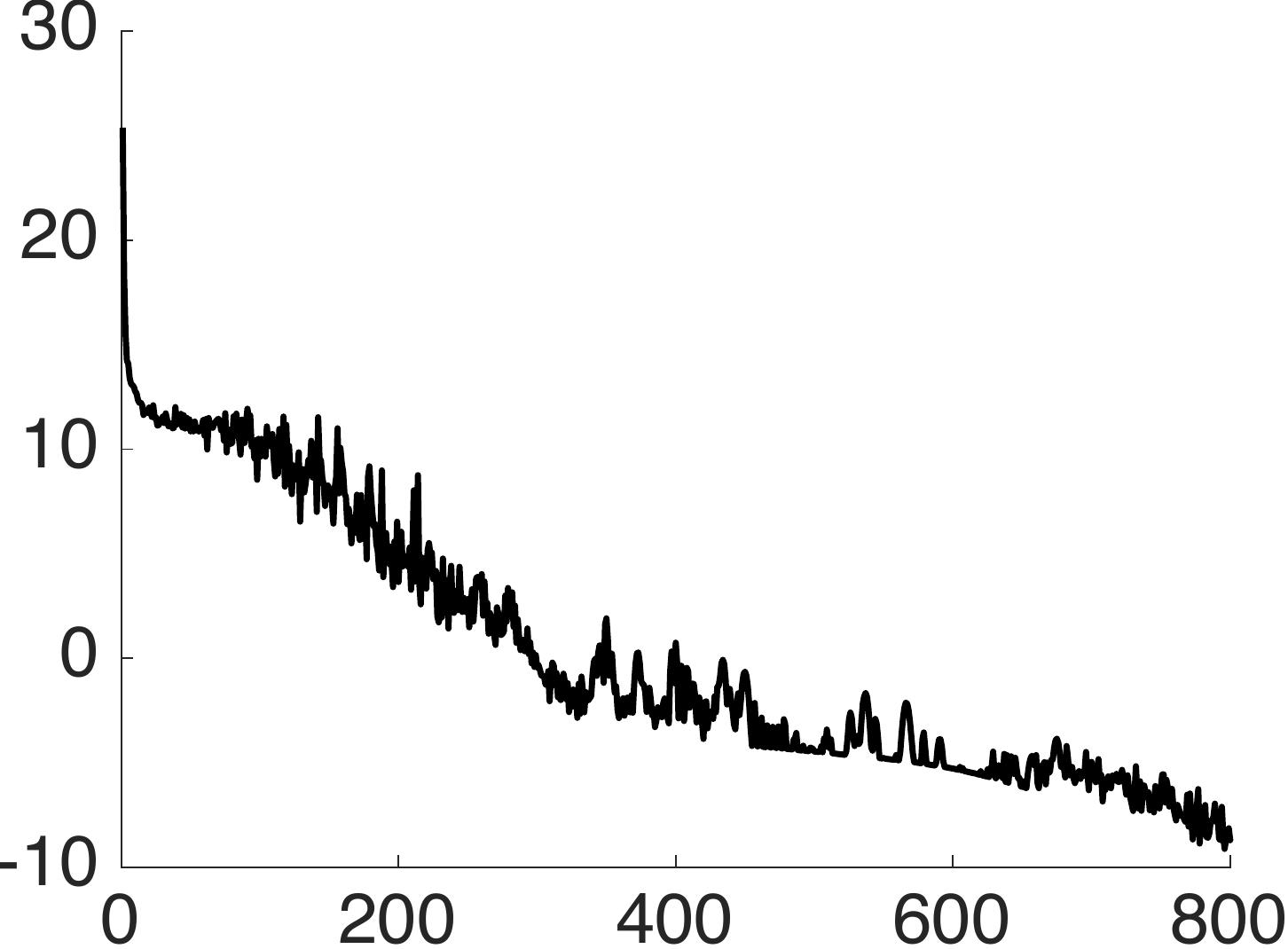}
            \includegraphics[width=\linewidth]{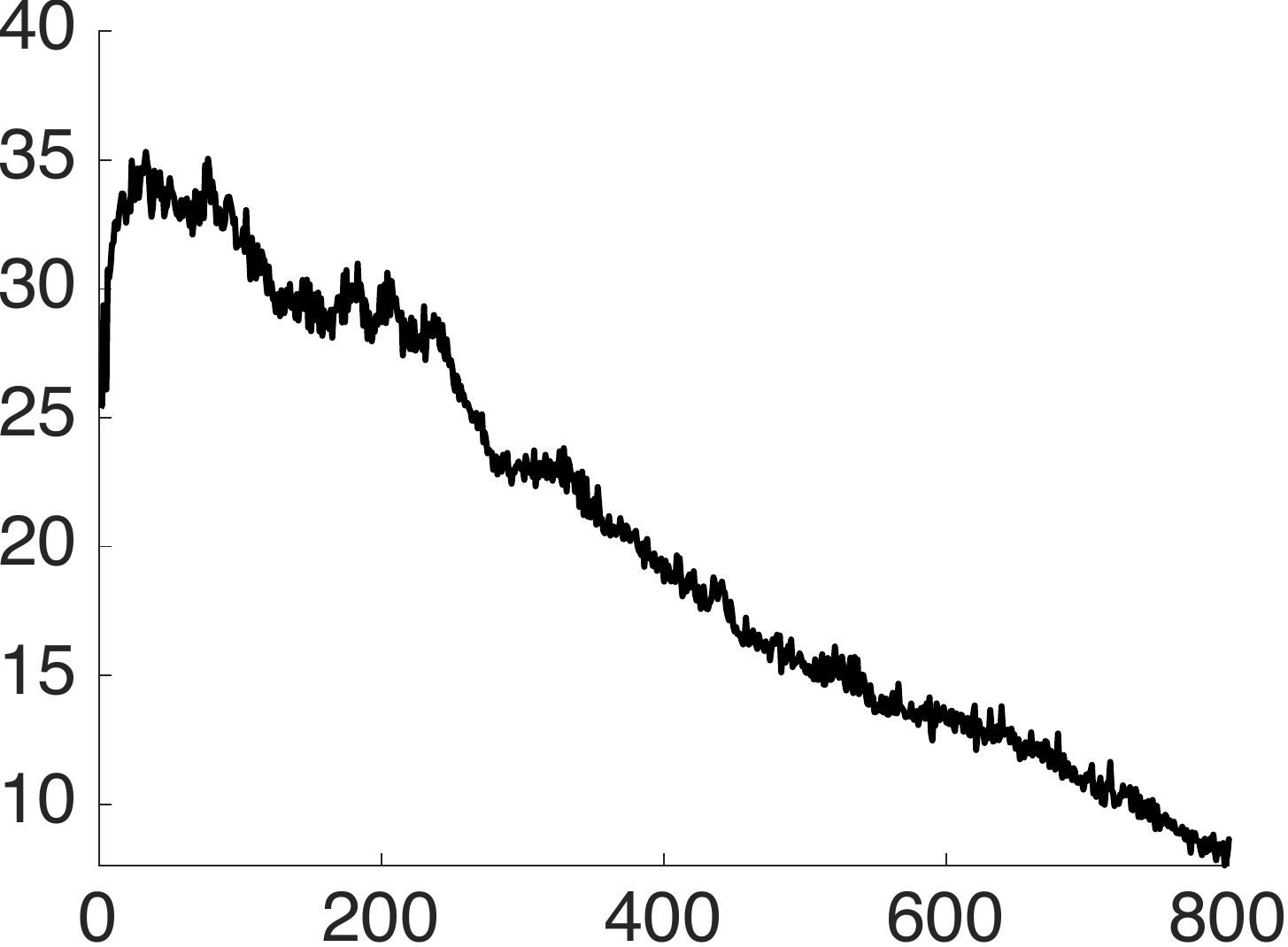}
            \caption{TV}\end{subfigure}
        
        \caption{Logarithm of the gradient norm as a function of L-BFGS iteration with different priors. Top row: SPDE prior. Middle row: the first order difference priors. Bottom row: the second order difference prior.}
        \label{others-stats}
    \end{figure}

    
    \FloatBarrier

    \section{Conclusion}
    \label{sec:conc}

    The objective of this work was to provide a comparison of multiple Cauchy Markov random field priors, for the task of addressing Bayesian inverse problems with examples in one-dimensional and two-dimensional deconvolutions. Cauchy random field priors are of interest as they are aimed to recover non-smooth, or rough, features of images and functions. The priors we tested included SPDE Cauchy prior, Cauchy sheet prior and Cauchy difference priors. Our results show that the isotropic Cauchy difference priors solve the issue of the  coordinate axis dependency of their anisotropic counterparts. We also illustrated properties  of the Cauchy SPDE prior.  \textcolor{black}{The presented Cauchy priors allow to encode the existing knowledge of spatial features that the standard hand-crafted random field priors cannot offer, such as spikiness and discontinuities, but the heavy-tailedness comes with a price of slow convergence of the numerical estimators.}  
    
    In the numerical experiments,  we provided both the MAP and CM estimators.  We evaluated the CM estimators of the posteriors using three MCMC algorithms, which have been  designed to sample high-dimensional distributions, and assessed their performance using Gelman-Rubin diagnostics. The MCMC algorithms did not significantly differ in the sampling performance, since all they seem to experience severe chain mixing issues with the posteriors having the Cauchy SPDE, Cauchy sheet or higher order Cauchy difference priors in when the posterior consisted of thousands of components in spatial dimensions of two. \textcolor{black}{It can be argued that the proposed isotropic version of the Cauchy difference prior for our 2D experiments, worked the best}.
    As the focus of this work was primarily computational, there are interesting directions to consider both on the analysis and numerical side. These are summarized in the following bullet points.
    \begin{itemize}
        \item It would be of interest to provide a unification between the SPDE and non-SPDE Cauchy priors in the context of Bayesian inversion. The task would include a generalization of such priors, and whether analytical results such as well-posedness, or some posterior convergence analysis would hold. Inherently, the task is difficult, as mathematically the priors are quite different.  A clearer understanding of the Cauchy sheet priors prior would be required.  Challenges therefore  remain.
        \item As our experiments were specific to deconvolution, the next natural step would be to consider an extension to more advanced numerical models. Such examples would include partial differential equations based on impedance tomography or models arising in geophysical sciences.
        \item One of the priors we introduced was the Cauchy SPDE prior, based on \cite{LRL11}. The discretization we used for was a centered finite difference method. We would like to  consider a finite-element discretization for the SPDE prior and for the difference priors, if possible. Numerically, it would be interesting to see how the discretization could alter the sampling of the Cauchy random field priors, and  how  existing FEM theory could be used to to derive convergence analysis. 
        \item Finally, it would be of interest to see if the Cauchy priors can be generated using neural networks. Most of the work in the field has been for Gaussian priors, and a natural extension  would thus be to consider $\alpha$-stable Cauchy neural network priors. 
    \end{itemize}

    \section*{Acknowledgements}
    
    The authors thank Dr Sari Lasanen and Prof. Heikki Haario for useful discussions. JS and LR acknowledge Academy of Finland project funding (grant numbers 334816 and 336787).
    NKC is supported by KAUST baseline funding.
    
    \printbibliography

    \section*{Pseudocodes for the random walk Metropolis MCMC algorithms}
    
    \begin{algorithm}
        \caption{Blockwise Metropolis-Hastings algorithm with adaptation.}
        \label{scam}
        \begin{algorithmic}[1] 
            \State \textbf{Input:} Target probability density function $\pi$; no. samples $N$; no. adaptation cycles $N_a$;
            vector $\textbf{V}$ of indices  for each block, so that $\textrm{dim}(\mathbf{V}) = N_b$. 
            \State \textbf{Output:} Chain of samples  $[\mathbf{u}_{N_a},\mathbf{u}_{N_a +1},\cdots \mathbf{u}_N]$.
            \State Set $\mathbf{u}_1=\mathbf{u}_0$.
            \For{$2\leq i \leq N$} 
            \For{$1\leq k \leq N_\textrm{b}$} 
            \State Set $j = \mathbf{V}_{k} $.
            \State Set $\mathbf{u}_p = \mathbf{u}_{i-1,j} $.
            \State  Generate  $\mathbf{r}\sim\mathcal{N}(0,\mathbf{I}_j)$.
            \State Set $\mathbf{u}_r = \mathbf{u}_{p} + \mathbf{Q}_j\mathbf{r}$.
            \State Generate  $z \sim \textrm{Unif}[0,1]$.
            \If{$z\leq\frac{\pi(\mathbf{u}_r|\mathbf{u}_{i-1,-j})}{\pi(\mathbf{u}_p|\mathbf{u}_{i-1,-j})}$}  
            \State Set $\mathbf{u}_{i,j}=\mathbf{u}_r$. 
            \Else 
            \State Set $\mathbf{u}_{i,j}=\mathbf{u}_{p}$.
            \EndIf
            \If{$i<N_a$}  
            \State Set $\mathbf{Q}_j=\textrm{chol}\left( \frac{2.38^2}{\textrm{dim}(j)} \left( \textrm{cov}(\mathbf{u}_{1,j}, \mathbf{u}_{2,j} \dots, \mathbf{u}_{i,j}  ) + \mathbf{I}_j \delta \right )\right)$.
            \EndIf
            \EndFor
            \EndFor\label{aq}
        \end{algorithmic}
    \end{algorithm}
    
    \begin{algorithm}
        \caption{Repelling-Attracting Metropolis with adaptation.}
        \label{ram}
        \begin{algorithmic}[1] 
            \State \textbf{Input:} Target probability density function $\pi$; no. samples $N$; no. adaptation cycles $N_a$;
            vector $\textbf{V}$ of indices  for each block, so that $\textrm{dim}(\mathbf{V}) = N_b$. 
            \State \textbf{Output:} Chain of samples  $[\mathbf{u}_{N_a},\mathbf{u}_{N_a +1},\cdots \mathbf{u}_N]$.
            \State Set $\mathbf{u}_1=\mathbf{u}_0$ and $\mathbf{w}_1=\mathbf{u}_0$.
            \For{$2\leq i \leq N$} 
            \For{$1\leq k \leq N_\textrm{b}$} 
            \State Set index $j = \mathbf{V}_{k} $.
            \State Set $\mathbf{u}_p = \mathbf{u}_{i-1,j} $, set $\mathbf{w}_p = \mathbf{w}_{i-1,j} $.
            
            \State  Generate  $\mathbf{r}\sim\mathcal{N}(0,\mathbf{I}_j)$, set $\mathbf{u}' = \mathbf{u}_{p} + \mathbf{Q}_j\mathbf{r}$.
            \State Generate  $z_1 \sim \textrm{Unif}[0,1]$.
            \While{$z_1\leq\frac{\pi(\mathbf{u}_p|\mathbf{u}_{i-1,-j})}{\pi(\mathbf{u}'|\mathbf{u}_{i-1,-j})}$}
            \State  Generate  $\mathbf{r}\sim\mathcal{N}(0,\mathbf{I}_j)$, set $\mathbf{u}' = \mathbf{u}_{p} + \mathbf{Q}_j\mathbf{r}$.
            \State Generate  $z_1 \sim \textrm{Unif}[0,1]$.
            \EndWhile
            
            \State  Generate  $\mathbf{r}\sim\mathcal{N}(0,\mathbf{I}_j)$, set $\mathbf{u}^* = \mathbf{u}' + \mathbf{Q}_j\mathbf{r}$.
            \State Generate  $z_2 \sim \textrm{Unif}[0,1]$.
            \While{$z_2\leq\frac{\pi(\mathbf{u}^*|\mathbf{u}_{i-1,-j})}{\pi(\mathbf{u}'|\mathbf{u}_{i-1,-j})}$}
            \State  Generate  $\mathbf{r}\sim\mathcal{N}(0,\mathbf{I}_j)$, set $\mathbf{u}^* = \mathbf{u}' + \mathbf{Q}_j\mathbf{r}$.
            \State Generate  $z_2 \sim \textrm{Unif}[0,1]$.
            \EndWhile
            
            \State  Generate  $\mathbf{r}\sim\mathcal{N}(0,\mathbf{I}_j)$, set $\mathbf{w}^* = \mathbf{u}^* + \mathbf{Q}_j\mathbf{r}$.
            \State Generate  $z_3 \sim \textrm{Unif}[0,1]$.
            \While{$z_3\leq\frac{\pi(\mathbf{u}^*|\mathbf{u}_{i-1,-j})}{\pi(\mathbf{w}^*|\mathbf{u}_{i-1,-j})}$}
            \State  Generate  $\mathbf{r}\sim\mathcal{N}(0,\mathbf{I}_j)$, set $\mathbf{w}^* = \mathbf{u}^* + \mathbf{Q}_j\mathbf{r}$.
            \State Generate  $z_3 \sim \textrm{Unif}[0,1]$.
            \EndWhile

            \State Generate  $z_4 \sim \textrm{Unif}[0,1]$.
            \If{$z_4\leq\frac{\pi(\mathbf{u}^*|\mathbf{u}_{i-1,-j}) \min\left(1,\pi(\mathbf{u}_p|\mathbf{u}_{i-1,-j})/\pi(\mathbf{w}_p|\mathbf{u}_{i-1,-j}) \right) }{\pi(\mathbf{u}_p|\mathbf{u}_{i-1,-j}) \min\left(1,\pi(\mathbf{u}^*|\mathbf{u}_{i-1,-j})/\pi(\mathbf{w}^*|\mathbf{u}_{i-1,-j}) \right) }$}  
            \State Set $\mathbf{u}_{i,j}=\mathbf{u}^*$, set $\mathbf{w}_{i,j}=\mathbf{w}^*$. 
            \Else 
            \State Set $\mathbf{u}_{i,j}=\mathbf{u}_{p}$,  set $\mathbf{w}_{i,j}=\mathbf{w}_p$.
            \EndIf
            \If{$i<N_a$}  
            \State Set $\mathbf{Q}_j=\textrm{chol}\left(\frac{2.38^2}{\textrm{dim}(j)}  \left( \frac{1}{2} \textrm{cov}(\mathbf{u}_{1,j}, \mathbf{u}_{2,j} \dots, \mathbf{u}_{i,j}  ) + \mathbf{I}_j \delta\right )\right)$.
            \EndIf
            \EndFor
            \EndFor\label{qa}
        \end{algorithmic}
    \end{algorithm}
    
    \begin{algorithm}
        \caption{No U-Turn Sampler with Hamiltonian Monte Carlo.}
        \label{nuts}
        \begin{algorithmic}[1] 
            \State \textbf{Input:} Target probability density function $\pi$; no. samples $N$; no. adaptation cycles $N_a$.
            \State \textbf{Output:} Chain of samples  $[\mathbf{u}_{N_a},\mathbf{u}_{N_a +1},\cdots \mathbf{u}_N]$.
            \State Set $\mathbf{u}_1=\mathbf{u}_0$ and $\mathbf{M}=\mathbf{M}_0$.
            \For{$2\leq i \leq N$} 
            \State Generate $\mathbf{p} \sim \mathcal{N}(0,\mathbf{M}^{-1})$.
            \State Set $\mathcal{T} = (\mathbf{u}_{i-1},\mathbf{p}) $, $\mathbf{u}_\text{l} = \mathbf{u}_{i-1}$, $\mathbf{u}_\text{r} = \mathbf{u}_{i-1}$, $\mathbf{p}_\text{l} = \mathbf{p}$, $\mathbf{p}_\text{r} = \mathbf{p}$, $n = 0$.
            \While {}
            \State Generate $d\sim \{-1,1\}$.
            \If{$d=1$}  
            \State Set $\mathcal{T}_{\text{n}}, \_,\_, \mathbf{u}_\text{r}, \mathbf{p}_\text{r},\text{q} =  \text{ContinueTree}(\mathbf{u}_{\text{r}},\mathbf{p}_{\text{r}},d,n,\mathbf{M},\epsilon)$.
            \Else
            \State Set $\mathcal{T}_{\text{n}}, \mathbf{u}_\text{l}, \mathbf{p}_\text{l}, \_,\_, \text{q} =  \text{ContinueTree}(\mathbf{u}_{\text{l}},\mathbf{p}_{\text{l}},d,n,\mathbf{M},\epsilon)$.
            \EndIf           
            \If{ $\text{q}\lor \text{U-Turned}(\mathbf{p}_\text{l}, \mathbf{p}_\text{r}, \mathcal{T}$)}  
            \State \textbf{break}
            \EndIf       
            \State Set $\mathcal{T} = \mathcal{T} \cup \mathcal{T}_n $, $n = n +1$.    
            \EndWhile
            \State Sample $\mathbf{u}_i $ from $ \{ \mathbf{u}_k \}_{k=\_}^{k=+}$ with weights proportional to $\frac{ \exp{\left( -H(\mathbf{u}_k,\mathbf{p}_k )  \right)}}{\sum_k \exp{\left( -H(\mathbf{u}_k,\mathbf{p}_k )  \right)}} $.
            \If{ $i < N_a$)}  
            \State  Set $\mathbf{M}=\text{NewMetric}([\mathbf{u}_{1},\mathbf{u}_{2},\cdots \mathbf{u}_i])$,  $\epsilon = \text{AdaptStepSize}(\epsilon,\mathcal{T})$.
            \EndIf 
            \EndFor\label{q65a}
            
            \Procedure{ContinueTree}{$\mathbf{u}^*,\mathbf{p}^*,d,n,\mathbf{M},\epsilon$}
            \If{ $n=0$ }  
            \State Set $\mathbf{u}^*,\mathbf{p}^* = \text{OneLeapFrogStepToDir}(\mathbf{u}^*, \mathbf{p}^*,d,\mathbf{M},\epsilon)$.
            \State Return $(\mathbf{u}^*, \mathbf{p}^*), \mathbf{u}^*, \mathbf{p}^*,\mathbf{u}^*, \mathbf{p}^*$, false.
            \Else
            \State Set $\mathcal{T}^*, \mathbf{u}^*_{\text{l}}, \mathbf{p}^*_{\text{l}},\mathbf{u}^*_{\text{r}}, \mathbf{p}^*_{\text{r}},\text{q}^*$ = ContinueTree($\mathbf{u}^*,\mathbf{p}^*,d,n-1,\mathbf{M},\epsilon$).
            \If{$d=1$}  
            \State Set $\mathcal{T}_{\text{n}}^{**}, \_,\_, \mathbf{u}^*_\text{r}, \mathbf{p}^*_\text{r},\text{q}^{**} =  \text{ContinueTree}(\mathbf{u}^*_{\text{r}},\mathbf{p}^*_{\text{r}},d,n-1,\mathbf{M},\epsilon)$.
            \Else
            \State Set $\mathcal{T}_{\text{n}}^{**}, \mathbf{u}_\text{l}^*, \mathbf{p}^*_\text{l}, \_,\_,\text{q}^{**} =  \text{ContinueTree}(\mathbf{u}^*_{\text{l}},\mathbf{p}^*_{\text{l}},d,n-1,\mathbf{M},\epsilon)$.
            \EndIf
            \State Set $\mathcal{T}^* = \mathcal{T}^* \cup \mathcal{T}^{**} $, $\text{q}^* =\text{q}^* \lor \text{U-Turned}(\mathcal{T}^*,\mathbf{p}^*_\text{l},\mathbf{p}^*_\text{r})$.
            \State Return $\mathcal{T}^*, \mathbf{u}^*_{\text{l}}, \mathbf{p}^*_{\text{l}},\mathbf{u}^*_{\text{r}}, \mathbf{p}^*_{\text{r}}, \text{q}^*$.
            \EndIf  
            \EndProcedure
            
            \Procedure{U-Turned}{$\mathbf{p}_{\text{l}}, \mathbf{p}_{\text{r}}, \mathcal{T}$}
            \If{$ \left( \mathbf{M}^{-1}\mathbf{p}_{\text{l}} \cdot \left(\sum_{i\in\mathcal{T}} {\mathcal{T}_i}[\mathbf{p}] \right) <0 \right) \lor \left( \mathbf{M}^{-1}\mathbf{p}_{\text{r}} \cdot \left(\sum_{i\in\mathcal{T}} {\mathcal{T}_i}[\mathbf{p} \right)] <0 \right)$}  
            \State Return true.
            \Else
            \State Return false.
            \EndIf   
            \EndProcedure
        \end{algorithmic}
    \end{algorithm}

\end{document}